\documentclass[12pt]{article}

\usepackage{titlesec}
\usepackage{psfrag,graphicx,verbatim,array,multicol,multirow,arydshln,palatino,enumerate}
\usepackage{amsmath,amssymb,amsfonts,bm,rotating}
\usepackage{pstricks,pst-node}
\usepackage{epsfig}
\usepackage{longtable}
\usepackage{harvard}
\usepackage{hyperref}
\hypersetup{colorlinks=true,linkcolor=black}

\titleformat{\paragraph}[hang]{\normalfont\bfseries}{\theparagraph}{0pt}{}

\citationmode{abbr}
\begin{document}

\sloppy 

\begin{center}

\vspace*{1in}

{\Large \bf  The Design of Observational Longitudinal Studies\\[10ex]}

\begin{large}
{Xavier Basaga\~na$^1$, Donna Spiegelman$^{1,2}$}
\vskip3mm

{$^1$Department of Biostatistics \\[1ex]
Harvard School of Public Health} \\[2ex]

{$^2$Department of Epidemiology \\[1ex]
Harvard School of Public Health} \\[2ex]

\end{large}

\end{center}

\newpage

\setlength{\baselineskip}{22pt}  
 
\section*{Summary}

This paper considers the design of observational longitudinal studies with a continuous response and a binary time-invariant exposure, where, typically, the exposure is unbalanced, the mean response in the two groups differs at baseline and the measurement times might not be the same for all participants. We consider group differences that are constant and those that increase linearly with time. We study power, number of study participants ($N$) and number of repeated measures ($r$), and provide formulas for each quantity when the other two are fixed, for compound symmetry, damped exponential and random intercepts and slopes covariances. When both $N$ and $r$ can be chosen by the investigator, we study the optimal combination for maximizing power subject to a cost constraint and minimizing cost for fixed power. Intuitive parameterizations are used for all quantities. All calculations are implemented in freely available software.
\vspace{24pt}

\numberwithin{equation}{section}

\section{Introduction} 
\label{introp1}

Sample size and power calculation in longitudinal studies with continuous outcomes that compare two groups have been considered previously \cite{Yi:2002,Schouten:1999,Galbraith:2002,Frison:1992,Frison:1997,Dawson:1993,Raudenbush:2001,Overall:1994,Hedeker:1999,Jung:2003,Schlesselman:1973,Liu:1997,Kirby:1994,Rochon:1998}. These publications have based their formulas on several different test statistics, designed to maximize power over several typical hypotheses which arise in longitudinal studies. Most of this previous work was motivated by the design of clinical trials. In an observational setting, study design calculations must be based on tests which remain valid when baseline response levels of the exposed and unexposed differ and when the exposure prevalence is not 0.5. Although most of the aforementioned formulas can be applied to a non-randomized setting, in-depth investigation of the formulas in settings relevant in observational research is lacking. For example, a study of the behavior of study power as the exposure prevalence deviates from 0.5 in longitudinal designs has not appeared previously. In clinical trials, the time scale of interest is usually time from randomization and the repeated measures are scheduled at a common set of times for all participants -- therefore, all the sample size formulas were based on this case. Here, we consider situations where time in the study is not the time variable of interest, but rather age, time since exposure or other measures of time of importance. Design of longitudinal studies is complex, involve the a priori specification of up to ten parameters about which investigators may have little information unless pilot data are available. We therefore formulated intuitive parameterizations to our formulas, using percent changes for the specification of effects, and intuitive covariance parameters for three covariance structures in order to facilitate widespread use in applications. 

In addition to exposure prevalence, we studied in detail the effect of the following factors on power: the number of repeated measures; the length of follow-up, the frequency of measurement; the use of age as the time metameter instead of time since randomization. The effect of these parameters on the required number of participants when the number of repeated measures is fixed was also studied. We studied the effect of the covariance parameters on the required number of repeated measures when the number of participants is fixed. Additionally, when the number of study participants ($N$) and the number of repeated measures ($r$) can be controlled by the investigator, their optimal combination for maximizing the power to detect a group difference subject to a cost constraint was derived. \citeasnoun{Cochran:1977} and \citeasnoun{Raudenbush:1997} examined this problem in the special cases of the alternative hypothesis of a group difference that is constant over time and under compound symmetry. \citeasnoun{Snijders:1993} developed the methodology to obtain the optimal number of participants and measurements, $(N_{opt} ,r_{opt} )$, subject to a cost constraint, under compound symmetry (CS) and random slopes (RS) covariance structures, for both a group difference that is constant over time (CMD) and for a group by time interaction  (LDD). The model upon which they based these developments explicitly separates the between- from the within- subjects effects (B\&W), and optimal designs are different than those given in this paper which follow the modeling approaches most commonly used in epidemiology (models \eqref{cmdp1}-\eqref{ldddummyp1}). In this paper, we will briefly address how design considerations differ for the B\&W model from the models considered here, as relevant, and in addition, we extend results to the damped exponential (DEX) covariance and settings where subjects are observed at different times (e.g. when baseline ages vary). Finally, we study in detail for the first time the effect of all the parameters on the resulting optimal combination.

This paper is structured as follows. In section~\ref{notationp1}, we present intuitively parameterized models for the two alternative hypotheses commonly considered in longitudinal studies, and the test statistics that will serve as the basis for power and sample size calculation for each of them. We show that some of the test statistics that have been considered previously are biased or less efficient in observational (non-randomized) studies, when the expected value of the baseline measures is not equal in the two exposure groups. In section~\ref{Nrfixedp1}, we derive formulas for power and sample size for an arbitrary covariance matrix and study the effect of exposure prevalence. Using an intuitive formulation for the parameters of interest, as well as for the nuisance parameters, we provide the formulas under compound symmetry and study the effect of the covariance between repeated measurements on power. Then, we extend the formulas to other covariance structures by incorporating additional intuitively defined parameters, formulated in a manner that is intended to be accessible to non-statistical investigators and enables the use of existing pilot or published data and when unavailable, intuitive hunches. The effects of departure from compound symmetry on power, sample size and number of repeated measures are studied. In section~\ref{optNr}, we provide methods to find the optimal combination of number of participants and number of repeated measures for maximizing power under a fixed budgetary constraint. In section~\ref{examplep1}, we explore aspects of the design of an epidemiological study of the effects of cigarette smoking on lung function, based on publicly available data that we will treat as a pilot study. In section~\ref{softp1}, we compare the functionality of currently available software for longitudinal study design, discuss their limitations, and introduce our comprehensive software for the observational longitudinal design setting. Finally, in section~\ref{conclp1}, we summarize our findings on the many different factors that need to be taken into account when planning a longitudinal observational study of a time-invariant exposure. 

\section{Notation and Preliminary Results}
\label{notationp1}

Consider the case where there are two groups, the exposed and the non-exposed. Let $N$ be the total sample size and $p_e $ be the prevalence of exposure. Let $Y_{ij}$ be a normally distributed outcome of interest for the measurement taken at the $j{\text{th}}$ time $(j = 1, \ldots ,r)$ for the $i{\text{th}}$ $(i = 1, \ldots ,N)$ participant, and let $k_i $ $(k_i  = 0,1)$ be the exposure group for subject $i$. We consider studies that obtain repeated measures every $s$ time units, as is the usual design in epidemiologic studies. Thus, the total length of follow-up is $\tau  = s\,r
$. For example, a study that follows participants every 6 months $(s = 6)$ for five years ($\tau  = 12*5 = 60$ months) would have 11 measures per participant, one at baseline plus $r = 10$ repeated measures. In epidemiology, there is often interest in the variation of the response by age and not by time in the study. Since participants enter the study at different ages, each participant has a different set of ages, ${\mathbf{t}}_i $, over which they are observed. Since each measurement is taken every $s$ units, the vector of times is fully defined by the initial time (age at entry) $t_{i0} $, and then ${\mathbf{t}}_i ^\prime   = \;(t_{i0} ,\;t_{i0}  + s,\;t_{i0}  + 2s,\; \ldots ,\;t_{i0}  + r\,s)$. When $V\left( {t_0 } \right) = 0
$, where $V\left( {t_0 } \right)$ is the variance of the primary time metameter of the analysis at baseline, as when using time since enrollment in the study as the time variable of interest, all participants have the same time vector.
We assume a linear form for the mean, $\mathbb{E}\left[ {{\mathbf{Y}}_i } \right] = {\mathbf{X}}_i {\mathbf{{\rm B}}}\;(i = 1, \ldots ,N)$, where ${\mathbf{X}}_i $ is the covariate matrix for participant $i$, and ${\mathbf{{\rm B}}}$ is the vector of unknown regression coefficients relating the conditional mean of ${\mathbf{Y}}_i $ to its corresponding covariates; and $Var\left[ {{\mathbf{Y}}_i |{\mathbf{X}}_i } \right] = {\mathbf{\Sigma }}_i \;\;(i = 1, \ldots ,N)$, where ${\mathbf{\Sigma }}_i 
$ is the $(r + 1) \times (r + 1)$ residual covariance matrix assumed equal for all participants. Note that ${\mathbf{\Sigma }}_i $ can be any valid covariance matrix, and can include terms associated with between-subjects variability as well as within-subjects variation. The generalized least squares (GLS) estimator of ${\mathbf{{\rm B}}}$ is 
$$
{\mathbf{\hat B}} = \left( {\sum\limits_{i = 1}^N {{\mathbf{X'}}_i {\mathbf{\Sigma }}_i^{ - 1} {\mathbf{X}}_i } } \right)^{ - 1} \left( {\sum\limits_{i = 1}^N {{\mathbf{X'}}_i {\mathbf{\Sigma }}_i^{ - 1} {\mathbf{Y}}_i } } \right).
$$
One way to circumvent the problem of the design matrix not being known a priori in an observational study is to use the asymptotic variance of this estimator. Other possible approaches are discussed in section~\ref{conclp1}. The asymptotic variance of ${\mathbf{\hat B}}$ is $\frac{1} {N}{\mathbf{\Sigma }}_{\rm B} $, where  
\begin{equation}
\label{sigmaBp1}
{\mathbf{\Sigma }}_{\rm B}  = \left( {\mathbb{E}_X \left[ {{\mathbf{X'}}_i {\mathbf{\Sigma }}_i^{ - 1} {\mathbf{X}}_i } \right]} \right)^{ - 1}. 
\end{equation}
and, provided ${\mathbf{\Sigma }}_i $ does not depend on the covariates, this covariance matrix can be fully specified by knowing the first and second order moments of the covariate distribution, and not their full distribution \cite{Tu:2004}.  We assume that the prevalence of exposure is $p_e $, the variance of the initial time is $V\left( {t_0 } \right)$ and the correlation between exposure and initial time is $\rho _{\operatorname{e} ,t_0 } $. We also assume that the variance of the initial time is the same in the two exposure groups.  

Our sample size and power equations are based on the Wald test for the coefficient of interest. Thus, our test statistic has the canonical form
\begin{equation}
\label{testp1}
T = \frac{{\sqrt {N\,} {\mathbf{c'\hat {\rm B}}}}}
{{\sqrt {{\mathbf{c'\Sigma }}_{\rm B} {\mathbf{c}}} }},
\end{equation}
where ${\mathbf{c}}$ is a $(g + 1) \times 1$ vector, where $g$ is the number of explanatory variables in the model, with a one and $g$ zeros isolating the particular component of ${\mathbf{{\rm B}}}$ that is of interest. Our models will be written with the coefficient of interest always the last one, so  ${\mathbf{c}}$ will be of the form ${\mathbf{c}} = \left( {0, \ldots ,0,1} \right)$.

\begin{figure}
  \centering 
  \includegraphics[width=6in]{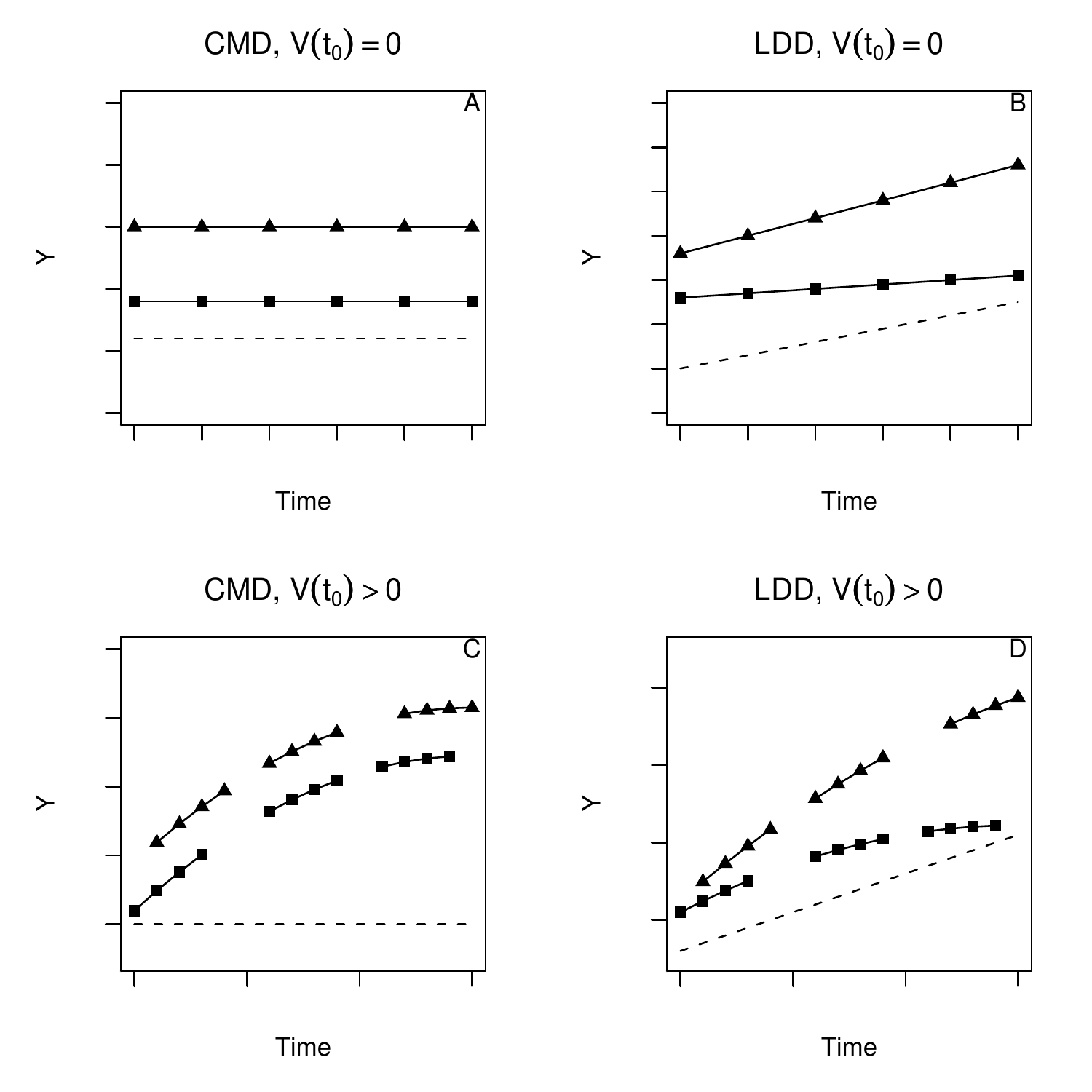}
  \caption{Possible patterns of response under the alternative hypotheses considered in this paper. In A and B, all participants have six measurements at the same time points. In C and D, the graph shows six participants (three exposed and three unexposed) with a total of four measurements each. The different lines represent unexposed ($\blacksquare$), exposed ($\blacktriangle$) and the difference between exposed and unexposed ({-~-~-}).}
  \label{patternsp1}
\end{figure}
	
Power and sample size calculations depend on the alternative hypothesis under consideration. Two patterns have been commonly considered in past literature, and both are quite relevant for applications in epidemiology. We first consider a constant mean difference (CMD) \cite{Frison:1992} between exposed and unexposed (figure~\ref{patternsp1}A for $V(t_0 ) = 0$ and figure~\ref{patternsp1}C for $V(t_0 ) > 0$). The CMD alternative hypothesis in the context of clinical trials assumes that the baseline means of the two groups are equal, and then the post-baseline means have a constant difference with respect to time. In observational studies, there is nothing special about baseline, and we need to allow the two groups to differ at baseline as well, as would usually be the case in observational studies.  This situation can be modeled as
\begin{equation}
\label{cmdp1}
\mathbb{E}\left( {Y_{ij} |X_{ij} } \right) = \beta _0  + \beta _1 t_{ij}  + \beta _2 k_i 
\end{equation}
		
if the effect of time can be considered linear. Our formulas will be based on model~\eqref{cmdp1}, however conclusions will extend to the more general model
\begin{equation}
\label{cmdfp1}
\mathbb{E}\left( {Y_{ij} |X_{ij} } \right) = \beta _0  + f(t_{ij} ;{\mathbf{\beta }}_1 ) + \beta _2 k_i, 
\end{equation}

where $f(t_{ij} ;{\mathbf{\beta }}_1 )$ is an arbitrary function of time, only if time and exposure can be considered independent. The resulting response profiles for the exposed and unexposed are parallel. The null hypothesis of interest is $H_0 :\beta _2  = 0$ vs. the alternative, $H_0 :\beta _2  \ne 0$, and serves as a basis for the test of whether the two response profiles coincide or not. When all participants are observed at the same set of time points, we can include indicators for the $(r + 1)$ time points and write model~\eqref{cmdfp1} as
\begin{equation}
\label{cmddummyp1}
\mathbb{E}\left( {Y_{ij} |X_{ij} } \right) = \mu _{0.0}  + \mu _{0.1}  +  \cdots  + \mu _{0.r}  + \beta _2 k_i, 
\end{equation}
where $\mu _{0.0} ,\mu _{0.1} , \cdots ,\mu _{0.r} $ give the means at times $t_0 , \ldots ,\,t_r $ in the unexposed group, and $\beta _2 $ is the constant difference in response between exposed and unexposed. 

In the second pattern, linearly divergent differences (LDD), the effect of exposure varies linearly with time (figure~\ref{patternsp1}B for $V(t_0 ) = 0$ and figure~\ref{patternsp1}D for $V(t_0 ) > 0$) \cite{Frison:1997}. In clinical trials, the mean of the two groups at baseline are assumed to be equal. Here, we allow for a baseline difference, as this would usually be the case in observational studies. In the simplest version of LDD, the effect of time is linear in both groups (figure~\ref{patternsp1}B) and can be modeled as 
\begin{equation}
\label{lddp1}
\mathbb{E}\left( {Y_{ij} |X_{ij} } \right) = \gamma _0  + \gamma _1 t_{ij}  + \gamma _2 k_i  + \gamma _3 \left( {t_{ij}  \times k_i } \right),
\end{equation}
although we can allow the relationship between response and time in the unexposed to be more general, of form 
\begin{equation}
\label{lddfp1}
\mathbb{E}\left( {Y_{ij} |X_{ij} } \right) = \gamma _0  + f\left( {t_{ij} ;{\mathbf{\gamma }}_1 } \right) + \gamma _2 k_i  + \gamma _3 \left( {t_{ij}  \times k_i } \right),
\end{equation}
where $f\left( {t_{ij} ;{\mathbf{\gamma }}_1 } \right)$ is now a function of time that includes a linear term but is otherwise arbitrary. The formulas we derive, however, will be valid for model~\eqref{lddfp1} only if exposure and time can be considered independent. When all participants are observed at the same set of time points, one can use the model 
\begin{equation}
\label{ldddummyp1}
\mathbb{E}\left( {Y_{ij} |X_{ij} } \right) = \mu _{0.0}  + \mu _{0.1}  +  \cdots  + \mu _{0.r}  + \gamma _2 k_i  + \gamma _3 (k_i  \times t_j ),
\end{equation}
which accommodates any shape over time. The test of interest is thus $H_0 :\gamma _3  = 0$ vs. the alternative $H_A :\gamma _3  \ne 0
$, where $\gamma _3 $ is the difference in the rates of change of the response over time between the exposed and unexposed, per a single unit of time. Note that if the null hypothesis is accepted there can still be a constant difference at baseline, which persists over time, between the exposed and unexposed, as in CMD. That is, we test whether the two response profiles are parallel or not, or, in other words, whether the effect of time is the same in exposed and unexposed. 

Models \eqref{lddp1}-\eqref{ldddummyp1} assume that the cross-sectional and longitudinal effects of time coincide. Models that separate the cross-sectional (between-subjects) and the longitudinal (within-subjects) effects have been developed (B\&W) \cite{Diggle:2002,Fitzmaurice:2004,Neuhaus:1998,Ware:1990}. Applying those models and allowing different effects for the exposed and the unexposed, one can fit the following model to the data,
\begin{equation}
\label{lddbwp1}
\begin{gathered}
  \mathbb{E}\left( {Y_{ij} |X_{ij} } \right) = \eta _0  + \eta _1 t_{i0}  + \eta _2 \left( {t_{ij}  - t_{i0} } \right) + \eta _3 k_i  + \eta _4 \left( {k_i  \times t_{i0} } \right) + \eta _5 \left( {k_i  \times \left( {t_{ij}  - t_{i0} } \right)} \right) =  \\ 
  \eta _0  + \eta '_1 t_{i0}  + \eta _2 t_{ij}  + \eta _3 k_i  + \eta '_4 \left( {k_i  \times t_{i0} } \right) + \eta _5 \left( {k_i  \times t_{ij} } \right) \\ 
\end{gathered}, 
\end{equation}
where $\eta _1 $ is the cross-sectional effect of time in the unexposed, $\eta _4 
$ is the difference in the cross-sectional effect of time between the exposed and the unexposed, $\eta _2 $ is the longitudinal effect of time in the unexposed, and $\eta _5 $ is the difference in the longitudinal effect of time between the exposed and the unexposed. The hypothesis to be tested is $H_0 :\eta _5  = 0$ vs. the alternative $H_A :\eta _5  \ne 0$. When there is no confounding, which in this context means that the exposed and unexposed do not differ with respect to the distribution of unmeasured risk factors, and when $V(t_0 ) = 0$, the distribution of time is the same among the exposed and unexposed, and $\eta _5  = \gamma _3 $. Otherwise, they are different, and $\eta _5 $ will be the parameter of interest in epidemiology. Another possibility is to fit a model for the differences from one visit to the next, 
\begin{equation}
\label{ldddiffp1}
\mathbb{E}\left( {Y_{i,j}  - Y_{i,j - 1} |X_{ij} } \right) = \lambda _0  + \lambda _1 k_i.
\end{equation}		
We prove in Appendix~\ref{apvar4p1} that $s\hat \lambda _1  = \hat \eta _5 $ and $s^2 Var\left( {\hat \lambda _1 } \right) = Var\left( {\hat \eta _5 } \right)$, so inferences based on  $\hat \lambda _1 $ or $\hat \eta _5 $ are equivalent. As relevant, we will discuss the impact on design when \eqref{lddbwp1} or \eqref{ldddiffp1} is to be used instead of \eqref{lddp1}-\eqref{ldddummyp1}. 
 
Other test statistics have been previously proposed for these settings, including those which adjust for the baseline response, such as ANCOVA, SLAIN and SLANC \cite{Frison:1992,Frison:1997}. Because differences at baseline among exposed and unexposed commonly occur in observational studies, the properties of the test statistics are different in observational studies compared to randomized trials (Appendix~\ref{apancova}). Although ANCOVA, SLAIN and SLANC are valid for the CMD hypothesis, they are less powerful than the test statistic on which we based our design calculations. Under the LDD hypothesis, ANCOVA, SLAIN and SLANC are all biased (Appendix~\ref{apancova}). In the absence of additional model covariates, with all participants observed at the same time points, a two-stage estimator, where a regression of the response vs. time is performed for each participant, and in a second stage, these $N$ independent estimates of the slopes are regressed on exposure, is algebraically identical to the estimating $\gamma _3 $ from model~\eqref{lddp1} by OLS (Appendix~\ref{ap2stage}). It turns out that if we can assume a compound symmetry (CS) covariance structure or a random slopes (RS) covariance structure, where random effects are assumed both for the intercept and the time slope, the two stage estimator, and, the OLS estimator are equivalent to the GLS estimator of $\gamma _3 $ (Appendix~\ref{ap2stage}). This result does not hold for damped exponential (DEX) correlation structure (Appendix~\ref{ap2stage}). The two-stage approach and GLS approach do not coincide when participants are observed at different times. Since $\hat \beta _2 $ and $\hat \gamma _3 $ in this paper are GLS and therefore are the best linear unbiased estimates for their respective models \eqref{cmdp1}-\eqref{cmddummyp1} and \eqref{lddp1}-\eqref{ldddummyp1} \cite{Searle:1971}, other valid options for the testing of the LDD hypothesis, such as comparing the maximum change over the exposed to the maximum change in the unexposed \cite{Koh-Banerjee:2003}, are inefficient and will not be considered further.

The only parameters of the models given above that are needed for power calculations are $\beta _2 $, $\gamma _3 $ and $\eta _5 $. Typically, it is difficult to provide a priori values for these parameters that are realistic and well justified and even more difficult to supply realistic and well justified values for their variance. Therefore, we reparameterized the key parameters of the models above in terms of quantities more likely to be known to the investigator, available from published papers, or easily calculated in pilot data. These parameters are: 
\begin{enumerate}
\item the mean response at baseline (or at the mean initial time) in the unexposed group $(\mu _{00} )$, where $\mu _{00}  = \mathbb{E}\left( {Y_{i0} |k_i  = 0} \right),\,{\kern 1pt} \,i = 1, \cdots ,N$. 
\item the percent difference between exposed and unexposed groups ($p_1$) at baseline (or at the mean initial time), where 
$$p_1  = \frac{{\mathbb{E}\left( {Y_{i0} |k_i  = 1} \right) - \mathbb{E}\left( {Y_{i0} |k_i  = 0} \right)}}
{{\mathbb{E}\left( {Y_{i0} |k_i  = 0} \right)}},\;\,i = 1, \cdots ,N.
$$\item the percent change from baseline (or from the mean initial time)  to end of follow-up (or to the mean final time) in the unexposed group ($p_2 $), where  
$$p_2  = \frac{{\mathbb{E}\left( {Y_{i\tau } |k_i  = 0} \right) - \mathbb{E}\left( {Y_{i0} |k_i  = 0} \right)}}
{{\mathbb{E}\left( {Y_{i0} |k_i  = 0} \right)}},\;i = 1, \cdots ,N.
$$For situations where $\tau $ is not fixed, $p_2 $ is defined at time $s$ instead of at time $\tau $.
\item the percent difference between the change from baseline (or from the mean initial time)  to end of follow-up (or mean final time) in the exposed group and the unexposed group ($p_3 $), where 
$$p_3  = \frac{{\mathbb{E}\left( {Y_{i\tau }  - Y_{i0} |k_i  = 1} \right) - \mathbb{E}\left( {Y_{i\tau }  - Y_{i0} |k_i  = 0} \right)}}
{{\mathbb{E}\left( {Y_{i\tau }  - Y_{i0} |k_i  = 0} \right)}},\;\,i = 1, \cdots ,N.
$$When $p_2  = 0$, $p_3 $ will be defined as the percent change from baseline (or from the mean initial time)  to the end of follow-up (or to the mean final time) in the exposed group, 
$$
p_3  = \frac{{\mathbb{E}\left( {Y_{i\tau } |k_i  = 1} \right) - \mathbb{E}\left( {Y_{i0} |k_i  = 1} \right)}}
{{\mathbb{E}\left( {Y_{i0} |k_i  = 1} \right)}},\;i = 1, \cdots ,N.
$$
For situations where $\tau $ is not fixed, $p_3 $ is defined at time $s$ instead of at time $\tau $.
\item the residual variance of the response given the covariates, $\sigma ^2  = Var\left( {Y_{ij} |X_{ij} } \right)
$. Note importantly that this parameter is not equal to the marginal, cross-sectional variance of ${\mathbf{Y}}$. It can be approximated by the variance of the response among the unexposed at baseline, a quantity that may be available from the literature or pilot data, or if age is the time variable of interest, within a reasonably narrow age group. If only a marginal response variance is available over a range of ages and exposure levels, as will often be the case in epidemiology, the investigator can approximate the residual variance by multiplying it by the quantity $1 - R^2 $, where $R^2 $ is the assumed proportion of the marginal variance of the response variable that is explained by the model to be fit when the study is conducted \cite{Snijders:1994}, here one of models \eqref{cmdp1}-\eqref{ldddummyp1}. Typically, in epidemiology, $R^2 $ ranges from 0.10 to 0.30 or so. Under CMD, the parameter of interest is $\beta _2  = p_1 \mu _{00} $, and under LDD, it is $\gamma _3  = \eta _5  = {p_2 p_3 \mu _{00} }/{\tau }$ or, when $p_2  = 0$, it is $\gamma _3  = \eta _5  = {(1 + p_1 )p_3 \mu _{00} }/ {\tau }$. Hence, when CMD is of interest, the investigator needs to specify the alternative through two parameters, $p_1 $ and $\mu _{00} $, and when LDD is of interest, four parameters are needed, $p_2 $, $p_3 $, $\mu _{00} $, and $\tau $. In our experience, investigators can readily provide values, or ranges of values, for these parameters, while it is difficult to directly obtain values for $\beta _2 $, $\gamma _3 $, or $\eta _5 $. It is even more difficult, if not impossible, to obtain a priori values for $Var(\hat \beta _2 )$, $Var(\hat \gamma _3 )$ or $Var(\hat \eta _5 )$, since these quantities depend on $\sigma ^2 $, along with other parameters. 
\end{enumerate}

\section{Power and Sample Size when either $N$ or $r$ is Fixed}
\label{Nrfixedp1}

\subsection{General case when ${\mathbf{\Sigma }}_i  = {\mathbf{\Sigma }}\;\forall i$}
\label{generalp1}

We assume that the covariance matrix is the same for all participants, i.e. ${\mathbf{\Sigma }}_i~=~{\mathbf{\Sigma }}\;\forall i$. Section~\ref{rsp1} will consider a particular where this is not true. The general power formula associated with the test statistic, $T$, is 
\begin{equation}
\label{powerp1}
\Phi \left[ {\frac{{\sqrt {N\,} \left| {\left( {{\mathbf{c'{\rm B}}}} \right)_{H_A } } \right|}}
{{\sqrt {{\mathbf{c'\Sigma }}_{\rm B} {\mathbf{c}}} }} - z_{1 - \alpha /2} } \right],
\end{equation}
where $\left( {{\mathbf{c'{\rm B}}}} \right)_{H_A } $ is the value of the regression parameter vector under the alternative (i.e. $\beta _2 $ for CMD and $\gamma _3 $ for LDD), $\alpha $ is the significance level, ${\mathbf{\Sigma }}_{\rm B} $ is defined in equation~\eqref{sigmaBp1}, and $z_p $ and $\Phi \left(  \cdot  \right)$ are the $p{\text{th}}$ quantile and the cumulative density of a standard normal, respectively. From \eqref{powerp1}, it is clear that power will increase as the number of participants, $N$, increases. From equation~\eqref{sigmaBp1} we can see that the matrix ${\mathbf{\Sigma }}_{\rm B} $ depends both on the inverse of the residual covariance matrix, ${\mathbf{\Sigma }}^{ - 1} $, and the covariate matrices ${\mathbf{X}}_i $. Let $v_{jj'} $ be the $(j,j'){\text{th}}$ element of ${\mathbf{\Sigma }}^{ - 1} $, and let 
$$
{\mathbf{A}} = \left( {\begin{array}{*{20}c}
   {\sum\limits_{j = 0}^r {\sum\limits_{j' = 0}^r {v_{jj'} } } } & {\sum\limits_{j = 0}^r {\sum\limits_{j' = 0}^r {jv_{jj'} } } }  \\
   {\sum\limits_{j = 0}^r {\sum\limits_{j' = 0}^r {jv_{jj'} } } } & {\sum\limits_{j = 0}^r {\sum\limits_{j' = 0}^r {jj'v_{jj'} } } }  \\

 \end{array} } \right).
$$
Then, under CMD, we show that ${\mathbf{c'\Sigma }}_{\rm B} {\mathbf{c}}$ is (Appendix~\ref{apvar1p1}) 
\begin{equation}
\label{varcmdp1}
{\mathbf{c'\Sigma }}_{\rm B} {\mathbf{c}} = \frac{{s^2 \det ({\mathbf{A}}) + \left( {\sum\limits_{j = 0}^r {\sum\limits_{j' = 0}^r {v_{jj'} } } } \right)^2 V\left( {t_0 } \right)}}
{{p_e (1 - p_e )\left( {\sum\limits_{j = 0}^r {\sum\limits_{j' = 0}^r {v_{jj'} } } } \right)\left[ {s^2 \det ({\mathbf{A}}) + \left( {\sum\limits_{j = 0}^r {\sum\limits_{j' = 0}^r {v_{jj'} } } } \right)^2 \left( {1 - \rho _{\operatorname{e} ,t_0 }^2 } \right)V\left( {t_0 } \right)} \right]}}.
\end{equation}
If either $V\left( {t_0 } \right)~=~0$, i.e. all participants enter the study at the same time, or $\rho _{\operatorname{e} ,t_0 }  = 0$, i.e. exposure and initial time are uncorrelated, this formula reduces to
\begin{equation}
\label{varcmdvt00p1}
{\mathbf{c'\Sigma }}_{\rm B} {\mathbf{c}} = \left[ {p_e (1 - p_e )\left( {\sum\limits_{j = 0}^r {\sum\limits_{j' = 0}^r {v_{jj'} } } } \right)} \right]^{ - 1} 
\end{equation}
(Appendix~\ref{apvarp1}). Under LDD, we have 
\begin{equation}
\label{varlddp1}
{\mathbf{c'\Sigma }}_{\rm B} {\mathbf{c}} = \frac{{\left( {\sum\limits_{j = 0}^r {\sum\limits_{j' = 0}^r {v_{jj'} } } } \right)}}
{{p_e (1 - p_e )\left[ {s^2 \det ({\mathbf{A}}) + \left( {1 - \rho _{\operatorname{e} ,t_0 }^2 } \right)V(t_0 )\left( {\sum\limits_{j = 0}^r {\sum\limits_{j' = 0}^r {v_{jj'} } } } \right)^2 } \right]}}
\end{equation}
(Appendix~\ref{apvar2p1}). Under model~\eqref{lddbwp1}, the B\&W model, or model~\eqref{ldddiffp1} we show that 
$$
{\mathbf{c'\Sigma }}_{\rm B} {\mathbf{c}} = \frac{{\left( {\sum\limits_{j = 0}^r {\sum\limits_{j' = 0}^r {v_{jj'} } } } \right)}}
{{p_e (1 - p_e )s^2 \det ({\mathbf{A}})}}
$$
(Appendix~\ref{apvar3p1}), i.e. the formula for ${\mathbf{c'\Sigma }}_{\rm B} {\mathbf{c}}$ is the same as in the LDD case when $V(t_0 ) = 0$. Therefore, the results for LDD when $V(t_0 ) = 0$ apply to model~\eqref{lddbwp1} and model~\eqref{ldddiffp1} and will not be presented in a separate section. When the follow-up period, $\tau $, is fixed and the time points are equidistant, there are instances where ${\mathbf{c'\Sigma }}_{\rm B} {\mathbf{c}}$, under LDD and when $V(t_0 ) = 0$, is the same for $r = 1$ and $r = 2$. On pure efficiency grounds, in these situations, it is never cost-effective to add only one additional measure. In Appendix~\ref{apr1r2}, we derive a condition on the matrix ${\mathbf{\Sigma }}$ that needs to hold for this situation to occur, and we show that this will be the case for the three covariance structures considered in this paper. Of course, this result assumes that the interaction term is linear, and does not consider that the third measure is needed to assess the validity of this assumption. This counter-intuitive result is due to the fact that with equidistant points, the additional measure would be taken at half the follow up period, which is the mean of the time vector. This is similar to the fact that, in simple linear regression, adding an observation whose value for the explanatory variable is the explanatory variable mean produces no change on the variance of the slope.  

The power formula depends on $N$, $r$ and $s$, producing a discrete three-dimensional surface of constant power. Fixing $r$ and $s$, the formula for the required sample size in $N$ to achieve a pre-specified power $\pi $ is 
\begin{equation}
\label{Np1}
N = \frac{{\left( {{\mathbf{c'\Sigma }}_{\rm B} {\mathbf{c}}} \right)\,\left( {z_\pi   + z_{1 - \alpha /2} } \right)^2 }}
{{({\mathbf{c'{\rm B}}}_{H_A } )^2 }}.
\end{equation}

\begin{figure}
\centering 
\includegraphics{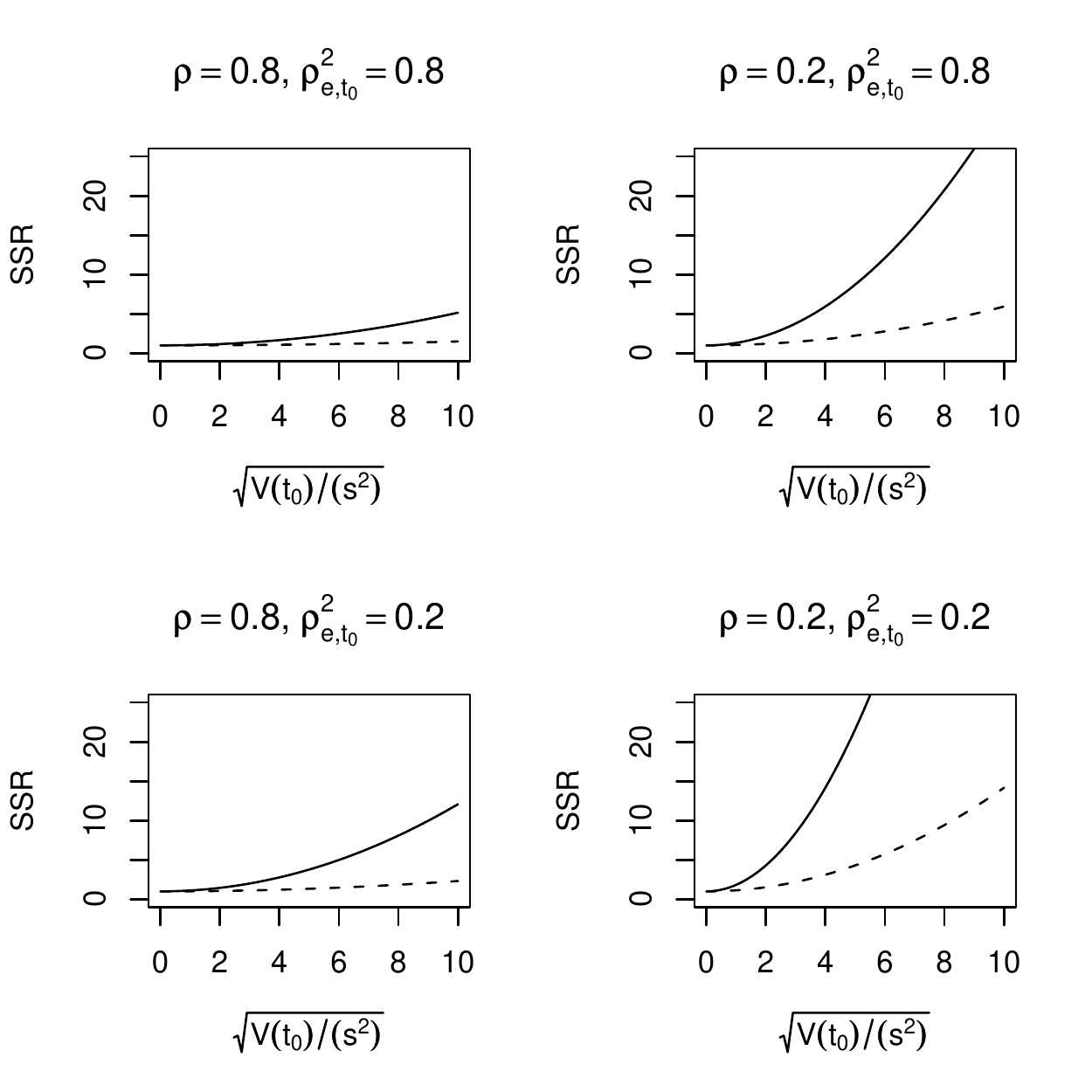}
\caption{Ratio of required sample sizes (SSR) to achieve the same power comparing the $V(t_0 ) = 0$ case to $V(t_0 ) > 0$ under LDD and CS $ \left( {SSR = \frac{{N_{V(t_0 ) = 0} }} {{N_{V(t_0 )} }}} \right) $. The groups are: $r = 2$ ({---}{---}) and $r = 5$ (-~-~-).}
\label{ssrvt0p1}
\end{figure}

For some parameters, namely $p_e $, $s$, $V\left( {t_0 } \right)$ and $\rho _{\operatorname{e} ,t_0 } $, their effect on ${\mathbf{c'\Sigma }}_{\rm B} {\mathbf{c}}$, and by virtue of \eqref{powerp1} and \eqref{Np1}, on power and sample size with $r$ fixed, can be derived for a general covariance structure. Unlike randomized clinical trials, in observational studies, $p_e $ is rarely 0.5, which is the value of $p_e $ that maximizes power. For other values of $p_e $ the sample size obtained for $p_e  = 0.5$ needs to be multiplied by $\frac{{0.5^2 }}
{{p_e (1 - p_e )}}$. For example, for $p_e  = 0.2$ and $r$ fixed we need 56\% more participants to achieve the same power than for $p_e  = 0.5$, and for $p_e  = 0.1$, the sample size is multiplied by 2.7. It can also be derived from equations \eqref{varcmdp1} and \eqref{varlddp1} that, unless ${\mathbf{\Sigma }}
$ is an explicit function of $s$ (as in the DEX covariance structure models considered in section~\ref{dexp1}), increasing $s$ reduces ${\mathbf{c'\Sigma }}_{\rm B} {\mathbf{c}}$ and therefore increases power and reduces the required sample size for both CMD and LDD. However, if either $V\left( {t_0 } \right) = 0
$ or $\rho _{\operatorname{e} ,t_0 }  = 0$, $s$ does not play a role in the CMD case. The effect of  $V(t_0 )$, the baseline variance in the primary time scale, is different under CMD and LDD. Under CMD, as this parameter increases, keeping $\rho _{\operatorname{e} ,t_0 } $ and all other parameters fixed, so does ${\mathbf{c'\Sigma }}_{\rm B} {\mathbf{c}}$ and therefore power decreases, since 
$$
\frac{{\delta {\mathbf{c'\Sigma }}_{\rm B} {\mathbf{c}}}}
{{\delta V(t_0 )}} > 0.
$$
The decrease in power will be larger as $\rho _{\operatorname{e} ,t_0 } $ departs from zero. When $\rho _{\operatorname{e} ,t_0 }  = 0$, inspection of \eqref{varcmdp1} shows readily that the power is the same as that for the $V(t_0 ) = 0$ case (see \eqref{varcmdvt00p1}). Unlike the CMD case, under LDD, as $V(t_0 )$ increases the power also increases, as is easily seen upon examination of \eqref{varlddp1}, which depends on $V(t_0 )$ in the denominator only. Increasing $V(t_0 )$ increases the range of the regressor, $t$, which is known in simpler regression problems to increase the power of the study to detect a non-zero regression slope. Apparently, this result extends to interaction terms that are a function of a continuous variable, as well. The gain in power due to $V(t_0 )$ is largest at $\rho _{\operatorname{e} ,t_0 }  = 0$, and it vanishes when $\rho _{\operatorname{e} ,t_0 }  = 1$ or $\rho _{\operatorname{e} ,t_0 }  =  - 1$, in which case power is equivalent to the $V(t_0 ) = 0$ case. It only makes sense to examine the effect of $\rho _{\operatorname{e} ,t_0 } $ when $V(t_0 ) > 0$, since if $V(t_0 ) = 0$ then $t_{i0} $ is constant for all participants and $\rho _{\operatorname{e} ,t_0 } $ is zero. Figure~\ref{ssrvt0p1} shows the gain in efficiency of having $V(t_0 ) > 0
$ compared to $V(t_0 ) = 0$ for the CS case by showing the ratio of required sample size to achieve the same power in every case. The gains in efficiency can be very large for small $r$, large $\frac{{\sqrt {V(t_0 )} }}{s}$, small $\rho $ and small $\rho _{\operatorname{e} ,t_0 } $.

Sometimes, $N$ is fixed and the interest is in finding the minimum $r$ to achieve a certain power. This problem has not been examined in any detail in previous literature. It may arise, for example, when an existing cross-sectional study is to form the basis of a new longitudinal one or when there are a fixed number of participants available, (e.g. nurses who returned a baseline questionnaire in 1989). For some covariance structures, an explicit formula for $r$ as a function of $N$, $\pi $ and $s$ can also be obtained, and for other structures, $r$ can be obtained only numerically. It is shown that the minimum $r$ for fixed $N$ is obtained when $p_e  = 0.5$ (Appendix~\ref{appe}). Since no other global results are available, we will therefore consider this problem in the next section. 

Sections \ref{csp1}-\ref{rsp1} provide power and sample size formulas for particular covariance structures. Then, for the two alternative hypotheses and for each covariance structure, we assess the effect of $r$ and the covariance parameters on power. Two scenarios will be considered when studying the effect of $r$ on power. First, the frequency of measurements, $s$, is fixed. For example, participants might visit the clinic every 6 months, and increasing the number of repeated measures increases duration of follow-up, $\tau $. In the second situation, the length of follow-up, $\tau $, is fixed, for example, to 5 years, and increasing the number of repeated measures involves increasing the frequency of measurement, $s$. Formulas for $r$ as a function of $N$, $\pi $ and $s$ or $\tau $ will be given when closed-form solutions exist. The effect of the covariance parameters on $r$ will also be studied. 

\subsection{Compound symmetry (CS)}
\label{csp1}

The simplest residual covariance structure that can be assumed for longitudinal data is compound symmetry. The residual covariance matrix is fully defined using two parameters: $\sigma ^2$ and $\rho $. The first one is the variance of the response given the covariates, which under CS is assumed to be constant over time, and was defined in section~\ref{notationp1}. The second parameter, $\rho $, is the correlation between two measurements from the same participant. Under CS, it is also the reliability coefficient, or intraclass correlation coefficient, 
$$
\rho  = \frac{{\sigma _{between}^2 }}
{{\sigma _{within}^2  + \sigma _{between}^2 }},
$$
where $\sigma _{between}^2 $ and $\sigma _{within}^2 $ are the between- and within-subject variance, respectively, and $\sigma _{within}^2  + \sigma _{between}^2  = \sigma ^2 $. Typically, $\rho $ is unavailable at the time a study is designed, and power or sample size will be calculated over a range. Under CS, then, the covariance matrix is 
$$
Var({\mathbf{Y}}_i |{\mathbf{X}}_i ) = {\mathbf{\Sigma }}_{(r + 1) \times (r + 1)}  = \sigma ^2 \left( {\begin{array}{*{20}c}
   1 & \rho  &  \cdots  & \rho   \\
   \rho  & 1 &  \ddots  &  \vdots   \\
    \vdots  &  \ddots  &  \ddots  & \rho   \\
   \rho  &  \ldots  & \rho  & 1  \\

 \end{array} } \right).
$$ 

\subsubsection{CMD}
\label{cscmdp1}

\begin{sidewaystable}
	\centering
	\caption{Numerator and denominator of the test statistic $T$
 (equation~\eqref{testp1}) to obtain power and sample size for several correlation structures. For LDD, it is assumed here that $V(t_0 ) = 0$.}
\bigskip
\begin{tabular}{p{57pt}p{20pt}p{248pt}p{260pt}}
\hline
 \parbox{57pt}{\centering {\small Pattern,}} & 
  \parbox{20pt}{\centering {${\mathbf{\Sigma }}$}} & 
   \multicolumn{2}{c}{\parbox{509pt}{\centering ${\mathbf{c'\Sigma }}_{\rm B} {\mathbf{c}}$}} \\
   
\cline{3-4}    \parbox{57pt}{\centering {\small $$\left( {{\mathbf{c'{\rm B}}}} \right)_{H_A }$$}} & & \parbox{248pt}{\centering Fixed $s$} &
 \parbox{260pt}{\centering  Fixed $\tau$} \\
  \hline
  
 \parbox{57pt}{\centering {\small \multirow{2}{57pt}{ {\centering CMD, $$ \beta_2  = p_1 \mu _{00} $$}}}} & 
  \parbox{20pt}{\centering {\footnotesize CS}} & 
   \parbox{248pt}{\centering {\small $${\frac{{\sigma ^2 (1 + r\rho )}}
{{{\kern 1pt} p_e (1 - p_e )(r + 1)}} }^{(1)}$$}} & 
    \parbox{260pt}{\centering {\small $${\frac{{\sigma ^2 (1 + r\rho )}}
{{{\kern 1pt} p_e (1 - p_e )(r + 1)}} }^{(1)}$$}} \\
  & \parbox{20pt}{\centering {\footnotesize AR(1)}} & 
 \parbox{248pt}{\centering {\small $${\frac{{\sigma ^2 (1 + \rho ^s )}}
{{{\kern 1pt} p_e (1 - p_e )(1 + r + \rho ^s  - r\rho ^s )}}}^{(2)}$$}} &
  \parbox{260pt}{\centering {\small $${\frac{{\sigma ^2 (1 + \rho ^{\tau /r} )}}
{{{\kern 1pt} p_e (1 - p_e )(1 + r + \rho ^{\tau /r}  - r\rho ^{\tau /r} )}}}^{(2)}$$}} \\
\hdashline 

\parbox{57pt}{\centering {\small \multirow{3}{57pt}{{\centering LDD,$$\gamma _3  = \frac{{p_2 p_3 \mu _{00} }}
{\tau } $$}}}} &
  \parbox{20pt}{\centering {\footnotesize CS}} & 
   \parbox{248pt}{\centering {\small $${\frac{{12\sigma ^2 (1 - \rho ){\kern 1pt} }}
{{{\kern 1pt} p_e (1 - p_e )\;s^2 \,r\,(r + 1)(r + 2)}}}^{(3)}$$}} & 
   \parbox{260pt}{\centering {\small $${\frac{{12\sigma ^2 (1 - \rho ){\kern 1pt} r}}
{{{\kern 1pt} p_e (1 - p_e )\;\tau ^2 \,(r + 1)(r + 2)}}}^{(3)}$$}} \\
   
  & \parbox{20pt}{\centering {\footnotesize AR(1)}} & 
  \parbox{248pt}{\centering {\small $${\frac{{12\sigma ^2 {\kern 1pt} (1 - \rho ^{2s} )\,\left[ {\,r\,s^2 p_e (1 - p_e )} \right]^{ - 1} }}
{{{\kern 1pt} \;\,(2 + r(r + 3) + 8\rho ^s  - 2r^2 \rho ^s  + (r - 2)(r - 1)\rho ^{2s} )}}}^{(4)}$$}} & 
   \parbox{260pt}{\centering {\small $${\frac{{12\sigma ^2 {\kern 1pt} (1 - \rho ^{2\tau /r} )\,r\,\left[ {\,\tau ^2 p_e (1 - p_e )} \right]^{ - 1} }}
{{{\kern 1pt} \;\,(2 + r(r + 3) + 8\rho ^{\tau /r}  - 2r^2 \rho ^{\tau /r}  + (r - 2)(r - 1)\rho ^{2\tau /r} )}}}^{(4)}$$}} \\
   
   & \parbox{20pt}{\centering {\footnotesize RS}} & 
 \parbox{248pt}{\centering {\small 
\begin{multline*}
 \left( {\frac{{12\sigma ^2 (1 - \rho _{t_0 } )}}
{{ s^2 p_e (1 - p_e )\,}}} \right) \\
\left( {\frac{1}
{{\,r(r + 1)(r + 2)}} + \left( {\frac{{\rho _{b_1 ,s,\tilde r} }}
{{1 - \rho _{b_1 ,s,\tilde r} }}} \right)\frac{1}
{{\tilde r(\tilde r + 1)(\tilde r + 2)}}} \right)^{(5)}
\end{multline*}
}} & 
  \parbox{260pt}{\centering {\small 
\begin{multline*}
\left( {\frac{{12\sigma ^2 (1 - \rho _{t_0 } )}}
{{ \tau ^2 p_e (1 - p_e )\,}}} \right) \\ 
\left( {\frac{r}
{{\,(r + 1)(r + 2)}} + \left( {\frac{{\rho _{b_1 ,\tau ,\tilde r} }}
{{1 - \rho _{b_1 ,\tau ,\tilde r} }}} \right)\frac{{\tilde r}}
{{\,(\tilde r + 1)(\tilde r + 2)}}} \right)^{(5)}
\end{multline*}
}} \\
\hline
  \parbox{585pt}{\footnotesize $^{(1)}$ \cite{Bloch:1986}} \\
  \parbox{585pt}{\footnotesize $^{(2)}$ Appendix~\ref{aplimitscmdar1}} \\
  \parbox{585pt}{\footnotesize $^{(3)}$ \cite{Diggle:2002,Dawson:1998,Frison:1997,Hedeker:1999,Kirby:1994,Jung:2003,Yi:2002}} \\
  \parbox{585pt}{\footnotesize $^{(4)}$ Appendix~\ref{aplimitslddar1}} \\
  \parbox{585pt}{\footnotesize $^{(5)}$ \cite{Fitzmaurice:2004,Galbraith:2002,Raudenbush:2001,Schlesselman:1973,Yi:2002}} \\
   
\end{tabular}
\vspace{2pt}

	\label{table1p1}
\end{sidewaystable}

Table~\ref{table1p1} shows the necessary terms to plug in to equation~\eqref{powerp1} to obtain the power formula for the case $V\left( {t_0 } \right)~=~0$. Under this same scenario, the formula for power as a function of $N$ and $r$ was previously given by \citeasnoun{Bloch:1986}. For the cases with $V\left( {t_0 } \right) > 0$ the equations do not simplify a great deal and interested readers should use formulas \eqref{varcmdp1} or \eqref{varlddp1} directly. For any $V\left( {t_0 } \right)$, the power to detect a difference increases as either $N$ or $r$ increase, but while by increasing $N$ power can get arbitrarily close to one, by increasing $r$ the maximum power that can be reached (when $r \to \infty $) (Appendix~\ref{aplimitscmdcs}) is 
$$
\Phi \left[ {\frac{{\sqrt {N{\kern 1pt} p_e {\kern 1pt} (1 - p_e )\,} \left| {\beta _2 } \right|}}
{{\sqrt {\sigma ^2 \rho } }} - z_{1 - \alpha /2} } \right].
$$
It has been shown that, when $V\left( {t_0 } \right) = 0,$
as the correlation, $\rho $, increases,  the power to detect a difference decreases \cite{Hedeker:1999}. This is not necessarily the case when $V\left( {t_0 } \right) > 0$. For example, when $N = 50$, $r = 2$, $\sigma ^2  = 1$, $V(t_0 ) = 20$ and $\rho _{\operatorname{e} ,t_0 }  = 0.7$, the variance for $\rho  = 0.8$ is 0.29 and for $\rho  = 0.9$ it is 0.25. Plugging in the corresponding values in Table~\ref{table1p1} to equation~\eqref{Np1} one obtains an equation for sample size for $V\left( {t_0 } \right)~=~0$. The equation for the minimum value of $r$ which achieves a specified particular power $\pi $,  with $N$ fixed under CS, CMD and $V\left( {t_0 } \right)~=~0$ is 
\begin{equation}
\label{rcmdcsp1}
r = \frac{{\beta _2^2 N\,p_e (1 - p_e ) - \left( {z_\pi   + z_{1 - \alpha /2} } \right)^2 \sigma ^2 }}
{{\left( {z_\pi   + z_{1 - \alpha /2} } \right)^2 \sigma ^2 \rho  - \beta _2^2 N\,p_e (1 - p_e )}}.
\end{equation}
As noted before, the desired power cannot always be reached by increasing the number of repeated measures, so equation \eqref{rcmdcsp1} will not always have a positive solution. The effect of the intraclass correlation, $\rho $, on $r$ depends on the additional parameters. If $\left( {z_\pi   + z_{1 - \alpha /2} } \right)^2 \sigma ^2  > \beta _2^2 N\,p_e (1 - p_e )$, then $r$ increases as $\rho $ increases, otherwise $r$ decreases as $\rho $ increases (Appendix~\ref{apefrho}). Our program also computes the required $r$ for the case where $V(t_0 ) > 0$.

\subsubsection{LDD}
\label{cslddp1}

Here, 
\begin{equation}
\label{varlddcsp1}
{\mathbf{c'\Sigma }}_{\rm B} {\mathbf{c}} = \frac{{12\sigma ^2 (1 - \rho ){\kern 1pt} (1 + r\rho )}}
{{N{\kern 1pt} p_e (1 - p_e )(r + 1)\;\left( {r\,(r + 2)(1 + r\rho )\,s^2  + 12(1 - \rho )(1 - \rho _{\exp ,t_0 }^2 )V(t_0 )} \right)}}.
\end{equation}
For $V(t_0 ) = 0$, this variance was given previously \cite{Diggle:2002,Dawson:1998,Frison:1997,Hedeker:1999,Kirby:1994,Jung:2003,Yi:2002}  (Table~\ref{table1p1}). The power formula can be obtained plugging in \eqref{varlddcsp1} into \eqref{powerp1}.

Power is a monotone function of $r$, and the limit of power when $r$ goes to infinity is one, both when $s$ is fixed and when $\tau $ is fixed (Appendix~\ref{aplimitslddcs}). Thus, any pre-specified power can be achieved by increasing the number of repeated measures. The effect of $\rho $ on power depends on a complicated fashion on $r$, $s$, $V(t_0 )
$ and $\rho _{\operatorname{e} ,t_0 }^2 $. However, when all participants are observed at the same time points ($V(t_0 ) = 0$), then increasing $\rho $ will always increase power, the opposite effect that it has under CMD \cite{Hedeker:1999}. 

Fixing $r$ and $s$, an expression for the required number of participants, $N$, is readily obtained using \eqref{Np1} and \eqref{varlddcsp1}. With $N$ and $s$ fixed, a closed form solution for $r$ is not available. Our program (see Section~\ref{softp1}) can be used to calculate the required $r$ for this case. As noted before, any pre-specified power can be reached by increasing $r$. If $V(t_0 ) = 0$, then as the intraclass correlation, $\rho $, increases, the required $r$ decreases, both when $s$ is fixed and when $\tau $ is fixed (in the latter case, provided $r > 1
$) (appendices A.5.2-A.5.3). When $V(t_0 ) > 0$, the effect of $\rho $ on $r$ is not necessarily monotone.

\subsection{Damped exponential}
\label{dexp1}

In this section, we consider a covariance structure that generalizes CS as a particular case. Following \citeasnoun{Munoz:1992}, the damped exponential (DEX) covariance matrix can be expressed as 
\begin{equation}
\label{DEXp1}
{\mathbf{\Sigma }} = \sigma ^2 \left( {\begin{array}{*{20}c}
   1 & {\rho ^{s^\theta  } } & {\rho ^{(2s)^\theta  } } &  \cdots  & {\rho ^{(rs)^\theta  } }  \\
   {\rho ^{s^\theta  } } & 1 & {\rho ^{s^\theta  } } &  \ddots  &  \vdots   \\
   {\rho ^{(2s)^\theta  } } & {\rho ^{s^\theta  } } & 1 &  \ddots  & {\rho ^{(2s)^\theta  } }  \\
    \vdots  &  \ddots  &  \ddots  &  \ddots  & {\rho ^{s^\theta  } }  \\
   {\rho ^{(rs)^\theta  } } &  \cdots  & {\rho ^{(2s)^\theta  } } & {\rho ^{s^\theta  } } & 1  \\

 \end{array} } \right).
\end{equation}
Under this covariance model,  $\rho $ is now the correlation between two measures from the same participant separated by one time unit. The correlation between two consecutive measures is $\rho ^{s^\theta  } $ and the correlation between two measurements of the same participant decreases as their separation in time increases. If the correlation between two measurements separated by $s$ units, $\rho _s $, is known, the correlation per one unit is $\rho  = \rho _s^{s^{ - \theta } } $. The parameter $\theta  \in \left[ {0,\,1} \right]$ controls the degree of attenuation of the correlation over time. This covariance structure includes compound symmetry when $\theta  = 0$ and AR(1) covariance structure when $\theta  = 1$. Thus, an investigator can vary the value of $\theta $ to determine the sensitivity of sample size and power calculations to departures of this sort from compound symmetry. For example, in a study on pulmonary function loss, the damping coefficient, $\theta $, was 0.48, and in a study on CD4 cell count in HIV infected subjects, it was 0.35 \cite{Munoz:1992}. Under DEX, ${\mathbf{c'\Sigma }}_{\rm B} {\mathbf{c}}$ cannot be simplified to a simpler expression since the inverse of a DEX matrix is a complicated expression, and there is no general expression for all values of $r$. Design calculations can be performed using our program. However, when $\theta  = 1$, that is, AR(1), a simple expression is obtained (Table~\ref{table1p1}).

\subsubsection{CMD}
\label{dexcmdp1}

\begin{figure}
  \centering 
  \includegraphics[width=6in]{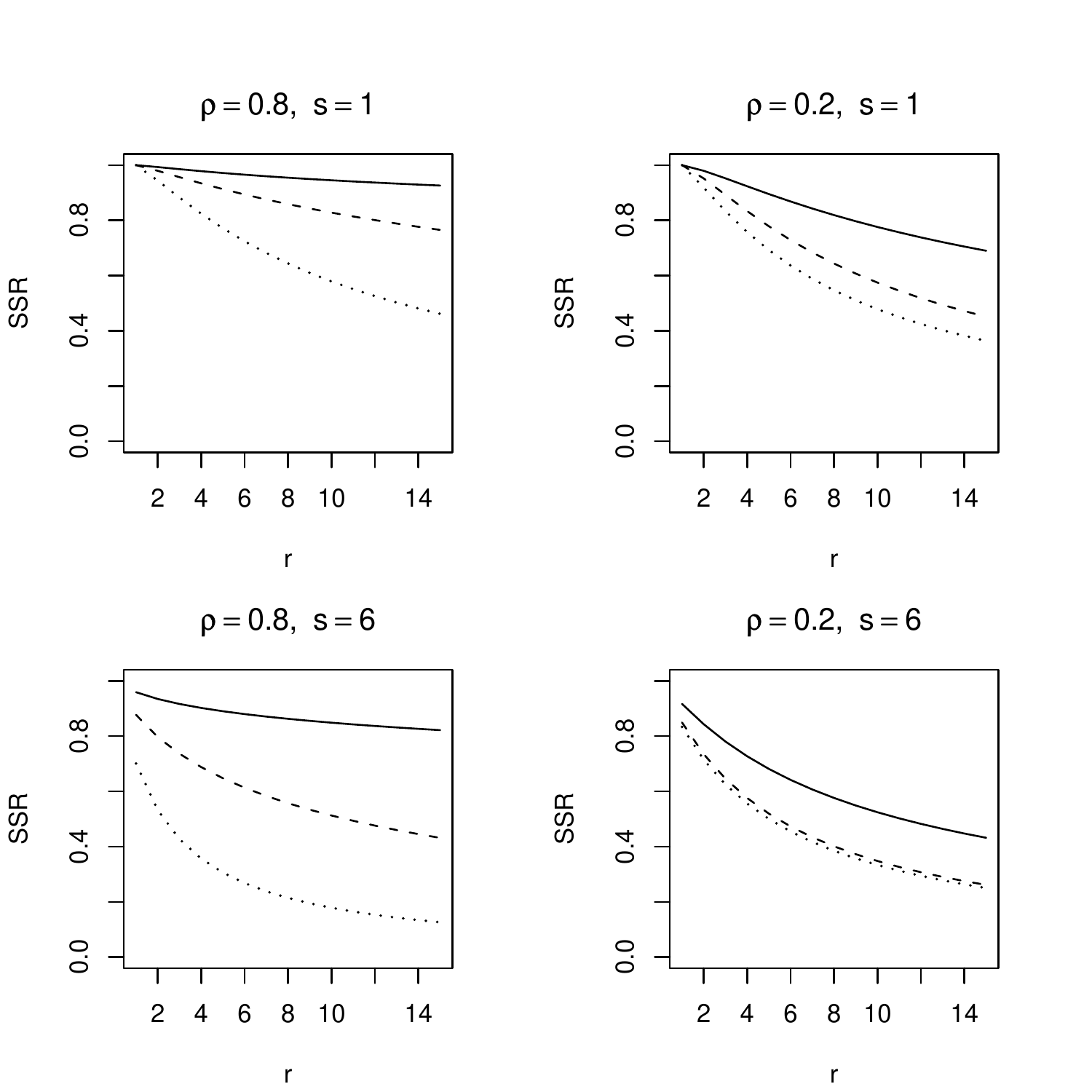}
  \caption{Ratio of required sample sizes (SSR) to achieve the same power comparing DEX and CS (  = 0) for fixed values of r, $\rho$, $\theta$, and $V(t_0 ) = 0$ under CMD $ \left( {SSR = \frac{{N_{DEX} }} {{N_{CS} }}} \right) $. The groups are: $\theta  = 0.2$ ({---}{---}), $\theta  = 0.5$ (-~-~-), and $\theta  = 1$ ($\cdots\cdots$).}
  \label{ssrcmddexp1}
\end{figure}

As noted before, with DEX, formulas do not have closed form, but the computations can be performed with our program. The formula for the variance of $\hat \beta _2 
$ under AR(1) and $V(t_0 ) = 0$ is given in Table~\ref{table1p1}. With $N$ fixed, under DEX and CMD, power increases as $r$ increases. For small values of $\theta $, we observed in some cases that by increasing $r$ to large values, the limit of the power when $r$ tends to infinity was not one, as in the CS case. However, we observed in some cases that when the frequency of measurements, $s$, is fixed, as $\theta $ gets large the limit of power as $r$ goes to infinity gets closer to one. With AR(1) covariance, we proved that the limit of power is one, so any pre-specified power can be reached by increasing the number of repeated measures (Appendix~\ref{aplimitscmdar1}). When the follow-up period, $\tau $, is fixed, this limit of power is not one. For example, when $V(t_0 ) = 0$ this limit is  
$$
\Phi \left[ {\frac{{\sqrt {N{\kern 1pt} p_e {\kern 1pt} (1 - p_e )(2 - \tau \log \rho )\,} \left| {\beta _1 } \right|}}
{{\sqrt {2\sigma ^2 } }} - z_{1 - \alpha /2} } \right]
$$
(Appendix~\ref{aplimitscmdar1}). Regarding the influence of the covariance parameters, the power to detect an exposure effect under CMD decreases as the correlation $\rho $ increases, as in the CS case, if $V\left( {t_0 } \right) = 0$; otherwise the relationship with $\rho $ is not always monotone. When $V\left( {t_0 } \right) = 0$, power increases as $\theta $ increases provided $s > 1$ unit (if $s < 1$ unit the correlation between two measures separated by one unit will be larger under DEX than under CS). The effect of departures from CS, i.e. $\theta  > 0$, on power and sample size can be assessed by computing the asymptotic relative efficiency (ARE) of the test statistic under CS and under DEX. Since the numerator of the test statistic is the same in both cases, the ARE is equivalent to the variance ratio, which in turn is equivalent to the inverse of the ratio of required sample sizes to achieve the same power \cite{Dawson:1993}. Figure~\ref{ssrcmddexp1} shows the percent reduction in the required sample size at a fixed power, when the covariance structure is DEX compared to CS under CMD for fixed $s$ and $V\left( {t_0 } \right) = 0$. Similar graphs are obtained for the fixed $\tau $ case. The reduction can be considerable and it is bigger when one takes many repeated measurements and the time between measurements $s$ is large.

The required $r$ to achieve a power $\pi $ when $N$ is fixed can be computed with our program. To assess the effect of $\theta $ on the required $r$, we computed it over a grid of values of the parameters, restricting to $N \in \left[ {400,2000} \right]$, $p_1  \in \left[ {0.1,0.2} \right]$, ${\sigma  \mathord{\left/
 {\vphantom {\sigma  {\mu _{00}  \in \left[ {0.5,2} \right]}}} \right.
 \kern-\nulldelimiterspace} {\mu _{00}  \in \left[ {0.5,2} \right]}}$, $\tau  \in \left[ {2,60} \right]$ and $V\left( {t_0 } \right) = 0$. Similar to the effect of $\theta $ on power and number of participants, we observed in this region of the parameter space that the required $r$ decreased as $\theta $ increased, provided $s > 1$ unit.

\subsubsection{LDD}
\label{dexlddp1}

\begin{figure}
  \centering 
  \includegraphics[width=6in]{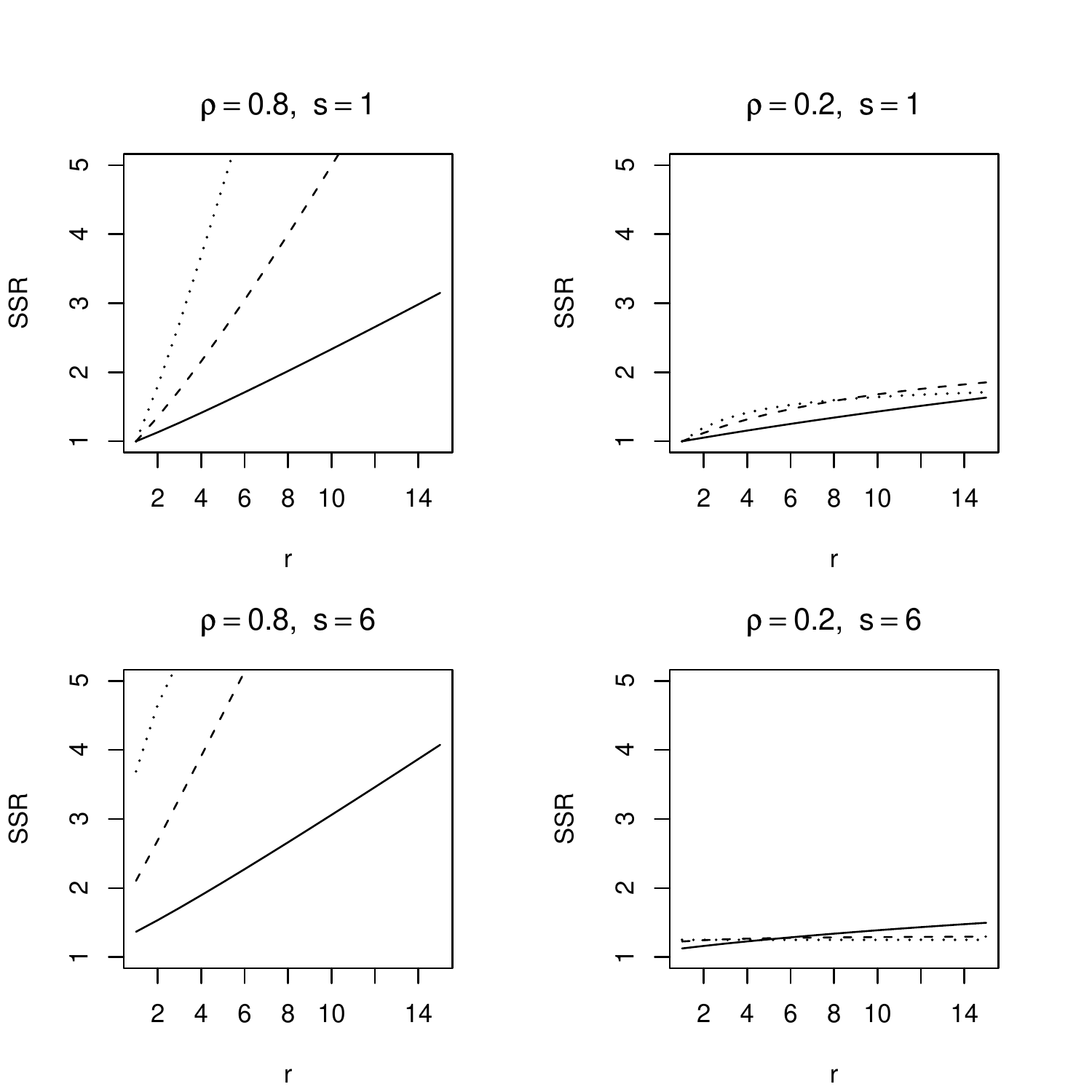}
  \caption{Ratio of required sample sizes (SSR) to achieve the same power comparing DEX and CS ($\theta~=~0$) for fixed values of $r$, $\rho$ and $\theta$, and under LDD with $V(t_0 ) = 0$ $ \left( {SSR = \frac{{N_{DEX} }}
{{N_{CS} }}} \right) $. The groups are $\theta  = 0.2$ ({---}), $\theta  = 0.5$
 (-~-~-), and $\theta  = 1$ ($\cdots\cdots$).}
  \label{ssrldddexp1}
\end{figure}

The formula for ${\mathbf{\Sigma }}_{\rm B} $ does not have a general simple expression for all values of $\theta $ and $r$, and therefore neither do the formulas for power or for $N$ with fixed $r$. For AR(1) and $V(t_0 ) = 0$, Table~\ref{table1p1} shows the formula for the variance of $\hat \gamma _3 $ needed by equation~\eqref{powerp1} to obtain the power of a study. Under LDD and DEX, we observed through a grid search over a wide range of the design space that the power to detect an effect increases as $r$ increases. We computed the limit of power when $r$ goes to infinity for the AR(1) case. For fixed $s$, this limit goes to one (Appendix~\ref{aplimitslddar1}). Therefore, any pre-specified power can be reached by increasing the number of repeated measures. For fixed $\tau $, this is not the case, and for example when $V(t_0 ) = 0$ the limit is (Appendix~\ref{aplimitslddar1}) 
$$
\Phi \left[ {\frac{{\sqrt {N{\kern 1pt} p_e {\kern 1pt} (1 - p_e )\left[ {12 + \tau \left( {\log \rho } \right)\left( {\tau \log \rho  - 6} \right)} \right]\,} \left| {\gamma _3 } \right|}}
{{\sqrt { - 24\sigma ^2 \log \rho } }} - z_{1 - \alpha /2} } \right].
$$
The effect of $\rho $ and $\theta $ on power depend on each other and it is not always monotone, even when $V(t_0 ) = 0$. We computed power for a grid of values for the parameters, restricting  $r \leqslant 15$ and $s \leqslant 6$ units, and observed that when $V(t_0 ) = 0$, as $\theta $ departs from 0, the power starts to decline, but it can increase again when $\theta $ approaches 1 for large values of $s$ and $r$ (data not shown). However, in the range of values we investigated, power was maximized at $\theta  = 0$ (i.e. CS). When $V(t_0 ) > 0$, power is not necessarily maximized at $\theta  = 0$ when $\frac{{V(t_0 )}}{{s^2 }}$ is large, around 20 or larger, and $\theta $ is near one, coupled with small values of $r$. Figure~\ref{ssrldddexp1} shows the increase in the required number of participants to achieve a certain power when there is covariance decay compared to CS, for LDD with $V(t_0 ) = 0$ and fixed $s$. For the case of fixed $\tau $, a similar pattern was observed.  The increase in $N$as $\theta $increases can be quite large when the intraclass correlation, $\rho $, is high and $r$ is large. For example, if the true covariance is AR(1) ($\theta  = 1$) and $\rho $ is large, one may have to enroll more than three times more participants than if the true covariance is CS, i.e. ($\theta  = 0$). 

The required $r$ to achieve a power $\pi $ when $N$ is fixed can be computed with our program. Computing the required $r$ for several values of the parameters we observed that the effect of $\theta $ on $r$ is not necessarily monotone. In general, $r$ increases as $\theta $ departs from 0, but it may decrease again as $\theta $ gets larger.

\subsection{Random intercepts and slopes}
\label{rsp1}

In this section, we consider another generalization of CS. We consider the covariance structure obtained when an additional random effect ($b_{1i} $) associated with time is assumed (i.e. random intercepts and slopes, denoted RS), leading to model 
$$
Y_{ij}  = \beta _0  + f(t_{ij} ;{\mathbf{\beta }}_1 ) + \beta _2 k_i  + b_{0i}  + t_{ij} b_{1i}  + e_{ij} 
$$ 
under CMD and 
$$
Y_{ij}  = \gamma _0  + f\left( {t_{ij} ,{\mathbf{\gamma }}_1 } \right) + \gamma _2 k_i  + \gamma _3 \left( {t_{ij}  \times k_i } \right) + b_{0i}  + t_{ij} b_{1i}  + e_{ij} 
$$
under LDD, where $f\left( {t_{ij} ,{\mathbf{\gamma }}_1 } \right)$ is a function of time that includes a linear term and is otherwise arbitrary. In mixed models notation, the residual covariance matrix is often written as ${\mathbf{\Sigma }}_i  = {\mathbf{Z}}_i {\mathbf{DZ'}}_i  + \sigma _{within}^2 {\mathbf{I}}$, where ${\mathbf{Z}}_i $ contains a subset of columns of the design matrix for participant $i$, and ${\mathbf{D}}$ is the covariance matrix of the random effects (e.g. \citeasnoun{Fitzmaurice:2004}, p. 199). Here, the matrix ${\mathbf{Z}}_i $ contains a column of ones and the column of times for participant $i$, and 
$$
{\mathbf{D}} = \left( {\begin{array}{*{20}c}
   {\sigma _{b_0 }^2 } & {\rho _{b_0 b_1 } \sigma _{b_0 } \sigma _{b_1 } }  \\
   {\rho _{b_0 b_1 } \sigma _{b_0 } \sigma _{b_1 } } & {\sigma _{b_1 }^2 }  \\
 \end{array} } \right),
$$
where $\sigma _{b_0 }^2 $ and $\sigma _{b_1 }^2 $ are the variance of the random effect associated with the intercept and slope, respectively, and $\rho _{b_0 b_1 } $ is the correlation between them. When there is only a random effect associated with the intercept, the matrix ${\mathbf{Z}}_i $ contains only a column of ones, and the resulting matrix, ${\mathbf{\Sigma }}_i$, follows a CS structure. Likewise, when $\sigma _{b_1 }^2  = 0$, RS reduces to CS. If $V(t_0 ) = 0$ then ${\mathbf{\Sigma }}_i  = {\mathbf{\Sigma }}
$, i.e. the covariance matrix is the same for all participants, as it was the case when CS or DEX was assumed, even when $V(t_0 ) > 0$ for CS and DEX. The RS covariance structure is heteroscedastic, with the residual variance of the responses (the diagonal elements of ${\mathbf{\Sigma }}$) assumed to change as a quadratic function of time, with positive curvature $\sigma _{b_1 }^2 $. In addition, this covariance structure assumes that the correlation between repeated measures changes with time and with increasing duration between visits - in either scenario, it can either increase or decrease. When pilot data are available, the parameters of ${\mathbf{D}}$ can be estimated and used directly as inputs into our program to perform design calculations. 

Often, however, longitudinal pilot data are not available, and a more intuitive parameterization is needed so that investigators can propose plausible values on which to base designs. To make the parameters more intuitive, we defined $\sigma _{t_0 }^2  = \sigma _{within}^2  + \sigma _{b_0 }^2$ as the residual variance at baseline (or at the mean initial time). Then, we define 
$$
\rho _{t_0 }  = \frac{{\sigma _{b_0 }^2 }}
{{\sigma _{within}^2  + \sigma _{b_0 }^2 }}
$$
as the reliability coefficient at baseline (or at the mean initial time), i.e. the percentage of residual variance at baseline that is due to between-subject variation. One additional parameter is needed, to fix the between-subjects variance in slopes. Following a parameterization proposed for characterizing the relative variability in slopes from several studies compared to their within-study variance in the context of meta-analysis \cite{Takkouche:1999}, we defined the slope reliability as the percentage of variation in the estimated coefficient $\hat \gamma _3 $ that is due to between-subjects variation. When $s$ is fixed, we define $\rho _{b_1 ,s,\tilde r} 
$ as the slope reliability with $\tilde r$ repeated measures, where $\tilde r$ is a hypothetical or trial value of $r$. When $V(t_0 ) = 0$, this quantity is 
$$
\rho _{b_1 ,s,\tilde r}  = \frac{{\sigma _{b_1 }^2 s^2 \tilde r(\tilde r + 1)(\tilde r + 2)}}
{{12(1 - \rho _{t_0 } )\sigma _{t_0 }^2  + \sigma _{b_1 }^2 s^2 \tilde r(\tilde r + 1)(\tilde r + 2)}}.
$$
For the case of fixed $\tau $, we define the equivalent quantity 
$$
\rho _{b_1 ,\tau ,\tilde r}  = \frac{{\sigma _{b_1 }^2 \tau ^2 (\tilde r + 1)(\tilde r + 2)}}
{{12\tilde r(1 - \rho _{t_0 } )\sigma _{t_0 }^2  + \sigma _{b_1 }^2 \tau ^2 (\tilde r + 1)(\tilde r + 2)}}.$$
The variance matrix can now be expressed in terms of these new intuitive, parameters $\sigma _{t_0 }^2 $, $\rho _{t_0 } \, \in \,[0,\,1]
$, $\rho _{b_0 b_1 } \, \in \,[ - 1,\,1]$, and $\rho _{b_1 ,s,\tilde r}  \in [0,1]$ or $\rho _{b_1 ,\tau ,\tilde r}  \in [0,1]$. In the case of fixed $\tau $, the covariance matrix can be expressed as
\begin{equation}
\label{sigmaRSp1}
{\mathbf{\Sigma }}_i  = \sigma _{t_0 }^2 \left( {{\mathbf{Z}}_i \left( {\begin{array}{*{20}c}
   {\rho _{t_0 } } & {\rho _{b_0 b_1 } \sqrt {\frac{{12\rho _{t_0 } (1 - \rho _{t_0 } )\tilde r}}
{{\tau ^2 (\tilde r + 1)(\tilde r + 2)}}\left( {\frac{{\rho _{b_1 ,\tau ,\tilde r} }}
{{1 - \rho _{b_1 ,\tau ,\tilde r} }}} \right)} }  \\
   {} & {\frac{{12(1 - \rho _{t_0 } )\tilde r}}
{{\tau ^2 (\tilde r + 1)(\tilde r + 2)}}\left( {\frac{{\rho _{b_1 ,\tau ,\tilde r} }}
{{1 - \rho _{b_1 ,\tau ,\tilde r} }}} \right)}  \\

 \end{array} } \right){\mathbf{Z'}}_i  + (1 - \rho _{t_0 } ){\mathbf{I}}} \right),
\end{equation}
and for the fixed $s$ case, one just needs to substitute $\rho _{b_1 ,\tau ,\tilde r} 
$ by $\rho _{b_1 ,s,\tilde r} $ and $\tau $ by $s\tilde r$. If the value of $r$ is known a priori, $\tilde r$ will take the value of $r$. Otherwise, for design problems where $r$ is not fixed (i.e. when finding $r$ for fixed $N$, or when finding $(N_{opt} ,r_{opt} )$), the investigator needs to provide the slope reliability together with a trial value of $\tilde r$ associated with it, and then find $r$ or $(N_{opt} ,r_{opt} )$. In the calculations that follow, $\tilde r$ will act as a constant. If the value of $r$ that solves the design problem is different from the one used to define the initial $\rho _{b_1 ,s,\tilde r} $ or $\rho _{b_1 ,\tau ,\tilde r} $, the investigator should recalculate $\rho _{b_1 ,s,\tilde r} $ or $\rho _{b_1 ,\tau ,\tilde r} $ with the new value of  $r$ to ascertain that the resulting values of $\rho _{b_1 ,s,r} $ or $\rho _{b_1 ,\tau ,r} $ are realistic. Our software automatically recalculates $\rho _{b_1 ,s,r} $ or $\rho _{b_1 ,\tau ,r} $ with the value of $r$ that is the solution to the design problem. In the figures shown in this paper and, otherwise, when grid searches were performed, we chose $\tilde r = 5$ and  $\rho _{b_1 ,s,r = 5} $ or $\rho _{b_1 ,\tau ,r = 5} $ are used.

\subsubsection{CMD}
\label{rscmdp1}

Formula~\eqref{varcmdvt00p1} under RS and CMD results in a complex formula for ${\mathbf{c'\Sigma }}_{\rm B} {\mathbf{c}}$ which we do not provide here. However, in practice, a RS correlation structure will be usually not fitted under CMD. For particular cases, calculations can be performed with our program by entering the intuitive parameters or by using formula \eqref{varcmdvt00p1} directly. Unlike the analogous CS and DEX scenarios, when $V(t_0 ) > 0$, formula \eqref{varcmdvt00p1} cannot be used because it is based on all participants having the same covariance matrix ${\mathbf{\Sigma }}$. When $V(t_0 ) > 0$ under RS, ${\mathbf{\Sigma }}_i 
$ is different for each participant (equation \eqref{sigmaRSp1}). We will still compute ${\mathbf{\Sigma }}_{\rm B} $ as $\mathbb{E}_X^{ - 1} \left( {{\mathbf{X'}}_i {\mathbf{\Sigma }}_i^{ - 1} {\mathbf{X}}_i } \right)$ as we did in section~\ref{notationp1}, but this calculation will now require correctly specifying the full distribution of $\left( {k_i ,t_{0i} } \right)$ and not just the first two moments. In this paper and in the software (section~\ref{softp1}), we assume that $t_0 $ is normally distributed within each exposure group, with the same variance, $V(t_0 )$, but a different mean that will depend on $\rho _{\operatorname{e} ,t_0 } $, and that $k_i $ follows a Bernoulli with probability $p_e $. We then compute $\mathbb{E}_X^{ - 1} \left( {{\mathbf{X'}}_i {\mathbf{\Sigma }}_i^{ - 1} {\mathbf{X}}_i } \right)$ by numerical integration (see Appendix~\ref{apvarRS} for more details). We assessed the sensitivity of results to the normality assumption for $t_0 $ by comparing to results obtained with $t_0 $ assumed to be uniform, a four-parameter Beta with several values for the shape parameters and lognormal with several values of the shape parameter, with the mean and variance of each distribution matched to the mean and variance of the normal case, over a grid of values of the covariance parameters. We found that the resulting variance depended on the distribution assumed, and depending on the values of the other parameters, the distributions we considered can provide variances that smaller or larger than the normal case, with no clear pattern. So, the results given in this paper for CMD, RS and $V(t_0 ) > 0 $ will rely on the times being normally distributed. 

The limit of the power when $r$ goes to infinity and $V(t_0 ) = 0$ is (Appendix~\ref{aplimitscmdrs}) $$
\Phi \left[ {\frac{{\sqrt {N{\kern 1pt} p_e {\kern 1pt} (1 - p_e )\,} \left| {\beta _1 } \right|}}
{{\sqrt {\sigma _{t_0 }^2 \rho _{t_0 } (1 - \rho _{b_0 b_1 }^2 )} }} - z_{1 - \alpha /2} } \right].
$$
The effect of the covariance parameters $\rho _{t_0 } $, $\rho _{b_0^{} b_1 } $, and $\rho _{b_1 ,s,r = 5} $ or $\rho _{b_1 ,\tau ,r = 1} $ on power, number of participants, and number of repeated measures is not monotone and depends upon the values of more than one parameter. Through a grid search, we found that for the same value of $\rho _{t_0 } $, power can be either larger or smaller than in the CS case. Similarly, through a grid search, we found that the effect of $V(t_0 )$ on power did not follow a monotone pattern throughout our grid search.

\subsubsection{LDD}
\label{rslddp1}

When $V(t_0 ) = 0$, power does not depend upon the parameter $\rho _{b_0 b_1 }^{} $. Table~\ref{table1p1} shows the terms needed to compute power and number of participants ($N$) for fixed $r$, using equations \eqref{powerp1} and \eqref{Np1}. These formulas are equivalent to those reported by \citeasnoun{Schlesselman:1973}, \citeasnoun{Raudenbush:2001}, \citeasnoun{Yi:2002}, \citeasnoun{Galbraith:2002} and \citeasnoun{Fitzmaurice:2004}. When $V(t_0 ) > 0$, formula \eqref{varlddp1} cannot be used because it requires that the response of all participants have the same covariance matrix, ${\mathbf{\Sigma }}$. As discussed above in the CMD case (section~\ref{rscmdp1}), under RS, when $V(t_0 ) > 0$, ${\mathbf{\Sigma }}_i $ is different for each participant. In this paper and in the software (section~\ref{softp1}), to compute ${\mathbf{\Sigma }}_{\rm B} $ we assumed that $t_0 $ is normally distributed within each exposure group, with the same variance $V(t_0 )$ and a different mean depending on $\rho _{\operatorname{e} ,t_0 } $, and that  $k_i $ follows a Bernoulli with probability $p_e $. We then computed ${\mathbf{\Sigma }}_{\rm B} $ as $\mathbb{E}_X^{ - 1} \left( {{\mathbf{X'}}_i {\mathbf{\Sigma }}_i^{ - 1} {\mathbf{X}}_i } \right)
$, by numerical integration (see Appendix~\ref{apvarRS} for details). As is section~\ref{rscmdp1} for the CMD case, we compared the results assuming normality for $t_0 $ to results assuming other distributions. The resulting variances were not materially different from the normal case for symmetric or moderately skewed distributions, but were greater than the normal case in situations where the distribution of $t_0 $ was very skewed. Since our program would then provide underestimates of the true variance, if it is believed that $t_0 $ is severely skewed, it might make sense to use formulas with $V(t_0 ) = 0
$, which does not require assumptions on the distribution of  $t_0 $ and appears to provide conservative estimates of the variance compared to the $V(t_0 ) > 0$ case, as observed through the grid searches over all the distributions of $t_0 $ we studied.

\begin{figure}
  \centering 
  \includegraphics[width=5in]{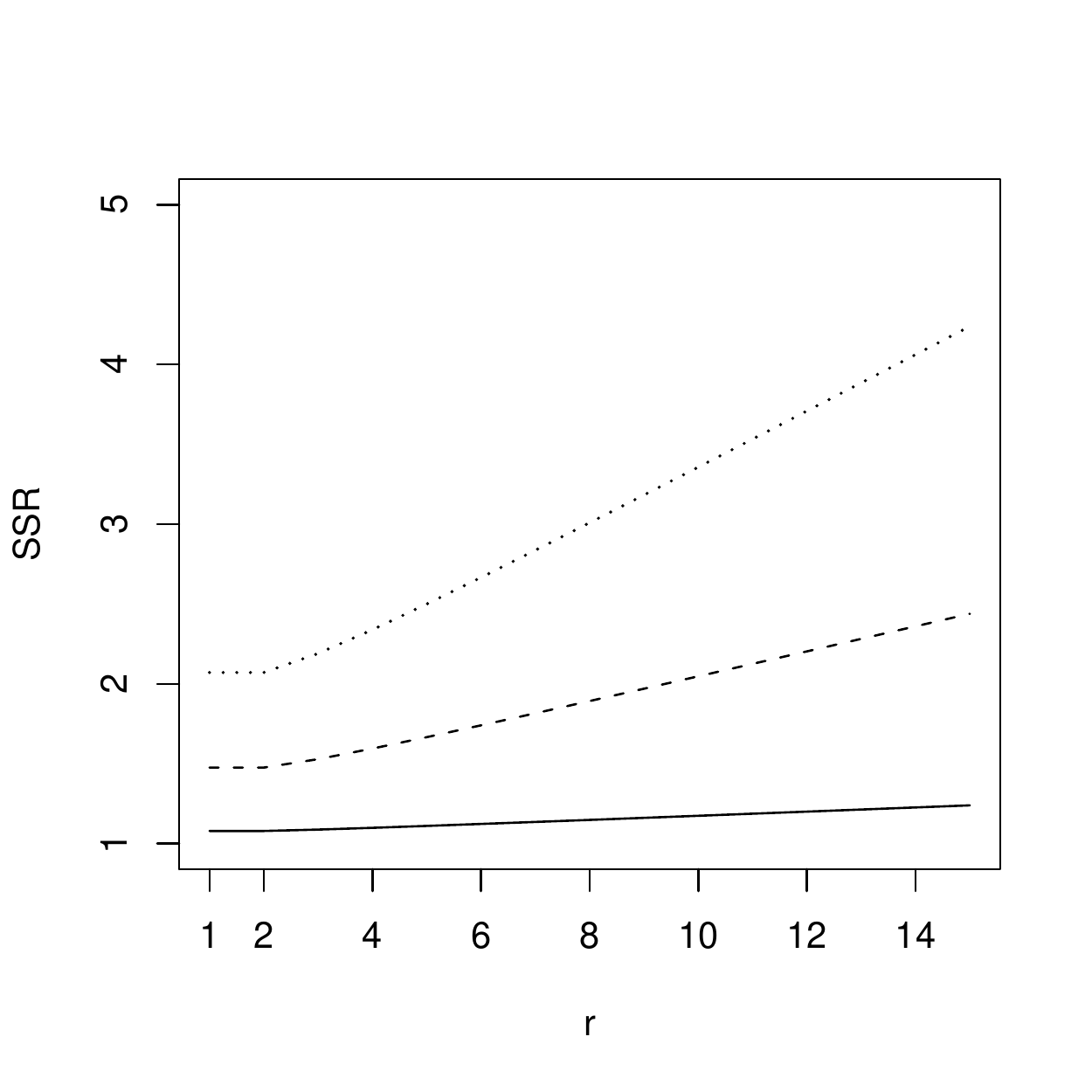}
  \caption{Ratio of required sample sizes (SSR) to achieve the same power comparing RS and CS ($\rho _{b_1 ,\tau ,r = 5}  = 0$) for fixed values of r, $\rho _{t_0 } $
 and $\rho _{b_1 ,\tau ,r} $ and $V(t_0 ) = 0$ under LDD $ \left( {SSR = \frac{{N_{RS} }} {{N_{CS} }}} \right) $. The groups are $\rho _{b_1 ,\tau ,r = 5}  = 0.1$  ({---}), $\rho _{b_1 ,\tau ,r = 5}  = 0.4$ (-~-~-), and $\rho _{b_1 ,\tau ,r = 5}  = 0.6$ ($\cdots\cdots$).}
  \label{ssrlddrsp1}
\end{figure}

Power is an increasing function of $r$. However, the limit of the power when $r$ goes to infinity is not one, so there can be instances where a pre-specified power cannot be achieved by just increasing $r$. When $V(t_0 ) = 0$ the limit is (Appendix~\ref{aplimitslddrs}) 
$$
\Phi \left[ {\frac{{\sqrt {N\;p_e {\kern 1pt} (1 - p_e )\,} \left| {\gamma _3 } \right|}}
{{\sqrt {\frac{{12\sigma _{t_0 } ^2 (1 - \rho _{t_0 } )\tilde r}}
{{\tau ^2 (\tilde r + 1)(\tilde r + 2)}}\left( {\frac{{\rho _{b_1 ,\tau ,\tilde r} }}
{{1 - \rho _{b_1 ,\tau ,\tilde r} }}} \right)} }} - z_{1 - \alpha /2} } \right]
$$
for the fixed $\tau $ case, and the equivalent expression substituting $\rho _{b_1 ,\tau ,\tilde r} $ by $\rho _{b_1 ,s,\tilde r} $ and $\tau $ by $s\tilde r$ for the fixed $s$ case.  
As in the CS case, the effect of $\rho _{t_0 } $ on power depends on other parameters in a complicated fashion, but when $V(t_0 )$= 0 increasing $\rho _{t_0 } $ always increases power. To examine the effect of $\rho _{b_1 ,s,\tilde r} $ or $\rho _{b_1 ,\tau ,\tilde r} $, that is, the effect of departures from compound symmetry towards a random slopes covariance structure, we calculated the ratio of sample sizes required to achieve the same power (ARE) comparing RS and CS. It is easily proven that more participants are required when either $\rho _{b_1 ,s,\tilde r} $ or $\rho _{b_1 ,\tau ,\tilde r} $ are greater than zero. Figure~\ref{ssrlddrsp1} shows the ARE as a function of $\rho _{b_1 ,\tau ,r = 5} $ when $V\left( {t_0 } \right) = 0$. It can be seen that the increase in number of participants can be quite large for large values of $r$ and $\rho _{b_1 ,\tau ,r = 5} $. This is, for fixed $\sigma _{within}^2 $, $\sigma _{b_0 }^2 $ and $\sigma _{b_1 }^2 $, as $r$ increases,  the within-subjects variance component of $Var\left( {\hat \gamma _3 } \right)$ is reduced, and it becomes very small for large values of $r$. Thus, the percentage of variance due to the between-subjects component is much greater, and the only way to reduce the between-subjects variance component is to recruit more subjects. When $V(t_0 ) > 0$, we computed power over a grid of values of the other parameters, with restrictions $r \leqslant 15$, $s \leqslant 6$ units, and $\frac{{\sqrt {V(t_0 )} }} {s} \leqslant 10
$ for the fixed $s$ case and $\tau  \leqslant 60$ units and $\frac{{\sqrt {V(t_0 )} }}
{\tau } \leqslant 10$ for the fixed $\tau $ case. Over this wide region of the parameter space, the power decreased as either $\rho _{b_1 ,s,r = 1} $ or $\rho _{b_1 ,\tau ,r = 1} $ increased, even when $V(t_0 ) > 0$.

With $N$ and $s$ fixed, there is no closed-form solution for the minimum value of $r$ to satisfy a specified power, but calculations to solve this non-linear equation can be performed with our program. As noted previously, there may be situations where the pre-specified power cannot be reached by simply increasing $r$. The effect of the covariance parameters on $r$ in this setting is not monotone. However, when $V(t_0 ) = 0$, we show that as the correlation $\rho _{t_0 } $ increases, the required $r$ decreases (appendices A.5.4-5). Conversely, as $\rho _{b_1 ,s,\tilde r} $ or $\rho _{b_1 ,\tau ,\tilde r} $ increase, the required $r$ increases, both when $s$ is fixed  and when $\tau $ is fixed (provided $r > 1$ for the latter case) (appendices A.5.6.1-2).

\section{Optimal Allocation}
\label{optNr}

\begin{sidewaystable}
	\centering
	\caption{Summary of the results in section~\ref{optNr}, CMD hypothesis.}
\bigskip
\begin{tabular}{p{57pt} p{218pt} p{153pt} p{188pt} }
\hline
 \parbox{57pt}{\raggedright } & 
 \parbox{218pt}{\centering \textbf{{\scriptsize $ r_{opt}$ when $V(t_0 ) = 0$}}} &
  \parbox{153pt}{\centering \textbf{{\scriptsize Characteristics *}}} &
   \parbox{188pt}{\centering \textbf{{\scriptsize $r_{opt} $when $V(t_0 ) \ne 0$*}}} \\
\hline

\parbox{57pt}{\raggedright \textbf{{\scriptsize CMD (Fixed $s$)}}} &
 \parbox{218pt}{\raggedright } & \parbox{153pt}{\centering } & 
  \parbox{188pt}{\centering } \\
\hline

\parbox{57pt}{\raggedright {\scriptsize ~~~~CS}} & 
 \parbox{218pt}{\centering $r_{opt}  = \sqrt {\frac{{(\kappa  - 1)(1 - \rho )}}
{\rho }}  - 1 $} & \parbox{153pt}{\centering {\scriptsize $r_{opt} $ decreases as $\rho $ increases}} & 
  \parbox{188pt}{\centering {\scriptsize  $r_{opt} $ not a monotone function of $V(t_0 )$}} \\
\hline

\parbox{57pt}{\raggedright {\scriptsize ~~~~DEX}} & \parbox{218pt}{\centering 
{\scriptsize $$r_{opt} \text{ greater than for CS } $$}} & 
 \parbox{153pt}{\centering {\scriptsize $r_{opt} $ decreases as $\rho $ increases \par
$r_{opt} $ increases as $\theta $ increases}} & 
  \parbox{188pt}{\centering {\scriptsize $r_{opt} $ not a monotone function of $V(t_0 )$ }} \\
\hline

\parbox{57pt}{\raggedright {\scriptsize ~~~~RS}} & 
 \parbox{218pt}{\centering 
{\scriptsize $$r_{opt} \text{ can be either greater or smaller than for CS }$$}} &
  \parbox{153pt}{\centering 
{\scriptsize {$r_{opt} $ not a monotone function of $\rho _{t_0 } $} {$r_{opt} 
$not a monotone function of $\rho _{b_1 ,s,r = 5} $ }}} & 
   \parbox{188pt}{\centering 
{\scriptsize $r_{opt} $ not a monotone function of $V(t_0 )$}} \\
\hline

\parbox{57pt}{\raggedright 
\textbf{{\scriptsize CMD (Fixed $\tau $)}}} & \parbox{218pt}{\raggedright } &
 \parbox{153pt}{\centering } & \parbox{188pt}{\centering } \\
\hline

\parbox{57pt}{\raggedright 
{\scriptsize ~~~~CS}} & \parbox{218pt}{\centering {\scriptsize $$r_{opt}  = \sqrt {\frac{{(\kappa  - 1)(1 - \rho )}}{\rho }}  - 1$$}} & \parbox{153pt}{\centering 
{\scriptsize $r_{opt} $decreases as $\rho $ increases}} & 
 \parbox{188pt}{\centering 
{\scriptsize  $r_{opt} $ not a monotone function of $V(t_0 )$, although $r_{opt} $ is only affected by $V(t_0 )$ for large values of $\rho _{\operatorname{e} ,t_0 } $ }} \\
\hline

\parbox{57pt}{\raggedright 
{\scriptsize ~~~~DEX}} & \parbox{218pt}{\raggedright \begin{itemize}
	\item {\scriptsize if $\tau $ small then $r_{opt} $ smaller than for CS *}
	\item {\scriptsize if $\tau $ large then $r_{opt} $ greater than for CS *}
\end{itemize} } & 
\parbox{153pt}{\centering {\scriptsize  $r_{opt} $ decreases as $\rho $ increases \par}
{\scriptsize $r_{opt} $ not a monotone function of $\theta $} } & 
  \parbox{188pt}{\centering {\scriptsize same as in the CS case}
} \\
\hline

\parbox{57pt}{\raggedright 
{\scriptsize ~~~~RS}
} & \parbox{218pt}{\centering 
{\scriptsize  $r_{opt} $ can be either greater or smaller than for CS}} & \parbox{153pt}{\centering {\scriptsize $r_{opt} $ not a monotone function of $\rho _{t_0 } $} {\scriptsize $r_{opt} $ not a monotone function of $\rho _{b_1 ,\tau ,r} $ }}  &
 \parbox{188pt}{\centering {\scriptsize $r_{opt} $ decreases as $V(t_0 )$ increases.}} \\
\hline

\end{tabular}

	\label{table2ap1}
\end{sidewaystable}

\begin{sidewaystable}
	\centering
	\caption{Summary of the results in section~\ref{optNr}, LDD hypothesis.}
\bigskip
\begin{tabular}{p{52pt} p{249pt} p{130pt} p{192pt}}
\hline
 \parbox{52pt}{\raggedright } & 
 \parbox{249pt}{\centering \textbf{{\scriptsize $r_{opt} $ when $V(t_0 ) = 0$}}} &
  \parbox{130pt}{\centering \textbf{{\scriptsize Characteristics *}}} &
   \parbox{192pt}{\centering \textbf{{\scriptsize $r_{opt} $when $V(t_0 ) \ne 0$*}}} \\
\hline

\parbox{52pt}{\raggedright \textbf{{\scriptsize LDD (Fixed $s$)}}} &
 \parbox{249pt}{\raggedright } & \parbox{130pt}{\centering } & \parbox{192pt}{\centering } \\
\hline

\parbox{52pt}{\raggedright 
{\scriptsize ~~~~CS}} & \parbox{249pt}{\centering {\scriptsize $r$ as large as possible}}& 
 \parbox{130pt}{\centering {\scriptsize $$\rho {\text{ does not affect }} r_{opt} $$}}  &
  \parbox{192pt}{\raggedright {\scriptsize  still 15 in our restricted space, except for large \par $V(t_0 )/s^2 $ and small $\rho _{\operatorname{e} ,t_0 } $ }} \\
\hline

\parbox{52pt}{\raggedright 
{\scriptsize ~~~~DEX}} & \parbox{249pt}{\centering {\scriptsize $$r \text{ as large as possible}$$}} &
 \parbox{130pt}{\centering {\scriptsize $\rho $ does not affect $r_{opt} $ \par} {\scriptsize  $\theta $ does not affect $r_{opt} $ }
} & \parbox{192pt}{\raggedright {\scriptsize $\theta $ does not affect $r_{opt} $}} \\
\hline

\parbox{52pt}{\raggedright {\scriptsize ~~~~RS}} & \parbox{249pt}{\centering 
{\scriptsize $r_{opt} $ solves $\kappa  = \frac{{r_{opt} ^2 \left( { - (3 + 2r_{opt} )\tilde r(\tilde r + 1)(\tilde r + 2) + (r_{opt}  + 1)^2 (r_{opt}  + 2)^2 \frac{{\rho _{b_1 ,s,\tilde r} }}
{{1 - \rho _{b_1 ,s,\tilde r} }}} \right)}}
{{\,\left( {2 + 6r_{opt}  + 3r_{opt} ^2 } \right)\tilde r(\tilde r + 1)(\tilde r + 2)}}$ }} &
 \parbox{130pt}{\centering {\scriptsize $\rho _{t_0 } $ does not affect $r_{opt} $}
{\scriptsize $r_{opt} $ decreases as $\rho _{b_1 ,s,r = 5} $ increases}} & \parbox{192pt}{\raggedright \begin{itemize}
	\item {\scriptsize $r_{opt} $ still smaller than CS }
	\item {\scriptsize $r_{opt} $ decreases as either $\rho _{b_1 ,s,r = 5} $ or $V(t_0)
 	$ increase}
\end{itemize}} \\
\hline

\parbox{52pt}{\raggedright \textbf{{\scriptsize LDD (Fixed $\tau $)}}} &
 \parbox{249pt}{\raggedright } & \parbox{130pt}{\centering } & \parbox{192pt}{\centering } \\
\hline

\parbox{52pt}{\raggedright {\scriptsize ~~~~CS}} & \parbox{249pt}{\raggedright \begin{itemize}
	\item {\scriptsize if $\kappa  < 5 $ then $r_{opt}  = 1 	$}
	\item {\scriptsize if $\kappa  > 5 $ then choose $r > \frac{{2(\kappa  + 1)}}
 	{{\kappa  - 5}}	$ }
\end{itemize}} & 
\parbox{130pt}{\centering {\scriptsize  $\rho $ does not affect $r_{opt} $ }} &
 \parbox{192pt}{\raggedright \begin{itemize}
	\item {\scriptsize can take repeated measures even with $\kappa  < 5$}
	\item {\scriptsize $r_{opt} $ not a monotone function of $V(t_0 ) $} 
	\end{itemize} } \\
\hline

\parbox{52pt}{\raggedright {\scriptsize ~~~~DEX}} 
& \parbox{249pt}{\centering 
{\scriptsize $r_{opt} $ is smaller or equal than for CS}} & 
 \parbox{130pt}{\centering {\scriptsize $r_{opt} $ decreases as $\theta $ increases}} {\scriptsize {$r_{opt} $ not a monotone function of $\theta $}} & 
  \parbox{192pt}{\centering {\scriptsize $r_{opt} $ not a monotone function of $V(t_0 )$}} \\
\hline

\parbox{52pt}{\raggedright {\scriptsize ~~~~RS}} &
 \parbox{249pt}{\raggedright 
  \begin{itemize}
	\item {\scriptsize if $\kappa  > 5	$, $r > \frac{{2(\kappa  + 1)}}
 	{{\kappa  - 5}}	$ and} 
 	\end{itemize}
 	{\scriptsize $$\rho _{b_1 ,\tau ,\tilde r}  < \frac{{\left[ { - 2(\kappa  + 1) + (\kappa  - 5)r} \right](\tilde r + 1)(\tilde r + 2)}}
 	{{6\tilde r(r + 1)(r + 2) + \left[ { - 2(\kappa  + 1) + (\kappa  - 5)r} \right](\tilde r + 1)(\tilde r + 2)}}	$$}
 	{\scriptsize ~~~~then $ r_{opt}  $ solves 
 	$$\kappa  = \frac{{r_{opt} (4 + 3r_{opt} )(\tilde r + 1)(\tilde r + 2) + \tilde r(r_{opt}  + 1)^2 (r_{opt}  + 2)^2 \frac{{\rho _{b_1 ,\tau ,\tilde r} }}
 	{{1 - \rho _{b_1 ,\tau ,\tilde r} }}}}
 	{{(r_{opt} ^2  - 2)(\tilde r + 1)(\tilde r + 2)}} $$}
	\begin{itemize}
	\item {\scriptsize Otherwise $r_{opt}  = 1 	$}
\end{itemize} } & 
\parbox{130pt}{\centering {\scriptsize  $\rho _{t_0 } $ does not affect $r_{opt} $ } \par
{\scriptsize $r_{opt} $ decreases as $\rho _{b_1 ,\tau ,\tilde r} $ increases}} &
 \parbox{192pt}{\raggedright \begin{itemize}
	\item {\scriptsize it	can be advisable to take repeated measures even with $\kappa  < 5	$ }
	\item {\scriptsize  	$r_{opt} $ decreases as $\rho _{b_1 ,\tau ,r = 5} $ increases. The effect of $V(t_0 )	$ is not monotone.} 
	\end{itemize} } \\
\hline

\parbox{623pt}{\raggedright {\scriptsize * Results derived from a grid search. The restrictions in the parameters where $r \leqslant 15$, $\kappa  \leqslant 40$, $s \leqslant 6$ units, and $\frac{{\sqrt {V(t_0 )} }}
{s} \leqslant 10$ for the fixed $s$ case and $\tau  \leqslant 60$ units, and $\frac{{\sqrt {V(t_0 )} }}
{\tau } \leqslant 10$ for the fixed $\tau $ case.}}

\end{tabular}

	\label{table2bp1}
\end{sidewaystable}

In planning a study, one often needs to consider cost. If the cost of recruiting a participant is $c_1 $ monetary units and the first measurement for each participant is $\kappa  \geqslant 1$ times more expensive than the rest, then the total cost of the study is 
\begin{equation}
\label{costp1}
COST = N{\kern 1pt} c_1 \left( {1 + {r \mathord{\left/
 {\vphantom {r \kappa }} \right.
 \kern-\nulldelimiterspace} \kappa }} \right).
\end{equation}
Then, when the budget is fixed at cost $C$,  we need to choose the combination $(N,r)$ that maximizes the power to detect the hypothesized effect, subject to this cost constraint. The imposition of the cost constraint determines a unique solution in $(N,r)$  that solves the optimization problem, unlike in section~\ref{Nrfixedp1}, where a discrete two-dimensional 'curve'  in  $(N,r)$ provides the desired power. Using a Lagrange multiplier, we maximize the power equation \eqref{powerp1} with respect to $r$, subject to constraint \eqref{costp1}. Once the optimal $r$ is obtained, it can be plugged into equation \eqref{costp1} to obtain the corresponding optimal $N$. The constrained problem reduces to the following unconstrained problem (Appendix~\ref{apconstraint})
\begin{equation}
\label{unconstrp1}
\mathop {Min}\limits_r \;(\kappa  + r)\left( {{\mathbf{c'\Sigma }}_{\rm B} {\mathbf{c}}} \right)
\end{equation}
It turns out that the value of $r_{opt} $ that maximizes power subject to a fixed cost is the same one that minimizes the cost of a study subject to a fixed power (Appendix~\ref{apconstraint}). Of course, $N_{opt} $ will be different, depending on the nature of the constraint. Since $r$ is not fixed by design, we must consider the two different scenarios as above:  when there is a fixed frequency of measurement (fixed $s$) and when there is a fixed follow-up time (fixed $\tau $). A summary of the results of this section is given in Tables \ref{table2ap1} and \ref{table2bp1}. The exposure prevalence, $p_e $, does not have any effect on $r_{opt} $ for any of the two constraints, and neither has an effect on $N_{opt} $ for the cost constraint problem. For the power constraint problem, $N_{opt} $ depends on $p_e $ as in the case of deriving $N$ for fixed $r$ described in section~\ref{generalp1}.

Throughout section~\ref{optNr}, when the analytical solutions could not be derived we computed the optimal $r$ for a grid of values of the other design parameters, with the restrictions $r \leqslant 15$, $\kappa  \leqslant 40$and $s \leqslant 6$ units  and $\frac{{\sqrt {V(t_0 )} }} {s} \leqslant 10$ for the fixed $s$ case, and $\tau  \leqslant 60$ units and $\frac{{\sqrt {V(t_0 )} }} {\tau } \leqslant 10$ for the fixed $\tau $ case. 

\subsection{Fixed frequency of measurement, $s$}
\label{optNrs}

In this section, the frequency of measurements is fixed, for example, to yearly visits. Then increasing the number of repeated measures, $r$, implies increasing the length of follow-up. If a correlation decay is assumed, e.g. under DEX, increasing $r$ will keep the correlation between adjacent measurements the same, but the first and last observations will be less correlated. Under RS, increasing the length of follow-up can either increase or decrease the variances of successive responses and their correlations with each other. 

\subsubsection{CMD}
\label{optNrscmd}

\begin{figure}
\centering 
\includegraphics[width=5in]{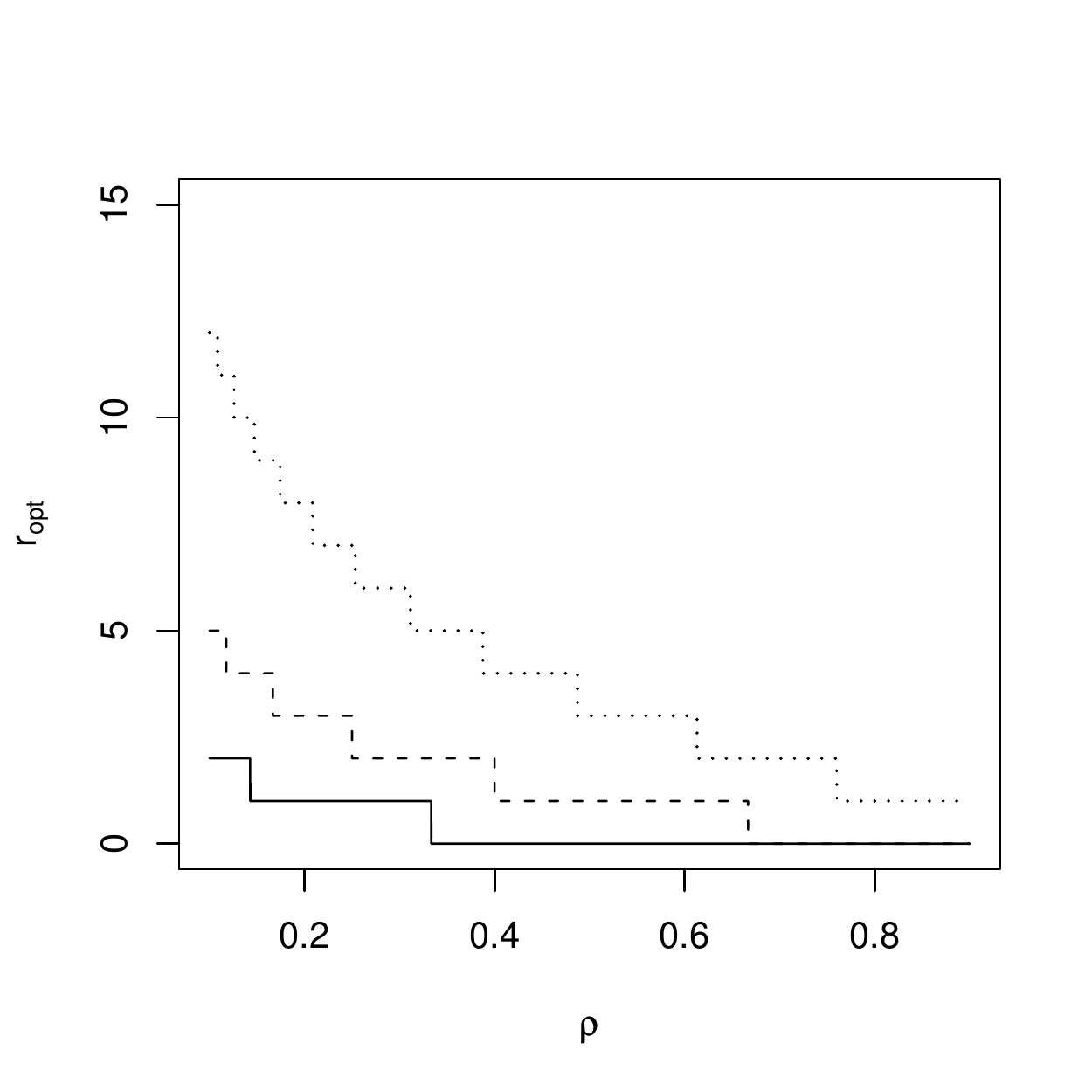}
\caption{Optimal number of repeated measures, ropt, as a function of  , under CMD, CS and $V(t_0 ) = 0$ for different cost ratios  ($\kappa  = 2$ ({---}{---}), $\kappa  = 5$ (-~-~-), and $\kappa  = 20$ ($\cdots\cdots$)).}
\label{roptcmdcsp1fig}
\end{figure}

\begin{sidewaysfigure}
  \centering 
  \includegraphics[width=7.5in]{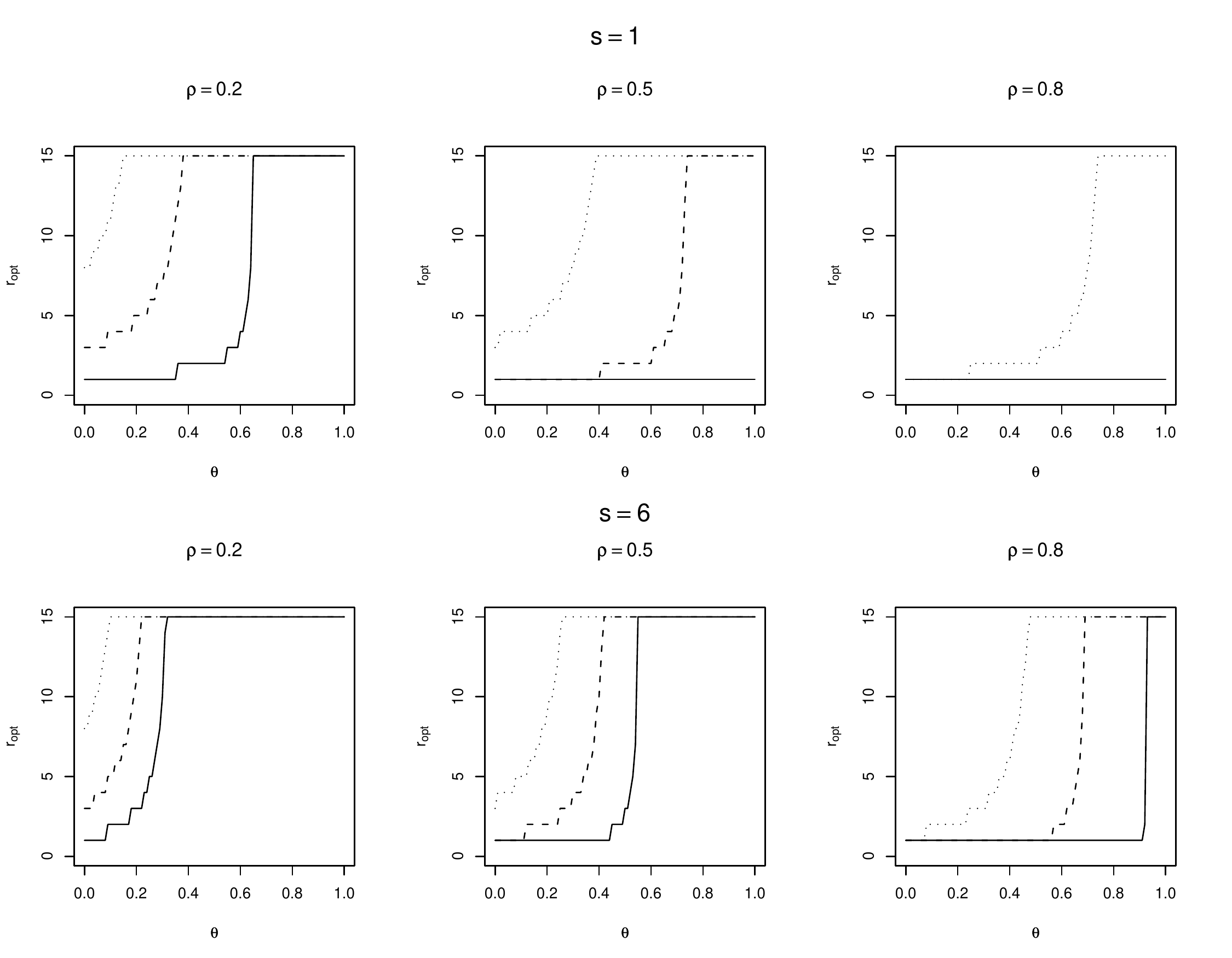}
  \caption{Optimal r as a function of $\theta $ under DEX, CMD, fixed frequency of measurement $s$ and $V(t_0 ) = 0$, for $r \in \left[ {0,15} \right]$ and different cost ratios ($\kappa  = 2$ ({---}{---}), $\kappa  = 5$ (-~-~-), and $\kappa  = 20$ ($\cdots\cdots$)).}
  \label{roptcmddexp1}
\end{sidewaysfigure}

For CMD under CS and $V(t_0 ) = 0$, the optimal $r$ is 
\begin{equation}
\label{roptcmdcsp1}
r_{opt}  = \sqrt {\frac{{(\kappa  - 1)(1 - \rho )}}
{\rho }}  - 1,
\end{equation}
and using the cost constraint \eqref{costp1}, 
$$
N_{opt}  = \frac{\kappa }
{{c_1 \,(\kappa  + r_{opt} )}}{\kern 1pt} COST.
$$
This result has been given in the context of cluster randomized trials \cite{Raudenbush:1997,Cochran:1977}. As can be seen in figure~\ref{roptcmdcsp1fig}, the greater the correlation between measurements of the same person, the smaller the optimal number of repeated measures; and the bigger the cost of the first measurement compared to the rest, the greater the optimal number of repeated measures. If all measurements have the same cost ($\kappa  = 1$), the optimal design takes no repeated measurements and recruits as many participants as the cost constraint allows. Large values of $\kappa $are needed to justify taking more than a small number of repeated measures. 
For the case where $V(t_0 ) > 0$ and $\rho _{\operatorname{e} ,t_0 }  \ne 0$, we computed the optimal $r$ for a range of values of the other design parameters and observed that the effect of $\frac{{\sqrt {V(t_0 )} }} {s}$ on $r_{opt} $ was not monotone. In general, $r_{opt} $ increased as $\frac{{\sqrt {V(t_0 )} }} {s} $ increases, but it decreased again for large values of $\frac{{\sqrt {V(t_0 )} }}
{s}$, especially for cases with small $\kappa $.

With DEX covariance structure, there is no closed-form solution for the optimal $(N,r)$. Computing the optimal $r$ for a grid of values of the other design parameters, we observed that the optimal $r$ was larger for DEX than for the compound symmetry case ($\theta  = 0$), the remaining parameters being equal. Figure~\ref{roptcmddexp1} shows $r_{opt} $  as a function of $\theta $ for several cases with $V(t_0 ) = 0$. The effect of $V(t_0 ) > 0$ was very similar to its effect in the CS case, with $r_{opt}$ increasing with $\frac{{\sqrt {V(t_0 )} }} {s}$ but decreasing again for large values of $\frac{{\sqrt {V(t_0 )} }} {s}$.

With RS covariance structure, there is no closed form solution for the optimal $(N,r)$. We computed the optimal design for a grid of combinations of the other parameters. Few patterns appeared. The optimal $r$ increased as $\kappa $ increased, but the effects of $\rho _{t_0 } $, $\rho _{b_1 ,s,r = 5} $, $\rho _{b_0 b_1 } $ and $V(t_0 )$ strongly depended on values of the other parameters. Calculations for specific situations can be performed with our program.

\subsubsection{LDD}
\label{optNrsldd}

\begin{figure}
\centering 
\includegraphics[width=5in]{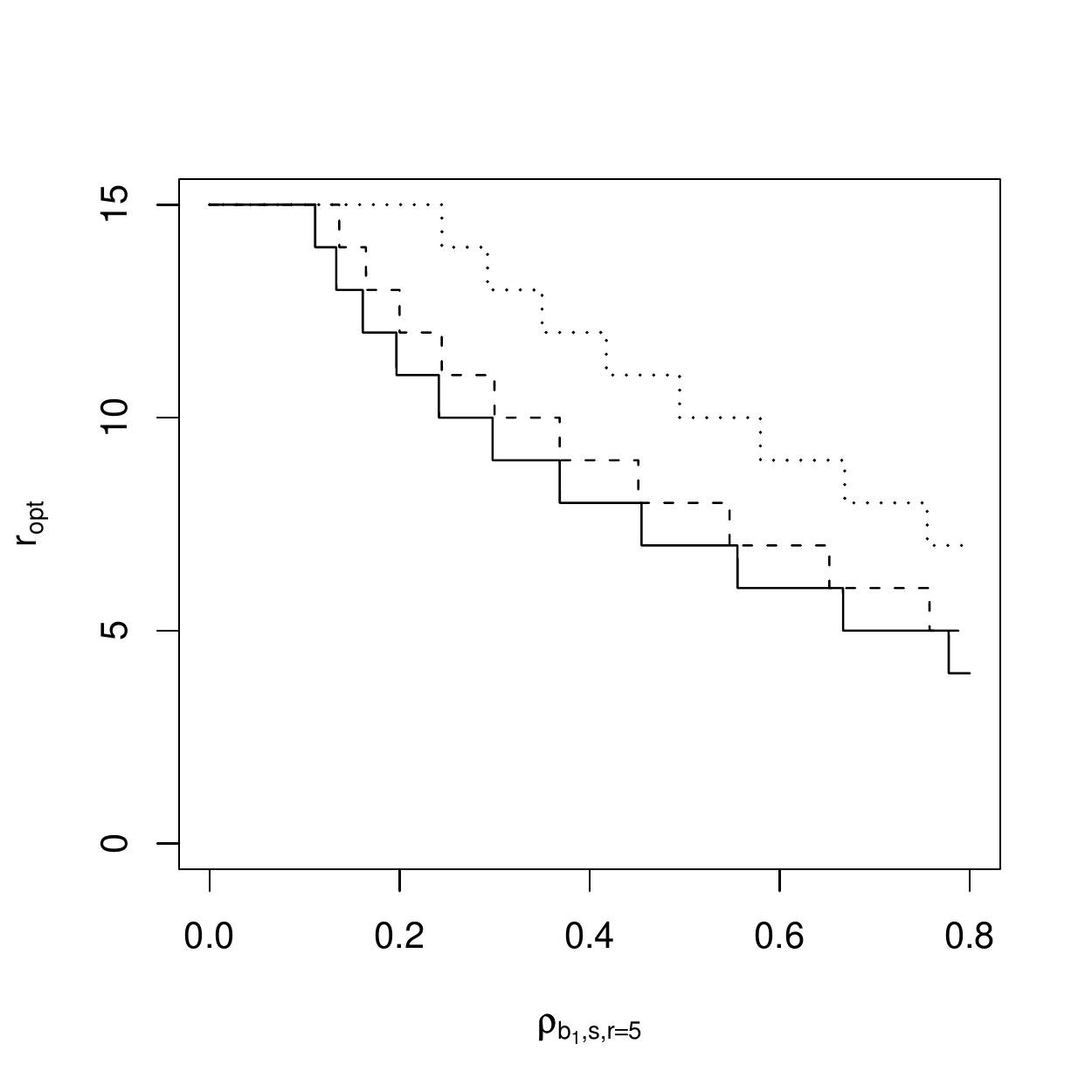}
\caption{Optimal number of repeated measures, ropt, under LDD and RS for fixed frequency of measurement assuming $V(t_0 ) = 0$ as a function of $\rho _{b_1 ,s,r = 5} $ for different cost ratios ($\kappa  = 2$ (---), $\kappa  = 5$ (-~-~-), and $\kappa  = 20$ ($\cdots\cdots$)).}
\label{roptlddrssp1}
\end{figure}

Under LDD, and with fixed frequency of measurements, $V(t_0 ) = 0$ and CS, it is always advisable, in terms of maximizing the power subject to cost constraint, to choose $r$ as large as possible, regardless of $\kappa $ (Appendix~\ref{apoptrlddcss}). In contrast, with $V(t_0 )~>~0$ and CS, under LDD, there are situations where choosing $r$ as large as possible is not the optimal design. We computed the optimal $r$ for a grid of combinations of the other parameters. The optimal was $r = 15$ for almost all cases investigated, except for a few combinations characterized mainly by large values of $\frac{{\sqrt {V(t_0 )} }} {s}$ (close to 10) and small $\rho_{\operatorname{e} ,t_0 } $. Using our program, $(N_{opt,} r_{opt} )$ can be obtained for specific cases.

For DEX, we computed the optimal $r$ over a grid of values of the parameters. We observed that, under DEX and $V(t_0 ) = 0$, we end up choosing the maximum $r$ (15 in our case), which agrees with the CS case. However, when  $V(t_0 ) > 0$, there were cases where choosing $r$ as large as possible did not give the highest power design at a fixed cost, in particular for large values of $\frac{{\sqrt {V(t_0 )} }} {s}$ (close to 10), small values of $\kappa $, large values of $\rho $  and small $\rho _{\operatorname{e} ,t_0 } $. 

RS behaves differently than CS and DEX since the length of follow-up has an effect on the variance of the observations. When $V(t_0 ) = 0$, for a given $\kappa $, $s$ and $\rho _{b_1 ,s,\tilde r} $ we show in Appendix~\ref{apoptrlddrss} that the optimal $r$ solves the equation 
$$
\kappa  = \frac{{r_{opt} ^2 \left( { - (3 + 2r_{opt} )\tilde r(\tilde r + 1)(\tilde r + 2) + (r_{opt}  + 1)^2 (r_{opt}  + 2)^2 \frac{{\rho _{b_1 ,s,\tilde r} }}
{{1 - \rho _{b_1 ,s,\tilde r} }}} \right)}}
{{\,\left( {2 + 6r_{opt}  + 3r_{opt} ^2 } \right)\tilde r(\tilde r + 1)(\tilde r + 2)}}.$$
Figure~\ref{roptlddrssp1} shows the optimal $r$ as a function of $\kappa $, and $\rho _{b_1 ,s,r = 5} $ when $V(t_0 ) = 0$. The optimal $r$ is smaller than for CS, where $\rho _{b_1 ,s,r = 5}  = 0$, and it decreases as $\rho _{b_1 ,s,r = 5} $ increases. 

When  $V(t_0 ) > 0$, with RS, the optimal $r$, $r_{opt} $, depends additionally on $\frac{{\sqrt {V(t_0 )} }} {s}$, $\rho _{b_0 b_1 } $ and $\rho _{\operatorname{e} ,t_0 } $. We computed $r_{opt} $ for a grid of values of the other parameters. Within the range of the parameter space investigated, the optimal $r$ decreased as either $\rho _{b_1 ,s,r = 5} $ or $\frac{{\sqrt {V(t_0 )} }} {s}$ increased, i.e. less repeated measures and more participants were needed as the variation of slopes between participants increased and as the variance of the baseline time variable increased. For specific cases, $r_{opt} $ can be computed with our program.

\subsection{Fixed follow-up period,  $\tau $}
\label{optNrtau}

Under this scenario, increasing $r$ involves increasing the frequency of measurements during a fixed time period, $\tau $. The interval between measurements, $s$, varies in this setting.

\subsubsection{CMD}
\label{optNrtaucmd}

\begin{sidewaysfigure}
  \centering 
  \includegraphics{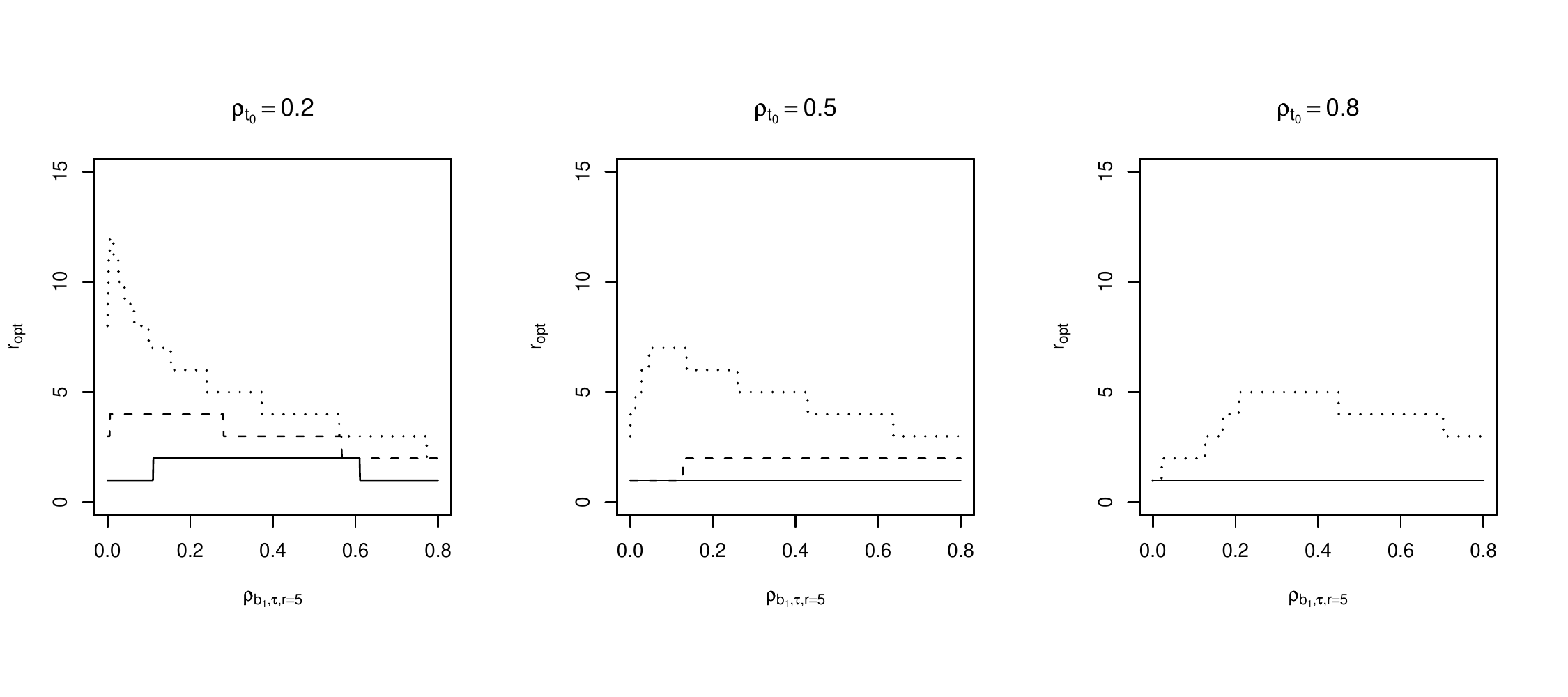}
  \caption{Optimal r as a function of $\rho _{b_1 ,\tau ,r = 5} $ under RS, CMD and fixed follow-up period $\tau  = 5$, for $r \in \left[ {0,15} \right]$, $\rho _{b_0 b_1 } \, =  - 0.5$, $V\left( {t_0 } \right) = 0$ and different cost ratios ($\kappa  = 2$ ({---}), $\kappa  = 5$ (-~-~-), and $\kappa  = 20$ ($\cdots\cdots$)).}
  \label{roptcmdrsp1}
\end{sidewaysfigure}

For CMD under CS and $V(t_0 ) = 0$, the problem is equivalent to the fixed frequency of measurement setting, since the correlations are not affected by the duration of follow-up. Therefore, the optimal $r$ is given by equation \eqref{roptcmdcsp1}. When $V(t_0 ) > 0$ and $\rho _{\operatorname{e} ,t_0 }  \ne 0$, we computed the optimal $r$ for a grid of values of the other parameters and observed a that $\frac{{\sqrt {V(t_0 )} }} {\tau }$  only changed $r_{opt} $ for large values of $\rho _{\operatorname{e} ,t_0 } $, and in that case the pattern is similar to the fixed $s$ case, where $r_{opt} $ increased as $\frac{{\sqrt {V(t_0 )} }}{\tau }$ separates from zero, but it decreased again for large values of $\frac{{\sqrt {V(t_0 )} }} {\tau }$. 

When there is a correlation decay ($\theta  > 0$), we computed $r_{opt} $ for a grid of values of the other parameters and found instances where the optimal $r$ was smaller than in the CS case when $\tau $ was small and bigger than the CS case when $\tau $ was large (data not shown). Increasing $\frac{{\sqrt {V(t_0 )} }}
{\tau }$ produced no changes in $r_{opt} $ in most situations. For RS, we investigated the dependency of $r_{opt} $ as a function of the parameters of the RS covariance structure using a grid of values of the other parameters. Few patterns appeared. We observed that the optimal $r$ increased with $\kappa $, the optimal being $r = 1$ or  $r = 2$ when $\kappa  = 2$, and ranging from 1 to 15 depending on the values of the other parameters when $\kappa  = 40$. Figure~\ref{roptcmdrsp1} shows the relationship between $r_{opt} $ and $\rho _{b_1 ,\tau ,r = 5} $ for several values of $\rho _{t_0 } $ and $\kappa $ and $V(t_0 ) = 0$. The effects of $\rho _{t_0 } $, $\rho _{b_1 ,\tau ,r = 5} $ and $\rho _{b_0 b_1 } $ strongly depended on the values of all other parameters. We observed that increasing $\frac{{\sqrt {V(t_0 )} }}
{\tau }$ reduced $r_{opt} $, whith  $r_{opt} $ being one for most cases with $\frac{{\sqrt {V(t_0 )} }} {\tau } > 3$.

\subsubsection{LDD}
\label{optNrtauldd}

\begin{figure}
\centering 
\includegraphics[width=5in]{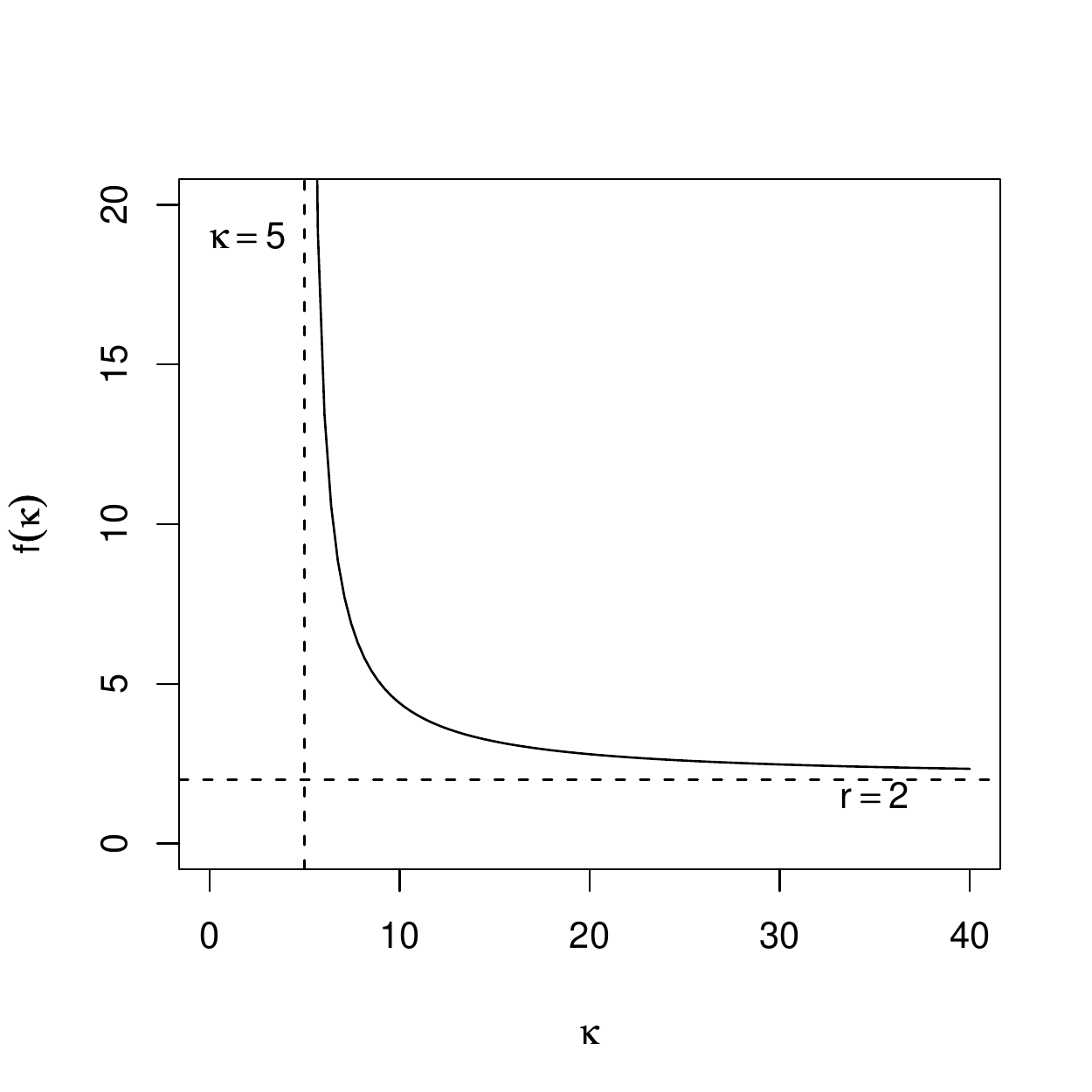}
\caption{Function $f(\kappa ) = \frac{{2(\kappa  + 1)}} {{\kappa  - 5}} $. For a particular $\kappa $, r needs to be greater than this function in order to have more power (for a given cost) than the power of the design with r=1 (under LDD and  CS).}
\label{fkappap1}
\end{figure}

Under CS with $V(t_0 ) = 0$, then $r_{opt}  = 1$ if $\kappa  < 5$. Recall that when LDD is the alternative hypothesis of interest, we need at least one repeated measure to identify the parameters. For $\kappa  > 5$, $r_{opt} $ is not one. Any combination of $(N,r)$ where  $r > \frac{{2(\kappa  + 1)}} {{\kappa  - 5}} $ would improve the power achieved by the combination where $r = 1$. The optimal would involve taking $r$ as large as possible, even though fewer participants would be recruited (Appendix~\ref{apoptrlddcstau}). If, for example, $\kappa  = 10$ then with $r \geqslant 5$ measures one would have more power for the same cost than with $r = 1$ (figure~\ref{fkappap1}). Note that large values of $\kappa $ are needed to justify taking $r$ = 3, 4 or 5, values that are common in many studies. 

\begin{sidewaysfigure}
  \centering 
  \includegraphics[width=8in]{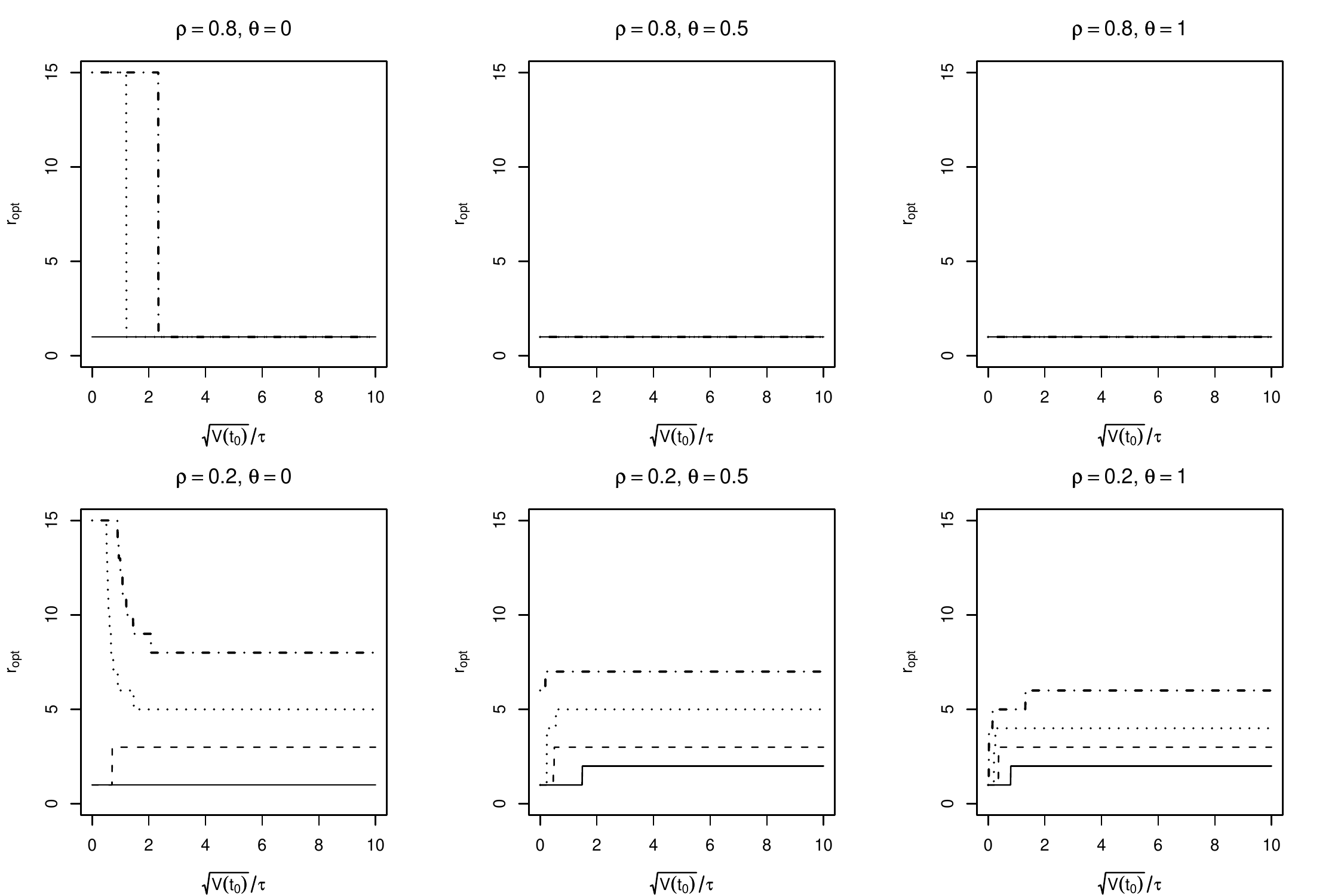}
  \caption{Optimal $r$ as a function of $\frac{{\sqrt {V(t_0 )} }} {\tau }$ under LDD, DEX, fixed $\tau  = 5$, $\rho _{\operatorname{e} ,t_0 }  = 0$. Lines indicate $\kappa  = 2$ ({---}), $\kappa  = 5$ (-~-~-), $\kappa  = 10$ ($\cdots\cdots$), $\kappa  = 20$ ($\cdot$~-~$\cdot$~-~$\cdot$~-).}
  \label{roptldddexvt0p1}
\end{sidewaysfigure}

When $V(t_0 ) > 0$ under CS covariance structure, we computed the optimal $r$ over a grid of values of the other parameters. We found that when $V(t_0 ) > 0$, $r_{opt} 
$ can be greater than one even when $\kappa  < 5$. In addition, the optimal $r$ rarely reached 15 except for small values of $\frac{{\sqrt {V(t_0 )} }} {\tau }$ coupled with large $\kappa $. So, for most of the combinations considered, the optimal $r$ was usually an intermediate values between one and fifteen. The effect of $\frac{{\sqrt {V(t_0 )} }} {\tau }$ on $r_{opt} $ was not monotone -- in some cases it was found to increase $r_{opt} $, and in others to decrease it. Figure~\ref{roptldddexvt0p1} when $\theta  = 0$ exemplified this for particular values of the parameters. For particular cases, the optimal $(N,r)$can be obtained with our program.

\begin{sidewaysfigure}
  \centering 
  \includegraphics[width=8in]{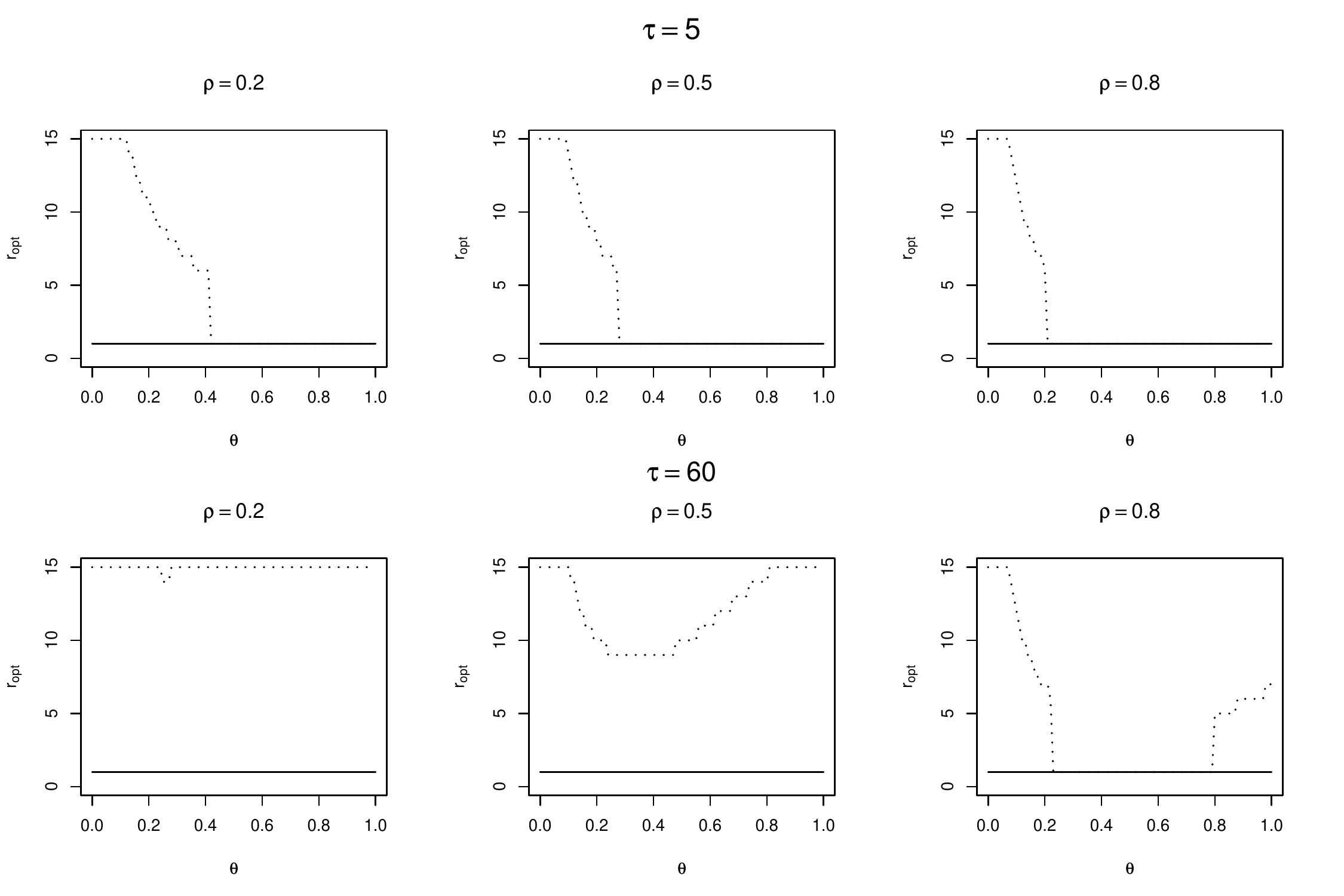}
  \caption{Optimal r as a function of $\theta $ under DEX, LDD and fixed follow-up period $\tau $, for $r \in \left[ {0,15} \right]$, $V\left( {t_0 } \right) = 0$ and different cost ratios ($\kappa  = 2$ ({---}), $\kappa  = 5$ (-~-~-), and $\kappa  = 20$ ($\cdots\cdots$)).}
  \label{roptldddexthetap1}
\end{sidewaysfigure}

Under DEX covariance structure, with $V(t_0 ) = 0$, the optimal value of $r$ that maximizes the power for a given cost also has a complicated expression. We computed the optimal over a grid of possible values of the other parameters, and observed that compared to CS, the optimal $r$ was smaller when $\theta  > 0$ (figure~\ref{roptldddexthetap1}). So, larger values of $\kappa $ are needed to justify taking the same number of repeated measures as would optimize the design under CS. We found in our grid search that the optimal $r$ increased as $\tau $ or $\kappa $ increased, and when $\rho $ decreased. However, the effect of $\theta $ was found to not be monotone for large values of $\tau $ (figure\ref{roptldddexthetap1}). The optimal $r$ and $N$ for different values of the parameters can be computed with our program.

Under DEX and $V(t_0 ) > 0$, we computed the optimal design for a grid of values of the other parameters. We found that the effect of $V(t_0 )$ on $r_{opt} $ was not monotone -- it both increased $r_{opt} $ and decreased it, depending on values of the other parameters. Figure~\ref{roptldddexvt0p1} exemplifies some cases for particular values of the parameters. When $V(t_0 ) > 0$, the optimal $r$ was not always smaller than in the otherwise analogous situation but under CS (i.e. when $\theta  = 0$). Our program can compute the optimal value for given values of the parameters.

Under RS covariance with $V(t_0 ) = 0$ involves the following condition must be met for the optimal $r$ to be greater that one, 
$$
\rho _{b_1 ,\tau ,\tilde r}  < \frac{{\left[ { - 2(\kappa  + 1) + (\kappa  - 5)r} \right](\tilde r + 1)(\tilde r + 2)}}
{{6\tilde r(r + 1)(r + 2) + \left[ { - 2(\kappa  + 1) + (\kappa  - 5)r} \right](\tilde r + 1)(\tilde r + 2)}}
$$
(Appendix~\ref{apoptrlddrstau}). Noting that this can only be true when the right hand side of the inequality is positive, we can deduce that $\kappa  > 5
$ and $r > \frac{{2(\kappa  + 1)}} 
{{\kappa  - 5}}$, as in the corresponding CS case (Appendix~\ref{apoptrlddrstau}). When the condition is met, then the optimal $r$ is the solution to the following equation 
$$
\kappa  = \frac{{r_{opt} (4 + 3r_{opt} )(\tilde r + 1)(\tilde r + 2) + \tilde r(r_{opt}  + 1)^2 (r_{opt}  + 2)^2 \frac{{\rho _{b_1 ,\tau ,\tilde r} }}
{{1 - \rho _{b_1 ,\tau ,\tilde r} }}}}
{{(r_{opt} ^2  - 2)(\tilde r + 1)(\tilde r + 2)}}.
$$

\begin{figure}
  \centering 
  \includegraphics[width=6in]{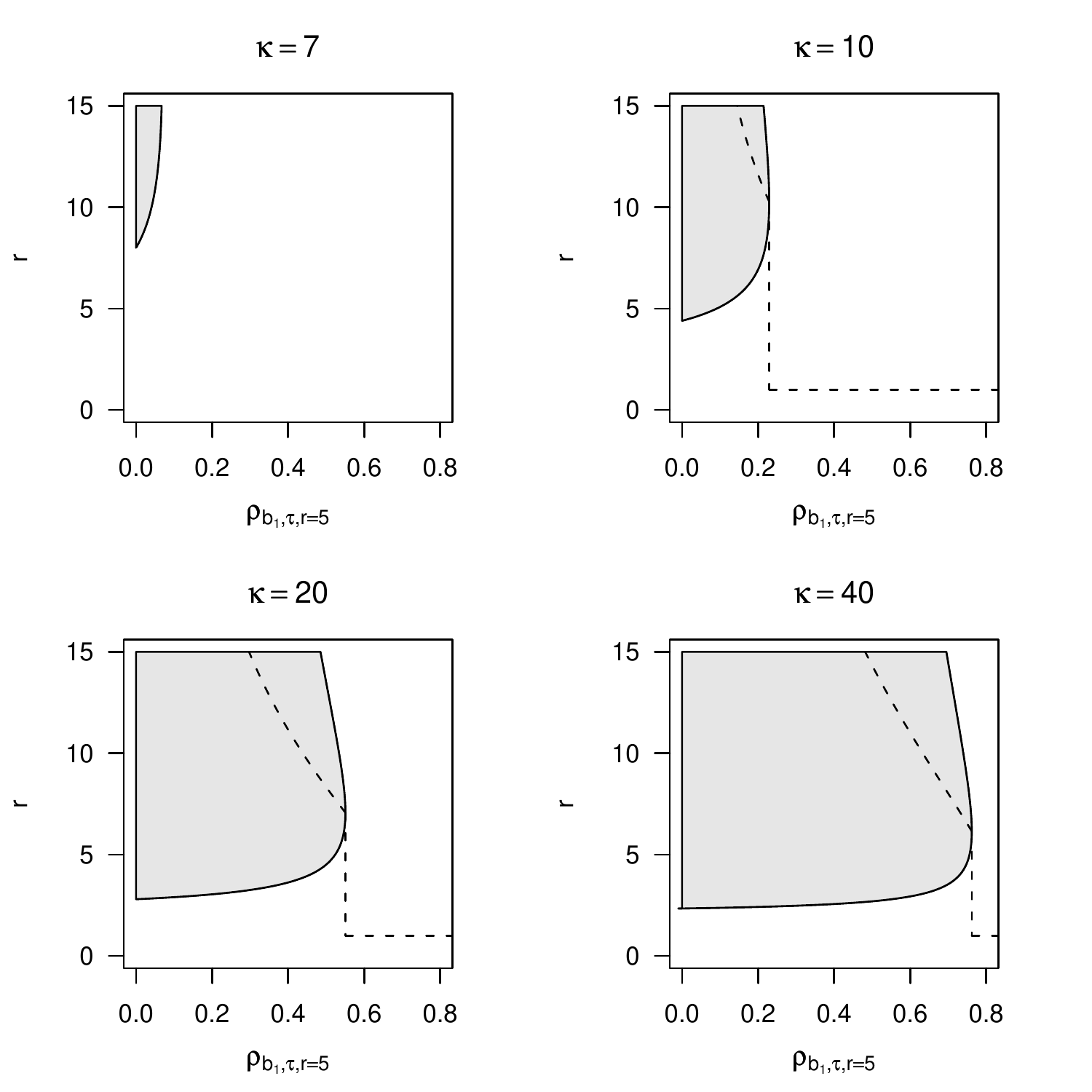}
  \caption{Optimal r (dashed line) and values of r that improve the power attained with $r = 1$ for a fixed cost and same $\kappa $ and $\rho _{b_1 ,\tau ,r = 5} $ (shaded area) under LDD, RS and fixed follow-up time, $\tau $}
  \label{roptlddrstaup1}
\end{figure}

Otherwise, the optimal is $r = 1$. Regions of the design space can be calculated for which at given values of $\rho _{b_1 ,\tau ,r = 5} $ and $\kappa $, $(N_{opt} ,r_{opt} )$  provides a design with more power than the design which takes $r = 1$. The shaded regions in Figure~\ref{roptlddrstaup1} show these regions in some examples. Note that if, for example, $\kappa  = 10$ and $\rho _{b_1 ,\tau ,r = 5}  = 0.3$, there are no values of $r$ that improve the power over that attained at $r = 1$. The optimal $r$ is also plotted in Figure~\ref{roptlddrstaup1}. For example, when $\kappa  = 20$ and $\rho _{b_1 ,\tau ,r = 5}  = 0.3$, the optimal $r$ is greater than 15. Our program can calculate the optimal value for a given $\kappa $ and $\rho _{b_1 ,\tau ,\tilde r} $.

\begin{sidewaysfigure}
  \centering 
  \includegraphics[width=8in]{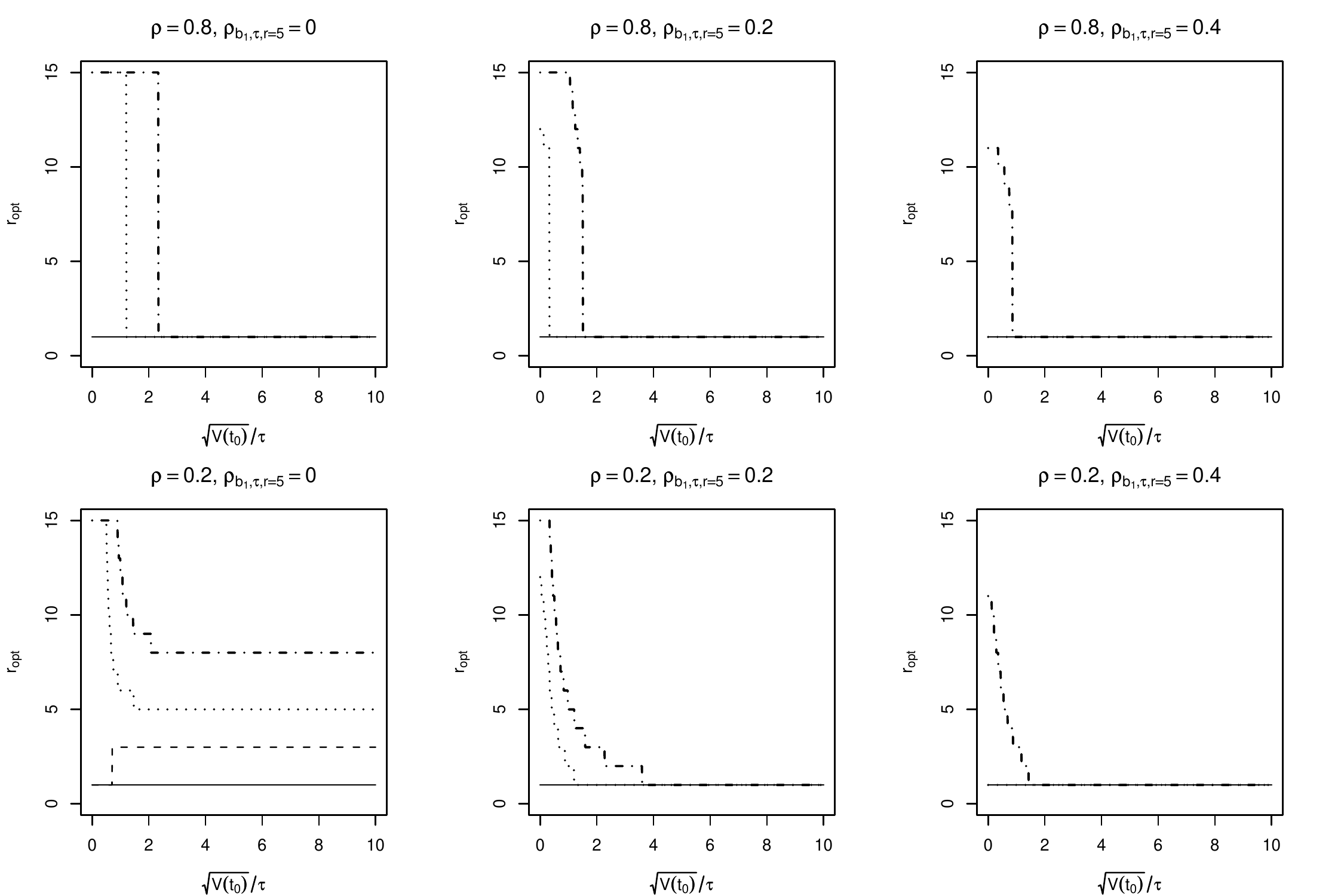}
  \caption{Optimal $r$ as a function of $\frac{{\sqrt {V(t_0 )} }} {\tau }$ under RS and fixed $\tau  = 5$, with $\rho _{b_0 ,b_1 }  =  - 0.5$ and $\rho_{\operatorname{e} ,t_0 }  = 0$. Lines indicate $\kappa  = 2$ ({---}), $\kappa  = 5$ (-~-~-), $\kappa  = 10$ ($\cdots\cdots$), $\kappa  = 20$ ($\cdot$~-~$\cdot$~-~$\cdot$~-).}
  \label{roptlddrsvt0p1}
\end{sidewaysfigure}

With RS and $V(t_0 ) > 0$, the optimal design depends on $\frac{{\sqrt {V(t_0 )} }}
{\tau }$, $\rho _{b_0 b_1 } $ and $\rho _{\operatorname{e} ,t_0 } $ as well. We computed the optimal design for a grid of values of the other parameters. As $\rho _{b_1 ,\tau ,r = 5} $ increased, $r_{opt} $ decreased, i.e. less repeated measurements and more participants were needed. The effect of $\frac{{\sqrt {V(t_0 )} }} {\tau }$ on $r_{opt} $ was observed to not be monotone -- it was found to both increase $r_{opt} $ and decrease it, depending on other values of the parameters. Figure~\ref{roptlddrsvt0p1} shows how $r_{opt} $ varies as a function of $\frac{{\sqrt {V(t_0 )} }} {\tau }$ in some particular cases cases. When  $V(t_0 ) > 0$, the optimal can be greater than one even when $\kappa  < 5$. Our program can compute the optimal value for given values of the parameters.

\section{Illustrative Example}
\label{examplep1}

To illustrate the methods used in this paper, we consider the subset of the Vlagtwedde- Vlaardingen study \cite{Rijcken:1987,Vanderlende:1981} that was made available on the website of a recent textbook on longitudinal analysis \cite{Fitzmaurice:2004} (\url{http://biosun1.harvard.edu/~fitzmaur/ala/}) as a pilot study, and use it as the basis of an investigation of options for the design of an expanded longitudinal study of the effect of smoking on lung function. Briefly, these pilot data consist of 133 men and women from rural Vlagtwedde, The Netherlands, aged 36 or older at baseline, who were followed every three years for up to 19 years for evaluation of their lung function, through spirometric measurement of forced expiratory volume (FEV$_1$). The exposure of interest, current smoking at baseline, was defined as smoking at least one cigarette per day at baseline. 

First, as discussed previously, the design of longitudinal study depends on up to nine parameters. To get a realistic idea about the likely range of design input parameters that apply to this study population and its anticipated extension, we fitted a linear model for FEV$_1$ as a function of smoking, time on study (in years) and the interaction of smoking with time to the pilot data, using CS, DEX and RS covariance structure assumptions. The estimated coefficients of this model assuming a DEX covariance structure model were $\hat \gamma _0  = 3.5086$, the average FEV$_1$ (liters)  at baseline among non-smokers, $\hat \gamma _1  =  - 0.2760$, the average yearly rate of decline of FEV$_1$ (liters)  among non-smokers, $\hat \gamma _2  =  - 0.0337$, the average difference in FEV$_1$ (liters)  at baseline between smokers and non-smokers, and $\hat \gamma _3  =  - 0.0045$, the average difference in the yearly rate of decline of FEV$_1$ (liters)  of smokers compared to non-smokers, corresponding to $\mu _{00}  = 3.5086$, the average FEV$_1$ (liters)  at baseline among non-smokers, $p_1  =  - 7.86\% $, the percent difference in FEV$_1$ (liters) between smokers and non-smokers at baseline, $p_2  =  - 18.2\% $, the percent change FEV$_1$ (liters)  from baseline to end of follow-up among non-smokers, and $p_3  = 13.35\% $, the percent difference between the change FEV$_1$ (liters)  from baseline to end of follow-up in smokers and non-smokers. 

Getting a good estimate of the residual variance from the model of interest at the design stage of a study is not easy. In order of increasingly likely accuracy, we suggest directly estimating $\sigma ^2 $ from longitudinal pilot data when available. Here, the data are available, and the estimated value of $\sigma ^2 $ from the regression of FEV$_1$ on baseline smoking status, time in years from start of the study, and their cross-product, was 0.3214 and 0.3179 under assumptions of CS and DEX, respectively. Under RS, $\sigma _{t_0 }^2 $ was 0.3400.  When longitudinal pilot data are not available, as will typically be the case, we suggest using cross-sectional pilot data if available. Here, the estimated value of $\sigma ^2 $ among the exposed only (since the majority are exposed) was 0.3403. If the time metameter for analysis is time since start of study, this value of the residual variance is likely to generate accurate design calculations. If the time metameter for analysis varies at the start of study, e.g. if time is age, it is best to estimate the residual variance over a restricted age range. The pilot data available do not permit estimation of the variance over a restricted age range - hence, in this example, the option is not available, as it often would not be in practice. Often at the design stage, variance values such as those discussed just above, over presumably comparable subjects but over a range of times, however time may be defined, and perhaps pooled across exposed and unexposed subjects, may be all that is available, from the literature or from pilot data. Here, the variance of FEV$_1$ using all of the measurements for all of the available subjects ($N = 133$) was 0.3837, and the analogous value given in the publication for 1607 subjects, pooled across gender, was 0.3740. Then, conservatively assuming that the ultimate model will explain no more than 10\% of this total variation, the investigator may use a value for $\sigma ^2 
$ of  $0.3837 \times 0.90 = 0.3453$, nearly identical to the analogous value obtained from the fit of the model to the pilot data. For the published value of the marginal variance, the projected value for the residual variance is 0.3366, also very similar to that obtained directly from the pilot data and to the one obtained from the model fit to the pilot data. When $N$ must be found subject to fixed power and a fixed number of measurements per person ($r$), it can be seen from equation \eqref{Np1} that the percent over- or under-estimation of  $N$ is directly proportional to the percent over- or under-estimation of $\sigma ^2 $. When power needs to be calculated as a function of $(N,r)$ or $r$ must be found subject to fixed power and a fixed number of participants ($N$), we can see from equations \eqref{powerp1} and \eqref{rcmdcsp1} that the effect of over- or under-estimation of $\sigma ^2 $ on design cannot be easily described. Interestingly, $(N_{opt} ,r_{opt} )$does not depend upon $\sigma ^2 $ when the design is constrained by a minimum acceptable power, only when it is constrained by a maximum cost.  

Values for other parameters characterizing the covariance structure are needed as well before design calculations can be conducted. If no pilot data are available to estimate them, it is suggested that sensitivity analysis be conducted over what is believed to be a realistic range. Here, we were able to estimate these values from the available longitudinal pilot data. Under the assumption of CS, $\rho $ was 0.857 and 0.896 under DEX, where $\theta $ was 0.18. Assuming RS, $\rho _{t_0 }$, was 0.877; $\rho _{b_1 ,s = 3,r = 6} $, the slope reliability for $r = 6$ measurements per participant was 0.36, indicating a moderate amount of between-subjects variation in slopes; alternatively, $\rho _{b_1 ,\tau  = 18,r = 1} $, the slope reliability at the end of  follow-up with $r = 1$, was 0.27; and $\rho _{b_0 ,b_1 } $ was -0.32. These RS covariance parameters correspond to $\sigma _{within}^2  = 0.0418$, $\sigma _{b_0 }^2  = 0.2982$,$\sigma _{b_1 }^2  = {\text{0}}{\text{.000095}}$ and $\sigma _{b_0 ,b_1 }  =  - 0.0017$. Finally, most of these pilot study participants were smokers, i.e. $p_e $, was $0.79$, and the published value of the standard deviation of  age at entry into the study was 10 years ($\sqrt {V(t_0 )} $) \cite{Rijcken:1987}.

\begin{table}
	\centering
	\caption{Minimum detectable effects in the pilot study $(N~=~133,\; r~=~6,\; \tau~=~18,\;$ $p_e  = 0.79,\, \mu _{00}  = 3.5086, p_2  =  - 18.2\% , {\kern 1pt} \,V(t_0 ) = 100,\rho _{\operatorname{e} ,t_0 }  = 0)$.}
    \bigskip 
\begin{tabular}{p{74pt}p{74pt}p{74pt}p{74pt}p{74pt}}
\hline
\parbox{74pt}{\raggedright Correlation} & \multicolumn{2}{c}{\parbox{149pt}{\centering CMD ($p_1$)}} & \multicolumn{2}{c}{\parbox{149pt}{\centering LDD ($p_3$)}} \\

\parbox{74pt}{\raggedright Power} & \parbox{74pt}{\centering 80\%} &
 \parbox{74pt}{\centering 90\%} & \parbox{74pt}{\centering 80\%} &
  \parbox{74pt}{\centering 90\%} \\
\hline

\parbox{74pt}{\raggedright CS$^{1}$} & \parbox{74pt}{\centering $\pm$ 9\%} &
 \parbox{74pt}{\centering $\pm$ 10\%} & \parbox{74pt}{\centering $\pm$ 22\%} &
  \parbox{74pt}{\centering $\pm$  25\%} \\
  
\parbox{74pt}{\raggedright DEX$^{2}$} & \parbox{74pt}{\centering $\pm$  9\%} &
 \parbox{74pt}{\centering $\pm$  10\%} & \parbox{74pt}{\centering $\pm$  26\%} &
  \parbox{74pt}{\centering $\pm$  30\%} \\
  
\parbox{74pt}{\raggedright RS$^{3 *}$} & \parbox{74pt}{\centering $\pm$  9\%} &
 \parbox{74pt}{\centering $\pm$  10\%} & \parbox{74pt}{\centering $\pm$  26\%} &
  \parbox{74pt}{\centering $\pm$  30\%} \\
\hline
\vspace{2pt}

\parbox{370pt} {\small $^1$ $\sigma ^2  = 0.3214$, $\rho  = 0.857$} \\
\parbox{370pt} {\small $^2$ $\sigma ^2  = 0.3179$, $\rho _1  = 0.896$, $\theta  = 0.18$
} \\
\parbox{370pt} {\small $^3$ $\sigma _{t_0 }^2  = 0.3400$, $\rho _{t_0 }  = 0.877$, $\rho _{b_1 ,s,r = 6}  = 0.36$, $\rho _{b_0 ,b_1 }  =  - 0.32$} \\
\parbox{370pt} {\small $^*$ $t_0 $ assumed normally distributed with variance $V(t_0 )$} \\
		
		\end{tabular}
	\label{tableeffectp1}
\end{table}

\begin{table}
	\centering
	\caption{Minimum number of participants ($N$) to detect a 10\% effect ($p_1~=~0.1$ or $p_3~=~0.1$) with 90\% power in the pilot study $(r = 6,\;\tau  = 18,\;p_e  = 0.79,\,\mu _{00}  = 3.5086,p_2  =  - 18.2\% )$.}
    \bigskip 

\begin{tabular}{p{29pt}p{33pt}p{31pt}p{35pt}p{31pt}p{35pt}p{31pt}p{35pt}p{31pt}}
\hline
\parbox{29pt}{\raggedright } & \multicolumn{4}{c}{\parbox{133pt}{\centering $V(t_0 ) = 0$}} & \multicolumn{4}{c}{\parbox{135pt}{\centering $V(t_0 ) = 100$}} \\

\parbox{29pt}{\raggedright } & \multicolumn{2}{c}{\parbox{65pt}{\centering CMD}} &
 \multicolumn{2}{c}{\parbox{67pt}{\centering LDD}} &
  \multicolumn{2}{c}{\parbox{67pt}{\centering CMD}} &
   \multicolumn{2}{c}{\parbox{67pt}{\centering LDD}} \\
   
\parbox{29pt}{\raggedright $\rho _{\operatorname{e} ,t_0 } $} &
 \parbox{33pt}{\centering 0} & \parbox{31pt}{\centering 0.8} &
  \parbox{35pt}{\centering 0} & \parbox{31pt}{\centering 0.8} &
   \parbox{35pt}{\centering 0} & \parbox{31pt}{\centering 0.8} &
    \parbox{35pt}{\centering 0} & \parbox{31pt}{\centering 0.8} \\
\hline

\parbox{29pt}{\raggedright CS$^{1}$} & \parbox{33pt}{\centering 151} &
 \parbox{31pt}{\centering 151} & \parbox{35pt}{\centering 918} &
  \parbox{31pt}{\centering 918} & \parbox{35pt}{\centering 151} &
   \parbox{31pt}{\centering 155} & \parbox{35pt}{\centering 863} &
    \parbox{31pt}{\centering 897} \\
    
\parbox{29pt}{\raggedright DEX$^{2}$} & \parbox{33pt}{\centering 144} & 
 \parbox{31pt}{\centering 144} & \parbox{35pt}{\centering 1330} &
  \parbox{31pt}{\centering 1330} & \parbox{35pt}{\centering 144} &
   \parbox{31pt}{\centering 152} & \parbox{35pt}{\centering 1215} &
    \parbox{31pt}{\centering 1286} \\
    
\parbox{29pt}{\raggedright RS$^{3 *}$} & \parbox{33pt}{\centering 144} &
 \parbox{31pt}{\centering 144} & \parbox{35pt}{\centering 1305} &
  \parbox{31pt}{\centering 1305} & \parbox{35pt}{\centering 147} &
   \parbox{31pt}{\centering 160} & \parbox{35pt}{\centering 1260} &
    \parbox{31pt}{\centering 1289} \\
\hline
\vspace{2pt}

\parbox{370pt} {\small $^1$ $\sigma ^2  = 0.3214$, $\rho  = 0.857$} \\
\parbox{370pt} {\small $^2$ $\sigma ^2  = 0.3179$, $\rho  = 0.896$, $\theta  = 0.18$} \\
\parbox{370pt} {\small $^3$ $\sigma _{t_0 }^2  = 0.3400$, $\rho _{t_0 }  = 0.877$, $\rho _{b_1 ,s,r = 6}  = 0.36$, $\rho _{b_0 ,b_1 }  =  - 0.32$} \\
\parbox{370pt} {\small $^*$ $t_0 $ assumed normally distributed with variance $V(t_0 )$.} \\

		\end{tabular}
	\label{tableNp1}
\end{table}

In Table~\ref{tableeffectp1}, we show the minimum detectable effects obtained for these trial parameter values under CMD and LDD for $r = 6$. In all the tables given in this section, we assumed, for the RS covariance structure, that $t_{0i} $ is normally distributed. As noted in Section~\ref{rslddp1}, we found that unless the distribution of $t_{0i} $ was extremely skewed, results would be quite insensitive to departures from this assumption. Under CMD, the minimum detectable effect was the same for the three alternate covariance structures, but under LDD, the patterns that allow for the covariance structure to vary with time on study had lower power and larger minimum detectable effects than under CS, as discussed in sections \ref{dexlddp1} and \ref{rslddp1}. We repeated the table assuming $\rho _{\operatorname{e} ,t_0 }  = 0.8$ and obtained almost identical results with the minimum detectable effects slightly higher. Suppose one wants to design a study with seven repeated measures ($r = 6$) taken every three years ($s = 3$). Using the parameters estimated from the pilot data, we computed the number of participants ($N$) needed to detect a 10\% difference in the parameter of interest ($p_1 $), the percent difference in exposure group means which is constant over time, and $p_3 $, the percent difference in the exposure group slopes characterizing their change over time, with 90\% power, for values of $V(t_0 )$ of zero and 100 and values of $\rho _{\operatorname{e} ,t_0 } $ of zero and 0.8. Results are presented in Table~\ref{tableNp1}. As expected, under CMD we need fewer participants when we have departures from CS, while under LDD departures from CS lead to having to recruit more participants. The departures from CS in these data did not appear to be large (e.g., $\theta  = 0.18$), but even this degree of decay had a considerable influence on the required sample sizes. This example suggests that it will often be important for investigators to consider even small departures from CS in their design calculations, and report the maximum departures from CS they are prepared to accommodate in their proposed study. We can also see in Table~\ref{tableNp1} a small increase in sample size under CMD when both $V(t_0 )$ and $\rho _{\operatorname{e} ,t_0 } $ are greater than zero. On the other hand, larger $V(t_0 )$ lead to reduction in sample under LDD, which is going to be maximum when $\rho _{\operatorname{e} ,t_0 }  = 0$.

Now suppose that neither $N$ nor $r$ are fixed by design and we have a budget of 15,000 monetary units (denoted without loss of generality as \$15,000) for CMD and \$100,000 monetary units for LDD, the cost of recruiting each participant and recording their first measurements is \$80 and the subsequent measures are $\kappa $ times cheaper. We must distinguish between two possible situations to proceed here: one in which the frequency of measurements is fixed, and the other, in which the follow-up time is fixed. We restricted consideration for reasons of feasibility to a maximum of $r = 10$ for the fixed frequency case, which would be equivalent to 30 years of follow-up, and to $r = 18$ for the fixed follow-up case, corresponding to one measurement per year. The optimal $(N,r)$ under CMD when $\kappa  = 5$ to detect $p_1  = 10\% $ was $(N,r) = (187,0)$ for all three covariance structures, i.e. the most powerful design for the least amount of money is a cross-sectional study that recruits as many participants as possible. With $\kappa  = 20$, the optimal design took one post-baseline measurement from 178 participants, and was again invariant to assumptions about the covariance structure. We obtained a small $r_{opt} $ even for large values of $\kappa $ because the correlation between measurements on the same participant is large, as seen previously in Figure~\ref{roptcmdcsp1fig} with CS. For LDD and $s = 3$, the optimal design to detect a 10\% difference between exposure group slopes was at the maximum feasible $r$,  here, $r = 10$, with $N = 146$ for all three covariance structures considered, fixed at the same values of the covariance parameters as in Table~\ref{tableNp1}. At this fixed cost of \$100,000, the power for $(N_{opt} ,r_{opt} ) = (10,416)$ differed depending on the assumed covariance structure, with a power of 99\% for CS, 88\% DEX with a small dampening coefficient ($\theta  = 0.18$) and 71\% for RS with a 69\% slope reliability  with $r = 10$ ($\rho _{b_1 ,s = 3,r = 10} $). With $r_{opt} $=10 and $s = 3$, the study is planned to be of 30 years duration and in many cases would not be a realistic choice. The optimal combination of $r$ and $N$ was the same when we assumed $V(t_0 ) = 0$, and the resulting power was only slightly smaller.

\begin{table}
	\centering
	\caption{Optimal design $(N_{opt} ,r_{opt} )$ to maximize power for a study to detect a 10\% difference in slopes ($p_3  = 10\% $) for fixed $\tau  = 18$ under LDD at a cost of no more than \$100,000 with $c_1  = \$ 80$ and $\rho _{\operatorname{e} ,t_0 }  = 0$.}
    \bigskip 

\begin{tabular}{p{43pt}p{69pt}p{61pt}p{61pt}p{61pt}p{61pt}}
\hline
\parbox{43pt}{\raggedright } & \parbox{69pt}{\raggedright } &
 \multicolumn{2}{c}{\parbox{122pt}{\centering $\kappa = 5$}} &
  \multicolumn{2}{c}{\parbox{123pt}{\centering $\kappa = 20$}} \\
  
\parbox{43pt}{\raggedright } & \parbox{69pt}{\raggedright } &
 \parbox{61pt}{\centering $(N,r)$} & \parbox{61pt}{\centering Power} &
  \parbox{61pt}{\centering $(N,r)$} & \parbox{61pt}{\centering Power} \\
\hline

\parbox{43pt}{\raggedright \multirow{2}{*}{CS}} & \parbox{69pt}{\raggedright $V(t_0 ) = 0$} & \parbox{61pt}{\centering (1041, 1)} & \parbox{61pt}{\centering 79\%} &
  \parbox{61pt}{\centering (657, 18)$^{4}$} & \parbox{61pt}{\centering 98\%} \\
  
 & \parbox{69pt}{\raggedright $V(t_0 ) = 100$} & \parbox{61pt}{\centering (1041, 1)}&
    \parbox{61pt}{\centering 83\%} & \parbox{61pt}{\centering (657, 18)} &
     \parbox{61pt}{\centering 99\%} \\
     
\parbox{43pt}{\raggedright \multirow{2}{*}{DEX}} & \parbox{69pt}{\raggedright $V(t_0 ) = 0$} & \parbox{61pt}{\centering (1041, 1)} & \parbox{61pt}{\centering 73\%} &
 \parbox{61pt}{\centering (925, 7)} & \parbox{61pt}{\centering 79\%} \\
 
 & \parbox{69pt}{\raggedright $V(t_0 ) = 100$} & \parbox{61pt}{\centering 
(1041, 1)} & \parbox{61pt}{\centering 77\%} & \parbox{61pt}{\centering (1190, 1)} &
 \parbox{61pt}{\centering 82\%} \\
 
\parbox{43pt}{\raggedright \multirow{2}{*}{RS}} & \parbox{69pt}{\raggedright $V(t_0 ) = 0$} & \parbox{61pt}{\centering (1041, 1)} & \parbox{61pt}{\centering 70\%} &
 \parbox{61pt}{\centering (757, 13)} & \parbox{61pt}{\centering 82\%} \\
 
 & \parbox{69pt}{\raggedright $V(t_0 ) = 100$*} & \parbox{61pt}{\centering 
(1041, 1)} & \parbox{61pt}{\centering 72\%} & \parbox{61pt}{\centering (781, 12)} &
 \parbox{61pt}{\centering 83\%} \\
 \hline
\vspace{2pt}

\parbox{370pt} {\small $^1$ $\sigma ^2  = 0.3214$, $\rho  = 0.857$} \\
\parbox{370pt} {\small $^2$ $\sigma ^2  = 0.3179$, $\rho  = 0.896$, $\theta  = 0.18$} \\
\parbox{370pt} {\small $^3$ $\sigma _{t_0 }^2  = 0.3400$, $\rho _{t_0 }  = 0.877$, $\rho _{b_1 ,s,r = 6}  = 0.36$, $\rho _{b_0 ,b_1 }  =  - 0.32$} \\
\parbox{370pt} {\small $^4$ Note that with $r = 18$, one measurement will be taken every year, three times more often than in the pilot study} \\
\parbox{370pt} {\small $^*$ $t_0 $ assumed normally distributed with variance $V(t_0 )$.} \\

\end{tabular}

	\label{tableNrp1}
\end{table}

\begin{figure}
\centering 
\includegraphics[width=6.5in]{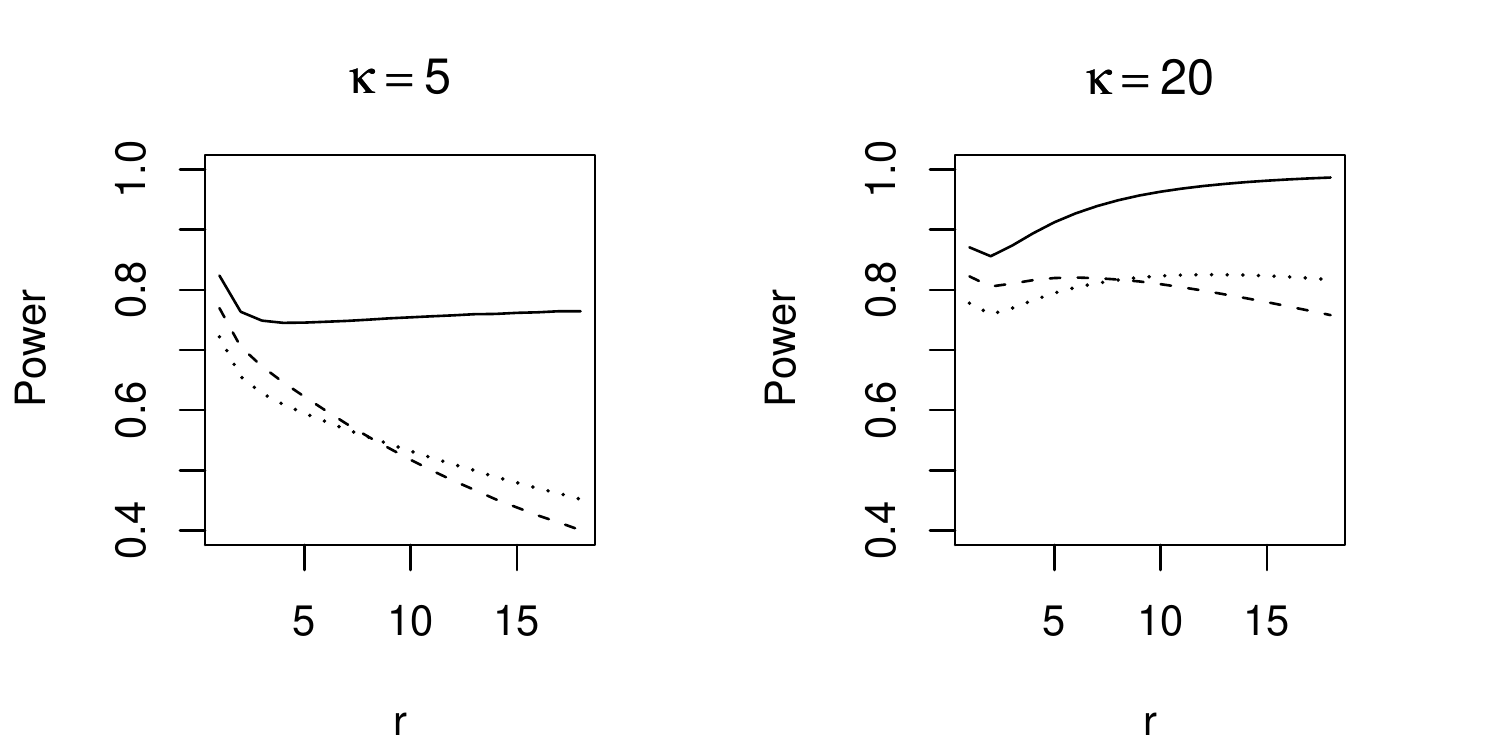}
\caption{Power as a function of r under LDD for $p_3  = 10\% $, fixed $\tau  = 18$
, $V(t_0 ) = 100$, $\rho _{\operatorname{e} ,t_0 }  = 0$ and a cost restriction of \$100,000. The values of the parameters are, for CS, $\sigma ^2  = 0.3214$, $\rho  = 0.857$; DEX, $\sigma ^2  = 0.3179$, $\rho  = 0.896$, $\theta  = 0.18$; RS, $\sigma_{t_0 }^2  = 0.3400$, $\rho _{t_0 }  = 0.877$, $\rho _{b_1 ,s,r = 6}  = 0.36
$, $\rho _{b_0 ,b_1 }  =  - 0.32$. The lines indicate CS ({---}{---}), DEX (-~-~-), and RS ($\cdots\cdots$).}
\label{figexamplep1}
\end{figure}

Table~\ref{tableNrp1} shows the optimal $(N,r)$ under LDD when the duration of the study is fixed at $\tau  = 18$, for $\kappa  = 5$ and $\kappa  = 20$ and for $V(t_0 ) = 0$ and $V(t_0 ) = 100$ and $\rho _{\operatorname{e} ,t_0 }  = 0$. We repeated the results with $\rho _{\operatorname{e} ,t_0 }  = 0.8$ and they were almost identical. The optimal design for $\kappa  = 5$ was at $r_{opt}  = 1$, as we observed in section~\ref{optNrtauldd} for $\kappa  \leqslant 5$, and this was independent of $V(t_0 )$, which only changed slightly the resulting power. For $\kappa  = 20$, the optimal design in $(N,r)$ varied considerably depending on the covariance structure, and within the same covariance structure it varied depending on the value of $V(t_0 )$. Figure~\ref{figexamplep1} shows how power varied as a function of $r_{opt} $  (and $N_{opt} $) for fixed study cost and different values of $\kappa $. When $\kappa $ is large, the optimal combination $(N,r)$ strongly depended on the covariance structure, and combinations that achieved a high power for one covariance structure were underpowered for others.

\section{Software}
\label{softp1}

\begin{sidewaystable}
	\centering
	\caption{Summary of the features of existing programs.}
\bigskip

\begin{tabular}{p{41pt}p{51pt}p{22pt}p{20pt}p{10pt}p{10pt}p{20pt}p{50pt}p{40pt}p{64pt}p{25pt}p{20pt}p{22pt}p{41pt}}
\hline

\parbox{437pt} {\tiny } \\

\parbox{41pt}{\raggedright {\small Software}} & 
 \parbox{51pt}{\centering {\small Reference}} & 
  \parbox{22pt}{\centering {\small CMD}} &  \parbox{20pt}{\centering {\small LDD}} &
   \parbox{10pt}{\centering {\small CS}} & \parbox{10pt}{\centering {\small RS}} &
    \parbox{20pt}{\centering {\small DEX}} & 
     \parbox{50pt}{\centering {\small $V\left({t_0 }\right)>0$}} &
      \parbox{40pt}{\centering {\small Exposure and time correlated}} &
       \parbox{64pt}{\centering {\small Optimal ($N$,$r$) for fixed cost and/or fixed power}} &
        \parbox{25pt}{\centering {\small Power}} & 
         \parbox{20pt}{\centering {\small $N$ for fixed $r$}} & 
          \parbox{22pt}{\centering {\small $r$ for fixed $N$}} &
           \parbox{41pt}{\centering {\small Minimum detectable effect}} \\

\parbox{437pt} {\tiny } \\

\hline

\parbox{437pt} {\tiny } \\

\parbox{41pt}{\raggedright {\small PINT}} & 
 \parbox{51pt}{\centering {\small Snijders (1993, 2003)}} & 
  \parbox{22pt}{\centering {\small $\surd$}} & 
   \parbox{20pt}{\centering {\small $\surd$}} &
    \parbox{10pt}{\centering {\small $\surd$}} &
     \parbox{10pt}{\centering {\small $\surd$}} & 
      \parbox{20pt}{\centering \textbf{{\small $\times{}$}}} &
       \parbox{50pt}{\centering \textbf{{\small \begin{math}\times{}\end{math}*}}} &
        \parbox{40pt}{\centering \textbf{{\small \begin{math}\times{}\end{math}}}} &
         \parbox{64pt}{\centering {\small $\surd$ (fixed cost)}} &
          \multicolumn{4}{c}{\parbox{109pt}{\centering {\small It computes the standard errors, $Var\left( {\hat \beta _2 } \right)$ and $Var\left( {\hat \gamma _3 } \right)$}}} \\

\parbox{437pt} {\tiny } \\
          
\parbox{41pt}{\raggedright {\small RMASS2}} &
 \parbox{51pt}{\centering {\small Hedeker (1999a, 1999b)}} &
  \parbox{22pt}{\centering {\small $\surd$}} &
   \parbox{20pt}{\centering {\small $\surd$}} &
    \parbox{10pt}{\centering {\small $\surd$}} &
     \parbox{10pt}{\centering {\small $\surd$}} &
      \parbox{20pt}{\centering {\small $\surd$}} & 
       \parbox{50pt}{\centering \textbf{{\small \begin{math}\times{}\end{math}}}} &
        \parbox{40pt}{\centering \textbf{{\small \begin{math}\times{}\end{math}}}} &
         \parbox{64pt}{\centering \textbf{{\small \begin{math}\times{}\end{math}}}} &
          \parbox{25pt}{\centering \textbf{{\small \begin{math}\times{}\end{math}}}}&
           \parbox{20pt}{\centering {\small $\surd$}} &
            \parbox{22pt}{\centering \textbf{{\small \begin{math}\times{}\end{math}}}} &
             \parbox{41pt}{\centering \textbf{{\small \begin{math}\times{}\end{math}}}
} \\

\parbox{437pt} {\tiny } \\

\parbox{41pt}{\raggedright {\small GEESIZE}} &
 \parbox{51pt}{\centering {\small Rochon (1998), Ziegler (2004) }} &
  \parbox{22pt}{\centering {\small $\surd$}} &
   \parbox{20pt}{\centering {\small $\surd$}} & 
    \parbox{10pt}{\centering {\small $\surd$}} &
     \parbox{10pt}{\centering \textbf{{\small \begin{math}\times{}\end{math}}}} &
      \parbox{20pt}{\centering {\small $\surd$}} & 
       \parbox{50pt}{\centering \textbf{{\small \begin{math}\times{}\end{math}}}} &
        \parbox{40pt}{\centering \textbf{{\small \begin{math}\times{}\end{math}}}} &
         \parbox{64pt}{\centering \textbf{{\small \begin{math}\times{}\end{math}}}} &
          \parbox{25pt}{\centering \textbf{{\small \begin{math}\times{}\end{math}}}}&
           \parbox{20pt}{\centering {\small $\surd$}} &
            \parbox{22pt}{\centering \textbf{{\small \begin{math}\times{}\end{math}}}
} &
             \parbox{41pt}{\centering \textbf{{\small \begin{math}\times{}\end{math}}}
} \\

\parbox{437pt} {\tiny } \\

\parbox{41pt}{\raggedright {\small OPTITXS}} &
 \parbox{51pt}{\centering {\small Basaga\~na and Spiegelman (2007)}} &
  \parbox{22pt}{\centering {\small $\surd$}} &
   \parbox{20pt}{\centering {\small $\surd$}} &
    \parbox{10pt}{\centering {\small $\surd$}} &
     \parbox{10pt}{\centering {\small $\surd$}} &
      \parbox{20pt}{\centering {\small $\surd$}} &
       \parbox{50pt}{\centering {\small $\surd$}} &
        \parbox{40pt}{\centering {\small $\surd$}} &
         \parbox{64pt}{\centering {\small $\surd$ (both constraints)}} &
          \parbox{25pt}{\centering {\small $\surd$}} &
           \parbox{20pt}{\centering {\small $\surd$}} &
            \parbox{22pt}{\centering {\small $\surd$}} &
             \parbox{41pt}{\centering {\small $\surd$}} \\

\parbox{437pt} {\tiny } \\

\hline
\vspace{2pt}

\parbox{400pt} {\small * Only considers B\&W model, which reduces to the $V(t_0 ) = 0$ case (Appendix~\ref{apvar3p1})} \\

\end{tabular}

	\label{tablesoft}
\end{sidewaystable}

Although others have provided public access software for longitudinal study design that is applicable to observational studies \cite{Hedeker:1999,Raudenbush:2005,Snijders:2003,Ziegler:2004}, nothing currently available comprehensively addresses all the cases considered in this paper, as they might arise in epidemiologic applications. Table~\ref{tablesoft} summarizes the features of the existing programs, as well as our own. Because program OD \cite{Raudenbush:1997,Raudenbush:2005,Raudenbush:2001} only supports 'balanced designs' where $p_e  = 0.5$, we do not consider it further here as such balance will rarely occur in an observational setting. Our program performs all the calculations described in this paper, and is publicly available, along with a user's guide, at the second author's website (\url{http://www.hsph.harvard.edu/faculty/spiegelman/optitxs.html}). The program runs in the R statistical package \cite{R:2006} which can be downloaded for free (\url{http://www.r-project.org}). Our program has an interactive user interface that queries the user for the optimal design scenario to consider, and for the relevant inputs for that scenario; no knowledge of R is required to run our program.  The program has modules to calculate power, number of participants when number of repeated measures is fixed, number of repeated measures when number of participants is fixed, minimum detectable difference and optimal number of repeated measures and participants under budget constraints. A demonstration of its use to compute the optimal $(N,r)$ under LDD and RS is shown in Appendix~\ref{apdemprogp1}.

\section{Conclusions}
\label{conclp1}

The power of a longitudinal study varies with several factors. Independent of assumptions about the covariance structure, power increases as the number of participants, the number of repeated measurements and the length of follow-up increase, and when the exposure prevalence approaches 0.5 from either direction. However, in many situations, we found that one cannot always achieve a pre-specified power solely by increasing the number of repeated measurements, since power sometimes reaches an asymptote. In practice, the length of follow-up or the time interval between successive measurements may be fixed or be restricted to a limited range of values. Sometimes, $N$ may be fixed (e.g. when planning longitudinal follow-up of an existing cross-sectional study) and $r$ needs to be determined. In other cases, $r$ may be fixed (e.g.  when the study is to be based on data collected during monthly clinic visits in an ongoing randomized trial of multivitamin and vitamin A supplementation among HIV-infected women in Dar es Salaam, Tanzania, for which funding can be obtained for no more than five years \cite{Villamor:2002} and $N$ will need to be determined. In still other situations, $(N,\,r)$ is fixed (e.g. in the Nurses' Health Study where the number of nurses, the interval between successive measurements and the duration of follow-up is all fixed \cite{Koh-Banerjee:2003}), and one may need to  determine power for a range of effect sizes given the data available. These problems can all be solved using the program developed in this paper to implement the calculations discussed. 

The power of a longitudinal study, the required number of participants and the required number of repeated measures all vary with assumptions about the covariance structure of the outcome variable as it evolves over time. For a difference between exposure groups that is constant over time (CMD), the lower the correlations between observations within the same participant, the more powerful the study. That is, a study with observations that are less correlated would have more power for the same number of participants and repeated measures; the study would need to recruit fewer participants to achieve the same power with a fixed number of repeated measures; and in most situations, a smaller number of repeated measures would be needed to achieve a pre-specified power for a fixed number of participants. If, in addition, there is a correlation decay compatible with DEX, the study would have more power for fixed $N$ and $r$ than a study with CS covariance, and it would require fewer participants and fewer repeated measures to achieve the same power. The effect of departures from CS compatible with RS cannot be summarized in a straightforward manner. 

When there is a linear interaction between time and exposure (LDD), higher correlations between repeated measurements within participants lead to increased power for the same $N$ and $r$, and require fewer participants and repeated measures to achieve the same power. We showed that departures from CS towards DEX or RS decreased power and increased the required number of participants and repeated measures, although the effect of $\theta $ on power and sample size was not monotone. These relationships under LDD may change when participants enter the study at different times (e.g. when age is the time variable of interest and participants begin follow-up at different ages). After providing tentative design parameters from pilot data, the literature, and intuition, investigators can perform a sensitivity analysis for departures from compound symmetry in their sample size or power calculations using our program. 
 
In practice, it is often the case that neither $N$ nor $r$ are fixed a priori. Then, there is an infinite set of combinations of $(N,\,r)$ that achieve the same power. However, by taking cost into account, one can find the optimal combination of $(N,\,r)
$ to maximize the power subject to a particular budget or to minimize the budget subject to a fixed power. We solved this problem when $s$ is fixed, where taking more repeated measures increases the length of follow-up, and when $\tau $ is fixed, where increasing $r$ reduces the frequency of measurement. For CMD, if all observations have the same cost, one would not take repeated measures, i.e. one will design a cross-sectional study recruiting as many participants as possible. If the subsequent measures of the same participant are cheaper than the first, it may be advisable to take repeated measures and recruit fewer participants. However, no repeated measures or just a small number of them are necessary when the correlation between measures of the same participant is large, as is often the case with longitudinal data. When deviations from CS exist, it becomes more advisable to take at least one repeated measure. Under LDD, if the follow-up period is not fixed, we showed that the optimal design has the maximum length of follow-up possible, except when a RS covariance structure is assumed, in which case the optimal combination of $(N,r)$  needs to be computed. If the follow-up period is fixed, we showed that the optimal design takes more than one repeated measure only when the subsequent measures are more than five times cheaper. When there are departures from CS, we showed that values of $\kappa $ around 10 or 20 are needed to justify taking 3 or 4 repeated measures. It must be pointed out that the optimal $(N,r)$ and the resulting power strongly depended on the covariance structure in the extensive cases we examined, and combinations that are optimal for one assumed covariance structure were often quite poor for another. For particular problems, investigators should perform sensitivity analysis for departures from the assumed values of covariance parameters. 

Of course, these recommendations for optimizing $(N,r)$ are based purely on considerations of power and cost. There are several other reasons to collect more repeated measurement than the optimal. For example, it might be useful to schedule regular visits during the whole follow-up period in order to minimize dropout and to have intermediate data for those who will eventually drop out. Another reason to collect many repeated measurements is when non-linearity of the response profile over time is anticipated. If time is mismodeled, and if time is not independent of exposure, i.e. if study participants do not adhere to visit schedules in a way that is associated with exposure, $\hat \beta _2 $ and $\hat \gamma _3 $ will be biased. Another situation where repeated measures are essential is when the exposure is time-varying, a situation that we will discuss in a subsequent paper and is found, for example, in crossover studies \cite{Jones:1989,Senn:2002}. 

We considered scenarios where the participants do not have the same initial time value, which can be the case, for example, when participants from different ages are recruited and age is the time variable of interest. This requires two new parameters to be provided, $V(t_0 )$ and $\rho _{\operatorname{e} ,t_0 } $. Under CMD, having $V(t_0 ) > 0$ reduces the power of the study, unless $\rho _{\operatorname{e} ,t_0 }  = 0$. Under LDD, however, as $V(t_0 )$ increases so does the power of the study, and the effect is stronger the closer $\rho _{\operatorname{e} ,t_0 } $ is to zero. This increase in power is due to a bigger range of observation of data used to estimate the time by exposure interaction term. The situation where $V(t_0 ) > 0$ resembles the accelerated longitudinal design \cite{Bell:1953}, where multiple age cohorts are sampled and followed. Then, the response profiles of the exposed and the unexposed are modeled. In some applications, when cohort or period effects are suspected to occur that cannot be adjusted for through standard individual-level covariate adjustments, or when between-subjects confounders (time-invariant confounders) are mis- or un-measured, B\&W models that separate within- and between-subjects effects can be fit \cite{Neuhaus:1998,Ware:1990}. If those models are to be used, the results for the LDD case with $V(t_0 ) = 0$ derived in this paper apply and the power gained from $V(t_0 ) > 0$  is lost.

We based our formulas on the asymptotic variance of the estimator, in which case only the first two moments of the covariates and not its full distribution need to be provided. Other approaches have been used. When power is the object of interest, calculations can be based upon the expected value of the non-centrality parameter over the distribution of the covariates ${\mathbf{X}}$ \cite[chapter 3]{Lachin:2000}, where the test uses the conditional variance of ${\mathbf{\hat {\rm B}}}$ given the covariates, i.e. computing 
$$
\Phi \left[ {\sqrt {\mathbb{E}_X \left\{ {\frac{{\left( {\left( {{\mathbf{c'{\rm B}}}} \right)_{H_A } } \right)^2 }}
{{{\mathbf{c}}'\left( {\sum\limits_{i = 1}^N {{\mathbf{X'}}_i {\mathbf{\Sigma }}_i^{ - 1} {\mathbf{X}}_i } } \right)^{ - 1} {\mathbf{c}}}}} \right\}}  - z_{1 - \alpha /2} } \right].
$$
When only exposure is random, this approach results in a $\left( {1 - \frac{1}
{N}} \right)$ correction over the approach used in this paper, which is negligible for large $N$, and equation \eqref{powerp1} becomes 
$$
\Phi \left[ {\frac{{\sqrt {N - 1\,} \left| {\left( {{\mathbf{c'{\rm B}}}} \right)_{H_A } } \right|}}
{{\sqrt {{\mathbf{c'\Sigma }}_{\rm B} {\mathbf{c}}} }} - z_{1 - \alpha /2} } \right]
$$
(Appendix~\ref{apvar1p1}). To find $N$ for fixed $r$ and power, or to find $r$ for fixed $N$ and power using Lachin's approach, one would need to solve the power equation provided above for the desired quantity. To find the optimal $(N,r)$ one would need to maximize the power equation given above subject to the cost constraint. Another approach would be to compute the expected value of the power formula over the distribution of ${\mathbf{X}}$, where again the conditional variance of ${\mathbf{\hat {\rm B}}}$ given the covariates is used,  i.e. 
$$
\mathbb{E}_X \left( {\Phi \left[ {\frac{{\sqrt {N\,} \left| {\left( {{\mathbf{c'{\rm B}}}} \right)_{H_A } } \right|}}
{{\sqrt {{\mathbf{c}}'\left( {\sum\limits_{i = 1}^N {{\mathbf{X'}}_i {\mathbf{\Sigma }}_i^{ - 1} {\mathbf{X}}_i } } \right)^{ - 1} {\mathbf{c}}} }} - z_{1 - \alpha /2} } \right]} \right)
$$
\cite{Glueck:2003}. Since the actual power for the particular sample that is finally recruited can be different from the average power, more conservative quantities such as the 0.025 quantile of power \cite{Glueck:2003} over the distribuition of ${\mathbf{X}}$ might be more appropriate, but the results require the development of a numerical algorithm for accurately computing these quantiles. Similarly, one can derive the required $N$ by taking the expected value of the sample size formula using the variance of ${\mathbf{\hat {\rm B}}}$ conditional on the covariates, 
$$
N = \frac{{\left( {{\mathbf{c}}'\left( {\sum\limits_{i = 1}^N {{\mathbf{X'}}_i {\mathbf{\Sigma }}^{ - 1} {\mathbf{X}}_i } } \right)^{ - 1} {\mathbf{c}}} \right)\,\left( {z_\pi   + z_{1 - \alpha /2} } \right)^2 }}
{{({\mathbf{c'{\rm B}}}_{H_A } )^2 }},
$$
over the distribution of ${\mathbf{X}}$.This leads to computing 
$$
\mathbb{E}_{\mathbf{X}} \left[ {{\mathbf{c}}'\left( {\sum\limits_{i = 1}^N {{\mathbf{X'}}_i {\mathbf{\Sigma }}^{ - 1} {\mathbf{X}}_i } } \right)^{ - 1} {\mathbf{c}}} \right],
$$
which, for the case where only exposure is random, is infinity, so this approach would not useful for this case. This approach would be hard to apply to find $r$, since no explicit functions for $r$ exist for most situations. To find the optimal $(N,r)$, one would need to maximize the power equation given above subject to the cost constraint. All these approaches should produce very similar results to the approach followed in this paper, with the advantage that with our approach the distribution of the covariates does not need to be specified and that simpler formulas are obtained. 

In this paper, we assumed that all study participants are observed at the same scheduled time points, even though their initial age can vary (i.e. only initial age is random), because this most closely resembles the experience typically encountered in epidemiological studies. \citeasnoun{Jung:2003} studied the effect of dispersed times around the scheduled time points, and found that even in the random measurement times case, the sample size calculated under fixed measurement time assumption had an empirical power very close to the nominal power. 

Another factor which could be considered when designing observational longitudinal studies is the dropout rate. Several studies looked at this issue \cite{Dawson:1998,Galbraith:2002,Hedeker:1999,Jung:2003,Yi:2002}   It was reassuring that \citeasnoun{Galbraith:2002} found in a simulation study that in a study of the power to detect group by exposure interactions (i.e. LDD) under RS, as long as the total percentage of lost to follow-up was no more than 30\%, then a study designed to achieve 90\% power ignoring dropout will generally achieve an actual power of at least 80\%. Another simple method consists of inflating the number of participants by a factor of $\frac{1} {{1 - f}}$, where $f$ is the anticipated fraction of lost to follow-up \cite[page 409]{Fitzmaurice:2004}. Further research is needed to assess how dropout would affect the optimal combination $(N,r)$, since the presence of dropout alters the inverse relationship between number of participants and number of time points at the same power \cite{Hedeker:1999}.

In summary, we provided formulas for power, sample size, number of repeated measures, and the optimal combination of participants and repeated measures are provided for longitudinal studies with a continuous response and a binary time-invariant exposure. Our results extend to the case where, for example, age is the time measure of interest, instead of time in the study. General results were derived for the effects of the parameters involved in the calculations whenever possible. The parameters are formulated in an intuitive way to facilitate the choice of appropriate values and to ease sensitivity analyses. Using our publicly available program, users can perform all the design calculations described in the paper and we encourage investigators to make use of this, rather than relying on generalizations which may or may not apply in a particular situation, given the complexity of what needs to be considered in exploring the optimal design.

\bibliography{thesisref}

@article{Bell:1953,
   Author = {Bell, R. Q.},
   Title = {Convergence: an accelerated longitudinal approach},
   Journal = {Child Dev},
   Volume = {24},
   Number = {2},
   Pages = {145-52},
   Year = {1953} }

@article{Bloch:1986,
   Author = {Bloch, D. A.},
   Title = {Sample size requirements and the cost of a randomized clinical trial with repeated measurements},
   Journal = {Stat Med},
   Volume = {5},
   Number = {6},
   Pages = {663-7},
   Year = {1986} }

@book{Cochran:1977,
   Author = {Cochran, William Gemmell},
   Title = {Sampling techniques},
   Publisher = {Wiley},
   Address = {New York},
   Edition = {3d},
   Year = {1977} }

@article{Dawson:1998,
   Author = {Dawson, J. D.},
   Title = {Sample size calculations based on slopes and other summary statistics},
   Journal = {Biometrics},
   Volume = {54},
   Number = {1},
   Pages = {323-30},
   Year = {1998} }

@article{Dawson:1993,
   Author = {Dawson, J. D. and Lagakos, S. W.},
   Title = {Size and power of two-sample tests of repeated measures data},
   Journal = {Biometrics},
   Volume = {49},
   Number = {4},
   Pages = {1022-32},
   Year = {1993} }

@book{Diggle:2002,
   Author = {Diggle, Peter and Heagerty, P. and Liang, K. Y. and Zeger, S.},
   Title = {Analysis of longitudinal data},
   Publisher = {Oxford University Press},
   Address = {Oxford; New York},
   Edition = {2nd},
   Series = {Oxford statistical science series 25},
   Year = {2002} }

@book{Fitzmaurice:2004,
   Author = {Fitzmaurice, Garrett M. and Laird, Nan M. and Ware, James H.},
   Title = {Applied longitudinal analysis},
   Publisher = {Wiley-Interscience},
   Address = {Hoboken, N.J.},
   Series = {Wiley series in probability and statistics},
   Year = {2004} }

@article{Frison:1997,
   Author = {Frison, L. J. and Pocock, S. J.},
   Title = {Linearly divergent treatment effects in clinical trials with repeated measures: efficient analysis using summary statistics},
   Journal = {Stat Med},
   Volume = {16},
   Number = {24},
   Pages = {2855-72},
   Year = {1997} }

@article{Frison:1992,
   Author = {Frison, L. and Pocock, S. J.},
   Title = {Repeated measures in clinical trials: analysis using mean summary statistics and its implications for design},
   Journal = {Stat Med},
   Volume = {11},
   Number = {13},
   Pages = {1685-704},
   Year = {1992} }

@article{Galbraith:2002,
   Author = {Galbraith, S. and Marschner, I. C.},
   Title = {Guidelines for the design of clinical trials with longitudinal outcomes},
   Journal = {Control Clin Trials},
   Volume = {23},
   Number = {3},
   Pages = {257-73},
   Year = {2002} }

@article{Glueck:2003,
   Author = {Glueck, D. H. and Muller, K. E.},
   Title = {Adjusting power for a baseline covariate in linear models},
   Journal = {Stat Med},
   Volume = {22},
   Number = {16},
   Pages = {2535-51},
   Year = {2003} }

@book{Graybill:1983,
   Author = {Graybill, Franklin A.},
   Title = {Matrices with applications in statistics},
   Publisher = {Wadsworth},
   Address = {Belmont, Calif.},
   Edition = {2nd},
   Year = {1983} }

@article{Hedeker:1999,
   Author = {Hedeker, D. and Gibbons, R. D. and Waternaux, C.},
   Title = {Sample size estimation for longitudinal designs with attrition: comparing time-related contrasts between two groups},
   Journal = {Journal of Educational and Behavioral Statistics},
   Volume = {24},
   Number = {1},
   Pages = {70-93},
      Year = {1999} }

@book{Jones:1989,
   Author = {Jones, Byron and Kenward, Michael G.},
   Title = {Design and analysis of cross-over trials},
   Publisher = {Chapman and Hall},
   Address = {London; New York},
   Edition = {1st},
   Series = {Monographs on statistics and applied probability; 34},
   Year = {1989} }

@article{Jung:2003,
   Author = {Jung, S. H. and Ahn, C.},
   Title = {Sample size estimation for GEE method for comparing slopes in repeated measurements data},
   Journal = {Stat Med},
   Volume = {22},
   Number = {8},
   Pages = {1305-15},
   Year = {2003} }

@article{Koh-Banerjee:2003,
   Author = {Koh-Banerjee, P. and Chu, N. F. and Spiegelman, D. and Rosner, B. and Colditz, G. and Willett, W. and Rimm, E.},
   Title = {Prospective study of the association of changes in dietary intake, physical activity, alcohol consumption, and smoking with 9-y gain in waist circumference among 16 587 US men},
   Journal = {American Journal of Clinical Nutrion},
   Volume = {78},
   Pages = {719-27},
   Year = {2003} }

@article{Kirby:1994,
   Author = {Kirby, A. J. and Galai, N. and Munoz, A.},
   Title = {Sample size estimation using repeated measurements on biomarkers as outcomes},
   Journal = {Control Clin Trials},
   Volume = {15},
   Number = {3},
   Pages = {165-72},
   Year = {1994} }

@book{Lachin:2000,
   Author = {Lachin, John M.},
   Title = {Biostatistical methods: the assessment of relative risks},
   Publisher = {Wiley},
   Address = {New York},
   Series = {Wiley series in probability and statistics},
   Year = {2000} }

@article{Liu:1997,
   Author = {Liu, G. and Liang, K. Y.},
   Title = {Sample size calculations for studies with correlated observations},
   Journal = {Biometrics},
   Volume = {53},
   Number = {3},
   Pages = {937-47},
   Year = {1997} }

@article{Munoz:1992,
   Author = {Munoz, A. and Carey, V. and Schouten, J. P. and Segal, M. and Rosner, B.},
   Title = {A parametric family of correlation structures for the analysis of longitudinal data},
   Journal = {Biometrics},
   Volume = {48},
   Number = {3},
   Pages = {733-42},
   Year = {1992} }

@article{Neuhaus:1998,
   Author = {Neuhaus, J. M. and Kalbfleisch, J. D.},
   Title = {Between- and within-cluster covariate effects in the analysis of clustered data},
   Journal = {Biometrics},
   Volume = {54},
   Number = {2},
   Pages = {638-45},
   Year = {1998} }

@article{Overall:1994,
   Author = {Overall, J. E. and Doyle, S. R.},
   Title = {Estimating sample sizes for repeated measurement designs},
   Journal = {Control Clin Trials},
   Volume = {15},
   Number = {2},
   Pages = {100-23},
   Year = {1994} }

@article{Puntanen:1989,
   Author = {Puntanen, S. and Styan, G. P. H.},
   Title = {The equality of the Ordinary Least Squares Estimator and the Best Linear Unbiased Estimator},
   Journal = {The American Statistician},
   Volume = {43},
   Number = {3},
   Pages = {153-61},
      Year = {1989} }

@manual{R:2006,
    Author = {{R Development Core Team}},
    Title = {R: A Language and Environment for Statistical Computing},
    Organization = {R Foundation for Statistical Computing},
    Address = {Vienna, Austria},
    Year = {2006}
}

@article{Raudenbush:1997,
   Author = {Raudenbush, S. W.},
   Title = {Statistical analysis and optimal design for cluster randomized trials},
   Journal = {Psychol Methods},
   Volume = {2},
   Number = {2},
   Pages = {173-85},
      Year = {1997} }

@article{Raudenbush:2005,
   Author = {Raudenbush, S. W. and Spybrook, J. and Xiao-Feng, L. and Congdon, R.},
   Title = {Optimal design for longitudinal and multilevel research},
   Journal = {http://www.ssicentral.com/otherproducts/othersoftware.html},
      Year = {2005} }

@article{Raudenbush:2001,
   Author = {Raudenbush, S. W. and Xiao-Feng, L.},
   Title = {Effects of study duration, frequency of observation, and sample size on power in studies of group differences in polynomial change},
   Journal = {Psychol Methods},
   Volume = {6},
   Number = {4},
   Pages = {387-401},
   Year = {2001} }

@article{Rijcken:1987,
   Author = {Rijcken, B. and Schouten, J. P. and Weiss, S. T. and Speizer, F. E. and van der Lende, R.},
   Title = {The relationship of nonspecific bronchial responsiveness to respiratory symptoms in a random population sample},
   Journal = {Am Rev Respir Dis},
   Volume = {136},
   Number = {1},
   Pages = {62-8},
   Year = {1987} }

@article{Rochon:1998,
   Author = {Rochon, J.},
   Title = {Application of GEE procedures for sample size calculations in repeated measures experiments},
   Journal = {Stat Med},
   Volume = {17},
   Number = {14},
   Pages = {1643-58},
   Year = {1998} }

@article{Schlesselman:1973,
   Author = {Schlesselman, J. J.},
   Title = {Planning a longitudinal study. II. Frequency of measurement and study duration},
   Journal = {J Chronic Dis},
   Volume = {26},
   Number = {9},
   Pages = {561-70},
   Year = {1973} }

@article{Schouten:1999,
   Author = {Schouten, H. J.},
   Title = {Planning group sizes in clinical trials with a continuous outcome and repeated measures},
   Journal = {Stat Med},
   Volume = {18},
   Number = {3},
   Pages = {255-64},
   Year = {1999} }

@book{Searle:1971,
   Author = {Searle, S. R.},
   Title = {Linear models},
   Publisher = {Wiley},
   Address = {New York,},
   Year = {1971} }

@book{Senn:2002,
   Author = {Senn, Stephen},
   Title = {Cross-over trials in clinical research},
   Publisher = {J. Wiley},
   Address = {Chichester, Eng.; New York},
   Edition = {2nd},
   Year = {2002} }

@article{Snijders:1994,
   Author = {Snijders, T.A.B and Bosker, R.J},
   Title = {Modeled variance in two-level models},
   Journal = {Sociological methods and research},
   Volume = {22},
   Number = {3},
   Pages = {342-63},
      Year = {1994} }

@article{Snijders:1993,
   Author = {Snijders, T.A.B. and Bosker, R.J.},
   Title = {Standard errors and sample sizes for two-level research},
   Journal = {Journal of Educational Statistics},
   Volume = {18},
   Number = {3},
   Pages = {237-259},
      Year = {1993} }

@article{Snijders:2003,
   Author = {Snijders, T.A.B.},
   Title = {PINT},
   Journal = {http://stat.gamma.rug.nl/snijders},
      Year = {20033} }

@article{Takkouche:1999,
   Author = {Takkouche, B. and Cadarso-Suarez, C. and Spiegelman, D.},
   Title = {Evaluation of old and new tests of heterogeneity in epidemiologic meta-analysis},
   Journal = {Am J Epidemiol},
   Volume = {150},
   Number = {2},
   Pages = {206-15},
   Year = {1999} }

@book{Timm:2002,
   Author = {Timm, Neil H.},
   Title = {Applied multivariate analysis},
   Publisher = {Springer},
   Address = {New York},
   Series = {Springer texts in statistics},
   Year = {2002} }

@article{Tu:2004,
   Author = {Tu, X. M. and Kowalski, J. and Zhang, J. and Lynch, K. G. and Crits-Christoph, P.},
   Title = {Power analyses for longitudinal trials and other clustered designs},
   Journal = {Stat Med},
   Volume = {23},
   Number = {18},
   Pages = {2799-815},
   Year = {2004} }

@article{Vanderlende:1981,
   Author = {Van der Lende, R. and Kok, T.J. and Peset, R. and Quanjer, P. H. and Schouten, J. P. and Orie, N.G.M.},
   Title = {Decreases in VC and FEV1 with time: Indicators for effects of smoking and air pollution},
   Journal = {Bulletin of European Physiopathology and Respiration},
   Volume = {17},
   Pages = {775-92},
      Year = {1981} }

@article{Villamor:2002,
   Author = {Villamor, E. and Msamanga, G. and Spiegelman, D. and Antelman, G. and Peterson, K. E. and Hunter, D. J. and Fawzi, W. W.},
   Title = {Effect of multivitamin and vitamin A supplements on weight gain during pregnancy among HIV-1-infected women},
   Journal = {Am J Clin Nutr},
   Volume = {76},
   Pages = {1082-90},
      Year = {2002} }

@article{Ware:1990,
   Author = {Ware, J. H. and Dockery, D. W. and Louis, T. A. and Xu, X. P. and Ferris, B. G., Jr. and Speizer, F. E.},
   Title = {Longitudinal and cross-sectional estimates of pulmonary function decline in never-smoking adults},
   Journal = {Am J Epidemiol},
   Volume = {132},
   Number = {4},
   Pages = {685-700},
   Year = {1990} }

@book{Wolfram:2005,
   Author = {{Wolfram Research Inc.}},
   Title = {Mathematica, version 5.2},
   Address = {Champaign, Illinois},
      Year = {2005} }

@article{Yi:2002,
   Author = {Yi, Q. and Panzarella, T.},
   Title = {Estimating sample size for tests on trends across repeated measurements with missing data based on the interaction term in a mixed model},
   Journal = {Control Clin Trials},
   Volume = {23},
   Number = {5},
   Pages = {481-96},
   Year = {2002} }

@article{Ziegler:2004,
   Author = {Ziegler, A.},
   Title = {GEESIZE},
   Journal = {http://www.imbs.uni-luebeck.de/pub/Geesize},
   Year = {2004} }

\section{Appendix}

\setcounter{tocdepth}{0} 
\setcounter{section}{0} 
\renewcommand{\thesection}{A.\arabic{section}}
\renewcommand{\theequation}{A.\arabic{equation}}
 
\section{Variance formulas}
\label{apvarp1}

\subsection{Proof of formulas \eqref{varcmdp1} and \eqref{varcmdvt00p1}}
\label{apvar1p1}

From equation~\eqref{sigmaBp1} we have ${\mathbf{\Sigma }}_{\rm B}  = \mathbb{E}^{ - 1} \left( {{\mathbf{X'}}_i {\mathbf{\Sigma }}_i^{ - 1} {\mathbf{X}}_i } \right)
$.  Now, 
\begin{multline*}
{\mathbf{X'}}_i {\mathbf{\Sigma }}^{ - 1} {\mathbf{X}}_i  =\\
 \left( {\begin{array}{*{20}c}
   1 &  \cdots  & 1 &  \cdots  & 1  \\
   {t_{i0} } &  \cdots  & {t_{i0}  + sj} &  \cdots  & {t_{i0}  + sr}  \\
   {k_i } &  \cdots  & {k_i } &  \cdots  & {k_i }  \\
 \end{array} } \right)\left( {\begin{array}{*{20}c}
   {\nu _{00} } &  \cdots  & {\nu _{0r} }  \\
    \vdots  &  \ddots  &  \vdots   \\
   {\nu _{r0} } &  \cdots  & {v_{rr} }  \\
 \end{array} } \right)\left( {\begin{array}{*{20}c}
   1 & {t_{i0} } & {k_i }  \\
    \vdots  &  \vdots  &  \vdots   \\
   1 & {t_{i0}  + sj} & {k_i }  \\
    \vdots  &  \vdots  &  \vdots   \\
   1 & {t_{i0}  + sr} & {k_i }  \\
 \end{array} } \right) = \\
 = \left( {\begin{array}{*{20}c}
   {\sum\limits_{j = 0}^r {\sum\limits_{j' = 0}^r {v_{jj'} } } } & {} & {}  \\
   { \begin{gathered} \biggl\{
  t_{i0} \sum\limits_{j = 0}^r {\sum\limits_{j' = 0}^r {v_{jj'} } }  +  \hfill \\
  \hfill s\sum\limits_{j = 0}^r {\sum\limits_{j' = 0}^r {jv_{jj'} } }  \biggr\}\\ 
\end{gathered}  } & { \begin{gathered} \biggl\{
  t_{i0}^2 \sum\limits_{j = 0}^r {\sum\limits_{j' = 0}^r {v_{jj'} } }  + st_{i0} \sum\limits_{j = 0}^r {\sum\limits_{j' = 0}^r {(j + j')v_{jj'} } }  \hfill \\
  \hfill + s^2 \sum\limits_{j = 0}^r {\sum\limits_{j' = 0}^r {jj'v_{jj'} } } \biggr\}  \\ 
\end{gathered}} & {}  \\
   {k_i \sum\limits_{j = 0}^r {\sum\limits_{j' = 0}^r {v_{jj'} } } } & {k_i \left( {t_{i0} \sum\limits_{j = 0}^r {\sum\limits_{j' = 0}^r {v_{jj'} } }  + s\sum\limits_{j = 0}^r {\sum\limits_{j' = 0}^r {jv_{jj'} } } } \right)} & {k_i ^2 \sum\limits_{j = 0}^r {\sum\limits_{j' = 0}^r {v_{jj'} } } }  \\
 \end{array} } \right)
\end{multline*}

Using $\bar t_0  = \mathbb{E}\left( {t_0 } \right)$, $p_e  = \mathbb{E}\left( k \right) = \mathbb{E}\left( {k^2 } \right)$ and 
$$
\rho _{\operatorname{e} ,t_0 }  = \frac{{\mathbb{E}\left( {kt_0 } \right) - p_e \bar t_0 }}
{{\sqrt {p_e (1 - p_e )} \sqrt {V\left( {t_0 } \right)} }},
$$
and assuming without loss of generality that $\bar t_0  = 0$ (this can be achieved by centering the initial time), which implies $\mathbb{E}\left( {t_0^2 } \right) = V(t_0 )$, we have that 
\begin{multline*}
\mathbb{E}\left( {{\mathbf{X'}}_i {\mathbf{\Sigma }}_i^{ - 1} {\mathbf{X}}_i } \right) =\\
{\small \left( {\begin{array}{*{20}c}
   {\sum\limits_{j = 0}^r {\sum\limits_{j' = 0}^r {v_{jj'} } } } & {} & {}  \\
   {s\sum\limits_{j = 0}^r {\sum\limits_{j' = 0}^r {jv_{jj'} } } } & {V\left( {t_0 } \right)\sum\limits_{j = 0}^r {\sum\limits_{j' = 0}^r {v_{jj'} } }  + s^2 \sum\limits_{j = 0}^r {\sum\limits_{j' = 0}^r {jj'v_{jj'} } } } & {}  \\
   {p_e \sum\limits_{j = 0}^r {\sum\limits_{j' = 0}^r {v_{jj'} } } } & {\left( {\rho _{\operatorname{e} ,t_0 } \sqrt {p_e (1 - p_e )} \sqrt {V\left( {t_0 } \right)} } \right)\sum\limits_{j = 0}^r {\sum\limits_{j' = 0}^r {v_{jj'} } }  + sp_e \sum\limits_{j = 0}^r {\sum\limits_{j' = 0}^r {jv_{jj'} } } } & {p_e \sum\limits_{j = 0}^r {\sum\limits_{j' = 0}^r {v_{jj'} } } }  \\
 \end{array} } \right)}
\end{multline*}
We are interested in the [3,3] component of the inverse of this matrix, which is 
\begin{multline*}
{\mathbf{c'\Sigma }}_{\rm B} {\mathbf{c}} = {\mathbf{c'}}\left( {\mathbb{E}\left( {{\mathbf{X'}}_i {\mathbf{\Sigma }}_i {\mathbf{X}}_i } \right)} \right)^{ - 1} {\mathbf{c}} = \\
\frac{{s^2 \det ({\mathbf{A}}) + \left( {\sum\limits_{j = 0}^r {\sum\limits_{j' = 0}^r {v_{jj'} } } } \right)^2 V(t_0 )}}
{{p_e (1 - p_e )\left( {\sum\limits_{j = 0}^r {\sum\limits_{j' = 0}^r {v_{jj'} } } } \right)\left[ {s^2 \det ({\mathbf{A}}) + \left( {\sum\limits_{j = 0}^r {\sum\limits_{j' = 0}^r {v_{jj'} } } } \right)^2 \left( {1 - \rho _{e,t_0 }^2 } \right)V(t_0 )} \right]}}
\end{multline*}
If either $V\left( {t_0 } \right)$ or $\rho _{\operatorname{e} ,t_0 } $ are zero then $$
{\mathbf{c'\Sigma }}_{\rm B} {\mathbf{c}} = \frac{1}
{{p_e (1 - p_e )\left( {\sum\limits_{j = 0}^r {\sum\limits_{j' = 0}^r {v_{jj'} } } } \right)}}.
$$
If we follow Lachin's approach \cite{Lachin:2000}, instead of using the asymptotic variance use the variance of ${\mathbf{\hat B}}$ conditional on the covariates, which is 
$$
\left( {\sum\limits_{i = 1}^N {{\mathbf{X'}}_i {\mathbf{\Sigma }}_i^{ - 1} {\mathbf{X}}_i } } \right)^{ - 1},
$$
and redefine ${\mathbf{\Sigma }}_{\rm B} $ as  
$$
\left( {\frac{1}
{N}\sum\limits_{i = 1}^N {{\mathbf{X'}}_i {\mathbf{\Sigma }}_i^{ - 1} {\mathbf{X}}_i } } \right)^{ - 1} 
$$
so that the test statistic is still 
$$
T = \frac{{\sqrt {N\,} {\mathbf{c'\hat {\rm B}}}}}
{{\sqrt {{\mathbf{c'\Sigma }}_{\rm B} {\mathbf{c}}} }}.
$$
Then, we would take the expected value of the non-centrality parameter under the alternative hypothesis over the distribution of ${\mathbf{X}}_i $, i.e. we would compute $\mathbb{E}\left[ {T^2 |H_1 } \right]$. If we assume that everyone is observed at the same set of time points, then the only random covariate is exposure. 
Thus, 
\begin{multline*}
\frac{1}
{N}\sum\limits_i {{\mathbf{X}}_i {\mathbf{\Sigma }}^{ - 1} {\mathbf{X'}}_i }  = \\
\left( {\begin{array}{*{20}c}
   {\sum\limits_{j = 0}^r {\sum\limits_{j' = 0}^r {v_{jj'} } } } & {} & {}  \\
   {s\sum\limits_{j = 0}^r {\sum\limits_{j' = 0}^r {jv_{jj'} } } } & {s^2 \sum\limits_{j = 0}^r {\sum\limits_{j' = 0}^r {jj'v_{jj'} } } } & {}  \\
   {\left( {\frac{{\sum\limits_i {k_i } }}
{N}} \right)\sum\limits_{j = 0}^r {\sum\limits_{j' = 0}^r {v_{jj'} } } } & {s\left( {\frac{{\sum\limits_i {k_i } }}
{N}} \right)\left( {\sum\limits_{j = 0}^r {\sum\limits_{j' = 0}^r {jv_{jj'} } } } \right)} & {\left( {\frac{{\sum\limits_i {k_i ^2 } }}
{N}} \right)\sum\limits_{j = 0}^r {\sum\limits_{j' = 0}^r {v_{jj'} } } }  \\
 \end{array} } \right),
\end{multline*}
and the [3,3] component of the inverse is 
$$
{\mathbf{c'\Sigma }}_{\rm B} {\mathbf{c}} = \left[ {\left( {\frac{{\sum\limits_i {k_i } }}
{N}} \right)\left( {1 - \frac{{\sum\limits_i {k_i } }}
{N}} \right)\left( {\sum\limits_{j = 0}^r {\sum\limits_{j' = 0}^r {v_{jj'} } } } \right)} \right]^{ - 1}.
$$
Then, 
$$
T^2  = \frac{{\hat \beta _2^2 }}
{{Var(\hat \beta _2 )}} = N\hat \beta _2^2 \left( {\frac{{\sum\limits_i {k_i } }}
{N}} \right)\left( {1 - \frac{{\sum\limits_i {k_i } }}
{N}} \right)\left( {\sum\limits_{j = 0}^r {\sum\limits_{j' = 0}^r {v_{jj'} } } } \right)
$$
and 
$$
\mathbb{E}\left[ {T^2 |H_1 } \right] = \mathbb{E}\left[ {N\beta _2^2 \left( {\frac{{\sum\limits_i {k_i } }}
{N}} \right)\left( {1 - \frac{{\sum\limits_i {k_i } }}
{N}} \right)\left( {\sum\limits_{j = 0}^r {\sum\limits_{j' = 0}^r {v_{jj'} } } } \right)} \right],
$$
where the expected value is taken over the distribution of $k_i $, so 
$$
\mathbb{E}\left[ {T^2 |H_1 } \right] = N\beta _2^2 \left( {\sum\limits_{j = 0}^r {\sum\limits_{j' = 0}^r {v_{jj'} } } } \right)\mathbb{E}\left[ {\left( {\frac{{\sum\limits_i {k_i } }}
{N}} \right)\left( {1 - \frac{{\sum\limits_i {k_i } }}
{N}} \right)} \right].
$$
Noticing that $Z = \sum\limits_i {k_i } $ is a Binomial variable we can work out the expected value, 
\begin{multline*}
\mathbb{E}\left[ {T^2 |H_1 } \right] = N\beta _2^2 \left( {\sum\limits_{j = 0}^r {\sum\limits_{j' = 0}^r {v_{jj'} } } } \right)\mathbb{E}\left[ {\left( {\frac{Z}
{N}} \right)\left( {1 - \frac{Z}
{N}} \right)} \right] \\
= N\beta _2^2 \left( {\sum\limits_{j = 0}^r {\sum\limits_{j' = 0}^r {v_{jj'} } } } \right)\left( {\frac{{\mathbb{E}\left( Z \right)}}
{N} - \frac{{\mathbb{E}\left( {Z^2 } \right)}}
{{N^2 }}} \right)\\
 = N\beta _2^2 \left( {\sum\limits_{j = 0}^r {\sum\limits_{j' = 0}^r {v_{jj'} } } } \right)\left( {p_e  - \frac{{Np_e (1 - p_e ) + N^2 p_e^2 }}
{{N^2 }}} \right)  \\
 = N\beta _2^2 \left( {\sum\limits_{j = 0}^r {\sum\limits_{j' = 0}^r {v_{jj'} } } } \right)\left( {\frac{{Np_e  - p_e  + p_e ^2  - Np_e ^2 }}
{N}} \right) \\ 
= \left( {N - 1} \right)\beta _2^2 \left( {\sum\limits_{j = 0}^r {\sum\limits_{j' = 0}^r {v_{jj'} } } } \right)p_e \left( {1 - p_e } \right).
\end{multline*}
The non-centrality parameter with the approach we followed in the paper is 
$$
N\beta _2^2 \left( {\sum\limits_{j = 0}^r {\sum\limits_{j' = 0}^r {v_{jj'} } } } \right)p_e \left( {1 - p_e } \right),
$$
so there is only a $\left( {1 - \frac{1} {N}} \right)$ correction compared with the one obtained with Lachin's method.

\subsection{Proof of formula~\eqref{varlddp1}}
\label{apvar2p1}

Following model~\eqref{lddp1}, and our derivations on Appendix~\ref{apvar1p1}, we now have that
\begin{multline*}
{\mathbf{X'}}_i {\mathbf{\Sigma }}^{ - 1} {\mathbf{X}}_i  = \\
\left( {\begin{array}{*{20}c}
   1 &  \cdots  & 1 &  \cdots  & 1  \\
   {t_{i0} } &  \cdots  & {t_{i0}  + sj} &  \cdots  & {t_{i0}  + sr}  \\
   {k_i } &  \cdots  & {k_i } &  \cdots  & {k_i }  \\
   {k_i t_{i0} } &  \cdots  & {k_i t_{i0}  + k_i sj} &  \cdots  & {k_i t_{i0}  + k_i sr}  \\
 \end{array} } \right)\left( {\begin{array}{*{20}c}
   {v_{00} } &  \cdots  & {v_{0r} }  \\
    \vdots  &  \ddots  &  \vdots   \\
   {v_{r0} } &  \cdots  & {v_{rr} }  \\
 \end{array} } \right) \\
 \left( {\begin{array}{*{20}c}
   1 & {t_{i0} } & {k_i } & {k_i t_{i0} }  \\
    \vdots  &  \vdots  &  \vdots  &  \vdots   \\
   1 & {t_{i0}  + sj} & {k_i } & {k_i t_{i0}  + k_i sj}  \\
    \vdots  &  \vdots  &  \vdots  &  \vdots   \\
   1 & {t_{i0}  + sr} & {k_i } & {k_i t_{i0}  + k_i sr}  \\
 \end{array} } \right)
\end{multline*}
and using the results in Appendix~\ref{apvar1p1} we only need to derive the components in the last row. We can derive $\mathbb{E}\left( {{\mathbf{X'}}_i {\mathbf{\Sigma }}_i^{ - 1} {\mathbf{X}}_i } \right)$, in which the [4,1] component is equivalent to the [3,2] and therefore it takes the value 
$$
\left( {\rho _{\operatorname{e} ,t_0 } \sqrt {p_e (1 - p_e )} \sqrt {V\left( {t_0 } \right)} } \right)\sum\limits_{j = 0}^r {\sum\limits_{j' = 0}^r {v_{jj'} } }  + sp_e \sum\limits_{j = 0}^r {\sum\limits_{j' = 0}^r {jv_{jj'} } }.
$$
The [4,2] component is 
$$
\mathbb{E}\left( {kt_0^2 } \right)\sum\limits_{j = 0}^r {\sum\limits_{j' = 0}^r {v_{jj'} } }  + 2s\left( {\rho _{\operatorname{e} ,t_0 } \sqrt {p_e (1 - p_e )} \sqrt {V\left( {t_0 } \right)} } \right)\sum\limits_{j = 0}^r {\sum\limits_{j' = 0}^r {jv_{jj'} } }  + s^2 p_e \sum\limits_{j = 0}^r {\sum\limits_{j' = 0}^r {jj'v_{jj'} } }, 
$$
the [4,3] component is 
$$
\left( {\rho _{\operatorname{e} ,t_0 } \sqrt {p_e (1 - p_e )} \sqrt {V\left( {t_0 } \right)} } \right)\sum\limits_{j = 0}^r {\sum\limits_{j' = 0}^r {v_{jj'} } }  + sp_e \sum\limits_{j = 0}^r {\sum\limits_{j' = 0}^r {jv_{jj'} } },
$$
and the [4,4] component is the same as the [4,2] component.
An expression for $\mathbb{E}\left( {kt_0^2 } \right) = p_e \mathbb{E}\left( {t_{0,k = 1}^2 } \right)$ in terms of the known parameters is needed. Since we assumed that $\bar t_0  = 0$, then $V\left( {t_0 } \right) = \mathbb{E}\left( {t_0^2 } \right) = (1 - p_e )\mathbb{E}\left( {t_{0,k = 0}^2 } \right) + p_e \mathbb{E}\left( {t_{0,k = 1}^2 } \right)$, which implies 
\begin{equation}
\label{eqA121}
\mathbb{E}\left( {t_{0,k = 0}^2 } \right) = \frac{{V\left( {t_0 } \right) - p_e \mathbb{E}\left( {t_{0,k = 1}^2 } \right)}}
{{1 - p_e }}.
\end{equation}
We have from Appendix~\ref{apvar1p1} that 
$$
\mathbb{E}\left( {kt_0 } \right) = p_e \bar t_{0,k = 1}  = \rho _{\operatorname{e} ,t_0 } \sqrt {p_e (1 - p_e )} \sqrt {V\left( {t_0 } \right)}, 
$$
therefore 
$$
\bar t_{0,k = 1}  = \rho _{\operatorname{e} ,t_0 } \sqrt {\frac{{(1 - p_e )}}
{{p_e }}} \sqrt {V\left( {t_0 } \right)} 
$$
and it can be deduced that 
$$
\bar t_{0,k = 0}  =  - \rho _{\operatorname{e} ,t_0 } \sqrt {\frac{{p_e }}
{{(1 - p_e )}}} \sqrt {V\left( {t_0 } \right)}.
$$
Then, 
$$
\left( {\bar t_{0,k = 1} } \right)^2  = \rho _{\operatorname{e} ,t_0 }^2 \frac{{(1 - p_e )}}
{{p_e }}V\left( {t_0 } \right)
$$
and 
$$
\left( {\bar t_{0,k = 0} } \right)^2  = \rho _{\operatorname{e} ,t_0 }^2 \frac{{p_e }}
{{(1 - p_e )}}V\left( {t_0 } \right).
$$
We assume that the variance of $t_0 $ is the same in exposed and unexposed, i.e. $V\left( {t_{0,k = 0} } \right) = V\left( {t_{0,k = 1} } \right)$. It follows that 
$$
V\left( {t_{0,k = 0} } \right) = V\left( {t_{0,k = 1} } \right) \Leftrightarrow \mathbb{E}\left( {t_{0,k = 0}^2 } \right) - \left( {\bar t_{0,k = 0} } \right)^2  = \mathbb{E}\left( {t_{0,k = 1}^2 } \right) - \left( {\bar t_{0,k = 1} } \right)^2.
$$
Plugging in expression~\eqref{eqA121} we obtain that 
$$
\mathbb{E}\left( {t_{0,k = 1}^2 } \right) = V\left( {t_0 } \right) + \rho _{\operatorname{e} ,t_0 }^2 \left( {\frac{{1 - 2p_e }}
{{p_e }}} \right)V\left( {t_0 } \right).
$$
Therefore, 
$$
\mathbb{E}\left( {kt_0^2 } \right) = p_e \mathbb{E}\left( {t_{0,k = 1}^2 } \right) = V\left( {t_0 } \right)\left[ {p_e  + \rho _{\operatorname{e} ,t_0 }^2 (1 - 2p_e )} \right].
$$
Now, plugging in this last expression in the formula for $\mathbb{E}\left( {{\mathbf{X'}}_i {\mathbf{\Sigma }}_i^{ - 1} {\mathbf{X}}_i } \right)$, and inverting the matrix, it can be derived that its [4,4] component is 
\begin{multline*}
{\mathbf{c'\Sigma }}_{\rm B} {\mathbf{c}} = {\mathbf{c'}}\left( {\mathbb{E}\left( {{\mathbf{X'}}_i {\mathbf{\Sigma }}_i^{ - 1} {\mathbf{X}}_i } \right)} \right)^{ - 1} {\mathbf{c}} = \\
\frac{{\left( {\sum\limits_{j = 0}^r {\sum\limits_{j' = 0}^r {v_{jj'} } } } \right)}}
{{p_e (1 - p_e )\left[ {s^2 \det ({\mathbf{A}}) + \left( {1 - \rho _{\operatorname{e} ,t_0 }^2 } \right)V(t_0 )\left( {\sum\limits_{j = 0}^r {\sum\limits_{j' = 0}^r {v_{jj'} } } } \right)^2 } \right]}}.
\end{multline*}
If $V(t_0 ) = 0$, then 
$$
{\mathbf{c'\Sigma }}_{\rm B} {\mathbf{c}} = \frac{{\left( {\sum\limits_{j = 0}^r {\sum\limits_{j' = 0}^r {v_{jj'} } } } \right)}}
{{p_e (1 - p_e )s^2 \det ({\mathbf{A}})}},
$$
and if $\rho _{\operatorname{e} ,t_0 }  = 0$ then 
$$
{\mathbf{c'\Sigma }}_{\rm B} {\mathbf{c}} = \frac{{\left( {\sum\limits_{j = 0}^r {\sum\limits_{j' = 0}^r {v_{jj'} } } } \right)}}
{{p_e (1 - p_e )\left[ {s^2 \det ({\mathbf{A}}) + V(t_0 )\left( {\sum\limits_{j = 0}^r {\sum\limits_{j' = 0}^r {v_{jj'} } } } \right)^2 } \right]}}.
$$
If we can assume that $t_0 $ and exposure are independent, then the formula we derived for the case $\rho _{\operatorname{e} ,t_0 }  = 0$ also applies to model \eqref{lddfp1}, which assumes a general form for the relationship between response and time in the unexposed but requires that a main effect of time is in the model, we can rewrite the model as 
$$
\mathbb{E}\left( {Y_{ij} |X_{ij} } \right) = \gamma _0  + \gamma _1 t_{ij}  + \alpha _1 f_1 \left( {t_{ij} } \right) +  \cdots  + \alpha _q f_q \left( {t_{ij} } \right) + \gamma _2 k_i  + \gamma _3 \left( {t_{ij}  \times k_i } \right),
$$
where $f_u \left( {t_{ij} } \right),\;u = 1, \ldots ,U$ are arbitrary functions of time. Since the $[m,q]$ term of the matrix $\mathbb{E}\left( {{\mathbf{X'}}_i {\mathbf{\Sigma }}^{ - 1} {\mathbf{X}}_i } \right)$ can be written as $\sum\limits_{j,j'} {v_{jj'} \mathbb{E}\left( {x_{ijm} x_{ij'q} } \right)} $, where $x_{kijm} $ is the value of the mth covariate for subject i from group k at time $t_j $, and exposure and time are independent, which implies 
$$
\mathbb{E}\left( {k_i f_u \left( {t_{ij'} } \right)} \right) = \mathbb{E}\left( {k_i } \right)\mathbb{E}\left( {f_u \left( {t_{ij'} } \right)} \right) = p_e \mathbb{E}\left( {f_u \left( {t_{ij'} } \right)} \right)\;\;\forall u,
$$
we have 
\begin{multline*}
\mathbb{E}\left( {{\mathbf{X'}}_i {\mathbf{\Sigma }}^{ - 1} {\mathbf{X}}_i } \right) = \\
\left( {\begin{array}{*{20}c}
   {\sum\limits_{j,j'} {v_{jj'} } } & {\sum\limits_{j,j'} {v_{jj'} \mathbb{E}\left( {t_{ij'} } \right)} } & {\sum\limits_{j,j'} {v_{jj'} \mathbb{E}\left( {f_1 \left( {t_{ij'} } \right)} \right)} } &  \cdots   \\
   {\sum\limits_{j,j'} {v_{jj'} \mathbb{E}\left( {t_{ij} } \right)} } & {\sum\limits_{j,j'} {v_{jj'} \mathbb{E}\left( {t_{ij} t_{ij'} } \right)} } & {\sum\limits_{j,j'} {v_{jj'} \mathbb{E}\left( {t_{ij} f_1 \left( {t_{ij'} } \right)} \right)} } &  \cdots   \\
   {\sum\limits_{j,j'} {v_{jj'} \mathbb{E}\left( {f_1 \left( {t_{ij} } \right)} \right)} } & {\sum\limits_{j,j'} {v_{jj'} \mathbb{E}\left( {t_{ij} f_1 \left( {t_{ij} } \right)} \right)} } & {\sum\limits_{j,j'} {v_{jj'} \mathbb{E}\left( {f_1 \left( {t_{ij} } \right)f_1 \left( {t_{ij'} } \right)} \right)} } &  \cdots   \\
    \vdots  &  \vdots  &  \vdots  & {}  \\
   {\sum\limits_{j,j'} {v_{jj'} \mathbb{E}\left( {f_V \left( {t_{ij'} } \right)} \right)} } & {\sum\limits_{j,j'} {v_{jj'} \mathbb{E}\left( {t_{ij} f_V \left( {t_{ij'} } \right)} \right)} } & {\sum\limits_{j,j'} {v_{jj'} \mathbb{E}\left( {f_1 \left( {t_{ij} } \right)f_V \left( {t_{ij'} } \right)} \right)} } &  \cdots   \\
   {p_e \sum\limits_{j,j'} {v_{jj'} } } & {p_e \sum\limits_{j,j'} {v_{jj'} \mathbb{E}\left( {t_{ij'} } \right)} } & {p_e \sum\limits_{j,j'} {v_{jj'} \mathbb{E}\left( {f_1 \left( {t_{ij} } \right)} \right)} } &  \cdots   \\
   {p_e \sum\limits_{j,j'} {v_{jj'} \mathbb{E}\left( {t_{ij} } \right)} } & {p_e \sum\limits_{j,j'} {v_{jj'} \mathbb{E}\left( {t_{ij} t_{ij'} } \right)} } & {p_e \sum\limits_{j,j'} {v_{jj'} \mathbb{E}\left( {t_{ij} f_1 \left( {t_{ij} } \right)} \right)} } &  \cdots   \\
 \end{array} } \right.  \\ 
\left. {\begin{array}{*{20}c}
    \cdots  & {\sum\limits_{j,j'} {v_{jj'} \mathbb{E}\left( {f_V \left( {t_{ij'} } \right)} \right)} } & {p_e \sum\limits_{j,j'} {v_{jj'} } } & {p_e \sum\limits_{j,j'} {v_{jj'} \mathbb{E}\left( {t_{ij'} } \right)} }  \\
    \cdots  & {\sum\limits_{j,j'} {v_{jj'} \mathbb{E}\left( {t_{ij} f_V \left( {t_{ij'} } \right)} \right)} } & {p_e \sum\limits_{j,j'} {v_{jj'} \mathbb{E}\left( {t_{ij'} } \right)} } & {p_e \sum\limits_{j,j'} {v_{jj'} \mathbb{E}\left( {t_{ij} t_{ij'} } \right)} }  \\
    \cdots  & {\sum\limits_{j,j'} {v_{jj'} \mathbb{E}\left( {f_1 \left( {t_{ij} } \right)f_V \left( {t_{ij'} } \right)} \right)} } & {p_e \sum\limits_{j,j'} {v_{jj'} \mathbb{E}\left( {f_1 \left( {t_{ij'} } \right)} \right)} } & {p_e \sum\limits_{j,j'} {v_{jj'} \mathbb{E}\left( {t_{ij} f_1 \left( {t_{ij'} } \right)} \right)} }  \\
   {} &  \vdots  &  \vdots  &  \vdots   \\
    \cdots  & {\sum\limits_{j,j'} {v_{jj'} \mathbb{E}\left( {f_V \left( {t_{ij} } \right)f_V \left( {t_{ij'} } \right)} \right)} } & {p_e \sum\limits_{j,j'} {v_{jj'} \mathbb{E}\left( {f_V \left( {t_{ij'} } \right)} \right)} } & {p_e \sum\limits_{j,j'} {v_{jj'} \mathbb{E}\left( {t_{ij} f_V \left( {t_{ij'} } \right)} \right)} }  \\
    \cdots  & {p_e \sum\limits_{j,j'} {v_{jj'} \mathbb{E}\left( {f_V \left( {t_{ij} } \right)} \right)} } & {p_e \sum\limits_{j,j'} {v_{jj'} } } & {p_e \sum\limits_{j,j'} {v_{jj'} \mathbb{E}\left( {t_{ij'} } \right)} }  \\
    \cdots  & {p_e \sum\limits_{j,j'} {v_{jj'} \mathbb{E}\left( {t_{ij} f_V \left( {t_{ij'} } \right)} \right)} } & {p_e \sum\limits_{j,j'} {v_{jj'} \mathbb{E}\left( {t_{ij} } \right)} } & {p_e \sum\limits_{j,j'} {v_{jj'} \mathbb{E}\left( {t_{ij} t_{ij'} } \right)} }  \\
 \end{array} } \right) \\
 = {\mathbf{M}} = \left( {\begin{array}{*{20}c}
   {\begin{array}{*{20}c}
   {\mathop {{\mathbf{M}}_1 }\limits_{(2 \times (V + 2))} }  \\
   {\mathop {{\mathbf{M}}_2 }\limits_{(V \times (V + 2))} }  \\
 \end{array} } & {\mathop {p_e {\mathbf{M'}}_1 }\limits_{((V + 2) \times 2)} }  \\
   {\mathop {p_e {\mathbf{M}}_1 }\limits_{(2 \times (V + 2))} } & {\mathop {p_e {\mathbf{M}}_4 }\limits_{(2 \times 2)} }  \\
 \end{array} } \right) 
\end{multline*}

Define the matrix 
$$
{\mathbf{Q}} = \left( {\begin{array}{*{20}c}
   1 & 0 &  \cdots  &  \cdots  &  \cdots  & 0  \\
   0 & 1 &  \ddots  & {} & {} &  \vdots   \\
    \vdots  &  \ddots  &  \ddots  &  \ddots  & {} &  \vdots   \\
   0 &  \cdots  & 0 & 1 &  \ddots  &  \vdots   \\
   { - p_e } & 0 &  \cdots  & 0 & 1 & 0  \\
   0 & { - p_e } &  \cdots  & 0 & 0 & 1  \\
 \end{array} } \right) = \left( {\begin{array}{*{20}c}
   {\mathop {\mathbf{I}}\limits_{((V + 2) \times (V + 2))} } & {\mathop {\mathbf{0}}\limits_{((V + 2) \times 2)} }  \\
   {\mathop {{\mathbf{Q}}_1 }\limits_{(2 \times (V + 2))} } & {\mathop {\mathbf{I}}\limits_{(2 \times 2)} }  \\
 \end{array} } \right),
$$
such that 
\begin{multline*}
{\mathbf{QMQ'}} = {\mathbf{B}} = \\
\left( {\begin{array}{*{20}c}
   {{\mathbf{B}}_1 } & {\begin{array}{*{20}c}
   {0\quad \quad } & 0  \\
   { \vdots \quad \quad } &  \vdots   \\
   {\quad \quad \quad \quad \quad 0\quad \quad \quad \quad \quad \quad \quad } & {\quad \quad \quad 0\quad \quad \quad }  \\
 \end{array} }  \\
   {\begin{array}{*{20}c}
   0 &  \cdots  & 0  \\
   0 &  \cdots  & 0  \\
 \end{array} } & {\begin{array}{*{20}c}
   {p_e (1 - p_e )\sum\limits_{jj'} {v_{jj'} } } & {p_e (1 - p_e )\sum\limits_{j,j'} {v_{jj'} \mathbb{E}\left( {t_{ij'} } \right)} }  \\
   {p_e (1 - p_e )\sum\limits_{jj'} {v_{jj'} \mathbb{E}\left( {t_{ij} } \right)} } & {p_e (1 - p_e )\sum\limits_{j,j'} {v_{jj'} \mathbb{E}\left( {t_{ij} t_{ij'} } \right)} }  \\
 \end{array} }  \\
 \end{array} } \right)  \\ 
 = \left( {\begin{array}{*{20}c}
   {{\mathbf{B}}_1 } & {\mathbf{0}}  \\
   {\mathbf{0}} & {{\mathbf{B}}_2 }  \\
 \end{array} } \right)
\end{multline*}
Since $\mathbb{E}\left( {t_{ij'} } \right) = \mathbb{E}\left( {t_0 } \right) + sj'$ and $\mathbb{E}\left( {t_{ij} t_{ij'} } \right) = \mathbb{E}\left( {t_0^2 } \right) + s(j + j')\mathbb{E}\left( {t_0 } \right) + s^2 jj'$, and assuming without loss of generality that $\mathbb{E}\left( {t_0 } \right) = 0$ and therefore $\mathbb{E}\left( {t_0^2 } \right) = V(t_0 )$, we have that 
$$
{\mathbf{B}}_2  = p_e (1 - p_e )\left( {\begin{array}{*{20}c}
   {\sum\limits_{j = 0}^r {\sum\limits_{j' = 0}^r {v_{jj'} } } } & {s\sum\limits_{j = 0}^r {\sum\limits_{j' = 0}^r {jv_{jj'} } } }  \\
   {s\sum\limits_{j = 0}^r {\sum\limits_{j' = 0}^r {jv_{jj'} } } } & {V\left( {t_0 } \right)\sum\limits_{j = 0}^r {\sum\limits_{j' = 0}^r {v_{jj'} } }  + s^2 \sum\limits_{j = 0}^r {\sum\limits_{j' = 0}^r {jj'v_{jj'} } } }  \\
 \end{array} } \right).
$$
We are interested in the [V+4,V+4] component of ${\mathbf{M}}^{ - 1} $, which corresponds to $NVar\left( {\hat \gamma _3 } \right)$. Now, 
$$
{\mathbf{M}}^{ - 1}  = {\mathbf{Q'{\rm B}}}^{ - 1} {\mathbf{Q}}  = \left( {\begin{array}{*{20}c}
   {{\mathbf{B}}_1^{ - 1}  + {\mathbf{Q}}_1 '{\mathbf{B}}_2^{ - 1} {\mathbf{Q}}_1 } & {{\mathbf{Q}}_1 '{\mathbf{B}}_2^{ - 1} }  \\
   {{\mathbf{B}}_2^{ - 1} {\mathbf{Q}}_1 } & {{\mathbf{B}}_2^{ - 1} }  \\
 \end{array} } \right),
$$
and 
\begin{multline*}
{\mathbf{B}}_2^{ - 1}  = \frac{1}
{{p_e (1 - p_e )\left[ {V(t_0 )\left( {\sum\limits_{j = 0}^r {\sum\limits_{j' = 0}^r {v_{jj'} } } } \right)^2  + s^2 \det ({\mathbf{A}})} \right]}} \\
\left( {\begin{array}{*{20}c}
   {V\left( {t_0 } \right)\sum\limits_{j = 0}^r {\sum\limits_{j' = 0}^r {v_{jj'} } }  + s^2 \sum\limits_{j = 0}^r {\sum\limits_{j' = 0}^r {jj'v_{jj'} } } } & { - s\sum\limits_{j = 0}^r {\sum\limits_{j' = 0}^r {jv_{jj'} } } }  \\
   { - s\sum\limits_{j = 0}^r {\sum\limits_{j' = 0}^r {jv_{jj'} } } } & {\sum\limits_{j = 0}^r {\sum\limits_{j' = 0}^r {v_{jj'} } } }  \\
 \end{array} } \right).
\end{multline*}
Thus, the [V+4,V+4] component of ${\mathbf{M}}^{ - 1} $ is 
$$
\frac{{\left( {\sum\limits_{j = 0}^r {\sum\limits_{j' = 0}^r {v_{jj'} } } } \right)}}
{{p_e (1 - p_e )\left[ {V(t_0 )\left( {\sum\limits_{j = 0}^r {\sum\limits_{j' = 0}^r {v_{jj'} } } } \right)^2  + s^2 \det ({\mathbf{A}})} \right]}}.
$$

If we follow Lachin's approach \cite{Lachin:2000}, instead of using the asymptotic variance use the variance of ${\mathbf{\hat B}}$ conditional on the covariates, which is 
$$
\left( {\sum\limits_{i = 1}^N {{\mathbf{X'}}_i {\mathbf{\Sigma }}_i^{ - 1} {\mathbf{X}}_i } } \right)^{ - 1},
$$
and redefine ${\mathbf{\Sigma }}_{\rm B} $ as  
$$
\left( {\frac{1}
{N}\sum\limits_{i = 1}^N {{\mathbf{X'}}_i {\mathbf{\Sigma }}_i^{ - 1} {\mathbf{X}}_i } } \right)^{ - 1} 
$$
so that the test statistic is still 
$$
T = \frac{{\sqrt {N\,} {\mathbf{c'\hat {\rm B}}}}}
{{\sqrt {{\mathbf{c'\Sigma }}_{\rm B} {\mathbf{c}}} }}.
$$
Then, we would take the expected value of the non-centrality parameter under the alternative hypothesis over the distribution of ${\mathbf{X}}_i $, i.e. we would compute $\mathbb{E}\left[ {T^2 |H_1 } \right]$. If we assume that everyone is observed at the same set of time points ($V(t_0 ) = 0$), then the only random covariate is exposure and we have 
\begin{multline*}
\frac{1}
{N}\sum\limits_i {{\mathbf{X}}_i {\mathbf{\Sigma }}^{ - 1} {\mathbf{X'}}_i }  = \\
\left( {\begin{array}{*{20}c}
   {\sum\limits_{j,j'}^{} {v_{jj'} } } & {} & {} & {}  \\
   {s\sum\limits_{j,j'}^{} {jv_{jj'} } } & {s^2 \sum\limits_{j,j'}^{} {jj'v_{jj'} } } & {} & {}  \\
   {\left( {\frac{{\sum\limits_i {k_i } }}
{N}} \right)\sum\limits_{j,j'}^{} {v_{jj'} } } & {s\left( {\frac{{\sum\limits_i {k_i } }}
{N}} \right)\sum\limits_{j,j'}^{} {jv_{jj'} } } & {\left( {\frac{{\sum\limits_i {k_i ^2 } }}
{N}} \right)\sum\limits_{j,j'}^{} {v_{jj'} } } & {}  \\
   {s\left( {\frac{{\sum\limits_i {k_i } }}
{N}} \right)\sum\limits_{j,j'}^{} {jv_{jj'} } } & {\left( {\frac{{s^2 \sum\limits_i {k_i } }}
{N}} \right)\sum\limits_{j,j'}^{} {jj'v_{jj'} } } & {\left( {\frac{{s\sum\limits_i {k_i } }}
{N}} \right)\sum\limits_{j,j'}^{} {jv_{jj'} } } & {\left( {\frac{{s^2 \sum\limits_i {k_i } }}
{N}} \right)\sum\limits_{j,j'}^{} {jj'v_{jj'} } }  \\
 \end{array} } \right)
\end{multline*}
and the [4,4] component of the inverse is 
$$
{\mathbf{c'\Sigma }}_{\rm B} {\mathbf{c}} = \frac{{\left( {\sum\limits_{j = 0}^r {\sum\limits_{j' = 0}^r {v_{jj'} } } } \right)}}
{{s^2 \det ({\mathbf{A}})\left( {\frac{{\sum\limits_i {k_i } }}
{N}} \right)\left( {1 - \frac{{\sum\limits_i {k_i } }}
{N}} \right)}}.
$$
Following the same steps as in Appendix~\ref{apvar1p1} we can derive that 
$$
\mathbb{E}\left[ {T^2 |H_1 } \right] = \frac{{\left( {N - 1} \right)\gamma _3^2 s^2 \det ({\mathbf{A}})p_e \left( {1 - p_e } \right)}}
{{\left( {\sum\limits_{j = 0}^r {\sum\limits_{j' = 0}^r {v_{jj'} } } } \right)}}.
$$
The non-centrality parameter with the approach we followed in the paper is 
$$
\frac{{N\gamma _3^2 s^2 \det ({\mathbf{A}})p_e \left( {1 - p_e } \right)}}
{{\left( {\sum\limits_{j = 0}^r {\sum\limits_{j' = 0}^r {v_{jj'} } } } \right)}},
$$
so there is only a $\left( {1 - \frac{1}{N}} \right)$ correction compared with the one obtained with Lachin's method.

\subsection{Proof that $N\,Var\left( {\hat \eta _5 } \right) = 
{\mathbf{c'\Sigma }}_{\rm B} {\mathbf{c}} = \frac{{\left( {\sum\limits_{j = 0}^r {\sum\limits_{j' = 0}^r {v_{jj'} } } } \right)}}
{{p_e (1 - p_e )s^2 \det ({\mathbf{A}})}}$ under model~\eqref{lddbwp1}.}
\label{apvar3p1}

From model \eqref{lddbwp1}, we have 
\begin{multline*}
{\mathbf{X'}}_i {\mathbf{\Sigma }}^{ - 1} {\mathbf{X}}_i  = \left( {\begin{array}{*{20}c}
   1 &  \cdots  & 1 &  \cdots  & 1  \\
   {t_{i0} } &  \cdots  & {t_{i0} } &  \cdots  & {t_{i0} }  \\
   0 &  \cdots  & {sj} &  \cdots  & {sr}  \\
   {k_i } &  \cdots  & {k_i } &  \cdots  & {k_i }  \\
   {k_i t_{i0} } &  \cdots  & {k_i t_{i0} } &  \cdots  & {k_i t_{i0} }  \\
   0 &  \cdots  & {k_i sj} &  \cdots  & {k_i sr}  \\
 \end{array} } \right)\left( {\begin{array}{*{20}c}
   {v_{00} } &  \cdots  & {v_{0r} }  \\
    \vdots  &  \ddots  &  \vdots   \\
   {v_{r0} } &  \cdots  & {v_{rr} }  \\
 \end{array} } \right) \\ 
 \left( {\begin{array}{*{20}c}
   1 & {t_{i0} } & 0 & {k_i } & {k_i t_{i0} } & 0  \\
    \vdots  &  \vdots  &  \vdots  &  \vdots  &  \vdots  &  \vdots   \\
   1 & {t_{i0} } & {sj} & {k_i } & {k_i t_{i0} } & {k_i sj}  \\
    \vdots  &  \vdots  &  \vdots  &  \vdots  &  \vdots  &  \vdots   \\
   1 & {t_{i0} } & {sr} & {k_i } & {k_i t_{i0} } & {k_i sr}  \\
 \end{array} } \right)
\end{multline*}
and we can deduce using the following results derived in appendices 1.1 and 1.2, i.e. 
$$
\begin{gathered}
\mathbb{E}\left( {t_0 } \right) = \bar t_0  = 0, \mathbb{E}\left( {t_0^2 } \right) = V\left( {t_0 } \right),  \mathbb{E}\left( k \right) = \mathbb{E}\left( {k^2 } \right) = p_e,  \\
 \mathbb{E}\left( {kt_0 } \right) = \rho _{\operatorname{e} ,t_0 } \sqrt {p_e (1 - p_e )} \sqrt {V\left( {t_0 } \right)}  = \overline {kt}, \\
 \mathbb{E}\left( {kt_0^2 } \right) = V\left( {t_0 } \right)\left[ {p_e  + \rho _{\operatorname{e} ,t_0 }^2 (1 - 2p_e )} \right] = \overline {kt^2 },
\end{gathered}
$$
that 
\begin{multline*}
\mathbb{E}\left( {{\mathbf{X'}}_i {\mathbf{\Sigma }}_i^{ - 1} {\mathbf{X}}_i } \right) = \\
\left( {\begin{array}{*{20}c}
   {\sum\limits_{j,j'}^{} {v_{jj'} } } & {} & {} & {} & {} & {}  \\
   0 & {V\left( {t_0 } \right)\sum\limits_{j,j}^{} {v_{jj'} } } & {} & {} & {} & {}  \\
   {s\sum\limits_{j,j'}^{} {jv_{jj'} } } & 0 & {s^2 \sum\limits_{j,j'}^{} {jj'v_{jj'} } } & {} & {} & {}  \\
   {p_e \sum\limits_{j,j'}^{} {v_{jj'} } } & {\overline {kt} \sum\limits_{j,j'}^{} {v_{jj'} } } & {sp_e \sum\limits_{j,j'}^{} {jv_{jj'} } } & {p_e \sum\limits_{j,j'}^{} {v_{jj'} } } & {} & {}  \\
   {\overline {kt} \sum\limits_{j,j'}^{} {v_{jj'} } } & {\overline {kt^2 } \sum\limits_{j,j'}^{} {v_{jj'} } } & {s\overline {kt} \sum\limits_{j,j'}^{} {jv_{jj'} } } & {\overline {kt} \sum\limits_{j,j'}^{} {v_{jj'} } } & {\overline {kt^2 } \sum\limits_{j,j'}^{} {v_{jj'} } } & {}  \\
   {sp_e \sum\limits_{j,j'}^{} {jv_{jj'} } } & {s\overline {kt} \sum\limits_{j,j'}^{} {jv_{jj'} } } & {s^2 p_e \sum\limits_{j,j'}^{} {jj'v_{jj'} } } & {sp_e \sum\limits_{j,j'}^{} {jv_{jj'} } } & {s\overline {kt} \sum\limits_{j,j'}^{} {jv_{jj'} } } & {s^2 p_e \sum\limits_{j,j'}^{} {jj'v_{jj'} } }  \\
 \end{array} } \right)
\end{multline*}
The [6,6] component of the inverse of this matrix is 
$$
{\mathbf{c'\Sigma }}_{\rm B} {\mathbf{c}} = \frac{{\left( {\sum\limits_{j = 0}^r {\sum\limits_{j' = 0}^r {v_{jj'} } } } \right)}}
{{p_e (1 - p_e )s^2 \det ({\mathbf{A}})}},
$$
as we derived in Appendix~\ref{apvar2p1} for the LDD case with $V(t_0 ) = 0$.

\subsection{Proof that $s\hat \lambda _1  = \hat \eta _5 $ and $s^2 Var\left( {\hat \lambda _1 } \right) = Var\left( {\hat \eta _5 } \right)$ from models \eqref{lddbwp1} and \eqref{ldddiffp1}}
\label{apvar4p1}

The GLS estimator has the expression 
$$
{\mathbf{\hat B}} = \left( {\sum\limits_{i = 1}^N {{\mathbf{X'}}_i {\mathbf{\Sigma }}^{ - 1} {\mathbf{X}}_i } } \right)^{ - 1} \left( {\sum\limits_{i = 1}^N {{\mathbf{X'}}_i {\mathbf{\Sigma }}^{ - 1} {\mathbf{Y}}_i } } \right),
$$
where ${\mathbf{X}}_i $ is the matrix of covariates for participant $i$. To derive $\hat \eta _5 $ from model~\eqref{lddbwp1} we only need the sixth row of 
$$
\left( {\sum\limits_{i = 1}^N {{\mathbf{X'}}_i {\mathbf{\Sigma }}^{ - 1} {\mathbf{X}}_i } } \right)^{ - 1},
$$
which we denote 
$$
\left[ {\left( {\sum\limits_{i = 1}^N {{\mathbf{X'}}_i {\mathbf{\Sigma }}^{ - 1} {\mathbf{X}}_i } } \right)^{ - 1} } \right]_{[6]} ,
$$
and then 
$$
\hat \eta _5  = \left[ {\left( {\sum\limits_{i = 1}^N {{\mathbf{X'}}_i {\mathbf{\Sigma }}^{ - 1} {\mathbf{X}}_i } } \right)^{ - 1} } \right]_{[6]} \left( {\sum\limits_{i = 1}^N {{\mathbf{X'}}_i {\mathbf{\Sigma }}^{ - 1} {\mathbf{Y}}_i } } \right),
$$
which we rewrite as 
$$
\hat \eta _5  = \left( {\sum\limits_{i = 1}^N {\left[ {\left( {\sum\limits_{i = 1}^N {{\mathbf{X'}}_i {\mathbf{\Sigma }}^{ - 1} {\mathbf{X}}_i } } \right)^{ - 1} } \right]_{[6]} {\mathbf{X'}}_i {\mathbf{\Sigma }}^{ - 1} {\mathbf{Y}}_i } } \right).
$$
Then, by calling 
$$
{\mathbf{c}}_\eta   = \left[ {\left( {\sum\limits_{i = 1}^N {{\mathbf{X'}}_i {\mathbf{\Sigma }}^{ - 1} {\mathbf{X}}_i } } \right)^{ - 1} } \right]_{[6]} {\mathbf{X'}}_i {\mathbf{\Sigma }}^{ - 1} 
$$
we have 
$$
\hat \eta _5  = \left( {\sum\limits_{i = 1}^N {{\mathbf{c}}_\eta  {\mathbf{Y}}_i } } \right).
$$
In Appendix~\ref{apvar3p1} we derived and expression for 
$$
\left( {\sum\limits_{i = 1}^N {{\mathbf{X'}}_i {\mathbf{\Sigma }}^{ - 1} {\mathbf{X}}_i } } \right)
$$
and from that we can derive 
\begin{multline*}
\left[ {\left( {\sum\limits_{i = 1}^N {{\mathbf{X'}}_i {\mathbf{\Sigma }}^{ - 1} {\mathbf{X}}_i } } \right)^{ - 1} } \right]_{[6]}  = 
 \frac{1} {{\det ({\mathbf{A}})p_e (1 - p_e )s}} \\
  \left( {\begin{array}{*{20}c}
   {p_e \sum\limits_{j = 0}^r {\sum\limits_{j' = 0}^r {jv_{jj'} } } ,} & {0,} & {\frac{{ - p_e }}
{s}\sum\limits_{j = 0}^r {\sum\limits_{j' = 0}^r {v_{jj'} } } ,} & { - \sum\limits_{j = 0}^r {\sum\limits_{j' = 0}^r {jv_{jj'} } ,} } & {0,} & {\frac{1}
{s}\sum\limits_{j = 0}^r {\sum\limits_{j' = 0}^r {v_{jj'} } } }  \\
 \end{array} } \right).
\end{multline*} 
For convenience, some terms can be rewritten in vector form. We define ${\mathbf{1}}$ as a $(r + 1) \times 1$ vector of ones, and ${\mathbf{t}}$ as a $(r + 1) \times 1$ matrix such that ${\mathbf{t'}} = (0,1,2, \ldots ,r)$, and then 
\begin{multline*}
\left[ {\left( {\sum\limits_{i = 1}^N {{\mathbf{X'}}_i {\mathbf{\Sigma }}^{ - 1} {\mathbf{X}}_i } } \right)^{ - 1} } \right]_{[6]}   = \frac{1}
{{\det ({\mathbf{A}})p_e (1 - p_e )s}} \\
\left( {\begin{array}{*{20}c}
   {p_e {\mathbf{t'\Sigma }}^{ - 1} {\mathbf{1}},} & {0,} & {\frac{{ - p_e }}
{s}{\mathbf{1'\Sigma }}^{ - 1} {\mathbf{1}},} & { - {\mathbf{t'\Sigma }}^{ - 1} {\mathbf{1}},} & {0,} & {\frac{1}
{s}{\mathbf{1'\Sigma }}^{ - 1} {\mathbf{1}}}  \\
 \end{array} } \right).
\end{multline*}
We can also derive 
$$
{\mathbf{X'}}_i {\mathbf{\Sigma }}^{ - 1}  = \left( {\begin{array}{*{20}c}
   1 &  \cdots  & 1 &  \cdots  & 1  \\
   {t_{i0} } &  \cdots  & {t_{i0} } &  \cdots  & {t_{i0} }  \\
   0 &  \cdots  & {sj} &  \cdots  & {sr}  \\
   {k_i } &  \cdots  & {k_i } &  \cdots  & {k_i }  \\
   {k_i t_{i0} } &  \cdots  & {k_i t_{i0} } &  \cdots  & {k_i t_{i0} }  \\
   0 &  \cdots  & {k_i sj} &  \cdots  & {k_i sr}  \\
 \end{array} } \right)\left( {\begin{array}{*{20}c}
   {v_{00} } &  \cdots  & {v_{0r} }  \\
    \vdots  &  \ddots  &  \vdots   \\
   {v_{r0} } &  \cdots  & {v_{rr} }  \\
 \end{array} } \right) = \left( {\begin{array}{*{20}c}
   {{\mathbf{1'\Sigma }}^{ - 1} }  \\
   {t_{i0} {\mathbf{1'\Sigma }}^{ - 1} }  \\
   {s{\mathbf{t'\Sigma }}^{ - 1} }  \\
   {k_i {\mathbf{1'\Sigma }}^{ - 1} }  \\
   {k_i t_{i0} {\mathbf{1'\Sigma }}^{ - 1} }  \\
   {sk_i {\mathbf{t'\Sigma }}^{ - 1} }  \\
 \end{array} } \right).
$$
Then, 
\begin{multline*}
{\mathbf{c}}_\eta   = \left[ {\left( {\sum\limits_{i = 1}^N {{\mathbf{X'}}_i {\mathbf{\Sigma }}^{ - 1} {\mathbf{X}}_i } } \right)^{ - 1} } \right]_{[6]} {\mathbf{X'}}_i {\mathbf{\Sigma }}^{ - 1}  = 
\frac{1} {{\det ({\mathbf{A}})p_e (1 - p_e )s}} \\
\left( {p_e {\mathbf{t'\Sigma }}^{ - 1} {\mathbf{11'\Sigma }}^{ - 1}  - p_e {\mathbf{t'\Sigma }}^{ - 1} {\mathbf{1'\Sigma }}^{ - 1} {\mathbf{1}} - k_i {\mathbf{t'\Sigma }}^{ - 1} {\mathbf{11'\Sigma }}^{ - 1}  + k_i {\mathbf{t'\Sigma }}^{ - 1} {\mathbf{1'\Sigma }}^{ - 1} {\mathbf{1}}} \right) \\
 = \frac{{( - p_e  + k_i )\left( {{\mathbf{1'\Sigma }}^{ - 1} {\mathbf{1}}} \right)}}
{{\det ({\mathbf{A}})p_e (1 - p_e )s}}{\mathbf{t'}}\left( {{\mathbf{\Sigma }}^{ - 1}  - {\mathbf{\Sigma }}^{ - 1} {\mathbf{1}}\left( {{\mathbf{1'\Sigma }}^{ - 1} {\mathbf{1}}} \right)^{ - 1} {\mathbf{1'\Sigma }}^{ - 1} } \right).
\end{multline*}
Now let us move to model~\eqref{ldddiffp1}. Define the $r \times (r + 1)$ matrix 
$$
{\mathbf{\Delta }} = \left( {\begin{array}{*{20}c}
   { - 1} & 1 & 0 &  \cdots  &  \cdots  & 0  \\
   0 & { - 1} & 1 & 0 &  \cdots  & 0  \\
   0 & 0 &  \ddots  &  \ddots  &  \ddots  &  \vdots   \\
    \vdots  & {} &  \ddots  &  \ddots  &  \ddots  & 0  \\
   0 &  \cdots  &  \cdots  & 0 & { - 1} & 1  \\

 \end{array} } \right).
$$
Note that ${\mathbf{\Delta Y}}_i $ contains the differences of the response from one visit to the next, so ${\mathbf{\Delta Y}}_i $ is the response variable in model (2.10). The covariance matrix of the response for model~\eqref{ldddiffp1} will then be ${\mathbf{\Delta \Sigma \Delta '}}$. Let us call ${\mathbf{Z}}$ the $r \times 2$ matrix of covariates for model~\eqref{ldddiffp1}, 
$$
{\mathbf{Z'}} = \left( {\begin{array}{*{20}c}
   1 &  \cdots  & 1  \\
   {k_i } &  \cdots  & {k_i }  \\
 \end{array} } \right)
$$
and ${\mathbf{X}}_{[3,6]} $ a $(r + 1) \times 2$ matrix containing the third and sixth column of ${\mathbf{X}}_i $ from model~\eqref{lddbwp1},
$$
{\mathbf{X'}}_{[3,6]}  = \left( {\begin{array}{*{20}c}
   0 & {} & {sj} & {} & {sr}  \\
   0 & {} & {k_i sj} & {} & {k_i sr}  \\
 \end{array} } \right).
$$
Then, it can be noted that $\frac{1} {s}{\mathbf{\Delta X}}_{[3,6]}  = {\mathbf{Z}}$. Therefore, the GLS estimate of $\lambda _1 $ can be written as 
\begin{multline*}
\hat \lambda _1  = \left[ {\left( {\sum\limits_{i = 1}^N {{\mathbf{Z'}}_i \left( {{\mathbf{\Delta \Sigma \Delta '}}} \right)^{ - 1} {\mathbf{Z}}_i } } \right)^{ - 1} } \right]_{[2]} \left( {\sum\limits_{i = 1}^N {{\mathbf{Z'}}_i \left( {{\mathbf{\Delta \Sigma \Delta '}}} \right)^{ - 1} {\mathbf{\Delta Y}}_i } } \right) =\\
\left( {\frac{1}
{s}\sum\limits_{i = 1}^N {\left[ {\left( {\frac{1}
{{s^2 }}\sum\limits_{i = 1}^N {\left( {{\mathbf{\Delta X}}_{[3,6]} } \right)^\prime  \left( {{\mathbf{\Delta \Sigma \Delta '}}} \right)^{ - 1} {\mathbf{\Delta X}}_{[3,6]} } } \right)^{ - 1} } \right]_{[2]} \left( {{\mathbf{\Delta X}}_{[3,6]} } \right)^\prime  \left( {{\mathbf{\Delta \Sigma \Delta '}}} \right)^{ - 1} {\mathbf{\Delta Y}}_i } } \right) \\
= \left( {\sum\limits_{i = 1}^N {s\left[ {\left( {\sum\limits_{i = 1}^N {{\mathbf{X'}}_{[3,6]} {\mathbf{\Delta '}}\left( {{\mathbf{\Delta \Sigma \Delta '}}} \right)^{ - 1} {\mathbf{\Delta X}}_{[3,6]} } } \right)^{ - 1} } \right]_{[2]} {\mathbf{X'}}_{[3,6]} {\mathbf{\Delta '}}\left( {{\mathbf{\Delta \Sigma \Delta '}}} \right)^{ - 1} {\mathbf{\Delta Y}}_i } } \right) \\
= \sum\limits_{i = 1}^N {{\mathbf{c}}_\lambda  {\mathbf{Y}}_i }.
\end{multline*}
Now, 
\begin{multline*}
\left( {\sum\limits_{i = 1}^N {{\mathbf{X'}}_{[3,6]} {\mathbf{\Delta '}}\left( {{\mathbf{\Delta \Sigma \Delta '}}} \right)^{ - 1} {\mathbf{\Delta X}}_{[3,6]} } } \right) \\
= s^2 \left( {\begin{array}{*{20}c}
   1 &  \cdots  & 1  \\
   {k_i } &  \cdots  & {k_i }  \\
 \end{array} } \right)_{(2 \times r)} \left( {{\mathbf{\Delta \Sigma \Delta '}}} \right)^{ - 1} _{(r \times r)} \left( {\begin{array}{*{20}c}
   1 & {k_i }  \\
    \vdots  &  \vdots   \\
   1 & {k_i }  \\
 \end{array} } \right)_{(r \times 2)} \\
  = s^2 \left( {{\mathbf{t'\Delta '}}\left( {{\mathbf{\Delta \Sigma \Delta '}}} \right)^{ - 1} {\mathbf{\Delta t}}} \right)\left( {\begin{array}{*{20}c}
   1 & {p_e }  \\
   {p_e } & {p_e }  \\
 \end{array} } \right),
\end{multline*}
so 
\begin{multline*}
\left[ {\left( {\sum\limits_{i = 1}^N {{\mathbf{X'}}_{[3,6]} {\mathbf{\Delta '}}\left( {{\mathbf{\Delta \Sigma \Delta '}}} \right)^{ - 1} {\mathbf{\Delta X}}_{[3,6]} } } \right)^{ - 1} } \right]_{[2]}  =\\
 \frac{1}
{{p_e (1 - p_e )s^2 \left( {{\mathbf{t'\Delta '}}\left( {{\mathbf{\Delta \Sigma \Delta '}}} \right)^{ - 1} {\mathbf{\Delta t}}} \right)}}\left( {\begin{array}{*{20}c}
   { - p_e } & 1  \\
 \end{array} } \right).
\end{multline*}
Now, by property B.3.5 of Seber (1984, page 536), 
$$
{\mathbf{\Delta '}}\left( {{\mathbf{\Delta \Sigma \Delta '}}} \right)^{ - 1} {\mathbf{\Delta }} = {\mathbf{\Sigma }}^{ - 1}  - {\mathbf{\Sigma }}^{ - 1} {\mathbf{1}}\left( {{\mathbf{1'\Sigma }}^{ - 1} {\mathbf{1}}} \right)^{ - 1} {\mathbf{1'\Sigma }}^{ - 1}.
$$
Then, 
$$
\frac{1}
{{\left( {{\mathbf{t'\Delta '}}\left( {{\mathbf{\Delta \Sigma \Delta '}}} \right)^{ - 1} {\mathbf{\Delta t}}} \right)}} = \frac{1}
{{\left( {{\mathbf{t'}}\left( {{\mathbf{\Sigma }}^{ - 1}  - {\mathbf{\Sigma }}^{ - 1} {\mathbf{1}}\left( {{\mathbf{1'\Sigma }}^{ - 1} {\mathbf{1}}} \right)^{ - 1} {\mathbf{1'\Sigma }}^{ - 1} } \right){\mathbf{t}}} \right)}}
$$
and with some algebra this expression equals 
$$
\frac{{\left( {{\mathbf{1'\Sigma }}^{ - 1} {\mathbf{1}}} \right)}}
{{\det ({\mathbf{A}})}}.
$$
So 
$$
\left[ {\left( {\sum\limits_{i = 1}^N {{\mathbf{X'}}_{[3,6]} {\mathbf{\Delta '}}\left( {{\mathbf{\Delta \Sigma \Delta '}}} \right)^{ - 1} {\mathbf{\Delta X}}_{[3,6]} } } \right)^{ - 1} } \right]_{[2]}  = \frac{{\left( {{\mathbf{1'\Sigma }}^{ - 1} {\mathbf{1}}} \right)}}
{{p_e (1 - p_e )s^2 \det ({\mathbf{A}})}}\left( {\begin{array}{*{20}c}
   { - p_e } & 1  \\
 \end{array} } \right).
$$
Now we need to derive ${\mathbf{X'}}_{[3,6]} {\mathbf{\Delta '}}\left( {{\mathbf{\Delta \Sigma \Delta '}}} \right){\mathbf{\Delta }}$, and by using Seber's property again we have 
$$
{\mathbf{X'}}_{[3,6]} {\mathbf{\Delta '}}\left( {{\mathbf{\Delta \Sigma \Delta '}}} \right){\mathbf{\Delta }} = {\mathbf{X'}}_{[3,6]} \left( {{\mathbf{\Sigma }}^{ - 1}  - {\mathbf{\Sigma }}^{ - 1} {\mathbf{1}}\left( {{\mathbf{1'\Sigma }}^{ - 1} {\mathbf{1}}} \right)^{ - 1} {\mathbf{1'\Sigma }}^{ - 1} } \right).
$$
So,
\begin{multline*}
{\mathbf{c}}_\lambda   = s\left[ {\left( {\sum\limits_{i = 1}^N {{\mathbf{X'}}_{[3,6]} {\mathbf{\Delta '}}\left( {{\mathbf{\Delta \Sigma \Delta '}}} \right)^{ - 1} {\mathbf{\Delta X}}_{[3,6]} } } \right)^{ - 1} } \right]_{[2]} {\mathbf{X'}}_{[3,6]} {\mathbf{\Delta '}}\left( {{\mathbf{\Delta \Sigma \Delta '}}} \right){\mathbf{\Delta }} = \\
 = \frac{{\left( {{\mathbf{1'\Sigma }}^{ - 1} {\mathbf{1}}} \right)}}
{{\det ({\mathbf{A}})sp_e (1 - p_e )}}\left( {\begin{array}{*{20}c}
   { - p_e } & 1  \\
 \end{array} } \right){\mathbf{X'}}_{[3,6]} \left( {{\mathbf{\Sigma }}^{ - 1}  - {\mathbf{\Sigma }}^{ - 1} {\mathbf{1}}\left( {{\mathbf{1'\Sigma }}^{ - 1} {\mathbf{1}}} \right)^{ - 1} {\mathbf{1'\Sigma }}^{ - 1} } \right) \\
 = \frac{{\left( {{\mathbf{1'\Sigma }}^{ - 1} {\mathbf{1}}} \right)}}
{{\det ({\mathbf{A}})p_e (1 - p_e )}}\left( {\begin{array}{*{20}c}
   { - p_e } & 1  \\
 \end{array} } \right)\left( {\begin{array}{*{20}c}
   {{\mathbf{t'}}}  \\
   {k_i {\mathbf{t'}}}  \\
 \end{array} } \right)\left( {{\mathbf{\Sigma }}^{ - 1}  - {\mathbf{\Sigma }}^{ - 1} {\mathbf{1}}\left( {{\mathbf{1'\Sigma }}^{ - 1} {\mathbf{1}}} \right)^{ - 1} {\mathbf{1'\Sigma }}^{ - 1} } \right) \\
 = \frac{{\left( { - p_e  + k_i } \right)\left( {{\mathbf{1'\Sigma }}^{ - 1} {\mathbf{1}}} \right)}}
{{\det ({\mathbf{A}})p_e (1 - p_e )}}{\mathbf{t'}}\left( {{\mathbf{\Sigma }}^{ - 1}  - {\mathbf{\Sigma }}^{ - 1} {\mathbf{1}}\left( {{\mathbf{1'\Sigma }}^{ - 1} {\mathbf{1}}} \right)^{ - 1} {\mathbf{1'\Sigma }}^{ - 1} } \right)
\end{multline*}
and we can observe that ${\mathbf{c}}_\lambda   = \frac{{{\mathbf{c}}_\eta  }}
{s} $ and therefore $s\hat \lambda _1  = \hat \eta _5 $ and $s^2 Var\left( {\hat \lambda _1 } \right) = Var\left( {\hat \eta _5 } \right)$.

\section{Bias and/or inefficiency of the ANCOVA, SLANC, SLAIN tests under CS}
\label{apancova}

These tests \cite{Frison:1992,Frison:1997} have the form 
$$
T = \frac{{N{\kern 1pt} p_e (1 - p_e )\left( {\bar S_1  - \bar S_0 } \right)^2 }}
{{{\mathbf{c'\Sigma c}}}},
$$
where $\bar S_k $ is exposure group $k$'s mean, $k = (0,1)$, of a summary measure, $S_i $, that is a linear combination of the repeated measures of each subject, $S_i  = {\mathbf{c'Y}}_i $. The vector ${\mathbf{c'}}$ defines the summary measures, which could be the within-subject mean of the repeated measures, the within-subject slope, or ANCOVA, SLANC and SLAIN. Let $n_k $ be the number of participants in exposure group $k$. Then, 
$$
\bar S_k  = \frac{{\sum\limits_{i = 1}^N {S_i I\left\{ {k_i  = k} \right\}} }}
{{n_k }} = {\mathbf{c'}}\left( {\frac{{\sum\limits_{i = 1}^N {{\mathbf{Y}}_i I\left\{ {k_i  = k} \right\}} }}
{{n_k }}} \right) = {\mathbf{c'\bar Y}}_k,
$$
where $I\left\{ {k_i  = k} \right\}$ is an indicator variable that takes the value one when $k_i  = k$ and zero otherwise, and ${\mathbf{\bar Y}}_k $ is the $(r + 1) \times 1$ vector of sample means for each time in group k. Thus, 
$$
T = \frac{{N{\kern 1pt} p_e (1 - p_e )\left( {{\mathbf{c'}}\left( {{\mathbf{\bar Y}}_1  - {\mathbf{\bar Y}}_0 } \right)} \right)^2 }}
{{{\mathbf{c'\Sigma c}}}}.
$$
The summary measure approach is appropriate when all subjects in the two exposure groups are observed in the same set of time points, i.e. when $V(t_0 ) = 0$. Otherwise, exposure group and time can be correlated and the summary measure approach would produce biased estimates of the effect. Clearly, $\mathbb{E}\left[ {\bar S_k } \right] = {\mathbf{c'\mu }}_k $, where ${\mathbf{\mu }}_k $ is the vector of true means for each time in group k, and $\mathbb{E}\left[ {\bar S_1  - \bar S_0 } \right] = {\mathbf{c'}}({\mathbf{\mu }}_1  - {\mathbf{\mu }}_0 )$. If 
$$
{\mathbf{c'}} = \left( {\frac{1}
{{r + 1}}, \cdots ,\frac{1}
{{r + 1}}} \right),
$$
we are testing the equality of the means of the two groups. The goal is to choose ${\mathbf{c'}}$ so that $\mathbb{E}\left( {T|H_0 } \right) = 0$ (valid) and for which the power of $T$ is at its maximum possible under $H_A $ (efficient). In this paper, interest is in two hypotheses which commonly arise in longitudinal studies, CMD and LDD. The CMD hypothesis specifies that the mean group differences are constant over time, or, equivalently, that the two response profiles are parallel, i.e. $({\mathbf{\mu }}_1  - {\mathbf{\mu }}_0 ) = p_1 \mu _{00} $. The LDD hypothesis specifies that the group mean differences are a linear function of time, or, equivalently, that there is a linear interaction between exposure and time, i.e. 
$$
{\mathbf{\mu }}_1  - {\mathbf{\mu }}_0  = (p_1  + \frac{{p_2 p_3 }}
{\tau }t_j )\mu _{00}.
$$
The parameter of interest under CMD is $p_1 $, and one wants to test $H_0 :p_1  = 0
$ vs. $H_A :p_1  \ne 0$. The parameter of interest under LDD is $p_3 $, and one wants to test $H_0 :p_3  = 0$ vs. $H_A :p_3  \ne 0$. To construct the relevant test statistic in each case, we need to identify a vector ${\mathbf{c'}}$ such that $H_0 :{\mathbf{c'}}({\mathbf{\mu }}_1  - {\mathbf{\mu }}_0 ) = 0$ vs. $H_A :{\mathbf{c'}}({\mathbf{\mu }}_1  - {\mathbf{\mu }}_0 ) \ne 0$ provides a test statistic of form $T$ that is the most powerful under the alternative hypothesis.

\subsection{Unbiasedness of the ANCOVA, SLAIN and SLANC test statistics under CMD, CS and $V(t_0 ) = 0$}
\label{apancovabias}

Under CMD, the expected value of the numerator of test statistics of the form discussed above, which include ANCOVA, SLAIN and SLANC, is 
$$
\mathbb{E}\left[ {{\mathbf{c'}}({\mathbf{\bar Y}}_1  - {\mathbf{\bar Y}}_0 )|\,H_0 } \right] = {\mathbf{c'}}\mathbb{E}\left[ {({\mathbf{\bar Y}}_1  - {\mathbf{\bar Y}}_0 )|\,H_0 } \right] = {\mathbf{c'}}\left( {\begin{array}{*{20}c}
   {p_1 \mu _{00} ,} &  \cdots  & {,p_1 \mu _{00} }  \\
 \end{array} } \right).
$$
Under $H_0 $, $p_1  = 0$ so $\mathbb{E}\left[ {{\mathbf{c'}}({\mathbf{\bar Y}}_1  - {\mathbf{\bar Y}}_0 )|\,H_0 } \right]$ is 0 and all vectors ${\mathbf{c'}}
$ produce unbiased estimators, including those given for ANCOVA, SLANC and SLAIN, for which the vectors ${\mathbf{c'}}$ under CS are 
$$
\begin{gathered}
{\mathbf{c'}} = \left( { - \rho ,\frac{1}
{r}, \cdots ,\frac{1}
{r}} \right), {\mathbf{c'}} = \frac{6}
{{r(r + 1)(r + 2)}}\left( { - \rho ,\;2 - r, \cdots ,\;2j - r, \cdots ,\;r} \right)
, \\
c_j  = \frac{{12j + 6\rho r(2j - r - 1)}}
{{r(r + 1)\left[ {\rho r(r - 1) + 2(2r + 1)} \right]}},
\end{gathered}
$$
for ANCOVA, SLANC and SLAIN, respectively \cite{Frison:1997}.

\subsection{Inefficiency of the ANCOVA, SLAIN and SLANC tests under CMD, CS and $V(t_0 ) = 0$}
\label{apancovainef}

\subsubsection{Inefficiency of the ANCOVA test under CMD and CS and $V(t_0 ) = 0$}
\label{apancovainef1}

\citeasnoun{Frison:1992} showed that, under CS, ANCOVA is the most powerful test for the CMD alternative hypothesis when the two groups have the same response at baseline, as would be the case in expectation in a randomized clinical trial, and when the post-baseline measures have a constant difference. However, in observational studies, where baseline has no special significance, the difference at baseline is assumed to be the same as the difference at the other time points. We will show here that the ANCOVA test is less powerful than the test we discussed in this manuscript, based on the GLS estimator, which is known to be the best linear unbiased estimator and therefore produces the most powerful test among those unbiased. 
	
To obtain the GLS estimator and its corresponding test statistic, we fit the model $Y_{ij}  = \beta _0  + \beta _1 k_i $, similarly to equation~\eqref{cmdp1} but without including the effect of time. Because $V(t_0 ) = 0$, exposure and time are independent and inclusion or exclusion of the time variable does not affect the estimation of $\beta _1 $, the parameter of interest under CMD. Reparameterizing using $\alpha _0  = \beta _0 $, $\alpha _1  = \beta _0  + \beta _1 $, and $\beta _1  = \alpha _1  - \alpha _0$, the model becomes $Y_{ij}  = \alpha _0 (1 - k_i ) + \alpha _1 k_i $. We derive the ${\mathbf{c}}$ vector for the test based upon the GLS estimator, and show that it is not equal to the ${\mathbf{c}}$ vector for the ANCOVA test. The design matrix for subject $i$ will contain a column of ones and a column of zeros if subject $i$ is unexposed, and a column of zeros and a column of ones if subject $i$ is unexposed. The GLS estimator for $\left( {\hat \alpha _0 ,\hat \alpha _1 } \right)$ is 
$$
\left( {\sum\limits_{i = 1}^N {\left( {{\mathbf{X'}}_i {\mathbf{\Sigma }}^{ - 1} {\mathbf{X}}_i } \right)} } \right)^{ - 1} \left( {\sum\limits_{i = 1}^N {{\mathbf{X'}}_i {\mathbf{\Sigma }}^{ - 1} {\mathbf{Y}}_i } } \right).
$$
Since, 
$$
\left( {\sum\limits_{i = 1}^N {\left( {{\mathbf{X'}}_i {\mathbf{\Sigma }}^{ - 1} {\mathbf{X}}_i } \right)} } \right)^{ - 1}  = \frac{1}
{{N\sum\limits_{j = 0}^r {\sum\limits_{j' = 0}^r {v_{jj'} } } }}\left( {\begin{array}{*{20}c}
   {\frac{1}
{{1 - p_e }}} & 0  \\
   0 & {\frac{1}
{{p_e }}}  \\
 \end{array} } \right),
$$
where $v_{jj'} $ is the $(j,j'){\text{th}}$ element of ${\mathbf{\Sigma }}^{ - 1} 
$, then 
$$
\hat \alpha _0  = \frac{1}
{{N(1 - p_e )\sum\limits_{j = 0}^r {\sum\limits_{j' = 0}^r {v_{jj'} } } }}\sum\limits_{i = 1}^{N(1 - p_e )} {\left( {1, \ldots ,1} \right){\mathbf{\Sigma }}^{ - 1} {\mathbf{Y}}_i }  = \frac{1}
{{\sum\limits_{j = 0}^r {\sum\limits_{j' = 0}^r {v_{jj'} } } }}\left( {1, \ldots ,1} \right){\mathbf{\Sigma }}^{ - 1} {\mathbf{\bar Y}}_0 
$$
and 
$$
\hat \alpha _1  = \frac{1}
{{Np_e \sum\limits_{j = 0}^r {\sum\limits_{j' = 0}^r {v_{jj'} } } }}\sum\limits_{i = N(1 - p_e ) + 1}^N {\left( {1, \ldots ,1} \right){\mathbf{\Sigma }}^{ - 1} {\mathbf{Y}}_i }  = \frac{1}
{{\sum\limits_{j = 0}^r {\sum\limits_{j' = 0}^r {v_{jj'} } } }}\left( {1, \ldots ,1} \right){\mathbf{\Sigma }}^{ - 1} {\mathbf{\bar Y}}_1,
$$
and 
$$
\beta _1  = \frac{1}
{{\sum\limits_{j = 0}^r {\sum\limits_{j' = 0}^r {v_{jj'} } } }}\left( {1, \ldots ,1} \right){\mathbf{\Sigma }}^{ - 1} {\mathbf{\bar Y}}_1  - \frac{1}
{{\sum\limits_{j = 0}^r {\sum\limits_{j' = 0}^r {v_{jj'} } } }}\left( {1, \ldots ,1} \right){\mathbf{\Sigma }}^{ - 1} {\mathbf{\bar Y}}_0.
$$
Therefore, 
$$
{\mathbf{c'}}_{GLS}  = \frac{1}
{{\sum\limits_{j = 0}^r {\sum\limits_{j' = 0}^r {v_{jj'} } } }}\left( {1, \ldots ,1} \right){\mathbf{\Sigma }}^{ - 1}. 
$$
This coincides with the result of \citeasnoun{Frison:1997} that the optimal ${\mathbf{c'}}$ is proportional $({\mathbf{\mu }}_1  - {\mathbf{\mu }}_0 )'{\mathbf{\Sigma }}^{ - 1} $. Since ANCOVA has a different vector ${\mathbf{c'}}$, ANCOVA could be, at best, as powerful as the GLS under CS.

Under  CMD and with a CS covariance specifically, we show that the test based upon the GLS estimator is explicitly more powerful than the ANCOVA test. The non-centrality parameter (NCP) for the test $T$ is 
$$
\lambda  = \frac{{\left[ {{\mathbf{c'}}({\mathbf{\mu }}_1  - {\mathbf{\mu }}_0 )} \right]^2 }}
{{{\mathbf{c'\Sigma c}}}}.
$$
For ANCOVA, the numerator of $\lambda $ is 
$$
\left[ {{\mathbf{c'}}({\mathbf{\mu }}_1  - {\mathbf{\mu }}_0 )} \right]^2  = \left[ {\left( { - \rho ,\frac{1}
{r}, \cdots ,\frac{1}
{r}} \right)\left( {\begin{array}{*{20}c}
   {p_1 \mu _{00} } &  \cdots  & {p_1 \mu _{00} }  \\
 \end{array} } \right)^\prime  } \right]^2  = 
\left( {p_1 \mu _{00} } \right)^2 \left( {1 - \rho } \right)^2. 
$$
The denominator of $\lambda _{ANCOVA} $ is 
\begin{multline*}
{\mathbf{c'\Sigma c}} = \sigma ^2 \left( { - \rho ,\frac{1}
{r}, \cdots ,\frac{1}
{r}} \right)\left( {\begin{array}{*{20}c}
   1 & \rho  &  \cdots  & \rho   \\
   \rho  & 1 &  \ddots  &  \vdots   \\
    \vdots  &  \ddots  &  \ddots  & \rho   \\
   \rho  &  \cdots  & \rho  & 1  \\
 \end{array} } \right)\left( { - \rho ,\frac{1}
{r}, \cdots ,\frac{1}
{r}} \right)^\prime   \\
= \sigma ^2 \left[ { - \rho ^2  + \frac{1}
{r}\left( {1 + (r - 1)\rho } \right)} \right]
\end{multline*}
The vector ${\mathbf{c'}}$ for the GLS approach is 
\begin{multline*}
{\mathbf{c'}} = \frac{{\sigma ^2 (1 + r\rho )}}
{{(r + 1)}}\left( {\frac{1}
{{\sigma ^2 (1 + r\rho )}},\frac{1}
{{\sigma ^2 (1 + r\rho )}}, \cdots ,\frac{1}
{{\sigma ^2 (1 + r\rho )}}} \right)  \\
= \left( {\frac{1}
{{r + 1}},\frac{1}
{{r + 1}},...,\frac{1}
{{r + 1}}} \right).
\end{multline*}
Thus, the numerator of the $\lambda _{GLS} $ is 
$$
\left[ {{\mathbf{c'}}({\mathbf{\mu }}_1  - {\mathbf{\mu }}_0 )} \right]^2  = \left[ {\left( {\frac{1}
{{r + 1}},\frac{1}
{{r + 1}}, \cdots ,\frac{1}
{{r + 1}}} \right)\left( {\begin{array}{*{20}c}
   {p_1 \mu _{00} ,} &  \cdots  & {,p_1 \mu _{00} }  \\
 \end{array} } \right)^\prime  } \right]^2 
 = \left( {p_1 \mu _{00} } \right)^2. 
$$
The denominator of $\lambda _{GLS} $ is 
\begin{multline*}
{\mathbf{c'\Sigma c}} = \sigma ^2 \left( {\frac{1}
{{r + 1}},\frac{1}
{{r + 1}}, \cdots ,\frac{1}
{{r + 1}}} \right)\left( {\begin{array}{*{20}c}
   1 & \rho  &  \cdots  & \rho   \\
   \rho  & 1 &  \ddots  &  \vdots   \\
    \vdots  &  \ddots  &  \ddots  & \rho   \\
   \rho  &  \cdots  & \rho  & 1  \\
 \end{array} } \right)\left( {\frac{1}
{{r + 1}},\frac{1}
{{r + 1}}, \cdots ,\frac{1}
{{r + 1}}} \right)^\prime  \\
 = \frac{{r\rho  + 1}}
{{r + 1}}.
\end{multline*}
The ratio of NCPs is 
$$
\frac{{\lambda _{GLS} }}
{{\lambda _{ANCOVA} }} = \frac{{\frac{{\left( {p_1 \mu _{00} } \right)^2 }}
{{\left( {\frac{{r\rho  + 1}}
{{r + 1}}} \right)}}}}
{{\frac{{\left( {p_1 \mu _{00} } \right)^2 \left( {1 - \rho } \right)^2 }}
{{ - \rho ^2  + \frac{1}
{r}\left( {1 + (r - 1)\rho } \right)}}}} = \frac{{r + 1}}
{{r(1 - \rho )}} > 1,
$$
proving that a test statistic based on the GLS approach is more powerful than ANCOVA under CMD with a  CS covariance matrix.

\subsubsection{Inefficiency of the SLANC and SLAIN tests under CMD }
\label{apancovainef2}

The vector ${\mathbf{c'}}$ for SLANC is defined by 
$$
{\mathbf{c'}} = \frac{6}
{{r(r + 1)(r + 2)}}\left( { - \rho ,\;2 - r, \cdots ,\;2j - r, \cdots ,\;r} \right),
$$
and for SLAIN by 
$$
c_j  = \frac{{12j + 6\rho r(2j - r - 1)}}
{{r(r + 1)\left[ {\rho r(r - 1) + 2(2r + 1)} \right]}}
$$
under CS \cite{Frison:1997}.  Since the vector ${\mathbf{c'}}$ are not equal to ${\mathbf{c'}}_{GLS}$, they are both inefficient. 

\subsection{Biasedness of the ANCOVA, SLAIN and SLANC test statistics under LDD }
\label{apancovaldd}

Under LDD, the expected value of the numerator of the ANCOVA test statistic is \begin{multline*}
\mathbb{E}\left[ {{\mathbf{c'}}({\mathbf{\bar Y}}_1  - {\mathbf{\bar Y}}_0 )|\,H_0 } \right] = {\mathbf{c'}}\mathbb{E}\left[ {({\mathbf{\bar Y}}_1  - {\mathbf{\bar Y}}_0 )|\,H_0 } \right] = \\
 = \left. {{\mathbf{c'}}\left( {\begin{array}{*{20}c}
   {(p_1  + \frac{{p_2 p_3 }}
{\tau }t_0 )\mu _{00} ,} &  \cdots  & {,(p_1  + \frac{{p_2 p_3 }}
{\tau }t_r )\mu _{00} }  \\
 \end{array} } \right)^\prime  } \right|_{H_0 :p_3  = 0}  \\
 = {\mathbf{c'}}\left( {\begin{array}{*{20}c}
   {p_1 \mu _{00} ,} &  \cdots  & {,p_1 \mu _{00} }  \\
 \end{array} } \right)^\prime   = 
\left( { - \rho ,\frac{1}
{r}, \cdots ,\frac{1}
{r}} \right)\left( {\begin{array}{*{20}c}
   {p_1 \mu _{00} ,} &  \cdots  & {,p_1 \mu _{00} }  \\
 \end{array} } \right)^\prime   = \\
 p_1 \mu _{00} \left( {1 - \rho } \right) \ne 0.
\end{multline*}
The expected value of the numerator of the  SLANC test statistic is
\begin{multline*}
\mathbb{E}\left[ {{\mathbf{c'}}({\mathbf{\bar Y}}_1  - {\mathbf{\bar Y}}_0 )|\,H_0 } \right] = {\mathbf{c'}}\mathbb{E}\left[ {({\mathbf{\bar Y}}_1  - {\mathbf{\bar Y}}_0 )|\,H_0 } \right]  \\ 
 = \left. {{\mathbf{c'}}\left( {\begin{array}{*{20}c}
   {(p_1  + \frac{{p_2 p_3 }}
{\tau }t_0 )\mu _{00} ,} &  \cdots  & {,(p_1  + \frac{{p_2 p_3 }}
{\tau }t_r )\mu _{00} }  \\
 \end{array} } \right)^\prime  } \right|_{H_0 :p_3  = 0} \\
  = {\mathbf{c'}}\left( {\begin{array}{*{20}c}
   {p_1 \mu _{00} ,} &  \cdots  & {,p_1 \mu _{00} }  \\
 \end{array} } \right)^\prime   = p_1 \mu _{00} {\mathbf{c'}}\left( {1, \ldots ,1} \right)\\
  = \frac{{6p_1 \mu _{00} }}
{{r(r + 1)(r + 2)}}\left( { - \rho  + \sum\limits_{j = 1}^r {\left( {2j - r} \right)} } \right) = \frac{{6p_1 \mu _{00} }}
{{r(r + 1)(r + 2)}}\left( {r - \rho } \right) \ne 0
\end{multline*}

The expected value of the numerator of the SLAIN test statistic is
\begin{multline*}
\mathbb{E}\left[ {{\mathbf{c'}}({\mathbf{\bar Y}}_1  - {\mathbf{\bar Y}}_0 )|\,H_0 } \right] = p_1 \mu _{00} {\mathbf{c'}}\left( {\begin{array}{*{20}c}
   1 &  \cdots  & 1  \\
 \end{array} } \right) \\
 = p_1 \mu _{00} \sum\limits_{j = 0}^r {\frac{{12j + 6\rho r(2j - r - 1)}}
{{r(r + 1)\left[ {\rho r(r - 1) + 2(2r + 1)} \right]}}} \\
 = \frac{{p_1 \mu _{00} }}
{{r(r + 1)\left[ {\rho r(r - 1) + 2(2r + 1)} \right]}}\sum\limits_{j = 0}^r {12j + 6\rho r(2j - r - 1)} \\
= \frac{{6p_1 \mu _{00} (1 - \rho )}}
{{\left[ {\rho r(r - 1) + 2(2r + 1)} \right]}} \ne 0
\end{multline*}

Thus, ANCOVA, SLAIN  and SLANC are all biased under the null and therefore not valid in observational studies under LDD.

\section{Proof that two-stage and GLS are equivalent approaches under CS or RS for $V(t_0 ) = 0$}
\label{ap2stage}

In the setting where all subjects are observed at the same set of time points, this appendix will proof:
\begin{itemize}
\item[(i)] {That the estimator of the difference of the rates of change in the two exposure groups obtained using the summary measure (two-stage) approach is algebraically equivalent to the estimator of $\gamma _3 $ obtained from fitting model~\eqref{lddp1} by OLS.}
\item[(ii)]{That when the covariance matrix ${\mathbf{\Sigma }}_i  = {\mathbf{\Sigma }}$ has a CS or RS structure, the estimators from model~\eqref{lddp1} obtained by OLS and GLS are algebraically equivalent. Given (i), this implies that the estimator from the summary measure approach is algebraically equivalent to the GLS estimator. We also show that this is not the case for DEX.} 
\end{itemize}

Given (i) and (ii), since the estimators from the summary measure (two-stage) approach, and GLS are the same linear combination of $\left( {{\mathbf{\bar Y}}_1  - {\mathbf{\bar Y}}_0 } \right)$, ${\mathbf{d'}}\left( {{\mathbf{\bar Y}}_1  - {\mathbf{\bar Y}}_0 } \right)$, once we assume a covariance structure for $Var\left[ {{\mathbf{Y}}_i |{\mathbf{X}}_i } \right] = {\mathbf{\Sigma }}_i $, the test statistic for the two methods is also equivalent and equal to 
$$
T = \frac{{{\mathbf{d'}}\left( {{\mathbf{\bar Y}}_1  - {\mathbf{\bar Y}}_0 } \right)}}
{{\sqrt {Var\left( {{\mathbf{d'}}\left( {{\mathbf{\bar Y}}_1  - {\mathbf{\bar Y}}_0 } \right)} \right)} }},
$$
where
\begin{multline*}
Var\left( {{\mathbf{d'}}\left( {{\mathbf{\bar Y}}_1  - {\mathbf{\bar Y}}_0 } \right)} \right) = {\mathbf{d'}}Var\left( {{\mathbf{\bar Y}}_1  - {\mathbf{\bar Y}}_0 } \right){\mathbf{d}} \\
= {\mathbf{d'}}\left( {\frac{1}
{{Np_e }}Var\left( {{\mathbf{Y}}_{i,k_i  = 1} } \right) + \frac{1}
{{N(1 - p_e )}}Var\left( {{\mathbf{Y}}_{i,k_i  = 0} } \right)} \right){\mathbf{d}} = \frac{{{\mathbf{d'\Sigma d}}}}
{{Np_e (1 - p_e )}}.
\end{multline*}

\subsection*{Proof of (i)}
\label{ap2stagei}

\subsubsection*{Summary measure (two-stage) approach}

Let ${\mathbf{Z}}_i $ be a $(r + 1) \times 2$ matrix that contains a column of ones and the column of times for participant $i$. Since all subjects are observed at the same set of time points then ${\mathbf{Z}}_i  = {\mathbf{Z}}$. Here, the summary measure is the subject-specific OLS slope associated with time from the regression of ${\mathbf{Y}}_i $ on ${\mathbf{Z}}_i  = {\mathbf{Z}}$. Let us call ${\mathbf{\hat \beta }}_i $, $i = 1, \ldots ,N$, the $(2 \times 1)$ vector containing the subject-specific intercept and slope of the regression, where ${\mathbf{\hat \beta }}_i  = \left( {{\mathbf{Z'Z}}} \right)^{ - 1} {\mathbf{Z'Y}}_i 
$. The subject-specific intercepts and slopes are averaged in each exposure group as follows, 
$$
{\mathbf{\hat \beta }}_k  = \frac{{\sum\limits_{i = 1}^N {\left( {{\mathbf{Z'Z}}} \right)^{ - 1} {\mathbf{Z'}}} {\mathbf{Y}}_i \,I\left\{ {k_i  = k} \right\}}}
{{n_k }} = \left( {{\mathbf{Z'Z}}} \right)^{ - 1} {\mathbf{Z'}}\frac{{\sum\limits_{i = 1}^N {{\mathbf{Y}}_i I\left\{ {k_i  = k} \right\}} }}
{{n_k }} = \left( {{\mathbf{Z'Z}}} \right)^{ - 1} {\mathbf{Z'\bar Y}}_k,
$$
where $I\left\{ {k_i  = k} \right\}$ is an indicator variable that takes the value one when $k_i  = k$ and zero otherwise; $n_k $ is the number of participants in exposure group $k$, $k = 0,1$; and ${\mathbf{\bar Y}}_k $ is the average of ${\mathbf{Y}}_i $ in group $k$. Since we are interested in the second component of ${\mathbf{\hat \beta }}_k $, the slope associated with time, we define $\bar S_k  = \left( {\left( {{\mathbf{Z'Z}}} \right)^{ - 1} {\mathbf{Z'}}} \right)_{(2)} {\mathbf{\bar Y}}_k $, where the subscript (2) indicates the second row of the matrix $\left( {{\mathbf{Z'Z}}} \right)^{ - 1} {\mathbf{Z'}}$. We are interested in the difference, which is $\left( {\bar S_1  - \bar S_0 } \right) = \left( {\left( {{\mathbf{Z'Z}}} \right)^{ - 1} {\mathbf{Z'}}} \right)_{(2)} \left( {{\mathbf{\bar Y}}_1  - {\mathbf{\bar Y}}_0 } \right)$.

\subsubsection*{OLS approach}

With the OLS approach, we fit all the data at the same time, using 
$$
\mathbb{E}\left( {Y_{ij} |X_{ij} } \right) = \gamma _0  + \gamma _1 t_{ij}  + \gamma _2 k_i  + \gamma _3 \left( {t_{ij}  \times k_i } \right),
$$
and our interest in on $\gamma _3 $. Reparameterizing, we can fit model 
$$
\mathbb{E}\left( {Y_{ij} |X_{ij} } \right) = \gamma _0^* \left( {1 - k_i } \right) + \gamma _1^* \left( {1 - k_i } \right)t_{ij}  + \gamma _2^* k_i  + \gamma _3^* k_i t_{ij} ,
$$
and our parameter of interest is now $\gamma _3  = \gamma _3^*  - \gamma _1^* $. The OLS estimator of the latter model can be derived as  
$$
{\mathbf{\hat \gamma }}^*  = \left( {{\mathbf{X'X}}} \right)^{ - 1} {\mathbf{X'Y}} = \left( {\sum\limits_{i = 1}^N {{\mathbf{X'}}_i {\mathbf{X}}_i } } \right)^{ - 1} \left( {\sum\limits_{i = 1}^N {{\mathbf{X'}}_i {\mathbf{Y}}_i } } \right),
$$
where ${\mathbf{X}}_i $ is the covariate matrix for subject $i$ and can be written as ${\mathbf{X}}_i  = \left( {\begin{array}{*{20}c}
   {\mathbf{Z}} & {\mathbf{0}}  \\
 \end{array} } \right)$ if participant $i$ is unexposed and ${\mathbf{X}}_i  = \left( {\begin{array}{*{20}c}
   {\mathbf{0}} & {\mathbf{Z}}  \\
 \end{array} } \right)$ if exposed. Then, 
$$
 \left( {\sum\limits_{i = 1}^N {{\mathbf{X'}}_i {\mathbf{X}}_i } } \right) = \left( {\begin{array}{*{20}c}
   {N(1 - p_e ){\mathbf{Z'Z}}} & {\mathbf{0}}  \\
   {\mathbf{0}} & {Np_e {\mathbf{Z'Z}}}  \\
 \end{array} } \right),
$$
$$\left( {\sum\limits_{i = 1}^N {{\mathbf{X'}}_i {\mathbf{X}}_i } } \right)^{ - 1}  = \frac{1}
{N}\left( {\begin{array}{*{20}c}
   {\frac{1}
{{(1 - p_e )}}\left( {{\mathbf{Z'Z}}} \right)^{ - 1} } & {\mathbf{0}}  \\
   {\mathbf{0}} & {\frac{1}
{{p_e }}\left( {{\mathbf{Z'Z}}} \right)^{ - 1} }  \\
 \end{array} } \right),
$$
and
\begin{multline*}
\left( {\sum\limits_{i = 1}^N {{\mathbf{X'}}_i {\mathbf{Y}}_i } } \right) = \left( {\begin{array}{*{20}c}
   {{\mathbf{Z'}}}  \\
   {\mathbf{0}}  \\
 \end{array} } \right)\left( {\sum\limits_{i = 1}^N {{\mathbf{Y}}_i I\left\{ {k_i  = 0} \right\}} } \right) + \left( {\begin{array}{*{20}c}
   {\mathbf{0}}  \\
   {{\mathbf{Z'}}}  \\
 \end{array} } \right)\left( {\sum\limits_{i = 1}^N {{\mathbf{Y}}_i I\left\{ {k_i  = 1} \right\}} } \right) \\
 = \left( {\begin{array}{*{20}c}
   {N(1 - p_e ){\mathbf{Z'\bar Y}}_0 }  \\
   {Np_e {\mathbf{Z'\bar Y}}_1 }  \\
 \end{array} } \right),
\end{multline*}
so 
$$
{\mathbf{\hat \gamma }}^*  = \left( {\begin{array}{*{20}c}
   {\left( {{\mathbf{Z'Z}}} \right)^{ - 1} {\mathbf{Z'\bar Y}}_0 }  \\
   {\left( {{\mathbf{Z'Z}}} \right)^{ - 1} {\mathbf{Z'\bar Y}}_1 }  \\
 \end{array} } \right).
$$
To compute $\hat \gamma _3  = \hat \gamma _3^*  - \hat \gamma _1^* $ we need to subtract the second from the fourth component, so $\hat \gamma _3  = \left( {\left( {{\mathbf{Z'Z}}} \right)^{ - 1} {\mathbf{Z'}}} \right)_{(2)} \left( {{\mathbf{\bar Y}}_1  - {\mathbf{\bar Y}}_0 } \right)$ as in the two-stage approach.

\subsection*{Proof of (ii)}
\label{ap2stageii}

A necessary and sufficient condition for the OLS and GLS estimators to be the same is ${\mathbf{HV}} = {\mathbf{VH}}$ \cite[condition Z5]{Puntanen:1989}, where ${\mathbf{H}}$ is the hat matrix ${\mathbf{H}} = {\mathbf{X}}({\mathbf{X'X}})^{ - 1} {\mathbf{X'}}$, ${\mathbf{X}}$ is our case the $N(r + 1) \times 4$ matrix of covariates based on model~\eqref{lddp1}, and ${\mathbf{V}}$ is the $N(r + 1) \times N(r + 1)$ covariance matrix of ${\mathbf{Y}}$, which is a block-diagonal matrix with the diagonal blocks equal to ${\mathbf{\Sigma }}$. As in the OLS derivation, we reparameterize the model as 
$$
\mathbb{E}\left( {Y_{ij} |X_{ij} } \right) = \gamma _0^* \left( {1 - k_i } \right) + \gamma _1^* \left( {1 - k_i } \right)t_{ij}  + \gamma _2^* k_i  + \gamma _3^* k_i t_{ij},
$$
and for convenience we sort ${\mathbf{X}}$ sot that the first $N(1 - p_e )$ participants are unexposed and therefore have ${\mathbf{X}}_i  = \left( {\begin{array}{*{20}c}
   {\mathbf{Z}} & {\mathbf{0}}  \\
 \end{array} } \right)$, and the following $Np_e $ are exposed and have ${\mathbf{X}}_i  = \left( {\begin{array}{*{20}c}
   {\mathbf{0}} & {\mathbf{Z}}  \\
 \end{array} } \right)$. As derived in the OLS case, 
$$
 ({\mathbf{X'X}})^{ - 1}  = \frac{1}
{N}\left( {\begin{array}{*{20}c}
   {\frac{1}
{{(1 - p_e )}}\left( {{\mathbf{Z'Z}}} \right)^{ - 1} } & {{\mathbf{0'}}}  \\
   {\mathbf{0}} & {\frac{1}
{{p_e }}\left( {{\mathbf{Z'Z}}} \right)^{ - 1} }  \\
 \end{array} } \right).
$$
Then, it can be derived that 
$$
{\mathbf{H}} = {\mathbf{X}}({\mathbf{X'X}})^{ - 1} {\mathbf{X'}} = \frac{1}
{N}\left( {\begin{array}{*{20}c}
   {{\mathbf{H}}_{11} } & {{\mathbf{0'}}}  \\
   {\mathbf{0}} & {{\mathbf{H}}_{22} }  \\
 \end{array} } \right),
$$
where ${\mathbf{H}}_{11} $ is a block matrix of $N(1 - p_e ) \times N(1 - p_e )$ blocks, each block being equal to $\frac{1}
{{1 - p_e }}{\mathbf{Z}}\left( {{\mathbf{Z'Z}}} \right)^{ - 1} {\mathbf{Z'}}$; and ${\mathbf{H}}_{22} $ is a block matrix with $Np_e  \times Np_e $ blocks, each block being equal to $\frac{1}
{{p_e }}{\mathbf{Z}}\left( {{\mathbf{Z'Z}}} \right)^{ - 1} {\mathbf{Z'}}$. Since ${\mathbf{V}}$ is block diagonal with the diagonal blocks equal to ${\mathbf{\Sigma }}
$, it follows that ${\mathbf{HV}}$ is going to be of the form 
$$
{\mathbf{HV}} = \frac{1}
{N}\left( {\begin{array}{*{20}c}
   {\left( {{\mathbf{HV}}} \right)_{11} } & {{\mathbf{0'}}}  \\
   {\mathbf{0}} & {\left( {{\mathbf{HV}}} \right)_{22} }  \\
 \end{array} } \right),
$$
where $\left( {{\mathbf{HV}}} \right)_{11} $ is a block matrix of $N(1 - p_e ) \times N(1 - p_e )$ blocks, each block being equal to $\frac{1}
{{1 - p_e }}{\mathbf{Z}}\left( {{\mathbf{Z'Z}}} \right)^{ - 1} {\mathbf{Z'\Sigma }}$; and $\left( {{\mathbf{HV}}} \right)_{22} $ is a block matrix with $Np_e  \times Np_e $ blocks, each block being equal to $\frac{1}
{{p_e }}{\mathbf{Z}}\left( {{\mathbf{Z'Z}}} \right)^{ - 1} {\mathbf{Z'\Sigma }}$. Similarly, we can derive that ${\mathbf{VH}}$ is of the form 
$$
{\mathbf{VH}} = \frac{1}
{N}\left( {\begin{array}{*{20}c}
   {\left( {{\mathbf{VH}}} \right)_{11} } & {{\mathbf{0'}}}  \\
   {\mathbf{0}} & {\left( {{\mathbf{VH}}} \right)_{22} }  \\
 \end{array} } \right),
$$
where $\left( {{\mathbf{VH}}} \right)_{11}$ is a block matrix of $N(1 - p_e ) \times N(1 - p_e )$ blocks, each block being equal to $\frac{1}
{{1 - p_e }}{\mathbf{\Sigma Z}}\left( {{\mathbf{Z'Z}}} \right)^{ - 1} {\mathbf{Z'}}$;  and $\left( {{\mathbf{VH}}} \right)_{22} $ is a block matrix with $Np_e  \times Np_e $ blocks, each block being equal to $\frac{1}
{{p_e }}{\mathbf{\Sigma Z}}\left( {{\mathbf{Z'Z}}} \right)^{ - 1} {\mathbf{Z'}}$. Clearly, then, proving that ${\mathbf{HV}} = {\mathbf{VH}}$ is equivalent to proving that ${\mathbf{Z}}\left( {{\mathbf{Z'Z}}} \right)^{ - 1} {\mathbf{Z'\Sigma }} = {\mathbf{\Sigma Z}}\left( {{\mathbf{Z'Z}}} \right)^{ - 1} {\mathbf{Z'}}$.

Next, we show that the ${\mathbf{Z}}\left( {{\mathbf{Z'Z}}} \right)^{ - 1} {\mathbf{Z'\Sigma }} = {\mathbf{\Sigma Z}}\left( {{\mathbf{Z'Z}}} \right)^{ - 1} {\mathbf{Z'}}$ holds for ${\mathbf{\Sigma }}$ having a CS or RS structure and therefore the OLS and GLS estimators are algebraically equivalent in those cases. We also show that the condition does not hold for DEX.

\subsubsection*{CS}

Under CS, ${\mathbf{\Sigma }} = \sigma ^2 \left( {\rho {\mathbf{11'}} + (1 - \rho ){\mathbf{I}}} \right)$, where ${\mathbf{I}}$ is the $(r + 1) \times (r + 1)$ identity matrix and ${\mathbf{1}}$ a $(r + 1) \times 1$ vector of ones. Then, 
\begin{multline*}
{\mathbf{Z}}\left( {{\mathbf{Z'Z}}} \right)^{ - 1} {\mathbf{Z'\Sigma }} = \sigma ^2 {\mathbf{Z}}\left( {{\mathbf{Z'Z}}} \right)^{ - 1} {\mathbf{Z'}}\left( {\rho {\mathbf{11'}} + (1 - \rho ){\mathbf{I}}} \right)  \\
 = \sigma ^2 \rho {\mathbf{Z}}\left( {{\mathbf{Z'Z}}} \right)^{ - 1} {\mathbf{Z'11'}} + \sigma ^2 (1 - \rho ){\mathbf{Z}}\left( {{\mathbf{Z'Z}}} \right)^{ - 1} {\mathbf{Z'}}.
\end{multline*}
Since ${\mathbf{Z}}\left( {{\mathbf{Z'Z}}} \right)^{ - 1} {\mathbf{Z'}}$ is a projection matrix in the subspace defined by columns of ${\mathbf{Z}}$, and the first column of ${\mathbf{Z}}$is ${\mathbf{1}}$, then ${\mathbf{Z}}\left( {{\mathbf{Z'Z}}} \right)^{ - 1} {\mathbf{Z'1}} = {\mathbf{1}}$ and ${\mathbf{Z}}\left( {{\mathbf{Z'Z}}} \right)^{ - 1} {\mathbf{Z'\Sigma }} = \sigma ^2 \rho {\mathbf{11'}} + \sigma ^2 (1 - \rho ){\mathbf{Z}}\left( {{\mathbf{Z'Z}}} \right)^{ - 1} {\mathbf{Z'}}$. Now, we derive an expression for 
\begin{multline*}
{\mathbf{\Sigma Z}}\left( {{\mathbf{Z'Z}}} \right)^{ - 1} {\mathbf{Z'}} = \sigma ^2 \left( {\rho {\mathbf{11'}} + (1 - \rho ){\mathbf{I}}} \right){\mathbf{Z}}\left( {{\mathbf{Z'Z}}} \right)^{ - 1} {\mathbf{Z'}} \\ 
= \sigma ^2 \rho {\mathbf{11'Z}}\left( {{\mathbf{Z'Z}}} \right)^{ - 1} {\mathbf{Z'}} + \sigma ^2 (1 - \rho ){\mathbf{Z}}\left( {{\mathbf{Z'Z}}} \right)^{ - 1} {\mathbf{Z'}}.
\end{multline*}
For the same reasoning used above, ${\mathbf{1'Z}}\left( {{\mathbf{Z'Z}}} \right)^{ - 1} {\mathbf{Z'}} = {\mathbf{1'}}$, and therefore ${\mathbf{\Sigma Z}}\left( {{\mathbf{Z'Z}}} \right)^{ - 1} {\mathbf{Z'}} = \sigma ^2 \rho {\mathbf{11'}} + \sigma ^2 (1 - \rho ){\mathbf{Z}}\left( {{\mathbf{Z'Z}}} \right)^{ - 1} {\mathbf{Z'}}$, which is the same expression we derived for ${\mathbf{Z}}\left( {{\mathbf{Z'Z}}} \right)^{ - 1} {\mathbf{Z'\Sigma }}$.

\subsubsection*{RS}

Under RS, ${\mathbf{\Sigma }} = {\mathbf{ZDZ'}} + \sigma _{within}^2 {\mathbf{I}}$. Then, 
\begin{multline*}
{\mathbf{Z}}\left( {{\mathbf{Z'Z}}} \right)^{ - 1} {\mathbf{Z'\Sigma }} = {\mathbf{Z}}\left( {{\mathbf{Z'Z}}} \right)^{ - 1} {\mathbf{Z'}}\left( {{\mathbf{ZDZ'}} + \sigma _{within}^2 {\mathbf{I}}} \right) \\
 = {\mathbf{Z}}\left( {{\mathbf{Z'Z}}} \right)^{ - 1} {\mathbf{Z'ZDZ'}} + \sigma _{within}^2 {\mathbf{Z}}\left( {{\mathbf{Z'Z}}} \right)^{ - 1} {\mathbf{Z'}} = {\mathbf{ZDZ'}} + \sigma _{within}^2 {\mathbf{Z}}\left( {{\mathbf{Z'Z}}} \right)^{ - 1} {\mathbf{Z'}}.
\end{multline*}
Now, we derive an expression for 
$$
{\mathbf{\Sigma Z}}\left( {{\mathbf{Z'Z}}} \right)^{ - 1} {\mathbf{Z'}} = \left( {{\mathbf{ZDZ'}} + \sigma _{within}^2 {\mathbf{I}}} \right){\mathbf{Z}}\left( {{\mathbf{Z'Z}}} \right)^{ - 1} {\mathbf{Z'}} = {\mathbf{ZDZ'}} + \sigma _{within}^2 {\mathbf{Z}}\left( {{\mathbf{Z'Z}}} \right)^{ - 1} {\mathbf{Z'}},
$$
which is the same expression we derived for ${\mathbf{Z}}\left( {{\mathbf{Z'Z}}} \right)^{ - 1} {\mathbf{Z'\Sigma }}$.

\subsubsection*{DEX}

A counterexample is enough to show that ${\mathbf{Z}}\left( {{\mathbf{Z'Z}}} \right)^{ - 1} {\mathbf{Z'\Sigma }} = {\mathbf{\Sigma Z}}\left( {{\mathbf{Z'Z}}} \right)^{ - 1} {\mathbf{Z'}}$ does not hold for DEX. With $r = 2$ then 
$$
{\mathbf{Z}}\left( {{\mathbf{Z'Z}}} \right)^{ - 1} {\mathbf{Z'}} = \left( {\begin{array}{*{20}c}
   {5/6} & {1/3} & { - 1/6}  \\
   {1/3} & {1/3} & {1/3}  \\
   { - 1/6} & {1/3} & {5/6}  \\
 \end{array} } \right).
$$
If we take $\sigma ^2  = 1$, $\rho  = 0.8$ and $\theta  = 1$ (AR(1) covariance structure) then 
$$
{\mathbf{\Sigma }} = \left( {\begin{array}{*{20}c}
   1 & {0.8} & {0.64}  \\
   {0.8} & 1 & {0.8}  \\
   {0.64} & {0.8} & 1  \\
 \end{array} } \right).
$$
Now, 
\begin{multline*}
{\mathbf{Z}}\left( {{\mathbf{Z'Z}}} \right)^{ - 1} {\mathbf{Z'\Sigma }} = \left( {\begin{array}{*{20}c}
   {5/6} & {1/3} & { - 1/6}  \\
   {1/3} & {1/3} & {1/3}  \\
   { - 1/6} & {1/3} & {5/6}  \\
 \end{array} } \right) 
  \left( {\begin{array}{*{20}c}
   1 & {0.8} & {0.64}  \\
   {0.8} & 1 & {0.8}  \\
   {0.64} & {0.8} & 1  \\
 \end{array} } \right) \\ 
 = \left( {\begin{array}{*{20}c}
   {0.993} & {0.866} & {0.633}  \\
   {0.813} & {0.866} & {0.813}  \\
   {0.633} & {0.866} & {0.993}  \\
 \end{array} } \right)
\end{multline*}
and 
\begin{multline*}
{\mathbf{\Sigma Z}}\left( {{\mathbf{Z'Z}}} \right)^{ - 1} {\mathbf{Z'}} = \left( {\begin{array}{*{20}c}
   1 & {0.8} & {0.64}  \\
   {0.8} & 1 & {0.8}  \\
   {0.64} & {0.8} & 1  \\
 \end{array} } \right)\left( {\begin{array}{*{20}c}
   {5/6} & {1/3} & { - 1/6}  \\
   {1/3} & {1/3} & {1/3}  \\
   { - 1/6} & {1/3} & {5/6}  \\
 \end{array} } \right) \\ 
 = \left( {\begin{array}{*{20}c}
   {0.993} & {0.813} & {0.633}  \\
   {0.866} & {0.866} & {0.866}  \\
   {0.633} & {0.813} & {0.993}  \\
 \end{array} } \right).
\end{multline*}
We can see that the the [2,1], [1,2], [3,2] and [2,3] components differ, so the condition does not hold.

\section{Proof that ${\mathbf{c'\Sigma }}_{\rm B} {\mathbf{c}}
$ is the same for $r = 1$ and $r = 2$ under LDD and $V(t_0 ) = 0$ with fixed follow-up period $\tau $ and equidistant time points.}
\label{apr1r2}

Formula~\eqref{varlddp1} for the $V(t_0 ) = 0$ case and expressed as a function of $\tau $ is 
$$
{\mathbf{c'\Sigma }}_{\rm B} {\mathbf{c}} = \frac{{\left( {\sum\limits_{j = 0}^r {\sum\limits_{j' = 0}^r {v_{jj'} } } } \right)r^2 }}
{{p_e (1 - p_e )\tau ^2 \det ({\mathbf{A}})}},
$$
where the term $v_{jj'} $ is the $[j,j']$ component of the inverse of ${\mathbf{\Sigma }}$ and 
$$
{\mathbf{A}} = \left( {\begin{array}{*{20}c}
   {\sum\limits_{j = 0}^r {\sum\limits_{j' = 0}^r {v_{jj'} } } } & {\sum\limits_{j = 0}^r {\sum\limits_{j' = 0}^r {jv_{jj'} } } }  \\
   {\sum\limits_{j = 0}^r {\sum\limits_{j' = 0}^r {jv_{jj'} } } } & {\sum\limits_{j = 0}^r {\sum\limits_{j' = 0}^r {jj'v_{jj'} } } }  \\
 \end{array} } \right) = \left( {\begin{array}{*{20}c}
   1 &  \cdots  & 1  \\
   0 &  \cdots  & r  \\
 \end{array} } \right){\mathbf{\Sigma }}\left( {\begin{array}{*{20}c}
   1 & 0  \\
    \vdots  &  \vdots   \\
   1 & r  \\
 \end{array} } \right).
$$
Let us call 
$$
{\mathbf{\Sigma }}_1  = \left( {\begin{array}{*{20}c}
   {\sigma _{11} } & {\sigma _{1\tau } }  \\
   {\sigma _{1\tau } } & {\sigma _{\tau \tau } }  \\
 \end{array} } \right)
$$
the covariance matrix for the case $r = 1$ and 
$$
{\mathbf{\Sigma }}_2  = \left( {\begin{array}{*{20}c}
   {\sigma _{11} } & {\sigma _{1,\tau /2} } & {\sigma _{1,\tau } }  \\
   {\sigma _{1,\tau /2} } & {\sigma _{\tau /2,\tau /2} } & {\sigma _{\tau /2,\tau } }  \\
   {\sigma _{1,\tau } } & {\sigma _{\tau /2,\tau } } & {\sigma _{\tau ,\tau } }  \\
 \end{array} } \right)
$$
for the case $r = 2$ and note that $\sigma _{11} $, $\sigma _{1,\tau } $ and $\sigma _{\tau ,\tau } $ are the same in the two matrices. Let us call ${\mathbf{A}}_1 $ the matrix ${\mathbf{A}}$ for the case $r = 1$ and ${\mathbf{A}}_2 $ for the case $r = 2$. Then, the expression ${\mathbf{c'\Sigma }}_{\rm B} {\mathbf{c}}$ will be the same for $r = 1$ and $r = 2$ if and only if 
$$
\frac{{{\mathbf{A}}_1 [1,1]}}
{{\det ({\mathbf{A}}_1 )}} = \frac{{4{\mathbf{A}}_2 [1,1]}}
{{\det ({\mathbf{A}}_2 )}}.
$$
We can now derive 
$$
{\mathbf{\Sigma }}_1^{ - 1}  = \frac{1}
{{\sigma _{11} \sigma _{\tau \tau }  - \sigma _{1\tau }^2 }}\left( {\begin{array}{*{20}c}
   {\sigma _{11} } & {\sigma _{1\tau } }  \\
   {\sigma _{1\tau } } & {\sigma _{\tau \tau } }  \\
 \end{array} } \right)
$$
and
\begin{multline*}
  {\mathbf{\Sigma }}_2^{ - 1}  = \frac{1}
{{ - 2\sigma _{1,\tau } \sigma _{1,\tau /2} \sigma _{\tau /2,\tau }  + \sigma _{1,\tau }^2 \sigma _{\tau /2,\tau /2}  + \sigma _{1,\tau /2}^2 \sigma _{\tau ,\tau }  + \sigma _{11} \left( {\sigma _{\tau /2,\tau }^2  - \sigma _{\tau /2,\tau /2} \sigma _{\tau ,\tau } } \right)}}  \\
  \left( {\begin{array}{*{20}c}
   {\sigma _{\tau /2,\tau }^2  - \sigma _{\tau /2,\tau /2} \sigma _{\tau ,\tau } } & {} & {}  \\
   {\sigma _{1,\tau /2} \sigma _{\tau ,\tau }  - \sigma _{1,\tau } \sigma _{\tau /2,\tau } } & {\sigma _{1,\tau }^2  - \sigma _{11} \sigma _{\tau ,\tau } } & {}  \\
   {\sigma _{1,\tau } \sigma _{\tau /2,\tau /2}  - \sigma _{1,\tau /2} \sigma _{\tau /2,\tau } } & {\sigma _{11} \sigma _{\tau /2,\tau }  - \sigma _{1,\tau } \sigma _{1,\tau /2} } & {\sigma _{1,\tau /2}^2  - \sigma _{11} \sigma _{\tau /2,\tau /2} }  \\
 \end{array} } \right)  
\end{multline*} 
Also, 
\begin{multline*} 
{\mathbf{A}}_1  = \frac{1}
{{\sigma _{11} \sigma _{\tau \tau }  - \sigma _{1\tau }^2 }}\left( {\begin{array}{*{20}c}
   1 & 1  \\
   0 & 1  \\
 \end{array} } \right)\left( {\begin{array}{*{20}c}
   {\sigma _{11} } & {\sigma _{1\tau } }  \\
   {\sigma _{1\tau } } & {\sigma _{\tau \tau } }  \\
 \end{array} } \right)\left( {\begin{array}{*{20}c}
   1 & 0  \\
   1 & 1  \\
 \end{array} } \right) \\
 = \frac{1}
{{\sigma _{11} \sigma _{\tau \tau }  - \sigma _{1\tau }^2 }}\left( {\begin{array}{*{20}c}
   {\sigma _{11}  - 2\sigma _{1\tau }  + \sigma _{\tau \tau } } & {\sigma _{11}  - \sigma _{1\tau } }  \\
   {\sigma _{11}  - \sigma _{1\tau } } & {\sigma _{11} }  \\
 \end{array} } \right),
\end{multline*}
$$
\det ({\mathbf{A}}_1 ) = \frac{1}
{{\sigma _{11} \sigma _{\tau \tau }  - \sigma _{1\tau }^2 }}
$$
and
$$
\frac{{\left( {\sum\limits_{j = 0}^1 {\sum\limits_{j' = 0}^1 {v_{jj'} } } } \right)}}
{{\det ({\mathbf{A}}_1 )}} = \frac{{{\mathbf{A}}_1 [1,1]}}
{{\det ({\mathbf{A}}_1 )}} = \sigma _{11}  - 2\sigma _{1\tau }  + \sigma _{\tau \tau }; 
$$
and
\begin{multline*}
  {\mathbf{A}}_2  = \frac{1}
{{ - 2\sigma _{1,\tau } \sigma _{1,\tau /2} \sigma _{\tau /2,\tau }  + \sigma _{1,\tau }^2 \sigma _{\tau /2,\tau /2}  + \sigma _{1,\tau /2}^2 \sigma _{\tau ,\tau }  + \sigma _{11} \left( {\sigma _{\tau /2,\tau }^2  - \sigma _{\tau /2,\tau /2} \sigma _{\tau ,\tau } } \right)}}  \\
  \left( {\begin{array}{*{20}c}
   1 & 1 & 1  \\
   0 & 1 & 2  \\
 \end{array} } \right) \\
 \left( {\begin{array}{*{20}c}
   {\sigma _{\tau /2,\tau }^2  - \sigma _{\tau /2,\tau /2} \sigma _{\tau ,\tau } } & {} & {}  \\
   {\sigma _{1,\tau /2} \sigma _{\tau ,\tau }  - \sigma _{1,\tau } \sigma _{\tau /2,\tau } } & {\sigma _{1,\tau }^2  - \sigma _{11} \sigma _{\tau ,\tau } } & {}  \\
   {\sigma _{1,\tau } \sigma _{\tau /2,\tau /2}  - \sigma _{1,\tau /2} \sigma _{\tau /2,\tau } } & {\sigma _{11} \sigma _{\tau /2,\tau }  - \sigma _{1,\tau } \sigma _{1,\tau /2} } & {\sigma _{1,\tau /2}^2  - \sigma _{11} \sigma _{\tau /2,\tau /2} }  \\
 \end{array} } \right) 
 \left( {\begin{array}{*{20}c}
   1 & 0  \\
   1 & 1  \\
   1 & 2  \\
 \end{array} } \right).
\end{multline*}
It can be derived that 
\begin{multline*}
  \frac{{{\mathbf{A}}_2 [1,1]}}
{{\det ({\mathbf{A}}_2 )}} = \frac{1}
{{\sigma _{11}  + 2\sigma _{1,\tau }  - 4\left( {\sigma _{1,\tau /2}  + \sigma _{\tau /2,\tau }  - \sigma _{\tau /2,\tau /2} } \right) + \sigma _{\tau ,\tau } }} \\
  \biggl\{ - \sigma _{1,\tau }^2  - \left( {\sigma _{1,\tau /2}  - \sigma _{\tau /2,\tau } } \right)^2  + 2\sigma _{1,\tau } \left( {\sigma _{1,\tau /2}  + \sigma _{\tau /2,\tau }  - \sigma _{\tau /2,\tau /2} } \right) + \\ 
  \left( {\sigma _{\tau /2,\tau /2}  - 2\sigma _{1,\tau /2} } \right)\sigma _{\tau ,\tau }  +   {\sigma _{11} \left( {\sigma _{\tau /2,\tau /2}  + \sigma _{\tau ,\tau }  - 2\sigma _{\tau /2,\tau } } \right)} \biggr\} 
\end{multline*} 
Then, 
$$
\frac{{{\mathbf{A}}_1 [1,1]}}
{{\det ({\mathbf{A}}_1 )}} = \frac{{4{\mathbf{A}}_2 [1,1]}}
{{\det ({\mathbf{A}}_2 )}}
$$
if and only if 
\begin{multline*}
  \sigma _{11}  - 2\sigma _{1\tau }  + \sigma _{\tau \tau }  = \frac{4}
{{\sigma _{11}  + 2\sigma _{1,\tau }  - 4\left( {\sigma _{1,\tau /2}  + \sigma _{\tau /2,\tau }  - \sigma _{\tau /2,\tau /2} } \right) + \sigma _{\tau ,\tau } }} \\
  \biggl\{  - \sigma _{1,\tau }^2  - \left( {\sigma _{1,\tau /2}  - \sigma _{\tau /2,\tau } } \right)^2  + 2\sigma _{1,\tau } \left( {\sigma _{1,\tau /2}  + \sigma _{\tau /2,\tau }  - \sigma _{\tau /2,\tau /2} } \right) + \\ 
  \left( {\sigma _{\tau /2,\tau /2}  - 2\sigma _{1,\tau /2} } \right)\sigma _{\tau ,\tau } +  {\sigma _{11} \left( {\sigma _{\tau /2,\tau /2}  + \sigma _{\tau ,\tau }  - 2\sigma _{\tau /2,\tau } } \right)} \biggr\},
\end{multline*} 
which with some algebra it reduces to $\sigma _{11}  - \sigma _{\tau \tau }  = 2\left( {\sigma _{1,\tau /2}  - \sigma _{\tau /2,\tau } } \right)$. So, ${\mathbf{c'\Sigma }}_{\rm B} {\mathbf{c}}$ will be the same for $r = 1$ and $r = 2$ if and only if $\sigma _{11}  - \sigma _{\tau \tau }  = 2\left( {\sigma _{1,\tau /2}  - \sigma _{\tau /2,\tau } } \right)$. We can check that for the covariance structures used in the paper, i.e. compound symmetry (CS) (section~\ref{csp1}), damped exponential (DEX) (section~\ref{dexp1}) and random intercepts and slopes (RS) (section~\ref{rsp1}) this condition is met. For CS, 
$$
{\mathbf{\Sigma }}_2  = \left( {\begin{array}{*{20}c}
   {\sigma _{11} } & {\sigma _{1,\tau /2} } & {\sigma _{1,\tau } }  \\
   {\sigma _{1,\tau /2} } & {\sigma _{\tau /2,\tau /2} } & {\sigma _{\tau /2,\tau } }  \\
   {\sigma _{1,\tau } } & {\sigma _{\tau /2,\tau } } & {\sigma _{\tau ,\tau } }  \\
 \end{array} } \right) = \sigma ^2 \left( {\begin{array}{*{20}c}
   1 & \rho  & \rho   \\
   \rho  & 1 & \rho   \\
   \rho  & \rho  & 1  \\
 \end{array} } \right),
$$
so $\sigma _{11}  - \sigma _{\tau \tau }  = \sigma ^2 \left( {1 - 1} \right) = 0
$ and $2\left( {\sigma _{1,\tau /2}  - \sigma _{\tau /2,\tau } } \right) = 2\sigma ^2 (\rho  - \rho ) = 0$ and the condition holds. For DEX, 
$$
{\mathbf{\Sigma }}_2  = \left( {\begin{array}{*{20}c}
   {\sigma _{11} } & {\sigma _{1,\tau /2} } & {\sigma _{1,\tau } }  \\
   {\sigma _{1,\tau /2} } & {\sigma _{\tau /2,\tau /2} } & {\sigma _{\tau /2,\tau } }  \\
   {\sigma _{1,\tau } } & {\sigma _{\tau /2,\tau } } & {\sigma _{\tau ,\tau } }  \\
 \end{array} } \right) = \sigma ^2 \left( {\begin{array}{*{20}c}
   1 & \rho  & {\rho ^{2^\theta  } }  \\
   \rho  & 1 & \rho   \\
   {\rho ^{2^\theta  } } & \rho  & 1  \\
 \end{array} } \right),
$$
so $\sigma _{11}  - \sigma _{\tau \tau }  = \sigma ^2 \left( {1 - 1} \right) = 0$ and $2\left( {\sigma _{1,\tau /2}  - \sigma _{\tau /2,\tau } } \right) = 2\sigma ^2 (\rho  - \rho ) = 0$ and the condition holds. For RS, 
\begin{multline*}
{\mathbf{\Sigma }}_2  = \left( {\begin{array}{*{20}c}
   {\sigma _{11} } & {\sigma _{1,\tau /2} } & {\sigma _{1,\tau } }  \\
   {\sigma _{1,\tau /2} } & {\sigma _{\tau /2,\tau /2} } & {\sigma _{\tau /2,\tau } }  \\
   {\sigma _{1,\tau } } & {\sigma _{\tau /2,\tau } } & {\sigma _{\tau ,\tau } }  \\
 \end{array} } \right) = \\
 {\footnotesize \left( {\begin{array}{*{20}c}
   {\sigma _{b_0 }^2  + \sigma _{within}^2 } & {} & {}  \\
   {\sigma _{b_0 }^2  + \rho _{b_0 b_1 } \sigma _{b_0 } \sigma _{b_1 } } & {\sigma _{b_0 }^2  + \sigma _{b_1 }^2  + 2\rho _{b_0 b_1 } \sigma _{b_0 } \sigma _{b_1 }  + \sigma _{within}^2 } & {}  \\
   {\sigma _{b_0 }^2  + 2\rho _{b_0 b_1 } \sigma _{b_0 } \sigma _{b_1 } } & {\sigma _{b_0 }^2  + 3\rho _{b_0 b_1 } \sigma _{b_0 } \sigma _{b_1 }  + 2\sigma _{b_1 }^2 } & {\sigma _{b_0 }^2  + 4\sigma _{b_1 }^2  + 4\rho _{b_0 b_1 } \sigma _{b_0 } \sigma _{b_1 }  + \sigma _{within}^2 }  \\
 \end{array} } \right)},
\end{multline*}
so  
$$
\sigma _{11}  - \sigma _{\tau \tau }  = \sigma _{b_0 }^2  + \sigma _{within}^2  - \sigma _{b_0 }^2  - 4\sigma _{b_1 }^2  - 4\rho _{b_0 b_1 } \sigma _{b_0 } \sigma _{b_1 }  - \sigma _{within}^2  =  - 4\sigma _{b_1 }^2  - 4\rho _{b_0 b_1 } \sigma _{b_0 } \sigma _{b_1 } 
$$
and
\begin{multline*}
2\left( {\sigma _{1,\tau /2}  - \sigma _{\tau /2,\tau } } \right) = 2\left( {\sigma _{b_0 }^2  + \rho _{b_0 b_1 } \sigma _{b_0 } \sigma _{b_1 }  - \sigma _{b_0 }^2  - 3\rho _{b_0 b_1 } \sigma _{b_0 } \sigma _{b_1 }  - 2\sigma _{b_1 }^2 } \right) \\
=  - 4\sigma _{b_1 }^2  - 4\rho _{b_0 b_1 } \sigma _{b_0 } \sigma _{b_1 } 
\end{multline*}
and the condition holds.

\section{Effect of $p_e $ on $r$}
\label{appe}

We write 
$$N = \frac{{\left( {{\mathbf{c'\Sigma }}_{\rm B} {\mathbf{c}}} \right)\,\left( {z_\pi   + z_{1 - \alpha /2} } \right)^2 }}
{{({\mathbf{c'{\rm B}}}_{H_A } )^2 }}
$$
as 
$$
N = \frac{{g({\mathbf{\eta }})f(r)\,\left( {z_\pi   + z_{1 - \alpha /2} } \right)^2 }}
{{p_e (1 - p_e )({\mathbf{c'{\rm B}}}_{H_A } )^2 }}
$$
where $g({\mathbf{\eta }})$ does not depend on $r$ or $p_e $. Then we can define $r$ implicitly as the value/s solving the equation $F(r) = 0$, where 
$$
F(r) = \frac{{g({\mathbf{\eta }})\,\left( {z_\pi   + z_{1 - \alpha /2} } \right)^2 }}
{{N({\mathbf{c'{\rm B}}}_{H_A } )^2 }} - \frac{{p_e (1 - p_e )}}
{{f(r)}}.
$$
Using implicit differentiation and differentiating both sides of $F(r) = 0$ we have 
$$
\frac{{\delta F(r)}}
{{\delta p_e }} = 0,
$$
from where we can derive $\frac{{\delta r}} {{\delta p_e }}$, 
$$
\frac{{\delta F(r)}}
{{\delta p_e }} = 0 \Leftrightarrow \frac{{(1 - 2p_e )f(r) - p_e (1 - p_e )f'(r)\frac{{\delta r}}
{{\delta p_e }}}}
{{\left[ {f(r)} \right]^2 }} = 0 \Leftrightarrow 
\frac{{\delta r}}
{{\delta p_e }} = \frac{{(1 - 2p_e )f(r)}}
{{p_e (1 - p_e )f'(r)}}.
$$
Then, to find the value of $p_e $ that minimizes $r$ we solve $\frac{{\delta r}}
{{\delta p_e }} = 0$, which results in the only root $p_e  = 0.5$. Since $(1 - 2p_e )$ is greater than zero for $p_e  < 0.5$ and smaller than zero for  $p_e  > 0.5$, $r$ has a maximum or a minimum at $p_e  = 0.5$. The sign of $\frac{{f(r)}}
{{f'(r)}}$ determines whether it is a maximum or a minimum. Since the variance ${\mathbf{c'\Sigma }}_{\rm B} {\mathbf{c}}$ is always positive so it is $f(r)$, and since the variance decreases as $r$ increases, $f'(r)$ is negative. Therefore $\frac{{f(r)}} {{f'(r)}}$ is negative and $\frac{{\delta r}}
{{\delta p_e }} < 0$ for $p_e  < 0.5$ and $\frac{{\delta r}}
{{\delta p_e }} > 0$ for $p_e  > 0.5$, implying that $r$ is minimum at $p_e  = 0.5$.

\section{Limit of ${\mathbf{c'\Sigma }}_{\rm B} {\mathbf{c}}$ when $r \to \infty $}
\label{aplimits}

\subsection{CMD, CS}
\label{aplimitscmdcs}

The inverse of a CS matrix has diagonal elements 
$$
\frac{1}
{{\sigma ^2 }}\frac{{1 + \rho (r - 2) - \rho ^2 (r - 1)}}
{{(1 - \rho )^2 \left( {1 + r\rho } \right)}}
$$
and off-diagonal elements 
$$
\frac{1}
{{\sigma ^2 }}\frac{{ - \rho }}
{{(1 - \rho )\left( {1 + r\rho } \right)}}.
$$
The sum of a row or a column of the inverse is 
$$
\frac{1}
{{\sigma ^2 }}\left( {\frac{{1 + \rho (r - 2) - \rho ^2 (r - 1)}}
{{(1 - \rho )^2 \left( {1 + r\rho } \right)}} - \frac{{r\rho }}
{{(1 - \rho )\left( {1 + r\rho } \right)}}} \right) = \frac{1}
{{\sigma ^2 \left( {1 + r\rho } \right)}}
$$
and therefore 
$$
\sum\limits_{j = 0}^r {\sum\limits_{j' = 0}^r {v_{jj'} } }  = \frac{{r + 1}}
{{\sigma ^2 (1 + r\rho )}}.
$$
Also, 
$$
\sum\limits_{j = 0}^r {\sum\limits_{j' = 0}^r {jv_{jj'} } }  = \sum\limits_{j = 0}^r {j\sum\limits_{j' = 0}^r {v_{jj'} } }  = \frac{{r(r + 1)}}
{{2\sigma ^2 \left( {1 + r\rho } \right)}}
$$
since $\sum\limits_{j' = 0}^r {v_{jj'} } $ is the sum of a row or column of the inverse. We can also derive 
$$
\sum\limits_{j = 0}^r {\sum\limits_{j' = 0}^r {jj'v_{jj'} } }  = \frac{{r(r + 1)(2 + r(4 + (r - 1)\rho ))}}
{{12\sigma ^2 (1 - \rho )\left( {1 + r\rho } \right)}}.
$$
Then, 
$$
\det ({\mathbf{A}}) = \left( {\sum\limits_{j = 0}^r {\sum\limits_{j' = 0}^r {v_{jj'} } } } \right)\left( {\sum\limits_{j = 0}^r {\sum\limits_{j' = 0}^r {jj'v_{jj'} } } } \right) - \left( {\sum\limits_{j = 0}^r {\sum\limits_{j' = 0}^r {jv_{jj'} } } } \right)^2  = \frac{{r(r + 1)^2 (r + 2)}}
{{12\sigma ^4 (1 - \rho )\left( {1 + r\rho } \right)}}.
$$
Plugging in all these expressions in to equation~\eqref{varcmdp1}, we have that under CMD and CS 
$$
{\mathbf{c'\Sigma }}_{\rm B} {\mathbf{c}} = \frac{{\sigma ^2 (1 + r\rho )\left( {r(r + 2)(1 + r\rho )s^2  + 12(1 - \rho )V\left( {t_0 } \right)} \right)}}
{{p_e (1 - p_e )(r + 1)\left( {r(r + 2)(1 + r\rho )s^2  + 12(1 - \rho )\left( {1 - \rho _{\operatorname{e} ,t_0 }^2 } \right)V\left( {t_0 } \right)} \right)}}.
$$
Then, using the highest order terms of $r$ on the numerator and denominator of ${\mathbf{c'\Sigma }}_{\rm B} {\mathbf{c}}$ we can derive that 
\begin{multline*}
\mathop {\lim }\limits_{r \to \infty } \;{\mathbf{c'\Sigma }}_{\rm B} {\mathbf{c}} = \mathop {\lim }\limits_{r \to \infty } \;\frac{{\sigma ^2 (1 + r\rho )\left( {r(r + 2)(1 + r\rho )s^2 } \right)}}
{{p_e (1 - p_e )(r + 1)\left( {r(r + 2)(1 + r\rho )s^2 } \right)}} = \\
\mathop {\lim }\limits_{r \to \infty } \frac{{\sigma ^2 (1 + r\rho )}}
{{p_e (1 - p_e )(r + 1)}}\; = \frac{{\sigma ^2 \rho }}
{{p_e (1 - p_e )}}.
\end{multline*}

\subsection{LDD, CS}
\label{aplimitslddcs}

Applying the results derived in Appendix~\ref{aplimitscmdcs} to equation~\eqref{varlddp1}, we can derive ${\mathbf{c'\Sigma }}_{\rm B} {\mathbf{c}}$ as 
$$
{\mathbf{c'\Sigma }}_{\rm B} {\mathbf{c}} = \frac{{12\sigma ^2 (1 - \rho ){\kern 1pt} (1 - r\rho )}}
{{N{\kern 1pt} p_e (1 - p_e )(r + 1)\left( {r(r + 2)(1 + r\rho )\,s^2  + 12(1 - \rho )\left( {1 - \rho _{\operatorname{e} ,t_0 }^2 } \right)V(t_0 )} \right)}}.
$$
Since the denominator is a polynomial of fourth degree of $r$ while the numerator is of first degree, then $\mathop {\lim }\limits_{r \to \infty } \;\;{\mathbf{c'\Sigma }}_{\rm B} {\mathbf{c}} = 0$.

\subsection{CMD, AR(1)}
\label{aplimitscmdar1}

\subsubsection*{Fixed $s$:}
\label{ar1fixeds}

The AR(1) covariance matrix is given by \eqref{DEXp1} with $\theta  = 1$, and its inverse is a tridiagonal matrix with the form 
$$
{\mathbf{\Sigma }}^{ - 1}  = \frac{1}
{{\left( {1 - \rho ^{2s} } \right)\sigma ^2 }}\left( {\begin{array}{*{20}c}
   1 & { - \rho ^s } & 0 & 0 &  \cdots  & 0  \\
   { - \rho ^s } & {1 + \rho ^{2s} } & { - \rho ^s } & 0 & {} & 0  \\
   0 & { - \rho ^s } & {1 + \rho ^{2s} } &  \ddots  &  \ddots  &  \vdots   \\
   0 & 0 &  \ddots  &  \ddots  & { - \rho ^s } & 0  \\
    \vdots  & {} &  \ddots  & { - \rho ^s } & {1 + \rho ^{2s} } & { - \rho ^s }  \\
   0 & 0 &  \cdots  & 0 & { - \rho ^s } & 1  \\
 \end{array} } \right)
$$
\cite[page 201]{Graybill:1983}. To use equation~\eqref{varcmdp1} we need to derive $\sum\limits_{j = 0}^r {\sum\limits_{j' = 0}^r {v_{jj'} } } $, $\sum\limits_{j = 0}^r {\sum\limits_{j' = 0}^r {jv_{jj'} } } $ and $\sum\limits_{j = 0}^r {\sum\limits_{j' = 0}^r {jj'v_{jj'} } } $. In can be easily shown that 
$$
\sum\limits_{j = 0}^r {\sum\limits_{j' = 0}^r {v_{jj'} } }  = \frac{{(1 + r + \rho ^s  - r\rho ^s )}}
{{\sigma ^2 {\kern 1pt} (1 + \rho ^s )}}.
$$
Also, 
$$
\sum\limits_{j = 0}^r {\sum\limits_{j' = 0}^r {j \cdot v_{jj'} } }  = \frac{{r\left( {1 - \rho ^s } \right)\left( {1 + r\left( {1 - \rho ^s } \right) + \rho ^s } \right)}}
{{2\left( {1 - \rho ^{2s} } \right)\sigma ^2 }}
$$
and
$$
\left( {\sum\limits_{j = 0}^r {\sum\limits_{j' = 0}^r {jj'v_{jj'} } } } \right) = \frac{r}
{{6\left( {1 - \rho ^{2s} } \right)\sigma ^2 }}\left( {1 + 4\rho ^s  + \rho ^{2s}  + 3r\left( {1 - \rho ^{2s} } \right) + 2r^2 \left( {1 - \rho ^s } \right)^2 } \right).
$$ 
If $V\left( {t_0 } \right) = 0$, we can use formula~\eqref{varcmdvt00p1} and 
$$
{\mathbf{c'\Sigma }}_{\rm B} {\mathbf{c}} = \frac{{\sigma ^2 (1 + \rho ^s )}}
{{{\kern 1pt} p_e (1 - p_e )(1 + r + \rho ^s  - r\rho ^s )}}
$$
as in Table~\ref{table1p1}. This formula has a polynomial of first order degree of $r$ in the denominator, and no terms involving $r$ in the numerator. Therefore,  $\mathop {\lim }\limits_{r \to \infty } \;\;{\mathbf{c'\Sigma }}_{\rm B} {\mathbf{c}} = 0$. If $V\left( {t_0 } \right) > 0$ the formula is very long ${\mathbf{c'\Sigma }}_{\rm B} {\mathbf{c}}$ and we used Mathematica \cite{Wolfram:2005} to get the formula and compute the limit, which was zero. Therefore, $\mathop {\lim }\limits_{r \to \infty } \;\;{\mathbf{c'\Sigma }}_{\rm B} {\mathbf{c}} = 0$ also for the case of $V\left( {t_0 } \right) > 0$.

\subsection*{Fixed $\tau $}
\label{ar1fixedtau}

For the fixed $\tau $ case we need to substitute $s$ by $\tau /r$. So, for the case of $V\left( {t_0 } \right) = 0$, 
$$
{\mathbf{c'\Sigma }}_{\rm B} {\mathbf{c}} = \frac{{\sigma ^2 (1 + \rho ^{\tau /r} )}}
{{{\kern 1pt} p_e (1 - p_e )(1 + r + \rho ^{\tau /r}  - r\rho ^{\tau /r} )}}.
$$
We want to compute 
\begin{multline*}
\mathop {\lim }\limits_{r \to \infty } \;\;{\mathbf{c'\Sigma }}_{\rm B} {\mathbf{c}} = \mathop {\lim }\limits_{r \to \infty } \;\frac{{\sigma ^2 (1 + \rho ^{\tau /r} )}}
{{{\kern 1pt} p_e (1 - p_e )(1 + r + \rho ^{\tau /r}  - r\rho ^{\tau /r} )}} = \\ \mathop {\lim }\limits_{r \to \infty } \;\frac{{\sigma ^2 (1 + \rho ^{\tau /r} )}}
{{{\kern 1pt} p_e (1 - p_e )\left[ {1 + \rho ^{\tau /r}  + r(1 - \rho ^{\tau /r} )} \right]}}.
\end{multline*}
By l'H\^{o}pital's rule it can be shown that $\mathop {\lim }\limits_{r \to \infty } \;r(1 - \rho ^{\tau /r} ) =  - \tau \log \rho $, and then 
$$
\mathop {\lim }\limits_{r \to \infty } \;\;{\mathbf{c'\Sigma }}_{\rm B} {\mathbf{c}} = \frac{{2\sigma ^2 }}
{{p_e (1 - p_e )\left[ {2 - \tau \log \rho } \right]^{} }}.
$$
If $V\left( {t_0 } \right) > 0$, we used Mathematica \cite{Wolfram:2005} to derive the limit, which in this case has a very complicated expression,
\begin{multline*}
 2\sigma ^2 \left( {\left( {\tau ^3  + 12V(t_0 )\tau } \right)\left( {\log (\rho )} \right)^2  - 6\left( {\tau ^2  + 4V(t_0 )} \right)\log (\rho ) + 12\tau } \right)\\
 \left[ {p_{\text{e}} (1 - p_{\text{e}} )(2 - \tau \log (\rho ))} \right]^{ - 1}  \\
\biggl\{  \left( {\tau ^3  + 12V(t_0 )\tau } \right)\left( {\log (\rho )} \right)^2  - 12V(t_0 )(\tau \log (\rho ) - 2)\rho _{e,t_0 }^2 \log (\rho ) \\
 - 6\left( {\tau ^2  + 4V(t_0 )} \right)\log (\rho ) + 12\tau  \biggr\}^{-1} 
\end{multline*}

\subsection{LDD, AR(1)}
\label{aplimitslddar1}

\subsubsection*{Fixed $s$}
\label{ar1lddfixeds}

Using the results from Appendix~\ref{aplimitscmdar1} and applying formula~\eqref{varlddp1} for the case $V\left( {t_0 } \right) = 0$, we can derive that 
$$
{\mathbf{c'\Sigma }}_{\rm B} {\mathbf{c}} = \frac{{12\sigma ^2 {\kern 1pt} (1 - \rho ^{2s} )\,\left[ {\,r\,s^2 p_e (1 - p_e )} \right]^{ - 1} }}
{{{\kern 1pt} \;\,(2 + r(r + 3) + 8\rho ^s  - 2r^2 \rho ^s  + (r - 2)(r - 1)\rho ^{2s} )}}
$$
as is shown in Table~\ref{table1p1}. Since the denominator is a polynomial of second degree of $r$ while the has no terms involving $r$, then $\mathop {\lim }\limits_{r \to \infty } \;\;{\mathbf{c'\Sigma }}_{\rm B} {\mathbf{c}} = 0
$. If $V\left( {t_0 } \right) > 0$, we used Mathematica \cite{Wolfram:2005} to derive that the limit was also zero.

\subsubsection*{Fixed $\tau $}
\label{ar1lddfixedtau}

For the fixed $\tau $ case we need to substitute $s$ by $\tau /r$. So, for the case of $V\left( {t_0 } \right) = 0$, we have 
$$
{\mathbf{c'\Sigma }}_{\rm B} {\mathbf{c}} = \frac{{12\sigma ^2 {\kern 1pt} (1 - \rho ^{2\tau /r} )\,r\,\left[ {\,\tau ^2 p_e (1 - p_e )} \right]^{ - 1} }}
{{{\kern 1pt} \;\,(2 + r(r + 3) + 8\rho ^{\tau /r}  - 2r^2 \rho ^{\tau /r}  + (r - 2)(r - 1)\rho ^{2\tau /r} )}},
$$
as shown in Table~\ref{table1p1}. This expression can be rewritten as
$$
\frac{{12\sigma ^2 {\kern 1pt} (1 - \rho ^{2{\tau  \mathord{\left/
 {\vphantom {\tau  r}} \right.
 \kern-\nulldelimiterspace} r}} )\,\left[ {p_e (1 - p_e )} \right]^{ - 1} }}
{{{\kern 1pt} \;\,\tau ^2 \left[ {r\left( {1 - \rho ^{{\tau  \mathord{\left/
 {\vphantom {\tau  r}} \right.
 \kern-\nulldelimiterspace} r}} } \right)^2  + \left( {3\left( {1 - \rho ^{2{\tau  \mathord{\left/
 {\vphantom {\tau  r}} \right.
 \kern-\nulldelimiterspace} r}} } \right)} \right) + \frac{1}
{r}\left( {2 + 8\rho ^{{\tau  \mathord{\left/
 {\vphantom {\tau  r}} \right.
 \kern-\nulldelimiterspace} r}}  + 2\rho ^{2{\tau  \mathord{\left/
 {\vphantom {\tau  r}} \right.
 \kern-\nulldelimiterspace} r}} } \right)} \right]}}.
$$
Then, to compute $\mathop {\lim }\limits_{r \to \infty } \;\;{\mathbf{c'\Sigma }}_{\rm B} {\mathbf{c}}$ we note that the limit of the numerator is $12\sigma ^2 \mathop {\lim }\limits_{r \to \infty } \;(1 - \rho ^{2{\tau  \mathord{\left/
 {\vphantom {\tau  r}} \right.
 \kern-\nulldelimiterspace} r}} ) = 0
$. In the denominator, the limit of last two terms is zero, and l'H\^{o}pital's rule can be used to derive that the limit of the first term is also zero. Thus, we apply l'H\^{o}pital's rule to derive the limit of ${\mathbf{c'\Sigma }}_{\rm B} {\mathbf{c}}
$, where the derivative of the numerator is 
$$
\frac{{24\sigma ^2 {\kern 1pt} \rho ^{2{\tau  \mathord{\left/
 {\vphantom {\tau  r}} \right.
 \kern-\nulldelimiterspace} r}} \tau \log \rho }}
{{r^2 }},
$$
the derivative of the first term of the denominator is 
$$
\tau ^2 \left[ {\left( {1 - \rho ^{{\tau  \mathord{\left/
 {\vphantom {\tau  r}} \right.
 \kern-\nulldelimiterspace} r}} } \right)^2  + \frac{{2\rho ^{{\tau  \mathord{\left/
 {\vphantom {\tau  r}} \right.
 \kern-\nulldelimiterspace} r}} \left( {1 - \rho ^{{\tau  \mathord{\left/
 {\vphantom {\tau  r}} \right.
 \kern-\nulldelimiterspace} r}} } \right)\tau \log \rho }}
{r}} \right],
$$
the derivative of the second term of the denominator is 
$$
3\tau ^2 \left[ {\frac{{2{\kern 1pt} \rho ^{2{\tau  \mathord{\left/
 {\vphantom {\tau  r}} \right.
 \kern-\nulldelimiterspace} r}} \tau \log \rho }}
{{r^2 }}} \right],
$$
and the derivative of the third term of the denominator is 
$$
2\tau ^2 \left[ {\frac{{ - 1 - 4\rho ^{{\tau  \mathord{\left/
 {\vphantom {\tau  r}} \right.
 \kern-\nulldelimiterspace} r}}  - \rho ^{{{2\tau } \mathord{\left/
 {\vphantom {{2\tau } r}} \right.
 \kern-\nulldelimiterspace} r}} }}
{{r^2 }} - \frac{{2\tau \log \rho \left( {2\rho ^{{\tau  \mathord{\left/
 {\vphantom {\tau  r}} \right.
 \kern-\nulldelimiterspace} r}}  + \rho ^{{{2\tau } \mathord{\left/
 {\vphantom {{2\tau } r}} \right.
 \kern-\nulldelimiterspace} r}} } \right)}}
{{r^3 }}} \right].
$$
Simplifying terms, 
\begin{multline*}
\mathop {\lim }\limits_{r \to \infty } \;\;{\mathbf{c'\Sigma }}_{\rm B} {\mathbf{c}} = \\
 {24\sigma ^2 {\kern 1pt} \rho ^{2{\tau  \mathord{\left/
 {\vphantom {\tau  r}} \right.
 \kern-\nulldelimiterspace} r}} \tau \log \rho \left[ {p_e (1 - p_e )} \right]^{ - 1} \tau ^{-2}} \\
 \biggl\{ r^2 \left( {1 - \rho ^{{\tau  \mathord{\left/
 {\vphantom {\tau  r}} \right.
 \kern-\nulldelimiterspace} r}} } \right)^2  + 2r\rho ^{{\tau  \mathord{\left/
 {\vphantom {\tau  r}} \right.
 \kern-\nulldelimiterspace} r}} \left( {1 - \rho ^{{\tau  \mathord{\left/
 {\vphantom {\tau  r}} \right.
 \kern-\nulldelimiterspace} r}} } \right)\tau \log \rho  + 3\left[ {2\rho ^{2{\tau  \mathord{\left/
 {\vphantom {\tau  r}} \right.
 \kern-\nulldelimiterspace} r}} \tau \log \rho } \right] \\
 + 2\left( { - 1 - 4\rho ^{{\tau  \mathord{\left/
 {\vphantom {\tau  r}} \right.
 \kern-\nulldelimiterspace} r}}  - \rho ^{2{\tau  \mathord{\left/
 {\vphantom {\tau  r}} \right.
 \kern-\nulldelimiterspace} r}} } \right) - \frac{{4\tau \log \rho \left( {2\rho ^{{\tau  \mathord{\left/
 {\vphantom {\tau  r}} \right.
 \kern-\nulldelimiterspace} r}}  + \rho ^{2{\tau  \mathord{\left/
 {\vphantom {\tau  r}} \right.
 \kern-\nulldelimiterspace} r}} } \right)}}
{r} \biggr\}^{-1}
\end{multline*}
Now the limit of the numerator is $24\sigma ^2 {\kern 1pt} \tau \log \rho $. In the denominator, we need to evaluate several terms. The limit of  $r^2 \left( {1 - \rho ^{{\tau  \mathord{\left/
 {\vphantom {\tau  r}} \right.
 \kern-\nulldelimiterspace} r}} } \right)^2 
$ can be obtained by applying l'H\^{o}pital's rule twice and it equals $\tau ^2 \left( {\log \rho } \right)^2 $. The limit of $2r\rho ^{{\tau  \mathord{\left/
 {\vphantom {\tau  r}} \right.
 \kern-\nulldelimiterspace} r}} \left( {1 - \rho ^{{\tau  \mathord{\left/
 {\vphantom {\tau  r}} \right.
 \kern-\nulldelimiterspace} r}} } \right)\tau \log \rho 
$ is obtained by applying l'H\^{o}pital's rule and it equals $ - 2\tau ^2 \left( {\log \rho } \right)^2 $. The limit of $3\left[ {2\rho ^{2{\tau  \mathord{\left/
 {\vphantom {\tau  r}} \right.
 \kern-\nulldelimiterspace} r}} \tau \log \rho } \right]$ is $3\left[ {2\tau \log \rho } \right]$, the limit of $2\left( { - 1 - 4\rho ^{{\tau  \mathord{\left/
 {\vphantom {\tau  r}} \right.
 \kern-\nulldelimiterspace} r}}  - \rho ^{2{\tau  \mathord{\left/
 {\vphantom {\tau  r}} \right.
 \kern-\nulldelimiterspace} r}} } \right)
$ is $ - 12$ and the limit of 
$$
\frac{{4\tau \log \rho \left( {2\rho ^{{\tau  \mathord{\left/
 {\vphantom {\tau  r}} \right.
 \kern-\nulldelimiterspace} r}}  + \rho ^{2{\tau  \mathord{\left/
 {\vphantom {\tau  r}} \right.
 \kern-\nulldelimiterspace} r}} } \right)}}
{r}
$$
is zero. Therefore, with some algebra we can deduce that 
$$
\mathop {\lim }\limits_{r \to \infty } \;\;{\mathbf{c'\Sigma }}_{\rm B} {\mathbf{c}} = \frac{{24\sigma ^2 {\kern 1pt} \log \rho }}
{{p_e (1 - p_e )\left[ { - 12\tau  + 6\tau ^2 \log \rho  - \tau ^3 \left( {\log \rho } \right)^2 } \right]}}.
$$
If $V\left( {t_0 } \right) > 0$, using the expression derived for fixed $s$ and substituting $s$ by $\tau /r$, we used Mathematica \cite{Wolfram:2005} to derive the limit, which has a complicated expression,
\begin{multline*}
 {24\sigma ^2 \log (\rho )\left[ {p_{\text{e}} (1 - p_{\text{e}} )} \right]} \\
\biggl\{  - \left( {\tau ^3  + 12V(t_0 )\tau } \right)\left( {\log (\rho )} \right)^2  + 12V(t_0 )(\tau \log (\rho ) - 2)\rho _{e,t_0 }^2 \log (\rho ) \\
+ 6\left( {\tau ^2  + 4V(t_0 )} \right)\log (\rho ) - 12\tau  \biggr\}^{-1}
\end{multline*}

\subsection{CMD, RS, $V(t_0 ) = 0$}
\label{aplimitscmdrs}

Since for the case $V(t_0 ) > 0$ we need to use numerical methods to compute ${\mathbf{c'\Sigma }}_{\rm B} {\mathbf{c}}$, we only compute the limits of ${\mathbf{c'\Sigma }}_{\rm B} {\mathbf{c}}$ for the case $V(t_0 ) = 0$. The covariance matrix of the repeated measurements is expressed as ${\mathbf{\Sigma }}_i  = {\mathbf{Z}}_i {\mathbf{DZ'}}_i  + \sigma _{within}^2 {\mathbf{I}}$, and since $V(t_0 ) = 0$ we have ${\mathbf{Z}}_i  = {\mathbf{Z}}$ and then we${\mathbf{\Sigma }}_i  = {\mathbf{\Sigma }} = {\mathbf{ZDZ'}} + \sigma _{within}^2 {\mathbf{I}}$. The matrix ${\mathbf{Z}}$ is $(r + 1) \times 2$ and contains a column of ones and the column of times ($sj,j = 0, \ldots ,r$). Note that formula~\eqref{varcmdp1} depends on 
$$
{\mathbf{A}} = \left( {\begin{array}{*{20}c}
   {\sum\limits_{j = 0}^r {\sum\limits_{j' = 0}^r {v_{jj'} } } } & {\sum\limits_{j = 0}^r {\sum\limits_{j' = 0}^r {jv_{jj'} } } }  \\
   {\sum\limits_{j = 0}^r {\sum\limits_{j' = 0}^r {jv_{jj'} } } } & {\sum\limits_{j = 0}^r {\sum\limits_{j' = 0}^r {jj'v_{jj'} } } }  \\
 \end{array} } \right)
$$
only through $s^2 \det ({\mathbf{A}})$. For convenience in this proof we define the new matrix 
$$
{\mathbf{\tilde A}} = \left( {\begin{array}{*{20}c}
   {\sum\limits_{j = 0}^r {\sum\limits_{j' = 0}^r {v_{jj'} } } } & {s\sum\limits_{j = 0}^r {\sum\limits_{j' = 0}^r {jv_{jj'} } } }  \\
   {s\sum\limits_{j = 0}^r {\sum\limits_{j' = 0}^r {jv_{jj'} } } } & {s^2 \sum\limits_{j = 0}^r {\sum\limits_{j' = 0}^r {jj'v_{jj'} } } }  \\
 \end{array} } \right)
$$
so that ${\mathbf{\tilde A}}$ is actually ${\mathbf{Z'\Sigma }}^{ - 1} {\mathbf{Z}}$. Note that $\det \left( {{\mathbf{\tilde A}}} \right) = s^2 \det ({\mathbf{A}})
$. Then, we have that  ${\mathbf{\tilde A}} = {\mathbf{Z'\Sigma }}^{ - 1} {\mathbf{Z}} = {\mathbf{Z'}}\left( {{\mathbf{ZDZ'}} + \sigma _{within}^2 {\mathbf{I}}} \right)^{ - 1} {\mathbf{Z}}$. Using the property 
$$
({\mathbf{ABA}}' + {\mathbf{C}})^{ - 1}  = {\mathbf{C}}^{ - 1}  - {\mathbf{C}}^{ - 1} {\mathbf{A}}\left( {{\mathbf{B}}^{ - 1}  + {\mathbf{A}}'{\mathbf{C}}^{ - 1} {\mathbf{A}}} \right)^{ - 1} {\mathbf{A}}'{\mathbf{C}}^{ - 1},
$$
which can be found in \cite[property 8, page 46]{Timm:2002}, we have that 
\begin{multline*}
\left( {{\mathbf{ZDZ'}} + \sigma _{within}^2 {\mathbf{I}}} \right)^{ - 1}  = \frac{1}
{{\sigma _{within}^2 }}{\mathbf{I}} - \frac{1}
{{\sigma _{within}^2 }}{\mathbf{IZ}}\left( {{\mathbf{D}}^{ - 1}  + {\mathbf{Z'}}\frac{1}
{{\sigma _{within}^2 }}{\mathbf{IZ}}} \right)^{ - 1} {\mathbf{Z'I}}\frac{1}
{{\sigma _{within}^2 }} \\
 = \frac{1}
{{\sigma _{within}^2 }}{\mathbf{I}} - \frac{1}
{{\sigma _{within}^4 }}{\mathbf{Z}}\left( {{\mathbf{D}}^{ - 1}  + \frac{1}
{{\sigma _{within}^2 }}{\mathbf{Z'Z}}} \right)^{ - 1} {\mathbf{Z'}}.
\end{multline*}
Now, 
\begin{multline*}
{\mathbf{Z'}}\left( {{\mathbf{ZDZ'}} + \sigma _{within}^2 {\mathbf{I}}} \right)^{ - 1} {\mathbf{Z}} \\
 = \frac{1}
{{\sigma _{within}^2 }}{\mathbf{Z'Z}} - \frac{1}
{{\sigma _{within}^4 }}{\mathbf{Z'Z}}\left( {{\mathbf{D}}^{ - 1}  + \frac{1}
{{\sigma _{within}^2 }}{\mathbf{Z'Z}}} \right)^{ - 1} {\mathbf{Z'Z}}
\end{multline*}
and using the property  
$$
{\mathbf{A}}^{ - 1}  - {\mathbf{A}}^{ - 1} \left( {{\mathbf{A}}^{ - 1}  + {\mathbf{B}}^{ - 1} } \right)^{ - 1} {\mathbf{A}}^{ - 1}  = ({\mathbf{A}} + {\mathbf{B}})^{ - 1},
$$
which can be found in \cite[property 6, page 46]{Timm:2002}, we have that \begin{multline*}
{\mathbf{\tilde A}} = {\mathbf{Z'}}\left( {{\mathbf{ZDZ'}} + \sigma _{within}^2 {\mathbf{I}}} \right)^{ - 1} {\mathbf{Z}} \\
 = \frac{1}
{{\sigma _{within}^2 }}{\mathbf{Z'Z}} - \frac{1}
{{\sigma _{within}^4 }}{\mathbf{Z'Z}}\left( {{\mathbf{D}}^{ - 1}  + \frac{1}
{{\sigma _{within}^2 }}{\mathbf{Z'Z}}} \right)^{ - 1} {\mathbf{Z'Z}} = \left( {\left( {{\mathbf{Z'Z}}} \right)^{ - 1} \sigma _{within}^2  + {\mathbf{D}}} \right)^{ - 1}. 
\end{multline*}
Now, 
\begin{multline*}
\left( {\left( {{\mathbf{Z'Z}}} \right)^{ - 1} \sigma _{within}^2  + {\mathbf{D}}} \right)^{ - 1}  = \\
\left( {\sigma _{within}^2 \left( {\begin{array}{*{20}c}
   {\frac{{2 + 4r}}
{{(r + 1)(r + 2)}}} & {\frac{{ - 6}}
{{s(r + 1)(r + 2)}}}  \\
   {\frac{{ - 6}}
{{s(r + 1)(r + 2)}}} & {\frac{{12}}
{{s^2 r(r + 1)(r + 2)}}}  \\
 \end{array} } \right) + \left( {\begin{array}{*{20}c}
   {\sigma _{b_0 }^2 } & {\rho _{b_0 b_1 } \sigma _{b_0 } \sigma _{b_1 } }  \\
   {\rho _{b_0 b_1 } \sigma _{b_0 } \sigma _{b_1 } } & {\sigma _{b_1 }^2 }  \\
 \end{array} } \right)} \right)^{ - 1} 
\end{multline*}
We computed this inverse using Mathematica \cite{Wolfram:2005}, and then using equation~\eqref{varcmdvt00p1} substituting $s^2 \det ({\mathbf{A}})
$ by $\det \left( {{\mathbf{\tilde A}}} \right)$ we derived the expression for ${\mathbf{c'\Sigma }}_{\rm B} {\mathbf{c}}$, which is 
$$
\frac{{\left( {\sigma _0^2  + \frac{{2(2r + 1)\sigma _w^2 }}
{{r^2  + 3r + 2}}} \right)\left( {\sigma _1^2  + \frac{{12\sigma _w^2 }}
{{\left( {r^3  + 3r^2  + 2r} \right)s^2 }}} \right) - \left( {\sigma _{01}  - \frac{{6\sigma _w^2 }}
{{\left( {r^2  + 3r + 2} \right)s}}} \right)^2 }}
{{(1 - p_e )p_e \left( {\sigma _1^2  + \frac{{12\sigma _w^2 }}
{{\left( {r^3  + 3r^2  + 2r} \right)s^2 }}} \right)}}.
$$
The limit of this expression is 
$$
\frac{{\sigma _0^2 \sigma _1^2  - \sigma _{01}^2 }}
{{p_e (1 - p_e )\sigma _1^2 }}.
$$
Equivalently, we derived the same results for the fixed $\tau $ case. This limit can be rewritten in terms of our paramterization as 
$$
\mathop {\lim }\limits_{r \to \infty } \;\;{\mathbf{c'\Sigma }}_{\rm B} {\mathbf{c}} = \frac{{\sigma _{t_0 }^2 \rho _{t_0 } \left( {1 - \rho _{01}^2 } \right)}}
{{p_e (1 - p_e )}}.
$$

\subsection{LDD, RS, $V(t_0 ) = 0$}
\label{aplimitslddrs}

When $V(t_0 ) = 0$, using equation \eqref{varlddp1}, we can derive ${\mathbf{c'\Sigma }}_{\rm B} {\mathbf{c}}$ by substituting $s^2 \det ({\mathbf{A}})$ by $\det \left( {{\mathbf{\tilde A}}} \right)$ to obtain, in terms of our parameterization, 
$$
{\mathbf{c'\Sigma }}_{\rm B} {\mathbf{c}} = \left( {\frac{{12\sigma ^2 (1 - \rho _{t_0 } )}}
{{{\kern 1pt} s^2 p_e (1 - p_e )\,}}} \right)\left( {\frac{1}
{{\,r(r + 1)(r + 2)}} + \left( {\frac{{\rho _{b_1 ,s,\tilde r} }}
{{1 - \rho _{b_1 ,s,\tilde r} }}} \right)\frac{1}
{{\tilde r(\tilde r + 1)(\tilde r + 2)}}} \right),
$$
as in Table~\ref{table1p1}. Then, it is easily derived that 
$$
\mathop {\lim }\limits_{r \to \infty } \;\;{\mathbf{c'\Sigma }}_{\rm B} {\mathbf{c}} = \left( {\frac{{12\sigma ^2 (1 - \rho _{t_0 } )}}
{{{\kern 1pt} s^2 p_e (1 - p_e )\,}}} \right)\left( {\frac{{\rho _{b_1 ,s,\tilde r} }}
{{1 - \rho _{b_1 ,s,\tilde r} }}} \right)\frac{1}
{{\tilde r(\tilde r + 1)(\tilde r + 2)}},
$$
and, equivalently, for the fixed $\tau $ case the limit is 
$$
\left( {\frac{{12\sigma ^2 (1 - \rho _{t_0 } )}}
{{{\kern 1pt} \tau ^2 p_e (1 - p_e )\,}}} \right)\left( {\frac{{\rho _{b_1 ,\tau ,\tilde r} }}
{{1 - \rho _{b_1 ,\tau ,\tilde r} }}} \right)\frac{{\tilde r}}
{{\,(\tilde r + 1)(\tilde r + 2)}}.
$$

\section{The effect of covariance parameters on the minimum $r$ for a fixed $N$, subject to power $\pi $}
\label{apeffectr}

\subsection{The effect of $\rho $ and $\rho _{t_0 } $}
\label{apefrho}

\subsubsection{CMD, CS, $V(t_0 ) = 0$}
\label{apefrhocmdcs}

From equation~\eqref{rcmdcsp1}, 
$$
r = \frac{{\beta _2^2 N\,p_e (1 - p_e ) - \left( {z_\pi   + z_{1 - \alpha /2} } \right)^2 \sigma ^2 }}
{{\left( {z_\pi   + z_{1 - \alpha /2} } \right)^2 \sigma ^2 \rho  - \beta _2^2 N\,p_e (1 - p_e )}}.
$$
Differentiating with respect to $\rho $, we get
$$
\frac{{\partial r}}
{{\partial \rho }} = \frac{{\left( {z_\pi   + z_{1 - \alpha /2} } \right)^2 \sigma ^2 \left( { - \beta _2^2 N\,p_e (1 - p_e ) + \left( {z_\pi   + z_{1 - \alpha /2} } \right)^2 \sigma ^2 } \right)}}
{{\left( {\left( {z_\pi   + z_{1 - \alpha /2} } \right)^2 \sigma ^2 \rho  - \beta _2^2 N\,p_e (1 - p_e )} \right)^2 }}.
$$
If $\left( {z_\pi   + z_{1 - \alpha /2} } \right)^2 \sigma ^2  > \beta _2^2 N\,p_e (1 - p_e )$, then $\frac{{\partial r}} {{\partial \rho }} > 0$, so r increases as $\rho
$ increases.
If $\left( {z_\pi   + z_{1 - \alpha /2} } \right)^2 \sigma ^2  < \beta _2^2 N\,p_e (1 - p_e )$, then $\frac{{\partial r}}{{\partial \rho }} < 0$, so r decreases as $\rho $ increases.

\subsubsection{LDD, CS, fixed s, $V(t_0 ) = 0$}
\label{apefrholddcs}

The minimum $r$ for fixed $N$ and fixed power, $\pi $, solves 
$$
N = \frac{{12\sigma ^2 (1 - \rho )\left( {z_\pi   + z_{1 - \alpha /2} } \right)^2 {\kern 1pt} }}
{{{\kern 1pt} \gamma _3^2 p_e (1 - p_e )\;s^2 \,r\,(r + 1)(r + 2)}},
$$
which was obtained plugging in the corresponding value of ${\mathbf{c'\Sigma }}_{\rm B} {\mathbf{c}} $ in Table~\ref{table1p1} into equation~\eqref{Np1}. Defining 
$$
F\left( {r,\,\rho } \right) = \frac{{N\gamma _3^2 p_e (1 - p_e )\;s^2 \,}}
{{12\sigma ^2 \left( {z_\pi   + z_{1 - \alpha /2} } \right)^2 }} - \frac{{(1 - \rho )}}
{{r\,(r + 1)(r + 2)}},
$$
the equation $F\left( {r,\,\rho } \right) = 0$ implicitly defines the function $r = f(\rho )$. Using implicit differentiation and taking into account that r is a function of $\rho $, $r(\rho )$, we obtain 
$$
\frac{{\partial r}}
{{\partial \rho }} = \frac{{ - r\,\left( {r + 1} \right)\left( {r + 2} \right)}}
{{(1 - \rho )(3r^2  + 6r + 2)}}.
$$
Since r is positive, the derivative is always negative, and r decreases as $\rho $ increases.

\subsubsection{LDD, CS, fixed $\tau $, $V(t_0 ) = 0$}
\label{apefrholddcstau}

The minimum $r$ for fixed $N$ and fixed power, $\pi $, solves 
$$
N = \frac{{12\sigma ^2 (1 - \rho )\left( {z_\pi   + z_{1 - \alpha /2} } \right)^2 {\kern 1pt} r}}
{{{\kern 1pt} \gamma _3^2 p_e (1 - p_e )\;\tau ^2 \,\,(r + 1)(r + 2)}},
$$
which was obtained plugging in the corresponding value of ${\mathbf{c'\Sigma }}_{\rm B} {\mathbf{c}}$ in Table~\ref{table1p1} into equation~\eqref{Np1}. Defining 
$$
F\left( {r,\,\rho } \right) = \frac{{N\gamma _3^2 p_e (1 - p_e )\;\tau ^2 \,}}
{{12\sigma ^2 \left( {z_\pi   + z_{1 - \alpha /2} } \right)^2 }} - \frac{{(1 - \rho )r}}
{{\,(r + 1)(r + 2)}},
$$
the equation $F\left( {r,\,\rho } \right) = 0$ implicitly defines the function $r = f(\rho )$. Using implicit differentiation and  taking into account that r is a function of $\rho $, we obtain 
$$
\frac{{\partial r}}
{{\partial \rho }} = \frac{{r\left( {r + 1} \right)\left( {r + 2} \right)}}
{{(1 - \rho )\left( { - r^2  + 2} \right)}}.
$$
If $r \geqslant 2$, then $\frac{{\partial r}}{{\partial \rho }} < 0$. So if we are taking at least two post-baseline measures, larger values of $\rho $ lead to smaller values of r to achieve the specified power. Since $\frac{r}{{\left( {r + 1} \right)\left( {r + 2} \right)}}$ is the same for $r = 1$ and $r = 2$, it is preferable to choose $r = 1$ since fewer measurements need to be collected. Therefore, the choice between $r = 1$ and $r = 2$ is not affected by $\rho $.

\subsubsection{LDD, RS, fixed s, $V(t_0 ) = 0$}
\label{apefrholddrss}

The minimum $r$ for fixed $N$ and fixed power, $\pi $, solves 
$$
N = \frac{{\left( {z_\pi   + z_{1 - \alpha /2} } \right)^2 \left( {\frac{{12\sigma ^2 (1 - \rho _{t_0 } )}}
{{{\kern 1pt} s^2 \,}}} \right)\left( {\frac{1}
{{\,r(r + 1)(r + 2)}} + \left( {\frac{{\rho _{b_1 ,s,\tilde r} }}
{{1 - \rho _{b_1 ,s,\tilde r} }}} \right)\frac{1}
{{\tilde r(\tilde r + 1)(\tilde r + 2)}}} \right)}}
{{{\kern 1pt} \gamma _3^2 p_e (1 - p_e )\;}},
$$
which was obtained by plugging in the corresponding value of ${\mathbf{c'\Sigma }}_{\rm B} {\mathbf{c}}
$ in Table~\ref{table1p1} into equation~\eqref{Np1}. Defining 
\begin{multline*}
F\left( {r,\,\rho _{t_0 } } \right) = \\
\frac{{N\gamma _3^2 p_e (1 - p_e )s^2 }}
{{12\sigma ^2 \left( {z_\pi   + z_{1 - \alpha /2} } \right)^2 }} - (1 - \rho _{t_0 } )\left( {\frac{1}
{{\,r(r + 1)(r + 2)}} + \left( {\frac{{\rho _{b_1 ,s,\tilde r} }}
{{1 - \rho _{b_1 ,s,\tilde r} }}} \right)\frac{1}
{{\tilde r(\tilde r + 1)(\tilde r + 2)}}} \right),
\end{multline*}
the equation $F\left( {r,\,\rho _{t_0 } } \right) = 0$ implicitly defines the function $r = f(\rho _{t_0 } )$. Using implicit differentiation and taking into account that r depends on $\rho _{t_0 } $, we obtain 
$$
\frac{{\partial r}}
{{\partial \rho _{t_0 } }} = \frac{{ - r\,\left( {r + 1} \right)\left( {r + 2} \right)\left[ {\rho _{b_1 ,s,\tilde r} r\,\left( {r + 1} \right)\left( {r + 2} \right) + \left( {1 - \rho _{b_1 ,s,\tilde r} } \right)\tilde r(\tilde r + 1)(\tilde r + 2)} \right]}}
{{\left( {1 - \rho _{b_1 ,s,\tilde r} } \right)\tilde r(\tilde r + 1)(\tilde r + 2)(1 - \rho _{t_0 } )\left( {3r^2  + 6r + 2} \right)}} < 0.
$$
Since the derivative is always negative when $r > 0$, r decreases as $\rho _{t_0 } $ increases.

\subsubsection{LDD, RS, fixed $\tau $, $V(t_0 ) = 0$}
\label{apefrholddrstau}

The minimum $r$ for fixed $N$ and fixed power, $\pi $, solves 
$$
N = \frac{{\left( {z_\pi   + z_{1 - \alpha /2} } \right)^2 \left( {\frac{{12\sigma ^2 (1 - \rho _{t_0 } )}}
{{{\kern 1pt} \tau ^2 \,}}} \right)\left( {\frac{r}
{{\,(r + 1)(r + 2)}} + \left( {\frac{{\rho _{b_1 ,\tau ,\tilde r} }}
{{1 - \rho _{b_1 ,\tau ,\tilde r} }}} \right)\frac{{\tilde r}}
{{\,(\tilde r + 1)(\tilde r + 2)}}} \right)}}
{{{\kern 1pt} \gamma _3^2 p_e (1 - p_e )\;}},
$$
which was obtained by plugging in the corresponding value of ${\mathbf{c'\Sigma }}_{\rm B} {\mathbf{c}}$ in Table~\ref{table1p1} into equation~\eqref{Np1}. Defining \begin{multline*}
F\left( {r,\,\rho } \right) = \\
\frac{{N\gamma _3^2 p_e (1 - p_e )\tau ^2 }}
{{12\sigma ^2 \left( {z_\pi   + z_{1 - \alpha /2} } \right)^2 }} - (1 - \rho _{t_0 } )\left( {\frac{r}
{{\,(r + 1)(r + 2)}} + \left( {\frac{{\rho _{b_1 ,\tau ,\tilde r} }}
{{1 - \rho _{b_1 ,\tau ,\tilde r} }}} \right)\frac{{\tilde r}}
{{\,(\tilde r + 1)(\tilde r + 2)}}} \right),
\end{multline*}
the equation $F\left( {r,\,\rho _{t_0 } } \right) = 0$ implicitly defines the function $r = f(\rho _{t_0 } )$. Using implicit differentiation, and taking into account that r depends on $\rho _{t_0 } $, we obtain 
$$
\frac{{\partial r}}
{{\partial \rho _{t_0 } }} = \frac{{\left( {r + 1} \right)\left( {r + 2} \right)\left[ {\rho _{b_1 ,\tau ,\tilde r} \tilde r\left( {r + 1} \right)\left( {r + 2} \right) + \left( {1 - \rho _{b_1 ,\tau ,\tilde r} } \right)(\tilde r + 1)(\tilde r + 2)r} \right]}}
{{\left( {1 - \rho _{b_1 ,\tau ,\tilde r} } \right)(\tilde r + 1)(\tilde r + 2)\left( {1 - \rho _{t_0 } } \right)(2 - r^2 )}}.
$$
If $r \geqslant 2$ then $\frac{{\partial r}} {{\partial \rho _{t_0 } }} < 0$. So if we are taking at least two post-baseline measures, larger values of $\rho _{t_0 } $ lead to smaller minimal values of r to achieve a certain power. Since $\frac{r}
{{\left( {r + 1} \right)\left( {r + 2} \right)}}$ is the same for $r = 1
$ and $r = 2$, the resulting power of both studies would be the same and it would be preferable to choose $r = 1$ since less measurements need to be collected. The choice between $r = 1$ and $r = 2$ is not affected by $\rho _{t_0 } $.

\subsection{The effect of $\rho _{b_1 ,s,\tilde r} $}
\label{apefrhob1}

\subsubsection{LDD, RS, fixed s, $V(t_0 ) = 0$}
\label{apefrhob1ldds}

The minimum $r$ for fixed $N$ and fixed power, $\pi $, solves 
$$
N = \frac{{\left( {z_\pi   + z_{1 - \alpha /2} } \right)^2 \left( {\frac{{12\sigma ^2 (1 - \rho _{t_0 } )}}
{{{\kern 1pt} s^2 \,}}} \right)\left( {\frac{1}
{{\,r(r + 1)(r + 2)}} + \left( {\frac{{\rho _{b_1 ,s,\tilde r} }}
{{1 - \rho _{b_1 ,s,\tilde r} }}} \right)\frac{1}
{{\tilde r(\tilde r + 1)(\tilde r + 2)}}} \right)}}
{{{\kern 1pt} \gamma _3^2 p_e (1 - p_e )\;}},
$$
which was obtained plugging in the corresponding value of ${\mathbf{c'\Sigma }}_{\rm B} {\mathbf{c}}$ in Table~\ref{table1p1} into equation~\eqref{Np1}. Defining 
\begin{multline*}
F\left( {r,\,\rho _{b_1 ,s,\tilde r} } \right) = \\
\frac{{N\gamma _3^2 p_e (1 - p_e )s^2 }}
{{12\sigma ^2 \left( {z_\pi   + z_{1 - \alpha /2} } \right)^2 (1 - \rho _{t_0 } )}} - \left( {\frac{1}
{{\,r(r + 1)(r + 2)}} + \left( {\frac{{\rho _{b_1 ,s,\tilde r} }}
{{1 - \rho _{b_1 ,s,\tilde r} }}} \right)\frac{1}
{{\tilde r(\tilde r + 1)(\tilde r + 2)}}} \right),
\end{multline*}
the equation $F\left( {r,\,\rho _{b_1 ,s,\tilde r} } \right) = 0$ implicitly defines the function $r = f(\rho _{b_1 ,s,\tilde r} )$. Using implicit differentiation, and taking into account that r depends on $\rho _{b_1 ,s,\tilde r} $, we obtain  
$$
\frac{{\partial r}}
{{\partial \rho _{b_1 ,s,\tilde r} }} = \frac{{r^2 (r + 1)^2 (r + 2)^2 }}
{{\tilde r(\tilde r + 1)(\tilde r + 2)(1 - \rho _{b_1 ,s,\tilde r} )^2 (3r^2  + 6r + 2)}} > 0.
$$
Since the derivative is always positive, r increases as $\rho _{b_1 ,s,\tilde r} $ increases.

\subsubsection{LDD, RS, fixed $\tau $, $V(t_0 ) = 0$}
\label{apefrhob1ldtau}

The minimum $r$ for fixed $N$ and fixed power, $\pi $, solves
$$
N = \frac{{\left( {z_\pi   + z_{1 - \alpha /2} } \right)^2 \left( {\frac{{12\sigma ^2 (1 - \rho _{t_0 } )}}
{{{\kern 1pt} \tau ^2 \,}}} \right)\left( {\frac{r}
{{\,(r + 1)(r + 2)}} + \left( {\frac{{\rho _{b_1 ,\tau ,\tilde r} }}
{{1 - \rho _{b_1 ,\tau ,\tilde r} }}} \right)\frac{{\tilde r}}
{{\,(\tilde r + 1)(\tilde r + 2)}}} \right)}}
{{{\kern 1pt} \gamma _3^2 p_e (1 - p_e )\;}},
$$
which was obtained plugging in the corresponding value of ${\mathbf{c'\Sigma }}_{\rm B} {\mathbf{c}}$ in Table~\ref{table1p1} into equation~\eqref{Np1}. Defining 
\begin{multline*}
F\left( {r,\,\rho _{b_1 ,\tau ,\tilde r} } \right) = \\
\frac{{N\gamma _3^2 p_e (1 - p_e )\tau ^2 }}
{{12\sigma ^2 \left( {z_\pi   + z_{1 - \alpha /2} } \right)^2 (1 - \rho _{t_0 } )}} - \left( {\frac{r}
{{\,(r + 1)(r + 2)}} + \left( {\frac{{\rho _{b_1 ,\tau ,\tilde r} }}
{{1 - \rho _{b_1 ,\tau ,\tilde r} }}} \right)\frac{{\tilde r}}
{{\,(\tilde r + 1)(\tilde r + 2)}}} \right),
\end{multline*}
the equation $F\left( {r,\,\rho _{b_1 ,s,\tilde r} } \right) = 0$ implicitly defines the function $r = f(\rho _{b_1 ,s,\tilde r} )$. Using implicit differentiation, and taking into account that r depends on $\rho _{b_1 ,\tau ,\tilde r} $, we obtain 
$$
\frac{\partial }
{{\partial \rho _{b_1 ,\tau ,\tilde r} }} = \frac{{\tilde r(r + 1)^2 (r + 2)^2 }}
{{\,(r^2  - 2)(\tilde r + 1)(\tilde r + 2)(1 - \rho _{b_1 ,\tau ,\tilde r} )^2 }}.
$$
If $r \geqslant 2$, $\frac{{\partial r}}
{{\partial \rho _{b_1 ,\tau ,\tilde r} }} > 0$. So if we are taking at least two post-baseline measurements, the effect of increasing $\rho _{b_1 ,\tau ,\tilde r} 
$ is to increase the minimum r needed to achieve a pre-specified power. Since $\frac{r}
{{\left( {r + 1} \right)\left( {r + 2} \right)}}$ is the same for $r = 1$ and $r = 2$, the resulting power of both studies would be the same and it is therefore preferable to choose $r = 1$ since less measurements need to be collected. The choice between $r = 1$ and $r = 2$ is not affected by $\rho _{b_1 ,\tau ,\tilde r} $.

\section{Calculation of the variance under RS and $V(t_0 ) > 0$ assuming $t_{0i} $ are normally distributed}
\label{apvarRS}

We need to derive 
$$
{\mathbf{\Sigma }}_{\rm B}  = \left( {\mathbb{E}\left( {{\mathbf{X'}}_i {\mathbf{\Sigma }}_i^{ - 1} {\mathbf{X}}_i } \right)} \right)^{ - 1}. 
$$
When ${\mathbf{\Sigma }}_i  = {\mathbf{\Sigma }}$ for all subjects, $\mathbb{E}\left( {{\mathbf{X'}}_i {\mathbf{\Sigma }}_i^{ - 1} {\mathbf{X}}_i } \right)$ can be computed exactly. This will happen when $V(t_0 ) = 0$, and then equations \eqref{varcmdvt00p1} and \eqref{varlddp1} for $V(t_0 ) = 0$ provide general expression for ${\mathbf{c'\Sigma }}_{\rm B} {\mathbf{c}}$ for CMD and LDD, respectively. However, if $V(t_0 ) > 0$ then ${\mathbf{\Sigma }}_i  \ne {\mathbf{\Sigma }}$ under RS. Specifically, ${\mathbf{\Sigma }}_i $ depends on $t_{0i} $, so we have ${\mathbf{\Sigma }}(t_{0i} )$. The formula for ${\mathbf{\Sigma }}_i $ under RS is ${\mathbf{\Sigma }}_i  = {\mathbf{Z}}_i {\mathbf{DZ'}}_i  + \sigma _{within}^2 {\mathbf{I}}$, where  
$$
{\mathbf{Z'}}_i  = \left( {\begin{array}{*{20}c}
   1 &  \cdots  &  \cdots  &  \cdots  & 1  \\
   {t_{0i} } &  \cdots  & {t_{0i}  + js} &  \cdots  & {t_{0i}  + rs}  \\
 \end{array} } \right)
$$
and 
$$
{\mathbf{D}} = \left( {\begin{array}{*{20}c}
   {\sigma _0^2 } & {\sigma _{01} }  \\
   {\sigma _{01} } & {\sigma _1^2 }  \\
 \end{array} } \right).
$$
At this point it is convenient to introduce a new matrix 
$$
{\mathbf{W}}_i  = \left( {\begin{array}{*{20}c}
   1 & 0 & {k_i } & 0  \\
   0 & 1 & 0 & {k_i }  \\
 \end{array} } \right).
$$
Note that under LDD, 
\begin{multline*}
{\mathbf{Z}}_i {\mathbf{W}}_i  = \left( {\begin{array}{*{20}c}
   1 & {t_{0i} }  \\
    \vdots  &  \vdots   \\
    \vdots  & {t_{0i}  + js}  \\
    \vdots  &  \vdots   \\
   1 & {t_{0i}  + rs}  \\
 \end{array} } \right)\left( {\begin{array}{*{20}c}
   1 & 0 & {k_i } & 0  \\
   0 & 1 & 0 & {k_i }  \\
 \end{array} } \right) = \\
 \left( {\begin{array}{*{20}c}
   1 & {t_{0i} } & {k_i } & {t_{0i} k_i }  \\
    \vdots  &  \vdots  &  \vdots  &  \vdots   \\
    \vdots  & {t_{0i}  + js} &  \vdots  & {\left( {t_{0i}  + js} \right)k_i }  \\
    \vdots  &  \vdots  &  \vdots  &  \vdots   \\
   1 & {t_{0i}  + rs} & {k_i } & {\left( {t_{0i}  + rs} \right)k_i }  \\
 \end{array} } \right) = {\mathbf{X}}_i.
\end{multline*}
Therefore, 
$$
{\mathbf{X}}_i ^\prime  {\mathbf{\Sigma }}_i^{ - 1} {\mathbf{X}}_i  = {\mathbf{W}}_i ^\prime  {\mathbf{Z}}_i ^\prime  {\mathbf{\Sigma }}_i^{ - 1} {\mathbf{Z}}_i {\mathbf{W}}_i  = {\mathbf{W}}_i ^\prime  {\mathbf{Z}}_i ^\prime  \left( {{\mathbf{Z}}_i {\mathbf{DZ'}}_i  + \sigma _{within}^2 {\mathbf{I}}} \right)^{ - 1} {\mathbf{Z}}_i {\mathbf{W}}_i.
$$
In Appendix~\ref{aplimitscmdrs} we derived , 
$$
{\mathbf{Z'}}\left( {{\mathbf{ZDZ'}} + \sigma _{within}^2 {\mathbf{I}}} \right)^{ - 1} {\mathbf{Z}} = \left( {\left( {{\mathbf{Z'Z}}} \right)^{ - 1} \sigma _{within}^2  + {\mathbf{D}}} \right)^{ - 1},
$$
so we can deduce ${\mathbf{X}}_i ^\prime  {\mathbf{\Sigma }}_i^{ - 1} {\mathbf{X}}_i  = {\mathbf{W}}_i ^\prime  \left( {\left( {{\mathbf{Z'Z}}} \right)^{ - 1} \sigma _{within}^2  + {\mathbf{D}}} \right)^{ - 1} {\mathbf{W}}_i 
$. Now, 
$$
\left( {{\mathbf{Z'}}_i {\mathbf{Z}}_i } \right) = \left( {\begin{array}{*{20}c}
   {r + 1} & {\left( {r + 1} \right)t_{0i}  + \frac{{s{\kern 1pt} r(r + 1)}}
{2}}  \\
   {\left( {r + 1} \right)t_{0i}  + \frac{{s{\kern 1pt} r(r + 1)}}
{2}} & {\left( {r + 1} \right)t_{0i} ^2  + st_{0i} r(r + 1) + \frac{{s^2 r(r + 1)(2r + 1)}}
{6}}  \\
 \end{array} } \right),
$$
and 
$$
\left( {{\mathbf{Z'}}_i {\mathbf{Z}}_i } \right)^{ - 1}  = \frac{1}
{{r(r + 1)(r + 2)s^2 }}\left( {\begin{array}{*{20}c}
   {2\left( {r\left( {1 + 2r} \right)s^2  + 6r{\kern 1pt} st_{0i}  + 6t_{0i} ^2 } \right)} & { - 6\left( {r{\kern 1pt} s + 2t_{0i} } \right)}  \\
   { - 6\left( {r{\kern 1pt} s + 2t_{0i} } \right)} & {12}  \\
 \end{array} } \right).
$$
Using this result we computed 
$$
\left( {\left( {{\mathbf{Z'Z}}} \right)^{ - 1} \sigma _{within}^2  + {\mathbf{D}}} \right)^{ - 1}  = \left( {\begin{array}{*{20}c}
   {a(t_{0i} )} & {c(t_{0i} )}  \\
   {c(t_{0i} )} & {d(t_{0i} )}  \\
 \end{array} } \right),
$$
where 
$$
a(t_{0i} ) = \frac{{\left( {\frac{{12\sigma _w^2 }}
{{r(r + 1)(r + 2)s^2 }} + \sigma _1^2 } \right)}}
{{\left( {\frac{{12\sigma _w^2 }}
{{r(r + 1)(r + 2)s^2 }} + \sigma _1^2 } \right)\left( {\sigma _0^2  + \frac{{2\sigma _w^2 \left( {r(1 + 2r)s^2  + 6rst_{0i}  + 6t_{0i} ^2 } \right)}}
{{r(r + 1)(r + 2)s^2 }}} \right) - \left( {\sigma _{01}  - \frac{{6\sigma _w^2 (rs + 2t_{0i} )}}
{{r(r + 1)(r + 2)s^2 }}} \right)^2 }},
$$ 
$$
c(t_{0i} ) = \frac{{\left( { - \sigma _{01}  + \frac{{6\sigma _w^2 (rs + 2t_{0i} )}}
{{r(r + 1)(r + 2)s^2 }}} \right)}}
{{\left( {\frac{{12\sigma _w^2 }}
{{r(r + 1)(r + 2)s^2 }} + \sigma _1^2 } \right)\left( {\sigma _0^2  + \frac{{2\sigma _w^2 \left( {r(1 + 2r)s^2  + 6rst_{0i}  + 6t_{0i} ^2 } \right)}}
{{r(r + 1)(r + 2)s^2 }}} \right) - \left( {\sigma _{01}  - \frac{{6\sigma _w^2 (rs + 2t_{0i} }}
{{r(r + 1)(r + 2)s^2 }}} \right)^2 }},
$$
$$
d(t_{0i} ) = \frac{{\left( {\sigma _0^2  + \frac{{2\sigma _w^2 \left( {r(1 + 2r)s^2  + 6rst_{0i}  + 6t_{0i} ^2 } \right)}}
{{r(r + 1)(r + 2)s^2 }}} \right)}}
{{\left( {\frac{{12\sigma _w^2 }}
{{r(r + 1)(r + 2)s^2 }} + \sigma _1^2 } \right)\left( {\sigma _0^2  + \frac{{2\sigma _w^2 \left( {r(1 + 2r)s^2  + 6rst_{0i}  + 6t_{0i} ^2 } \right)}}
{{r(r + 1)(r + 2)s^2 }}} \right) - \left( {\sigma _{01}  - \frac{{6\sigma _w^2 (rs + 2t_{0i} }}
{{r(r + 1)(r + 2)s^2 }}} \right)^2 }}.
$$
Pre- and post-multiplying by ${\mathbf{W}}_i $ we get 
\begin{multline*}
{\mathbf{X}}_i ^\prime  {\mathbf{\Sigma }}_i^{ - 1} {\mathbf{X}}_i  = {\mathbf{W}}_i ^\prime  \left( {\left( {{\mathbf{Z'Z}}} \right)^{ - 1} \sigma _{within}^2  + {\mathbf{D}}} \right)^{ - 1} {\mathbf{W}}_i  = \\
\left( {\begin{array}{*{20}c}
   {a(t_{0i} )} & {c(t_{0i} )} & {k_i a(t_{0i} )} & {k_i c(t_{0i} )}  \\
   {c(t_{0i} )} & {d(t_{0i} )} & {k_i c(t_{0i} )} & {k_i d(t_{0i} )}  \\
   {k_i a(t_{0i} )} & {k_i c(t_{0i} )} & {k_i a(t_{0i} )} & {k_i c(t_{0i} )}  \\
   {k_i c(t_{0i} )} & {k_i d(t_{0i} )} & {k_i c(t_{0i} )} & {k_i d(t_{0i} )}  \\
 \end{array} } \right).
\end{multline*} 
Now, 
$$
\mathbb{E}\left[ {{\mathbf{X}}_i ^\prime  {\mathbf{\Sigma }}_i^{ - 1} {\mathbf{X}}_i } \right] = \left( {\begin{array}{*{20}c}
   {\mathbb{E}\left[ {a(t_{0i} )} \right]} & {\mathbb{E}\left[ {c(t_{0i} )} \right]} & {\mathbb{E}\left[ {k_i a(t_{0i} )} \right]} & {\mathbb{E}\left[ {k_i c(t_{0i} )} \right]}  \\
   {\mathbb{E}\left[ {c(t_{0i} )} \right]} & {\mathbb{E}\left[ {d(t_{0i} )} \right]} & {\mathbb{E}\left[ {k_i c(t_{0i} )} \right]} & {\mathbb{E}\left[ {k_i d(t_{0i} )} \right]}  \\
   {\mathbb{E}\left[ {k_i a(t_{0i} )} \right]} & {\mathbb{E}\left[ {k_i c(t_{0i} )} \right]} & {\mathbb{E}\left[ {k_i a(t_{0i} )} \right]} & {\mathbb{E}\left[ {k_i c(t_{0i} )} \right]}  \\
   {\mathbb{E}\left[ {k_i c(t_{0i} )} \right]} & {\mathbb{E}\left[ {k_i d(t_{0i} )} \right]} & {\mathbb{E}\left[ {k_i c(t_{0i} )} \right]} & {\mathbb{E}\left[ {k_i d(t_{0i} )} \right]}  \\
 \end{array} } \right).
$$
To compute the expected values in the matrix, we need to know the distribution of $t_{0i} $ and the joint distribution of $\left( {t_{0i} ,k_i } \right)$. We assume that the distribution of $t_{0i} $ has mean zero and variance $V(t_0 )$ and that $k_i $ follows a Bernoulli distribution with probability of success $p_e $. Additionally, we assume that within each exposure group, $t_{0i} $ is normally distributed with the same variance. In Appendix~\ref{apvar2p1} we deduced the means of $t_{0i} $ in each of the exposure groups as 
$$
\mathbb{E}\left( {t_0 |k = 1} \right) = \rho _{\operatorname{e} ,t_0 } \sqrt {\frac{{(1 - p_e )}}
{{p_e }}} \sqrt {V\left( {t_0 } \right)} 
$$
and 
$$
\mathbb{E}\left[ {t_0 |k = 1} \right] =  - \rho _{\operatorname{e} ,t_0 } \sqrt {\frac{{p_e }}
{{(1 - p_e )}}} \sqrt {V\left( {t_0 } \right)}.
$$
Also, using the results from Appendix~\ref{apvar2p1}, we can deduce that the common variance in the two groups is 
\begin{multline*}
V\left( {t_0 |k = 1} \right) = \mathbb{E}\left( {t_0^2 |k = 1} \right) - \left[ {\mathbb{E}\left( {t_0 |k = 1} \right)} \right]^2  \\
 = \frac{{V\left( {t_0 } \right)\left[ {p_e  + \rho _{\operatorname{e} ,t_0 }^2 (1 - 2p_e )} \right]}}
{{p_e }} - \rho _{\operatorname{e} ,t_0 }^2 \frac{{(1 - p_e )}}
{{p_e }}V\left( {t_0 } \right) = V\left( {t_0 } \right)\left( {1 - \rho _{\operatorname{e} ,t_0 }^2 } \right).
\end{multline*}
Therefore, we can write 
\begin{multline*}
f(t_{0i} |k_i  = 1) = \\
\frac{1}
{{\sqrt {2\pi V\left( {t_0 } \right)\left( {1 - \rho _{\operatorname{e} ,t_0 }^2 } \right)} }}\exp \left[ {\frac{{ - 1}}
{{2V\left( {t_0 } \right)\left( {1 - \rho _{\operatorname{e} ,t_0 }^2 } \right)}}\left( {t_{0i}  - \rho _{\operatorname{e} ,t_0 } \sqrt {\frac{{(1 - p_e )}}
{{p_e }}} \sqrt {V\left( {t_0 } \right)} } \right)^2 } \right]
\end{multline*} 
and 
\begin{multline*}
f(t_{0i} |k_i  = 0) = \\
\frac{1}
{{\sqrt {2\pi V\left( {t_0 } \right)\left( {1 - \rho _{\operatorname{e} ,t_0 }^2 } \right)} }}\exp \left[ {\frac{{ - 1}}
{{2V\left( {t_0 } \right)\left( {1 - \rho _{\operatorname{e} ,t_0 }^2 } \right)}}\left( {t_{0i}  + \rho _{\operatorname{e} ,t_0 } \sqrt {\frac{{p_e }}
{{(1 - p_e )}}} \sqrt {V\left( {t_0 } \right)} } \right)^2 } \right].
\end{multline*}
We can write this conditional distribution as 
\begin{multline*}
f(t_{0i} |k_i ) = \frac{1}
{{\sqrt {2\pi V\left( {t_0 } \right)\left( {1 - \rho _{\operatorname{e} ,t_0 }^2 } \right)} }} \\
\exp \left[ {\frac{{ - 1}}
{{2V\left( {t_0 } \right)\left( {1 - \rho _{\operatorname{e} ,t_0 }^2 } \right)}}\left( {t_{0i}  - \rho _{\operatorname{e} ,t_0 } \sqrt {V\left( {t_0 } \right)} ( - 1)^{1 - k_i } (1 - p_e )^{k_i  - \frac{1}
{2}} p_e ^{\frac{1}
{2} - k_i } } \right)^2 } \right].
\end{multline*}
Then, the joint distribution is 
\begin{multline*}
f(t_{0i} ,k_i )$ $ = \frac{{p_e^{k_i } (1 - p_e )^{1 - k_i } }}
{{\sqrt {2\pi V\left( {t_0 } \right)\left( {1 - \rho _{\operatorname{e} ,t_0 }^2 } \right)} }} \\
\exp \left[ {\frac{{ - 1}}
{{2V\left( {t_0 } \right)\left( {1 - \rho _{\operatorname{e} ,t_0 }^2 } \right)}}\left( {t_{0i}  - \rho _{\operatorname{e} ,t_0 } \sqrt {V\left( {t_0 } \right)} ( - 1)^{1 - k_i } (1 - p_e )^{k_i  - \frac{1}
{2}} p_e ^{\frac{1}
{2} - k_i } } \right)^2 } \right].
\end{multline*}
The marginal distribution of $t_{0i} $ is a mixture of two normals, 
$$
f(t_{0i} ) = (1 - p_e )f(t_{0i} |k_i  = 0) + p_e f(t_{0i} |k_i  = 1).
$$
Then, we can derive 
$$
\mathbb{E}\left[ {a(t_{0i} )} \right] = \int {a(t_{0i} )f(t_{0i} )}  = (1 - p_e )\int {a(t_{0i} )f(t_{0i} |k_i  = 0)}  + p_e \int {a(t_{0i} )f(t_{0i} |k_i  = 1)}, 
$$
and equivalently for $\mathbb{E}\left[ {c(t_{0i} )} \right]$ and $\mathbb{E}\left[ {d(t_{0i} )} \right]$. We can also derive 
\begin{multline*}
\mathbb{E}\left[ {k_i a(t_{0i} )} \right] = \int {\int {k_i a(t_{0i} )f(t_{0i} ,k_i )dk_i dt_{0i} } }  = \sum\limits_{k_i  = 0,1} {\int {k_i a(t_{0i} )f(t_{0i} ,k_i )dt_{0i} } }  = \\
\int {a(t_{0i} )f(t_{0i} ,1)dt_{0i} }  =  \int {a(t_{0i} )f(t_{0i} ,1)dt_{0i} }  = \\
\int a(t_{0i} )\frac{{p_e }}
{{\sqrt {2\pi V\left( {t_0 } \right)\left( {1 - \rho _{\operatorname{e} ,t_0 }^2 } \right)} }} \\
\exp \left[ {\frac{{ - 1}}
{{2V\left( {t_0 } \right)\left( {1 - \rho _{\operatorname{e} ,t_0 }^2 } \right)}}\left( {t_{0i}  - \rho _{\operatorname{e} ,t_0 } \sqrt {V\left( {t_0 } \right)} (1 - p_e )^{\frac{1}
{2}} p_e ^{ - \frac{1}
{2}} } \right)^2 } \right]dt_{0i}   \\
 = p_e \int {a(t_{0i} )f(t_{0i} |k = 1)dt_{0i} },
\end{multline*}
and equivalently for $\mathbb{E}\left[ {k_i c(t_{0i} )} \right]$ and $\mathbb{E}\left[ {k_i d(t_{0i} )} \right]$.
Using the expressions derived here, our program computes the expectations numerically to obtain $\mathbb{E}\left[ {{\mathbf{X}}_i ^\prime  {\mathbf{\Sigma }}_i^{ - 1} {\mathbf{X}}_i } \right]$, then it inverts this matrix and extracts the [4,4] component, which will be $Var\left( {\hat \gamma _3 } \right)$. For CMD, the procedure is exactly the same but using the matrix 
$$
{\mathbf{W}}_i  = \left( {\begin{array}{*{20}c}
   1 & 0 & {k_i }  \\
   0 & 1 & 0  \\
 \end{array} } \right).
$$

\section{Proof that $r_{opt} $ is the same for both the cost constraint and the power constraint, and reduces to the solution to the unconstrained problem~\eqref{unconstrp1}, but $N_{opt} $ depends upon the constraint}
\label{apconstraint}

The power optimization problem is  
$$
\mathop {Max}\limits_r \;\Phi \left[ {\frac{{\sqrt N \left| {\left( {{\mathbf{c'B}}} \right)_{H_A } } \right|}}
{{\sqrt {{\mathbf{c'\Sigma }}_{\rm B} (r){\mathbf{c}}} }} - z_{1 - \alpha /2} } \right]{\text{ subject to }}COST = Nc_1  + \frac{{Nrc_1 }}
{\kappa }.
$$
The cost constraint, 
$$
N = \frac{{\kappa COST}}
{{c_1 (\kappa  + r)}},
$$
can be plugged in the optimization function to obtain the unconstrained problem 
$$
\mathop {Max}\limits_r \;\Phi \left[ {\frac{{\sqrt {\frac{{\kappa COST}}
{{c_1 (\kappa  + r)}}} \left| {\left( {{\mathbf{c'B}}} \right)_{H_A } } \right|}}
{{\sqrt {{\mathbf{c'\Sigma }}_{\rm B} (r){\mathbf{c}}} }} - z_{1 - \alpha /2} } \right].
$$
Since $\Phi $ is a monotone function, this is equivalent to 
$$
\mathop {Max}\limits_r \;\frac{{\sqrt {\frac{{\kappa COST}}
{{c_1 (\kappa  + r)}}} \left| {\left( {{\mathbf{c'B}}} \right)_{H_A } } \right|}}
{{\sqrt {{\mathbf{c'\Sigma }}_{\rm B} (r){\mathbf{c}}} }} - z_{1 - \alpha /2}.
$$
Removing positive constant terms with respect to $r$, it is equivalent to 
$$
\mathop {Max}\limits_r \;\frac{1}
{{(\kappa  + r){\mathbf{c'\Sigma }}_{\rm B} (r){\mathbf{c}}}},
$$
which is in turn equivalent to $\mathop {Min}\limits_r \;(\kappa  + r){\mathbf{c'\Sigma }}_{\rm B} (r){\mathbf{c}}$. Once $r_{opt} $ is found solving this minimization problem, $N_{opt} $ would be 
$$
N_{opt}  = \frac{{\kappa \,COST}}
{{c_1 \left( {\kappa  + r_{opt} } \right)}}.
$$
The cost optimization problem is
$$
\mathop {Min}\limits_r \;Nc_1  + \frac{{Nrc_1 }}
{\kappa }{\text{ subject to }}\Phi \left[ {\frac{{\sqrt N \left| {\left( {{\mathbf{c'B}}} \right)_{H_A } } \right|}}
{{\sqrt {{\mathbf{c'\Sigma }}_{\rm B} (r){\mathbf{c}}} }} - z_{1 - \alpha /2} } \right] = \pi.
$$
Noting that 
$$
Nc_1  + \frac{{Nrc_1 }}
{\kappa } = Nc_1 \left( {\frac{{\kappa  + r}}
{\kappa }} \right)
$$
and that from the power constraint 
$$
N = \frac{{\left( {{\mathbf{c'}}\,{\mathbf{\Sigma }}_{\rm B} (r)\,{\mathbf{c}}} \right)\left( {z_{1 - \alpha /2}  + z_\pi  } \right)^2 }}
{{\left( {\left( {{\mathbf{c'{\rm B}}}} \right)_{H_A } } \right)^2 }},
$$
this is equivalent to the unconstrained problem 
$$
\mathop {Min}\limits_r \;\frac{{({\mathbf{c'}}\,{\mathbf{\Sigma }}_{\rm B} (r)\,{\mathbf{c}})\left( {z_{1 - \alpha /2}  + z_\pi  } \right)^2 }}
{{\left( {\left( {{\mathbf{c'{\rm B}}}} \right)_{H_A } } \right)^2 }}c_1 \left( {\frac{{\kappa  + r}}
{\kappa }} \right).
$$
Removing positive constant terms with respect to $r$, the problem becomes $\mathop {Min}\limits_r \;\left( {\kappa  + r} \right)({\mathbf{c'}}\,{\mathbf{\Sigma }}_{\rm B} (r)\,{\mathbf{c}})$, which is equivalent to the minimization problem obtained before. Thus, given $\kappa $, ${\mathbf{c}}$ and ${\mathbf{\Sigma }}_{\rm B} (r)$, the same $r_{opt} $ maximizes power and minimizes cost. For the cost problem, once $r_{opt} $ is found solving the minimization problem, 
$$
N_{opt}  = \frac{{\left( {{\mathbf{c'\Sigma }}_{\rm B} (r_{opt} ){\mathbf{c}}} \right)\,\left( {z_\pi   + z_{1 - \alpha /2} } \right)^2 }}
{{({\mathbf{c'{\rm B}}}_{H_A } )^2 }}.
$$

\section{Optimal $r$ under LDD and fixed $s$, for CS}
\label{apoptrlddcss}

The optimal $r$ solves $\mathop {Min}\limits_r \;(\kappa  + r){\mathbf{c'\Sigma }}_{\rm B} {\mathbf{c}}$ (Appendix~\ref{apconstraint}). Plugging in the appropriate value for ${\mathbf{c'\Sigma }}_{\rm B} {\mathbf{c}}$ from Table~\ref{table1p1}, the problem under LDD, CS and fixed $s$ is 
$$
\mathop {Min}\limits_r \;(\kappa  + r)\frac{{12\sigma ^2 (1 - \rho )}}
{{p_e (1 - p_e )s^2 r(r + 1)(r + 2)}}.
$$
Removing positive constant terms with respect to $r$, this problem becomes 
$$
\mathop {Min}\limits_r \quad F(r) = \frac{{(\kappa  + r)}}
{{r\;(r + 1)(r + 2)}}.
$$
Since 
$$
\frac{{\partial F}}
{{\partial r}} = \frac{{ - 2\kappa  - 6\kappa r - 3r^2  - 3\kappa r^2  - 2r^3 }}
{{r^2 (r + 1)^2 (r + 2)^2 }} < 0\;\forall \kappa, 
$$
$F(r)$ decreases as $r$ increases, and $r_{opt}  \to \infty $ subject to the cost constraint.

\section{$r_{opt} $ under LDD, RS and fixed $s$}
\label{apoptrlddrss}

The optimal $r$ solves $\mathop {Min}\limits_r \;(\kappa  + r){\mathbf{c'\Sigma }}_{\rm B} {\mathbf{c}}$ (Appendix~\ref{apconstraint}). Plugging in the appropriate value for ${\mathbf{c'\Sigma }}_{\rm B} {\mathbf{c}}$ from Table~\ref{table1p1}, the problem under LDD, RS and fixed $s$ is 
$$
\mathop {Min}\limits_r \;(\kappa  + r)\left( {\frac{{12\sigma ^2 (1 - \rho _{t_0 } )}}
{{s^2 p_e (1 - p_e )}}} \right)\left( {\frac{1}
{{r(r + 1)(r + 2)}} + \left( {\frac{{\rho _{b_1 ,s,\tilde r} }}
{{1 - \rho _{b_1 ,s,\tilde r} }}} \right)\frac{1}
{{\tilde r(\tilde r + 1)(\tilde r + 2)}}} \right).
$$
Removing positive constant terms with respect to $r$, this problem becomes 
$$
\mathop {Min}\limits_r \quad G(r) = (\kappa  + r)\left( {\frac{1}
{{\,r(r + 1)(r + 2)}} + \left( {\frac{{\rho _{b_1 ,s,\tilde r} }}
{{1 - \rho _{b_1 ,s,\tilde r} }}} \right)\frac{1}
{{\tilde r(\tilde r + 1)(\tilde r + 2)}}} \right).
$$
The solution, $r_{opt} $, solves 
\begin{multline*}
\frac{{\partial G}}
{{\partial r}} = \left( {\frac{{\rho _{b_1 ,s,\tilde r} }}
{{1 - \rho _{b_1 ,s,\tilde r} }}} \right)\frac{1}
{{\tilde r(\tilde r + 1)(\tilde r + 2)}} + \frac{{ - 2\kappa  - 6\kappa r - 3r^2  - 3\kappa r^2  - 2r^3 }}
{{r^2 (r + 1)^2 (r + 2)^2 }} \\
= \left( {\frac{{\rho _{b_1 ,s,\tilde r} }}
{{1 - \rho _{b_1 ,s,\tilde r} }}} \right)\frac{1}
{{\tilde r(\tilde r + 1)(\tilde r + 2)}} + \frac{{\partial F}}
{{\partial r}} = 0,
\end{multline*}
where $\frac{{\partial F}} {{\partial r}}$ is the derivative of the objective function $F(r)$ for the analogous problem under compound symmetry (Appendix~\ref{apoptrlddcss}). We showed in Appendix~\ref{apoptrlddcss} that $\frac{{\partial F}}{{\partial r}}$ is always negative, and since 
$$
\frac{{\partial ^2 F}}
{{\partial r^2 }} = \frac{{2\left( {4\kappa  + 18\kappa r + 33\kappa r^2  + 7r^3  + 24\kappa r^3  + 9r^4  + 6\kappa r^4  + 3r^5 } \right)}}
{{r^3 \left( {r + 1} \right)^3 (r + 2)^3 }} \geqslant 0,
$$
$\frac{{\partial F}} {{\partial r}}$ is also an increasing function of $r$. In addition, $\mathop {\lim }\limits_{r \to \infty } \frac{{\partial F}}
{{\partial r}} = 0^ - $. Since $\frac{{\partial G}} {{\partial r}}$ is $\frac{{\partial F}} {{\partial r}}$ plus a constant, $\frac{{\partial G}}
{{\partial r}}$ will equal 0 at some interior point of $r$ between 1 and $\infty $. Since 
$$
\frac{{\partial ^2 G}}
{{\partial r^2 }} = \frac{{2\left( {4\kappa  + 18\kappa r + 33\kappa r^2  + 7r^3  + 24\kappa r^3  + 9r^4  + 6\kappa r^4  + 3r^5 } \right)}}
{{r^3 \left( {r + 1} \right)^3 (r + 2)^3 }} \geqslant 0
$$
for all $r > 0$, $G(r)$ is convex and the point that solves $\frac{{\partial G}}
{{\partial r}} = 0$ is a global minimum and therefore it is $r_{opt} $.
Now, 
$$
\frac{{\partial G}}
{{\partial r}} = 0 \Leftrightarrow \kappa  = \frac{{r_{opt} ^2 \left( { - (3 + 2r_{opt} )\tilde r(\tilde r + 1)(\tilde r + 2) + (r_{opt}  + 1)^2 (r_{opt}  + 2)^2 \frac{{\rho _{b_1 ,s,\tilde r} }}
{{1 - \rho _{b_1 ,s,\tilde r} }}} \right)}}
{{\,\left( {2 + 6r_{opt}  + 3r_{opt} ^2 } \right)\tilde r(\tilde r + 1)(\tilde r + 2)}}.
$$
Figure~\ref{roptlddrssp1} of the paper shows $r_{opt} $ for several values of $\kappa $ and $\rho _{b_1 ,s,\tilde r} $.

\section{$\left( {N_{opt} ,r_{opt} } \right)$ under LDD, CS and fixed $\tau $}
\label{apoptrlddcstau}

As shown in Appendix~\ref{apconstraint}, the optimal $r$ solves $\mathop {Min}\limits_r \;(\kappa  + r){\mathbf{c'\Sigma }}_{\rm B} {\mathbf{c}}$  Plugging in the appropriate value for ${\mathbf{c'\Sigma }}_{\rm B} {\mathbf{c}}$ from Table~\ref{table1p1}, the problem under LDD, CS and fixed $\tau $ is 
$$
\mathop {Min}\limits_r \;(\kappa  + r)\frac{{12\sigma ^2 (1 - \rho )r}}
{{p_e (1 - p_e )\tau ^2 (r + 1)(r + 2)}}.
$$
Removing positive constant terms with respect to $r$, this problem becomes 
$$
\mathop {Min}\limits_r \quad H(r) = \frac{{(\kappa  + r)\,r}}
{{(r + 1)(r + 2)}}.
$$
Taking derivatives with respect to $r$, $r_{opt} $ solves 
$$
\frac{{\partial H}}
{{\partial r}} = \frac{{(3 - \kappa )r^2  + 4r + 2\kappa }}
{{\left( {r + 1} \right)^2 (r + 2)^2 }} = 0.
$$
For $\kappa  < 3$ the derivative is positive. Therefore, when $\kappa  < 3
$ $H(r)$ increases with r and, consequently, the minimum is at $r =1$. If $\kappa  > 3
$, the derivative equals 0 at 
$$
r = \frac{{2 \pm \sqrt 2 \sqrt {2 - 3\kappa  + \kappa ^2 } }}
{{\kappa  - 3}},
$$
which gives a positive solution only at 
$$
r = \frac{{2 + \sqrt 2 \sqrt {2 - 3\kappa  + \kappa ^2 } }}
{{\kappa  - 3}}.
$$
Now, we need to check whether at this point there is a maximum or a minimum of $H(r)
$. The second derivative of $H(r)$ is 
$$
\frac{{\partial ^2 H}}
{{\partial r^2 }} = \frac{{2\left( {4 - 6\kappa  - 6\kappa r - 6r^2  - 3r^3  + \kappa r^3 } \right)}}
{{(r + 1)^3 (r + 2)^3 }}.
$$
We evaluated the second derivative at the point 
$$
r = \frac{{2 + \sqrt 2 \sqrt {2 - 3\kappa  + \kappa ^2 } }}
{{\kappa  - 3}}
$$
with Mathematica \cite{Wolfram:2005} and obtained 
$$
\left( {24 + \frac{{3\sqrt 2 }}
{{\sqrt {(\kappa  - 2)(\kappa  - 1)} }}} \right)\kappa  - 17\sqrt 2 \sqrt {(\kappa  - 2)(\kappa  - 1)}  - \frac{{7\sqrt 2 }}
{{\sqrt {(\kappa  - 2)(\kappa  - 1)} }} - 40.
$$
This expression can be proven to be negative for all $\kappa  > 3$. Therefore, $H(r)$ has a maximum at 
$$
r = \frac{{2 + \sqrt 2 \sqrt {2 - 3\kappa  + \kappa ^2 } }}
{{\kappa  - 3}},
$$
while we were looking for a minimum. Since this is the only local maximum or minimum of $H(r)$, the global minimum of $H(r)$ will be at $r = 1$ or at $r = \infty $ . The global minimum will be at $r = \infty$ if we can find a value of r such that 
$$
H(r) < H(1) = \frac{{(1 + \kappa )}}
{6}.
$$
With a little bit of algebra , we get 
$$
H(r) = \frac{{r(r + \kappa )}}
{{(r + 1)(r + 2)}} < \frac{{(1 + \kappa )}}
{6} \Leftrightarrow r^2 (\kappa  - 5) + r( - 3\kappa  + 3) + 2(\kappa  + 1) > 0,
$$
which has roots at $r = 1$ and ${\text{ }}r = \frac{{2(\kappa  + 1)}}
{{\kappa  - 5}}$. If $\kappa  < 5$, then ${\text{ }}r = \frac{{2(\kappa  + 1)}}
{{\kappa  - 5}} < 0$, outside of its valid range. The global minimum is then  $r_{opt}  = 1$, 
$$
N_{opt}  = \frac{{\kappa COST}}
{{c_1 (\kappa  + 1)}}
$$
or 
$$
N_{opt}  = \frac{{2\sigma ^2 (1 - \rho )\,\left( {z_\pi   + z_{1 - \alpha /2} } \right)^2 }}
{{\tau ^2 p_e (1 - p_e )\gamma _3^2 }}
$$
for the power maximization or cost minimization problems, respectively. If $\kappa  > 5$, then $r = \frac{{2(\kappa  + 1)}}{{\kappa  - 5}} > 1$ within the range, so taking $r$as large as possible subject to the cost constraint, 
$$
r_{opt}  = \kappa \left( {\frac{{COST}}
{{c_1 }} - 1} \right)
$$
and $N_{opt}  = 1$.  Under a power constraint, one would find the smallest $r$ that satisfies the power constraint and set $N_{opt}  = 1$. In reality, the investigator will set $r$ as large as is feasible, and then find $N_{opt} $ to satisfy the cost or power constraint.

\section{$\left( {N_{opt} ,r_{opt} } \right)$ under LDD, RS and fixed $\tau $}
\label{apoptrlddrstau}

The optimal $r$ solves $\mathop {Min}\limits_r \;(\kappa  + r){\mathbf{c'\Sigma }}_{\rm B} {\mathbf{c}}$ (Appendix~\ref{apconstraint}). Plugging in the appropriate value for ${\mathbf{c'\Sigma }}_{\rm B} {\mathbf{c}}$ from Table\ref{table1p1}, the problem under LDD, RS and fixed $\tau $ is 
$$
\mathop {Min}\limits_r \;(\kappa  + r)\left( {\frac{{12\sigma ^2 (1 - \rho _{t_0 } )}}
{{s^2 p_e (1 - p_e )}}} \right)\left( {\frac{r}
{{(r + 1)(r + 2)}} + \left( {\frac{{\rho _{b_1 ,\tau ,\tilde r} }}
{{1 - \rho _{b_1 ,\tau ,\tilde r} }}} \right)\frac{{\tilde r}}
{{(\tilde r + 1)(\tilde r + 2)}}} \right).
$$
Removing positive constant terms with respect to $r$, this problem becomes 
$$
\mathop {Min}\limits_r \quad I(r) = (\kappa  + r)\left( {\frac{r}
{{\,(r + 1)(r + 2)}} + \left( {\frac{{\rho _{b_1 ,\tau ,\tilde r} }}
{{1 - \rho _{b_1 ,\tau ,\tilde r} }}} \right)\frac{{\tilde r}}
{{\,(\tilde r + 1)(\tilde r + 2)}}} \right).
$$
Taking derivatives with respect to r, $r_{opt} $ solves 
\begin{multline*}
\frac{{\partial I}}
{{\partial r}} = \left( {\frac{{\rho _{b_1 ,\tau ,\tilde r} }}
{{1 - \rho _{b_1 ,\tau ,\tilde r} }}} \right)\frac{{\tilde r}}
{{\,(\tilde r + 1)(\tilde r + 2)}} + \frac{{(3 - \kappa )r^2  + 4r + 2\kappa }}
{{\left( {r + 1} \right)^2 (r + 2)^2 }} = \\
\left( {\frac{{\rho _{b_1 ,\tau ,\tilde r} }}
{{1 - \rho _{b_1 ,\tau ,\tilde r} }}} \right)\frac{{\tilde r}}
{{\,(\tilde r + 1)(\tilde r + 2)}} + \frac{{\partial H}}
{{\partial r}} = 0,
\end{multline*}
where $\frac{{\partial H}} {{\partial r}}$ is the derivative of the objective function $H(r)$ for the analogous problem under CS, given in Appendix~\ref{apoptrlddcstau}. There we showed that if $\kappa  < 3$ then $\frac{{\partial H}} {{\partial r}}$ was strictly positive for all $r$, and therefore so is $\frac{{\partial I}} {{\partial r}}$. Thus, if $\kappa  < 3$, $I(r)$ is minimized at $r_{opt}  = 1$. For $\kappa  > 3$, we know that $\frac{{\partial H}} {{\partial r}}$ is continuous, has only one root in the range of interest and it can be shown that $\mathop {\lim }\limits_{r \to \infty } \frac{{\partial H}} {{\partial r}} = 0^ -  $ and $\frac{{\partial H(1)}}
{{\partial r}} = \frac{{7 + \kappa }} {{36}}$. It can also be shown with Mathematica \cite{Wolfram:2005} that $\frac{{\partial ^2 H}} {{\partial r^2 }}$ has only one real root, $r*$. Therefore, $\frac{{\partial H}} {{\partial r}}$ is positive at $r = 1$, it crosses 0 at the root 
$$
r = \frac{{2 + \sqrt 2 \sqrt {2 - 3\kappa  + \kappa ^2 } }}
{{\kappa  - 3}},
$$
as shown in Appendix~\ref{apoptrlddcstau}, it has a minimum at the only root of  $\frac{{\partial ^2 H}}
{{\partial r^2 }}$ and it increases again towards zero, where it reaches an asymptote. Because of the form of $\frac{{\partial I}} {{\partial r}}$, it will have a similar shape, since it is equal to $\frac{{\partial H}} {{\partial r}}$ but moved upwards by a factor of  
$$
\left( {\frac{{\rho _{b_1 ,\tau ,\tilde r} }}
{{1 - \rho _{b_1 ,\tau ,\tilde r} }}} \right)\frac{{\tilde r}}
{{\,(\tilde r + 1)(\tilde r + 2)}}.
$$
Therefore, $\frac{{\partial I}}{{\partial r}}$ will have zero roots if 
$$
\left( {\frac{{\rho _{b_1 ,\tau ,\tilde r} }}
{{1 - \rho _{b_1 ,\tau ,\tilde r} }}} \right)\frac{{\tilde r}}
{{\,(\tilde r + 1)(\tilde r + 2)}} > \frac{{\partial H(r*)}}
{{\partial r}},
$$
or two roots otherwise. In the first case, when $\frac{{\partial I}} {{\partial r}}$ has zero roots, $\frac{{\partial I}} {{\partial r}}$ is always positive and therefore $I(r)$ increases as $r$ increases and the minimum of $I(r)$ is at $r_{opt}  = 1$. In the second case, $\frac{{\partial I}} {{\partial r}}$ has two roots, which solve 
$$
\kappa  = \frac{{r(4 + 3r)(\tilde r + 1)(\tilde r + 2) + \tilde r\left( {r + 1} \right)^2 (r + 2)^2 \left( {\frac{{\rho _{b_1 ,\tau ,\tilde r} }}
{{1 - \rho _{b_1 ,\tau ,\tilde r} }}} \right)}}
{{\,(r^2  - 2)(\tilde r + 1)(\tilde r + 2)}}.
$$
Also, $\frac{{\partial ^2 H}}
{{\partial r^2 }} = \frac{{\partial ^2 I}}
{{\partial r^2 }}$, and $\frac{{\partial ^2 H}}
{{\partial r^2 }}$ is continuous and it has only one root at $r*$.  $\frac{{\partial ^2 H}}
{{\partial r^2 }}$ is negative for $r < r*$ and positive for $r > r*$. Since $r*$ lies between the first and second roots of $\frac{{\partial I}} {{\partial r}}$, it can be concluded that the first root is a maximum of $I(r)$ and the second root is a minimum of $I(r)$. The function $I(r)$ has, therefore, two local minima, one at $r = 1$  and the other at the second root of $\frac{{\partial I}} {{\partial r}}$. To find out when the second root is the global minimum of $I(r)$ we need to solve $I(1) > I(r)
$, where 
$$
I(1) = (\kappa  + 1)\left( {\frac{1}
{6} + \left( {\frac{{\rho _{b_1 ,\tau ,\tilde r} }}
{{1 - \rho _{b_1 ,\tau ,\tilde r} }}} \right)\frac{{\tilde r}}
{{\,(\tilde r + 1)(\tilde r + 2)}}} \right)
$$
and 
$$
I(r) = (\kappa  + r)\left( {\frac{r}
{{\,(r + 1)(r + 2)}} + \left( {\frac{{\rho _{b_1 ,\tau ,\tilde r} }}
{{1 - \rho _{b_1 ,\tau ,\tilde r} }}} \right)\frac{{\tilde r}}
{{\,(\tilde r + 1)(\tilde r + 2)}}} \right).
$$
Provided $r > 2$, this is equivalent to 
$$
\rho _{b_1 ,\tau ,\tilde r}  < \frac{{\left[ { - 2(\kappa  + 1) + (\kappa  - 5)r} \right](\tilde r + 1)(\tilde r + 2)}}
{{6\tilde r(r + 1)(r + 2) + \left[ { - 2(\kappa  + 1) + (\kappa  - 5)r} \right](\tilde r + 1)(\tilde r + 2)}}.
$$
The condition only makes sense if $ - 2(\kappa  + 1) + (\kappa  - 5)r > 0$, which is equivalent to the conditions $\kappa  > 5$ and $r > \frac{{2\left( {\kappa  - 1} \right)}} {{\kappa  - 5}}$. Figure~\ref{roptlddrstaup1} of the paper shows this region for different values of $\kappa $and $\rho _{b_1 ,\tau ,\tilde r} $, together with a line for the optimal value.

\section{Demonstration of the program use}
\label{apdemprogp1}

This is the output of the program for the computation of the optimal combination of $(N, r)$ that minimizes the total cost of the study subject to achieving a fixed power under LDD and RS. For other examples and a user's guide, go to \url{http://www.hsph.harvard.edu/faculty/spiegelman/optitxs.html}.

\small{

\begin{verbatim}
> long.opt()

* By just pressing <Enter> after each question, the default value,
  shown between square brackets, will be entered.

* Press <Esc> to quit

Do you want to maximize power subject to a given cost (1) or to 
  minimize the total cost subject to a given power (2)[1]? 2

Enter the desired power (0<Pi<1) [0.8]: .8

Are you assuming the time between measurements (s) is fixed (1), 
  or the total duration of follow-up (tau) is fixed (2) [1]? 2

Enter the time of follow-up (tau) [1]: 18

Enter the exposure prevalence (pe) (0<=pe<=1) [0.5]: .79

Enter the variance of the time variable at baseline, V(t0) 
   (enter 0 if all participants begin at the same time) [0]: 100

Enter the correlation between the time variable at baseline and 
 exposure, rho_{e,t0} [0]: 0

Constant mean difference (1) or Linearly divergent difference (2)
 [1]: 2

Will you specify the alternative hypothesis on the absolute 
 (beta coefficient) scale (1) or the relative (percent) scale (2)
 [1]? 2

Enter mean response at baseline among unexposed (mu00) [10]: 3.5

Enter the percent change from baseline to end of follow-up among 
 unexposed (p2) (e.g. enter 0.10 for a 10% change) [0.1]: -.182

Enter the percent difference between the change from baseline to 
 end of follow-up in the exposed group and the unexposed group 
 (p3) (e.g. enter 0.10 for a 10% difference) [0.1]: .1

Which covariance matrix are you assuming: compound symmetry (1),
 damped exponential (2) or random slopes (3) [1]? 3

Enter (1) for standard notation (variance of residuals and random
  effects) or (2) for "reliability" notation [1]: 2

Enter the variance of the response given the assumed model 
 covariates at baseline (sigma2) [1]: .34

Enter the reliability coefficient at baseline (0<rho_t0<1) 
 [0.8]: .877

Enter the trial value of the number of measurements at which the
        slope reliability will be provided (\tilde r>0 ) [5]: 6

Enter the slope reliability for 6 repeated measurements
 (0<rho_{b1,s,\tilde r}<1  or 0<rho_{b1,tau,\tilde r}<1) 
 [0.1]: .364

Enter the correlation between the random effects of slope
 and intercept (-1<rho[b0,b1]<1) [0]: -.32

Enter the cost of the first observation of each subject (c1>0) 
 [80]: 80

Enter the ratio of costs between the first measure and the rest
 (kappa) [2]: 20

Cost optimization problem (min cost for a given power):
   Optimal r= 12 , Optimal N= 732 , Power= 0.8 ,Cost= 93696 

Slope reliability at r= 12 :  0.4818737

\end{verbatim}
}

\end{document}